\documentclass[11pt,a4paper]{article}
\pdfoutput=1 

\usepackage{graphicx}

\usepackage[utf8x]{inputenc}	
\usepackage[english]{babel}
\usepackage{paralist}
\usepackage{multirow}
\usepackage{bbm}
\usepackage{amsmath}
\numberwithin{equation}{section}
\usepackage{amssymb}
\usepackage{amsthm}
\usepackage{MnSymbol}
\usepackage{mathtools}		
\usepackage{dsfont}		
\usepackage{stmaryrd}		
\usepackage{cite}		
\usepackage[svgnames]{xcolor}
\usepackage{enumerate}
\usepackage{setspace}
\usepackage{booktabs}
\usepackage{tikz}
\usetikzlibrary{shapes}
\usetikzlibrary{positioning}
\usetikzlibrary{chains}
\usetikzlibrary{arrows,fit,decorations.pathreplacing, shapes.misc}
\usetikzlibrary{decorations.markings}
\tikzstyle{every picture}+=[remember picture]
\tikzstyle{na} = [baseline=-.5ex]

\usepackage{placeins}

\usepackage{todonotes}
\usepackage{tikz-3dplot}
\usepackage{pgfplots}
\usepackage[font=small,labelfont=bf]{caption}
\usepackage[font=small,labelfont=bf]{subcaption}
\usepackage[pdftex,breaklinks,colorlinks=true,urlcolor=RoyalBlue,linkcolor=blue,
citecolor=blue]{hyperref}

\usepackage{adjustbox}
\newcommand{\ra}[1]{\renewcommand{\arraystretch}{#1}}
\def\Tr{{\rm Tr}}

\newcommand{\HS}{\mathrm{HS}}
\newcommand{\HF}{\mathrm{H}}

\newcommand{\Ncal}{\mathcal{N}}

\newcommand{\Ocal}{\mathcal{O}}
\newcommand{\Vcal}{\mathcal{V}}
\newcommand{\Wcal}{\mathcal{W}}

\newcommand{\R}{\mathbb{R}}

\newcommand{\Z}{\mathbb{Z}}

\newcommand{\surm}{\mathrm{SU}}
\newcommand{\urm}{\mathrm{U}}

\newcommand{\sorm}{\mathrm{SO}}
\newcommand{\sprm}{\mathrm{Sp}}

\newcommand{\mfrak}{\mathfrak{m}}

\newcommand{\sh}{\operatorname{sh}}
\newcommand{\ch}{\operatorname{ch}}
\newcommand{\magQuiv}{\mathsf{Q}}
\newcommand{\Coulomb}{\mathcal{C}}
\newcommand{\Higgs}{\mathcal{H}}
\def\ns#1{
	\node[circle, draw, fill=white] at (#1){};
	\node[cross out, draw] at (#1){};
}
\def\dfive#1{
	\node[cross out, draw] at (#1){};
}

\usetikzlibrary{shapes.misc}
\tikzset{gauge/.style={inner sep=1mm,draw=none,fill=white,minimum size=2mm,circle, draw}}
\tikzset{flavour/.style={draw=none,minimum size=0.3mm,fill=white, regular polygon,regular polygon sides=4,draw}}
\tikzset{bd/.style={circle, draw=black, inner sep=0pt, fill=black, minimum size=2mm}}
\tikzset{gd/.style={circle, draw=green, inner sep=0pt, fill=green, minimum size=2mm}}
\tikzset{redCircle/.style={circle, very thick, draw=red, inner sep=0pt, minimum size=2mm}}
\tikzset{blackSquare/.style={regular polygon,regular polygon sides=4, very thick, draw=black, inner sep=0pt, minimum size=4mm}}
\tikzset{gauge1/.style={draw=none,minimum size=0.6cm,fill=white,circle, draw}}
\tikzset{gauge3/.style={draw=none,minimum size=0.35cm,fill=white,circle, draw}}
\tikzset{gauge5/.style={draw=none,minimum size=0.35cm,fill=white,circle, draw}}
\tikzset{crosses/.style={draw,circle,cross out}}
\tikzset{blank/.style={draw=none,minimum size=0.4cm,fill=none,circle, draw}}
\tikzset{flavor2/.style={draw=none,minimum size=0.4cm,fill=white,regular polygon sides=4,draw}}
\tikzset{flavour2/.style={draw=none,minimum size=0.4cm,fill=white,regular polygon sides=4,draw}}
\tikzset{flavorBlue/.style={draw=none,minimum size=0.4cm,fill=blue,regular polygon sides=4,draw}}
\tikzset{flavorRed/.style={draw=none,minimum size=0.4cm,fill=red,regular polygon sides=4,draw}}
\tikzset{none/.style={draw=none}}
\tikzset{flavourBlue/.style={draw=none,minimum size=0.4cm,fill=blue,regular polygon sides=4,draw}}
\tikzset{flavourRed/.style={draw=none,minimum size=0.4cm,fill=red,regular polygon sides=4,draw}}
\tikzset{none/.style={draw=none}}
\tikzset{redgauge/.style={draw=none,minimum size=0.4cm,fill=red,circle, draw}}
\tikzset{miniU/.style={draw=none,minimum size=0.1cm,fill=red,circle, draw}}
\tikzset{smallgauge1/.style={draw=none,minimum size=0.1cm,fill=white,circle, draw}}
\tikzset{blankflavor/.style={draw=none,minimum size=0.8mm,fill=none, regular polygon,regular polygon sides=4,draw}}
\tikzset{miniBlue/.style={draw=none,minimum size=0.1cm,fill=blue,circle, draw}}
\tikzset{gauge2/.style={draw=none,minimum size=0.35mm,fill=red,circle, draw}}
\tikzset{bluegauge/.style={draw=none,minimum size=0.4cm,fill=blue,circle, draw}}
\tikzset{flavor1/.style={draw=none,minimum size=0.35mm,fill=blue, regular polygon,regular polygon sides=4,draw}}
\tikzset{flavor0/.style={draw=none,minimum size=0.35mm,fill=white, regular polygon,regular polygon sides=4,draw}}
\tikzset{o5/.style={green,dotted}}
\tikzset{new edge style 0/.style={magenta}}
\tikzset{o3tildeplus/.style={dashed}}
\tikzset{o3plus/.style={dotted}}
\tikzset{smalldot/.style={draw=none,minimum size=0.1mm,fill=black, circle,draw}}
\tikzset{dotsize/.style={circle,fill,inner sep=1.5pt,draw}}
\tikzset{doubleguys/.style={double, double distance = 3pt}}
\tikzset{tripleguys/.style={triple}}
\tikzset{new edge style 1/.style={dashed}}
\tikzset{thickline/.style={line width=0.06cm}}
\tikzset{darke/.style={line width=0.3mm,black}}
\tikzset{brace/.style={decorate,decoration={brace,amplitude=10pt}}}
\tikzset{brace1/.style={decorate,decoration={brace,amplitude=10pt}}}
\tikzset{bd/.style={circle, draw,inner sep=3.5pt, fill=black}}
\pgfdeclarelayer{edgelayer}
\pgfdeclarelayer{nodelayer}
\usetikzlibrary{decorations.pathreplacing}
\pgfsetlayers{edgelayer,nodelayer,main}

\tikzset{gauge/.style={circle, draw,inner sep=3pt}}
\tikzset{gaugeSO/.style={circle, draw,inner sep=3pt,fill=red}}
\tikzset{gaugeSp/.style={circle, draw,inner sep=3pt,fill=blue}}
\tikzset{flavour/.style={regular polygon,regular polygon sides=4,inner
sep=3pt, draw}}
\tikzset{flavourSO/.style={regular polygon,regular polygon sides=4,inner
sep=3pt, draw,fill=red}}
\tikzset{flavourSp/.style={regular polygon,regular polygon sides=4,inner
sep=3pt, draw,fill=blue}}
\tikzset{defect/.style={circle, draw,inner sep=3pt,fill=black}}

\catcode`@=11
\allowdisplaybreaks

\begin{document}

\allowdisplaybreaks

\begin{titlepage}
\setcounter{page}{0}
%
\begin{center}

{\LARGE\bf
Magnetic quivers and line defects\\
---\\
On a duality between 3d $\Ncal=4$ unitary and orthosymplectic quivers
}

\vspace{15mm}

{\large Satoshi Nawata${}^{1}$},\
{\large Marcus Sperling${}^{2,3}$},\
{\large Hao Ellery Wang${}^{1}$},\  and \
{\large Zhenghao Zhong${}^{4}$}
\\[5mm]
\noindent ${}^{1}${\em Department of Physics and Center for Field Theory and Particle Physics, Fudan University, }\\
{\em  220, Handan Road, 200433 Shanghai, China}\\
{Email: {\tt snawata@gmail.com}} \\
{{\tt yukawahaow@gmail.com}}
\\[5mm]
\noindent ${}^{2}${\em Shing-Tung Yau Center, Southeast University} \\
{\em Xuanwu District, Nanjing, Jiangsu, 210096, China}\\
{Email: {\tt msperling@seu.edu.cn}}
\\[5mm]
\noindent ${}^{3}${\em Yau Mathematical Sciences Center, Tsinghua University}\\
{\em Haidian District, Beijing, 100084, China}
\\[5mm]
\noindent ${}^{4}${\em Theoretical Physics Group, The Blackett Laboratory, Imperial College London,}\\
{\em Prince Consort Road London, SW7 2AZ, UK}\\
{Email: {\tt zhenghao.zhong14@imperial.ac.uk}}
\\[5mm]
\vspace{15mm}

\begin{abstract}
Supersymmetric $\sprm(k)$ quantum chromodynamics with 8 supercharges in space-time dimensions 3 to 6 can be realised by two different Type II brane configurations in the presence of orientifolds. Consequently, two types of magnetic quivers describe the Higgs branch of the $\sprm(k)$ SQCD theory. This is a salient example of a general phenomenon: a given hyper-K\"ahler Higgs branch may admit several magnetic quiver constructions. It is then natural to wonder if these different magnetic quivers, which are described by 3d $\Ncal=4$ theories, are dual theories.
In this work, the unitary and orthosymplectic magnetic quiver theories are subjected to a variety of tests, providing evidence that they are IR dual to each other. For this, sphere partition function and supersymmetric indices are compared. Also, we study half BPS line defects and find interesting regularities from the viewpoints of exact results, brane configurations, and 1-form symmetry.

\end{abstract}

\end{center}

\end{titlepage}
{\baselineskip=3pt
{\small
\tableofcontents
}
}

%
\section{Introduction}
A long-standing challenge in quantum field theory is a strongly coupled phase, due to the breakdown of conventional techniques. One promising approach is known as dualities, which often describe the physical equivalence of a strongly coupled theory by a weakly coupled theory. Moreover, dualities are proposed based on the agreement of a few quantities computable in both involved
theories. Among them are the symmetries, dimensions, anomalies, and partition functions. Recently, as generalisations of conventional global symmetries, higher form symmetries have been understood as vital components in characterising a QFT beyond the level of local operators. The charged objects under higher form symmetries are extended defects, such as line defects for 1-form symmetries. As a duality should map conventional symmetries as well as higher form symmetries, a crucial test lies in the understanding of defects.

One instance of strong coupling phenomena is the following: Higgs branches of theories with 8 supercharges can change drastically at special points, contrary to a long-standing misbelief. These changes are due to new massless degrees of freedom arising from tensionless strings in 6d, massless gauge instantons in 5d, and Argyres-Douglas points in 4d.
Recently, magnetic quivers \cite{Cabrera:2018jxt,Cabrera:2019izd,Bourget:2019aer,Bourget:2019rtl,Cabrera:2019dob,Bourget:2020gzi,Bourget:2020asf,Closset:2020scj,Akhond:2020vhc,vanBeest:2020kou,Bourget:2020mez,VanBeest:2020kxw,Eckhard:2020jyr,Closset:2020afy,Akhond:2021knl,Bourget:2021xex,Akhond:2021ffo,vanBeest:2021xyt,Bourget:2021csg,Sperling:2021fcf} have been systematically introduced with the aim to uniformly address Higgs branches of theories with 8 supercharges in dimensions 4 to 6. For this an auxiliary quiver gauge theory $\mathsf{Q}$ is utilised such that its 3d $\Ncal=4$ Coulomb branch $\Coulomb$ provides a geometric description of the desired Higgs branch $\Higgs$ of a theory $T$ in a phase $P$:
\begin{align}
\Higgs\left(T,P\right) = \Coulomb\left(\magQuiv(T,P)\right) \,.
\end{align}
Prior to the systematic developments, 3d Coulomb branches had already been used to describe Higgs branches in 4d \cite{DelZotto:2014kka}, 5d \cite{Cremonesi:2015lsa,Ferlito:2017xdq}, and 6d \cite{Mekareeya:2017jgc,Hanany:2018uhm,Hanany:2018vph,Hanany:2018cgo}.

The study of Higgs branch moduli spaces, understood as symplectic singularities, or hyper-K\"ahler singularities, via magnetic quivers has uncovered a new phenomenon. Given a hyper-K\"ahler singularity $X$, there may exist several magnetic quiver constructions $\mathsf{Q}_i$, with $i=1,\ldots, n$, \textit{i.e.}\
\begin{align}
    X \cong \Coulomb(\magQuiv_i) \qquad \forall i \,.
    \label{eq:magQuiv_construction}
\end{align}
The individual magnetic quiver constructions usually differ in various aspects:
\begin{compactitem}
\item The nodes of a quiver encode dynamical 3d $\Ncal=4$ vector multiplets as well as background vector multiplets. The underlying gauge and flavour groups can be unitary groups $\urm(n)$, special unitary groups $\surm(n)$, orthogonal groups $\sorm(n)$, or symplectic groups $\sprm(n)$.
\item The links between the nodes entail the 3d $\Ncal=4$ hypermultiplets. Conventional links denote hypermultiplets in a bifundamental representation of the adjacent nodes. However, other matter contents are known: such are non-simply laced links, higher charge hypermultiplets, or fundamental-fundamental representations.
\end{compactitem}
Then, \eqref{eq:magQuiv_construction} raises an immediate question: what is the relation between the different magnetic quivers $\magQuiv_i$?  Do these auxiliary quivers, now understood as 3d $\Ncal=4$ quiver gauge theories, only have isomorphic Coulomb branches or might they enjoy a more fundamental relationship? In the 3d $\Ncal=4$ context, the notion of duality is understood as an IR-duality: two theories with different UV descriptions, which in many cases are Lagrangian, flow to the same IR conformal fixed point.
The first evidence for such a duality between $\magQuiv_i$ comes from the defining property $\Coulomb(\magQuiv_i) = \Coulomb(\magQuiv_j)$. Note that this proposed duality is therefore distinct from 3d mirror symmetry \cite{Intriligator:1996ex,Hanany:1996ie}.
Moreover, it is important to stress that 3d $\Ncal=4$ Coulomb branches are affected by quantum corrections, and $\Coulomb$ always refers to the fully quantum corrected moduli space \cite{Seiberg:1996nz} of the IR SCFT.

As a prototypical example, the two different mirror theories of 3d $\Ncal=4$ $\sprm(k)$ gauge theory with $N_f$ fundamental hypermultiplets enjoy an IR duality. One mirror is a $D$-type Dynkin quiver solely composed of unitary nodes \cite{Hanany:1999sj}, while the other mirror is a linear quiver chain of alternating orthogonal and symplectic nodes \cite{Feng:2000eq}. However, not all properties of the IR SCFT may be apparent in the UV description. This is famously known for the orthosymplectic\footnote{This denotes a quiver with alternating orthogonal and symplectic nodes.} mirror quiver, which lacks FI parameters such that the Coulomb branch global symmetry is not manifest.
It is intriguing to examine the expected IR duality further by considering extended operators, such as line defects. Since the $\sprm(k)$ theory and both of its mirror theories admit brane realisations, it is advantageous to consider line defects from the perspective of brane configurations \cite{Assel:2015oxa,Dey:2021jbf} as well as 1-form symmetry \cite{Gaiotto:2014kfa}.

Similarly, a 5d $\sprm(k)$ SQCD at infinite coupling admits two magnetic quiver constructions, corresponding to two different brane realisations. As in the 3d case, the two types are essentially a unitary magnetic quiver and an orthosymplectic magnetic quiver. Taking a purely 3d $\Ncal=4$ viewpoint, it is then reasonable to ask whether they are dual to each other.
Some first hints towards a duality between these magnetic quivers, understood as legitimate 3d theories, are the match of Coulomb branches and Higgs branches \cite{Bourget:2020xdz}.
This work aims to provide further evidence for a duality between $D$-type Dynkin quivers and linear orthosymplectic quivers. Probes for a potential duality between two 3d $\Ncal=4$ theories include:
\begin{compactenum}[(i)]
\item (Quaternionic) Dimensions of Higgs and Coulomb branches.
\item Global symmetries on Higgs and Coulomb branches, \textit{i.e.}\ the deformation parameters: masses and FI terms.
\item Higgs or Coulomb branch chiral rings, and their generating functions \cite{Benvenuti:2006qr,Feng:2007ur,Gray:2008yu,Cremonesi:2013lqa,Bullimore:2015lsa}.
\item Hasse diagram for Higgs and Coulomb branch \cite{Bourget:2019aer,Grimminger:2020dmg,Bourget:2021siw}.
\item Superconformal index \cite{Razamat:2014pta}.
\item A-twisted and B-twisted indices \cite[\S6]{Closset:2016arn}.
\item 3-sphere partition function \cite{Kapustin:2009kz}.
\item Extended operators, like line defects \cite{Assel:2015oxa,Dimofte:2019zzj,Dey:2021jbf}.
\end{compactenum}
Points (i)--(iv) are very much tailored to analyse the geometry of the hyper-K\"ahler moduli spaces. In contrast, the indices (v)--(vi) are tools suitable for studying the IR SCFTs. In fact, in certain limits, the indices reduce to the Higgs and Coulomb branch generating functions. Quite differently, the sphere partition function (vii) depends explicitly on the existence of deformation parameters in the UV description, but is generically easier to evaluate than indices.
In view of the exact partition functions (iii), (v)--(vii), only some are sensitive to the global structure of the gauge group in the UV description. This happens when the precise choice of magnetic lattice or cocharacter lattice is involved; for instance, in the monopole formula, the superconformal index, and the A-twisted index. Other quantities, like the sphere partition function or the Higgs branch Hilbert series, are only sensitive to the Lie algebra of the underlying gauge group.
As a more direct probe of the gauge group, extended operators like line defects are sensitive to the global structure \cite{Aharony:2013hda}. Put differently, the spectra of line defects depend on 1-form symmetries \cite{Gaiotto:2014kfa}, which are by definition sensitive to the gauge group and not just the gauge algebra. Hence, this paper provides a detailed study of the IR duality between unitary and orthosymplectic magnetic quivers by exact partition functions (v)--(vii), including line defects (viii).

The paper is organised as follows: the definitions and various aspects of suitable partition functions are recalled in \S\ref{sec:partition_fct}. Thereafter, the two different 3d mirror theories for 3d $\Ncal=4$ $\sprm(k)$ gauge theory with fundamental flavours are studied in \S\ref{sec:3d_Sp}. In particular, line operators are introduced and matched between the two types of mirror theories.
Moving on to 5d $\Ncal=1$ $\sprm(k)$ theories and their two types of magnetic quivers, \S\ref{sec:5d_Sp} starts by comparing partition functions. Next, Wilson line defects are included in both types of magnetic quivers, and an intriguing matching pattern is presented. At the end of the section, we demonstrate the identification of some 0-form symmetries, including that from gauging the 1-form symmetry, in the two types of magnetic quivers.
Lastly, we summarise this paper and discuss some open questions in \S\ref{sec:summary-open}. Several appendices provide computational evidence for statements made in the main body. Appendix \ref{app:brane} collects background material on relevant Type IIB brane configurations. Computational results on partition functions are provided in Appendix \ref{app:superconf_index} for the superconformal index, in Appendix \ref{app:twisted_indices} for twisted indices, and in Appendix \ref{app:sphere} for the sphere partition function.

%
\section{Exact partition functions}
\label{sec:partition_fct}

An exact partition function by supersymmetric localisation is a powerful tool to check dualities since it is independent of coupling constants.
To test 3d $\mathcal{N}=4$ unitary/orthosymplectic dualities, we use exact partition functions of three kinds: superconformal index, twisted indices and $S^3$ partition function.
The combination of these partition functions provides detailed information of 3d $\mathcal{N}=4$ quantum field theory, and confirms predictions from brane configurations in Type IIB theory.

Before discussing partition functions, we provide a lightning review on 3d $\Ncal=4$ theories. The superconformal symmetry of an $\Ncal=4 $ theory is $\mathrm{OSp}(4|4)$, whose bosonic generators comprise those of the conformal algebra in addition to those of the $\surm(2)_H\times \surm(2)_C$ R-symmetry. The SCFT has $G_H \times G_C$ global symmetry. If a UV Lagrangian description is available, the following data needs to be specified: (i) a gauge group $G$ and (ii) a representation $\mathcal{R}$ of the matter content.  The gauge group gives rise to a dynamical vector multiplet, while the matter content is given as terms of hypermultiplets transforming under $G$. The two complex scalars in the hypermultiplet transform as $[1]\otimes[0]$ under $\surm(2)_H\times \surm(2)_C$, whereas the three adjoint-valued real scalars in the vector multiplet transform as $[0]\otimes [2]$ under the R-symmetry. The IR global symmetry might not be manifest in the UV description. The Cartan generators of $G_H \times G_C$ correspond to deformation parameters. These are the masses, which are a Cartan subalgebra of $G_H$ and transform as $[0]\otimes [2]$ in the R-symmetry, and the FI-parameters, which belong to a Cartan subalgebra of $G_C$ and transform as $[2]\otimes [0]$ in the R-symmetry.

\subsection{Superconformal index}\label{sec:SCI}
The 3d $\mathcal{N}=4$ superconformal index \cite{Razamat:2014pta} is defined by
\begin{align}
  \mathbb{I}(\mathfrak{q},\mathfrak{t})={\operatorname{Tr}}(-1)^{F} \mathfrak{q}^{J+\frac{H+C}{4}} \mathfrak{t}^{H-C}  \,,
\end{align}
which is understood as the partition function of the theory on $S^1\times S^2$.
Here $F$ is the Fermion number, $J$ is the generator of the $\urm(1)_{J}$ rotational symmetry of the $S^2$ in the space-time, and $H$, $C$ stand for the Cartan generators of the $\surm(2)_{H}\times\surm(2)_{C}$ R-symmetry groups, respectively. Although fugacities of global symmetries can be turned on, we do not consider them throughout this paper. This index computes the graded dimensions of $\frac18$-BPS states in an IR SCFT. For instance, the mirror symmetry for unitary theories is studied by the superconformal index in \cite{Okazaki:2019ony}.

The contribution from a vector multiplet and an $\Ncal=2$ chiral multiplet under a representation $R$ of a gauge group $G$, respectively, reads off
\begin{subequations}
\begin{align}
Z_{\mathrm{vec}}(z,\mfrak)=&
\prod_{\alpha \in \Delta}\left(\frac{\mathfrak{q}^{\frac{1}{2}}}{\mathfrak{t}^2}\right)^{-|\alpha(\mfrak)|}  \left(1-\mathfrak{q}^{\frac{|\alpha(\mfrak)|}{2}} z^{\alpha}\right) \frac{\left(\mathfrak{t}^{2}\mathfrak{q}^{\frac{1+|\alpha(\mfrak)|}{2}} z^{\alpha} ;\mathfrak{q}\right)_{\infty}}{\left(\mathfrak{t}^{-2}\mathfrak{q}^{\frac{1+|\alpha(\mfrak)|}{2}}  z^{\alpha} ;\mathfrak{q}\right)_{\infty}} \\
Z_{\mathrm{chiral}}(z,\mfrak)=&\prod_{w\in R} \left(\frac{\mathfrak{q}^{\frac{1}{2}}}{\mathfrak{t}^2}\right)^{\frac{1}{4}|w(\mfrak)|} \frac{\left(\mathfrak{t}^{-1} \mathfrak{q}^{\frac{3}{4}+\frac{1}{2}|w(\mfrak)|} z^{-w} ;\mathfrak{q}\right)_{\infty}}{\left(\mathfrak{t} \mathfrak{q}^{\frac{1}{4}+\frac{1}{2}|w(\mfrak)|} z^{w} ;\mathfrak{q}\right)_{\infty}}~,
\end{align}
\end{subequations}
where $\Delta$ is the set of simple roots of a gauge group, and $\mfrak$ are magnetic fluxes on $S^2$ valued in the cocharacter lattice of the gauge group. Note that $(z;\mathfrak{q})_\infty\coloneqq\prod_{\ell=1}^\infty(1-z\mathfrak{q}^\ell)$ is the $\mathfrak{q}$-Pochhammer symbol. For unitary gauge groups, an $\Ncal=4$ hypermultiplet consists of two $\Ncal=2$ chiral multiplets with $R\oplus R^*$ so that
\begin{equation}
  Z_{\mathrm{hyp}}(z,\mfrak)\coloneqq Z_{\mathrm{chiral}}(z,\mfrak)Z_{\mathrm{chiral}}(z^{-1},\mfrak)=\prod_{w\in R} \left(\frac{\mathfrak{q}^{\frac{1}{2}}}{\mathfrak{t}^2}\right)^{\frac{1}{2}|w(\mfrak)|} \frac{\left(\mathfrak{t}^{-1} \mathfrak{q}^{\frac{3}{4}+\frac{1}{2}|w(\mfrak)|} z^{\mp w} ;\mathfrak{q}\right)_{\infty}}{\left(\mathfrak{t} \mathfrak{q}^{\frac{1}{4}+\frac{1}{2}|w(\mfrak)|} z^{\pm w} ;\mathfrak{q}\right)_{\infty}} \,,
\end{equation}
where $\mathfrak{q}$-Pochhammer symbols with repeated signs $\pm,\mp$ are all multiplied. Then, a superconformal index of a quiver gauge theory can be schematically expressed as the contour integral
\begin{equation}\label{SCI}
    \mathbb{I}(\mathfrak{q},\mathfrak{t})=\sum_{\mfrak}\int \prod_{\mathrm{gauge}} \frac{1}{\left|W_{G}\right|} \left[\frac{dz}{2\pi i z}\right]Z_{\mathrm{vec}}(z,\mfrak) \prod_{\mathrm{matter}} Z_{\mathrm{chiral}}(z,\mfrak)~.
\end{equation}
Employing the following change of variables
\begin{equation}\label{HC-limit}
\mathfrak{q} =t_h t_c~,\qquad \mathfrak{t}=\left(\frac{t_h}{t_c}\right)^{1 / 4}~,
\end{equation}
the limits $t_h\to0$ and $t_c\to0$ of the superconformal index lead to the Hilbert series of the Coulomb and Higgs branch of the 3d $\Ncal=4$ theory, respectively \cite{Razamat:2014pta}. In particular, the Higgs branch limit reduces to the Molien-Weyl formula \cite{Benvenuti:2006qr,Feng:2007ur,Gray:2008yu}, and the Coulomb branch limit reproduces the monopole formula \cite{Cremonesi:2013lqa}.

The superconformal index counts $\frac18$-BPS gauge-invariant local operators up to $(-1)^F$ in a 3d $\mathcal{N}=4$ SCFT. The other important operators are extended BPS operators such as Wilson and vortex line operators. The 3d mirror symmetry exchanges Wilson and vortex lines operators. In Type IIB theory, a Wilson line operator is realised as F1-strings suspended by D3 and D5-brane whereas a vortex line operator is realised as D1-branes suspended by D3 and NS5-branes \cite{Assel:2015oxa}. See Appendix \ref{app:brane} for details. Using IIB brane configurations, \cite{Assel:2015oxa} provides microscopic descriptions of vortex operators, and exactly evaluates expectation values of line operators in 3d $\Ncal=4$ theories by performing supersymmetric localisation. In 3d $\Ncal=4$ unitary/orthosymplectic duality, the exchange of Wilson and vortex line operators does not occur. Instead, we are interested in which gauge groups in dual quiver theories line operators of the same type are coupled.

\subsection{Twisted indices}
\label{sec:twisted_ind_def}
To assess whether a pair of unitary/orthosymplectic quiver theories is endowed with equivalent sets of line operators, we evaluate expectation values of line operators in a topologically twisted theory \cite{Rozansky:1996bq}.
In particular, we consider twisted partition functions \cite{Nekrasov:2014xaa,Gukov:2015sna,Benini:2015noa,Benini:2016hjo,Closset:2016arn,Gukov:2020lqm} on $S^1\times S^2$ with line operators where a topological twist is performed on $S^2$. A topological twist with the Cartan subgroup of $\surm(2)_{H}$ leads to an A-twisted index, whereas a twist with the Cartan subgroup of $\surm(2)_{C}$ leads to a B-twisted index. The A-twisted and B-twisted theories flow to a non-linear sigma model on the Coulomb and Higgs branches of the 3d $\Ncal=4$ theory, respectively. Hence, the A-twisted and B-twisted indices of $S^1\times S^2$ count the cohomology of the topological supercharge, which yields the Hilbert series of the Coulomb and Higgs branches, respectively \cite{Closset:2016arn,Gukov:2020lqm}. Since half-BPS vortex and Wilson lines preserve the topological supercharge of A-type and B-type, respectively, we evaluate their expectation values in the corresponding twisted index on  $S^1\times S^2$. In fact, the explicit expressions of the twisted indices (with line operators) are given as the Jeffrey-Kirwan contour integrals \cite[\S6]{Closset:2016arn} by performing the supersymmetric localisation. A vortex line expectation value in the A-twisted index is given by
\begin{align}\label{vortex}
\mathbb{I}^A= \sum_{\mfrak}  \oint_{\mathrm{JK}} \prod_{\mathrm{gauge}} \frac{1}{\left|W_{G}\right|}\left[\frac{d z}{2 \pi i z}\right] Z^{A}_{\mathrm{vec}}(z,\mfrak)  \prod_{\mathrm{matter}} Z^{A}_{\mathrm{chiral}}(z,\mfrak) ~\Vcal(z)~,
\end{align}
where the contributions from a vector multiplet and a chiral multiplet are
\begin{subequations}\label{A-twisted}
\begin{align}
  Z^{A}_{\text {vec}}(z,\mfrak)=&\left(t-t^{-1}\right)^{-\operatorname{rk}(G)} \prod_{\alpha \in  \Delta}\left(\frac{1-z^{\alpha}}{t-z^{\alpha} t^{-1}}\right)^{\alpha(\mfrak)+1}~,\\
Z^{A}_{\mathrm{chiral}}(z,\mfrak)=&\prod_{w \in R}\left(\frac{z^{\frac{w}2}t^{\frac12}}{1-z^{w}  t}\right)^{w(\mfrak)}~.
\end{align}
\end{subequations}
Here, we also sum over magnetic fluxes $\mfrak$ supported on $S^2$, which take values in the cocharacter lattices of the gauge group. The vortex loop contribution can be read off by evaluating a 1d $\Ncal=4$ supersymmetric quantum mechanics \cite{Assel:2015oxa}. For the $\urm(k)$ supersymmetric quantum mechanics with $N_1$ fundamental and $N_2$ anti-fundamental chiral multiplets
\begin{equation}\label{SQM}
\tikzset{every picture/.style={line width=0.75pt}} 
\begin{tikzpicture}
  \begin{pgfonlayer}{nodelayer}
    \node [style=gauge3,fill=black] (0) at (-5, -4+4) {};
    \node [style=flavour2] (1) at (-3.5, -4+4) {};
        \node [style=flavour2] (4) at (-6.5, -4+4) {};
    \node [style=none] (2) at (-5, -4.5+4) {$k$};
    \node [style=none] (3) at (-3.5, -4.5+4) {$N_2$};
      \node [style=none] (3) at (-6.5, -4.5+4) {$N_1$};
  \end{pgfonlayer}
  \begin{pgfonlayer}{edgelayer}
    \begin{scope}[decoration={markings,mark =at position 0.5 with {\arrow{stealth}}}]
  \draw[postaction={decorate},color=red,thick] (0)--(1) ;
    \draw[postaction={decorate},color=red,thick] (4)--(0);\end{scope}
  \end{pgfonlayer}
\end{tikzpicture}\end{equation}
the contribution to the twisted indices becomes
\begin{equation}
  \Vcal(z)= \sum_{k=\sum_{i = 1}^{N_{1}} k_{i}} \prod_{a=1}^{N_{1}}\left(\prod_{b \neq a}^{N_{1}} \frac{z_{a} t^{-1}-z_{b} t}{z_{a}-z_{b}} \left(\frac{1-z_{a} t}{z_{a}-t}\right)^{N_2}\right)^{k_{a}}~.
\end{equation}
In this paper, we focus only on $k=1$ for the sake of simplicity. Nonetheless, it is straightforward to generalise the results to arbitrary $k$.
To evaluate an A-twisted index of a generic $\Ncal=4$ non-Abelian quiver gauge theory as a sum over Bethe vacua  \cite{Nekrasov:2014xaa,Gukov:2015sna,Benini:2015noa,Benini:2016hjo,Closset:2016arn,Gukov:2020lqm}, we need to refine the formulas \eqref{A-twisted} by switching on all the deformation parameters of the theory such as FI parameters and real masses. Otherwise, solutions of Bethe ansatz equations would miss some of the vacua. However, with all the deformation parameters turned on, it is difficult to solve Bethe ansatz equations algebraically. In practice, this can be done only numerically in most cases. In Appendix \ref{app:A-twisted}, we instead perform the Jeffrey-Kirwan contour integral and sum over magnetic fluxes for A-twisted indices of $\sprm(1)$ SQCDs with vortex loops. Since this is a formidable task for a long or higher rank quiver gauge theory, we evaluate the B-twisted index with Wilson loop. A B-twisted index with Wilson loop is given by
\begin{equation}\label{Wilson}
\mathbb{I}^B= \oint_{\mathrm{JK}} \prod_{\mathrm{gauge}} \frac{1}{\left|W_{G}\right|}\left[\frac{d z}{2 \pi i z}\right] Z^{B}_{\mathrm{vec}}(z)  \prod_{\mathrm{matter}} Z^{B}_{\mathrm{chiral}}(z) ~\Wcal(z)~,
\end{equation}
where the contributions from a vector multiplet and a chiral multiplet are
\begin{subequations}
\begin{align}
  Z^{B}_{\text {vec}}(z)=&\left(t-t^{-1}\right)^{\operatorname{rk}(G)} \prod_{\alpha\in \Delta}\left(1-z^{\alpha}\right)\left(t-z^{\alpha} t^{-1}\right)~,\\
Z^{B}_{\mathrm{chiral}}(z)=&\prod_{w \in R}\frac{z^{\frac{w}2}t^{\frac12}}{1-z^{w}t}~.
\end{align}
\end{subequations}
The Wilson loop $\Wcal(z)$ amounts to an insertion of a character corresponding to its representation.
Here only the zero magnetic flux sector contributes to the B-twisted index so that we do not have to sum over the magnetic fluxes \cite{Closset:2016arn}. The B-twisted index can also be computed simply by expanding the integrand as a series and taking the $z$-independent part. Therefore, it is much easier to calculate the expectation value of a Wilson loop \eqref{Wilson} than that of a vortex loop \eqref{vortex}.

Here we briefly comment on magnetic lattices in \eqref{SCI} and \eqref{vortex}. As carefully studied in  \cite{Bourget:2020xdz}, in addition to the weight lattices, the lattices simultaneously shifted by a half need to be included for unframed orthosymplectic quivers in \S\ref{sec:5d_Sp}. This point is emphasised in \S\ref{sec:lattice} and \S\ref{sec:5d_Sp}.
\subsection{Sphere partition function}

As in \cite{Assel:2015oxa,Dey:2021jbf}, we can also use $S^3$ partition functions to evaluate the expectation value of line operators.
A $S^3$ partition function is evaluated by supersymmetric localisation in the pioneering work \cite{Kapustin:2009kz} as
\begin{equation}\label{S3}
  Z^{S^3}= \oint \prod_{\mathrm{gauge}} \frac{1}{\left|W_{G}\right|}\left[d s\right] Z_{\mathrm{vec}}^{S^3}(s)  \prod_{\mathrm{matter}} Z_{\mathrm{hyp}}^{S^3}(s,m) ~ Z_{\mathrm{FI}}(s, \xi)Z_{\mathrm{defect}}^{S^3}(s)~.
\end{equation}
A vector multiplet and a hypermultiplet contributes
\begin{subequations}
\begin{align}
Z^{S^3}_{\mathrm{vec}}(s)=&\prod_{\alpha\in \Delta} \operatorname{sh}(\alpha \cdot s)~,\\
Z^{S^3}_{\mathrm{hyp}}(s, {m})=&\prod_{w\in R} \frac{1}{\operatorname{ch}(w \cdot s-m)}~,
\end{align}
\end{subequations}
where $m$ is a mass parameter. Here we use the notation
$$\operatorname{sh}(x) \equiv 2 \sinh (\pi x)~, \qquad \operatorname{ch}(x) \equiv 2 \cosh (\pi x)~.$$
Whenever the gauge group has a $\urm(1)$ factor, the contribution from the FI parameter  $\xi$ is
\begin{equation}\label{FI}
Z_{\mathrm{FI}}(s, \xi)= e^{2 \pi i \xi\Tr(s)}~.
\end{equation}
Although an orthosymplectic quiver is not endowed with an FI parameter, we can turn on the \emph{unphysical} FI parameter \eqref{FI} as a regulator \cite{Benvenuti:2011ga} and compute its $S^3$ partition function by taking the limit $\xi\to0$.
As before, the Wilson loop is evaluated by inserting a character
\begin{equation}
Z_{\text {Wilson}}(s)=\sum_{w \in R} e^{2 \pi w \cdot s}\,.
\end{equation}
On the other hand, the vortex loop described by the 1d supersymmetric quantum mechanics in \eqref{SQM} results in the following contribution:
\begin{equation}
Z_{\text {vortex}}(s)=\sum_{\substack{k_{1}, \cdots, k_{N} \in\{0,1\} \\ k=\sum_{i = 1}^{N_{1}} k_{i}}}(-1)^{(N_1+N_2) k} \prod_{i<j}^{N_1} \frac{\operatorname{sh}\left[s_{i}-s_{j}+i\left(k_{i}-k_{j}\right) z\right]}{\operatorname{sh}\left[s_{i}-s_{j}\right]} \prod_{j=1}^{N_1} \prod_{a=1}^{N_2} \frac{\operatorname{ch}\left(s_{j}-m_{a}\right)}{\operatorname{ch}\left(s_{j}-m_{a}+i k_{j} z\right)}\,.
\end{equation}
For the sake of brevity, we focus only on $k=1$ in this paper.

\subsection{Comments on weight lattice and magnetic lattice}
\label{sec:lattice}
In some instances, the evaluation of exact partition functions entails a summation over the magnetic lattice (also referred to as cocharacter lattice) of the gauge group like \eqref{SCI} and \eqref{vortex}. Likewise, extended operators are sensitive to the weight lattice of the gauge group and not just the gauge algebra. Therefore, the weight lattice and the magnetic lattice need to be discussed.
For a review on the relevant lattice in 3d $\Ncal=4$ theories, the reader is referred to \cite{Goddard:1976qe,Kapustin:2005py,Hanany:2016ezz}, while the notation follows \cite{Bourget:2020xdz}. As two types of quivers appear, these are discussed in turn.

An important concept is the following: a quiver gauge theory can be understood as encoding a representation $\phi : G \to \mathrm{GL}(V)$ of a group $G$ into a finite-dimensional vector space $V$. For a quiver with gauge group $G$, the symmetry group $\mathrm{ker}\ \phi \subset G$ is a group that acts trivially on the matter content.
In view of \cite{Gaiotto:2014kfa}, $\mathrm{ker}\ \phi$ is an electric 1-form symmetry. Suppose $H \subset \mathrm{ker}\ \phi$ is a normal subgroup, and the gauge group is chosen as the quotient $G \slash H$. Taking the quotient with respect to $H$ or, equivalently, gauging the 1-form symmetry $H$ affects the spectrum of admissible line operators\cite{Aharony:2013hda,Gaiotto:2014kfa,Tong:2017oea}. For example, if a Wilson line transforms in a $G$ representation that is not $H$-invariant, then it is charged under the 1-form symmetry. Consequently, the representation is not admissible for $G\slash H$ or, put differently, the line defect does not survive gauging the 1-form symmetry. While 1-form symmetries of magnetic quivers are discussed in \cite{Bourget:2020xdz,Closset:2020afy,Closset:2020scj}, we tailor it below for the unitary and orthosymplectic quivers relevant in this paper.

\paragraph{Unitary quivers.}
If the 3d $\Ncal=4$ quiver gauge theory exhibits explicit flavour nodes, the group $\mathrm{ker}\ \phi$ is trivial and the weight lattice and magnetic weight lattice are well-known. For each $\urm(n)$ node, the weight lattice is $\Lambda_w^{\urm(n)}=\Z^n$, while the magnetic lattice is $\Lambda_w^{\urm(n)^\vee}=\Z^n$. For a product gauge group $G=\prod_i \urm(n_i)$, the lattice are given by
\begin{align}
    \Lambda_w^G = \oplus_i \Z^{n_i} \;, \qquad
    \Lambda_w^{G^\vee} = \oplus_i \Z^{n_i} \,.
\end{align}

If the quiver theory is composed of unitary gauge nodes, but does not contain any flavour nodes, the group $\mathrm{ker}\ \phi = \urm(1)$ is non-trivial and continuous. As elaborated in \cite{Bourget:2020xdz}, this continuous group leads to divergent partition functions, like the Coulomb branch Hilbert series or the superconformal index. The diagonally acting $\urm(1)$ can be removed in several ways.
\begin{compactitem}
\item If the quiver contains a $\urm(1)$ gauge node, we can simply turn this gauge node into a flavour node. The resulting quiver has an explicit flavour, and the above results apply. All theories considered in this paper fall into this class.
\item Generically, we may remove the diagonal $\urm(1)$ from any $\urm(n)$ node. However, the result is not equivalent to turning this node into $\surm(n)$ nor $\mathrm{PSU}(n)$, as extensively discussed in \cite{Bourget:2020xdz}. The choice of gauge group becomes $G= \left[\surm(n) \times \prod_i \urm(n_i) \right] \slash \Z_{n}^{\mathrm{diag}}$. The discrete group  originates from the centre $\Z_{n} = Z(\surm(n))$ and can then be embedded into each $\urm(n_i)$ factor, such that it defines a diagonally acting $\Z_{n}^{\mathrm{diag}}$ group.
\end{compactitem}

\paragraph{Orthosymplectic quivers.}
Turning to orthosymplectic quivers, the results of \cite{Bourget:2020xdz} show that $\mathrm{ker}\ \phi$ is trivial either if the quiver contains explicit flavour nodes or if the gauge group contains at least one $\sorm(2n+1)$ factor. In this case, the weight lattices and magnetic lattices of the different factors are as summarised in Table \ref{tab:lattice}. For the product group $G=\prod_i \sorm(2n_i+1) \times \prod_{j} \sprm(k_j) \times \prod_{l}\sorm(2h_j)$, the lattices are simply given by
\begin{subequations}
\begin{align}
    \Lambda_w^{G} &=
    \bigoplus_i \Lambda_w^{\sorm(2n_i+1)}
    \, \oplus \,
    \bigoplus_{j} \Lambda_w^{\sprm(k_j)}
    \, \oplus \,
    \bigoplus_{l} \Lambda_{w}^{\sorm(2h_j)} \,,\\
        \Lambda_w^{G^\vee} &=
        \bigoplus_i \Lambda_w^{\sorm(2n_i+1)^\vee}
        \, \oplus \,
        \bigoplus_{j} \Lambda_w^{\sprm(k_j)^\vee}
        \, \oplus  \,
        \bigoplus_{l} \Lambda_{w}^{\sorm(2h_j)^\vee}  \,.
\end{align}
\end{subequations}

If the orthosymplectic quiver does not admit flavours and does not contain $\sorm(2n+1)$ factors, the group $\mathrm{ker} \ \phi = \Z_2$ is non-trivial. It is then a \emph{choice} to remove this discrete group from the gauge group or not, as both choices lead to well-defined theories. However, in view of the unframed orthosymplectic magnetic quivers studied in \cite{Hanany:2018uhm,Hanany:2018cgo,Cabrera:2019dob,Bourget:2020gzi,Akhond:2020vhc,Akhond:2021knl,Bourget:2021xex}, the choice of removing the diagonal $\Z_2$ seems preferred from brane configurations. To be specific, consider the example
\begin{align}
      \raisebox{-.5\height}{
    \begin{tikzpicture}
        \node (g1) [gaugeSO,label=below:{\footnotesize{$\sorm(2n)$}}] {};
        \node (g2) [gaugeSp,right of =g1,label=below:{\footnotesize{$\sprm(k)$}}] {};
        \draw (g1)--(g2);
    \end{tikzpicture}
    }
    \label{eq:quiver_example}
\end{align}
such that $G= \sorm(2n) \times \sprm(k)$. Considering the quotient gauge group $G \slash \Z_2$ then leads to the following lattice:
\begin{subequations}
\label{eq:lattice_example}
\begin{align}
    \Lambda_w^{G\slash \Z_2} &=
    \left[ \Z_{\sum|\mathrm{even}}^{n} \oplus \Z_{\sum|\mathrm{even}}^{k} \right]
    \cup
    \left[ \Z_{\sum|\mathrm{odd}}^{n} \oplus \Z_{\sum|\mathrm{odd}}^{k} \right] \,, \\
    \Lambda_w^{(G\slash \Z_2)^\vee} &=
    \left[ \Z^{n} \oplus \Z^{k} \right]
    \cup
    \left[ \left(\Z+\tfrac{1}{2}\right)^{n} \oplus \left(\Z+\tfrac{1}{2}\right)^{k} \right]  \,,
\end{align}
\end{subequations}
see Table \ref{tab:lattice} for notation.
The magnetic lattice displays the established \emph{integer lattice} + \emph{half-integer lattice} structure, which enlarges the set of monopole operators contributing to the 3d quiver theory. In contrast, the weight lattice shows that the set of admissible representation has been diminished. In particular, the fundamental representations of $\sorm(2n)$ and $\sprm(k)$ are no longer allowed, as they reside in $\Z_{\sum|\mathrm{odd}}^{n}$ and   $\Z_{\sum|\mathrm{odd}}^{k}$, respectively. However, the bifundamental hypermultiplet in \eqref{eq:quiver_example} is still an admissible representation as it sits in $\Z_{\sum|\mathrm{odd}}^{n} \oplus \Z_{\sum|\mathrm{odd}}^{k}$.

\begin{table}[t]
\ra{1.5}
\centering
\begin{tabular}{llccc}
\toprule
  & group G   & $\sorm(2n+1)$ & $\sprm(n)$  & $\sorm(2n)$  \\ \midrule
 centre & $Z(G)$ & $1$ & $\Z_2$ & $\Z_2$ \\
 weight lattice of $G$ & $\Lambda_w^G$ & $\Z^n$ & $\Z^n$ & $\Z^n$ \\
 magnetic lattice of $G$ & $\Lambda_w^{G^\vee}$ & $\Z^n$ & $\Z^n$ & $\Z^n$ \\ \midrule
 weight lattice of $G\slash \Z_2$ & $\Lambda_w^{G\slash \Z_2}$ & NA & $\Z^n$ & $\Z^n$ \\
 magnetic lattice of $G\slash \Z_2$ & $\Lambda_w^{(G\slash \Z_2)^\vee}$ & NA &  $\Z^n_{\sum|\mathrm{even}}$ & $\Z^n_{\sum|\mathrm{even}}$ \\\bottomrule
\end{tabular}
\caption{Lattices associated to different gauge group factors. $Z(G)$ denotes the centre of a group $G$. $\Lambda_w^G$ stands for the \emph{weight lattice} of the group $G$, while $\Lambda_w^{G^\vee}$ is the \emph{magnetic lattice} of $G$ --- also known as weight lattice of the GNO dual group $G^\vee$ \cite{Goddard:1976qe}. $\Z^n$ denotes the standard integer lattice in $n$ dimensions. $\Z^n_{\sum|\mathrm{even}}$ stands for the set of points $(a_1,\ldots,a_n) \in \Z^n$ such that $\sum_i a_i = \mathrm{even}$.  Likewise, $\Z^n_{\sum|\mathrm{odd}}$ denotes the points $(a_1,\ldots,a_n) \in \Z^n$ such that $\sum_i a_i = \mathrm{odd}$.}
\label{tab:lattice}
\end{table}

\section{\texorpdfstring{3d $\sprm(k)$ SQCD: unitary and orthosymplectic mirror quivers}{3d Sp(k) SQCD: unitary and orthosymplectic mirror quivers}}
\label{sec:3d_Sp}

The goal of this section is two-fold. First, with brane perspective and exact partition functions, we investigate duality between unitary and orthosymplectic quivers, which are mirror to $\sprm(k)$ SQCDs. This investigation paves the way to study the duality of the magnetic quivers in \S\ref{sec:5d_Sp}. Second, we provide a detailed study of the brane dynamics and S-duality in Type IIB theory with O-plane by using exact partition functions.

Understanding D3-D5-NS5 brane dynamics in Type IIB theory \cite{Hanany:1996ie} is indispensable to study duality of 3d $\Ncal=4$ theories. At the same time, exact results in 3d $\Ncal=4$ theories uncover detailed information about brane systems in Type IIB theory \cite{Benvenuti:2011ga,Assel:2015oxa,Dey:2020hfe,Dey:2021jbf}. In this section, we study 3d $\Ncal=4$ mirror symmetry in the presence of an O-plane, including line operators, from this point of view. Also, we show the correspondence of Wilson lines between the unitary and orthosymplectic mirror quivers. The reader can refer to Appendix \ref{app:brane} for the rudiments and notations about 3d $\Ncal=4$ theories and Type IIB brane constructions.

\subsection{Duality of unitary and orthosymplectic mirror quivers}\label{sec:UOSp-3d}

It is well-known that Type IIB brane configurations can realise 3d $\Ncal=4$ low-energy theories with $\sprm(k)$ gauge group and $N_f$ fundamental matter fields. The Lagrangian description is conveniently summarised via a quiver diagram as follows:
\begin{align}\label{Spk-SQCDs}
      \raisebox{-.5\height}{
    \begin{tikzpicture}
	\node (g1) [gaugeSp,label=right:{$\sprm(k)$}] {};
	\node (f1) [flavourSO,above of=g1, label=right:{$\sorm(2N_f)$}] {};
	\draw (g1)--(f1);
	\end{tikzpicture}
    }
\end{align}
where the number of flavours is constrained by $N_f \geq 2k$. This condition ensures complete Higgsing such that the mirror theory can be derived from Type IIB configurations introduced in \cite{Hanany:1996ie,Feng:2000eq}.
A slightly more restrictive constraint is $N_f\geq 2k+1$, which ensures that the $\sprm(k)$ gauge node is \emph{good} in the sense of \cite{Gaiotto:2008ak}. For $N_f=2k+1$, the gauge group is balanced, see \eqref{eq:def_balance}, which implies an enhanced $\sorm(2)$ topological symmetry in the IR. Borrowing from \cite{Cremonesi:2013lqa}, the Coulomb branch Hilbert series for $\sprm(1)$ with $N_f$ flavours is $H=\mathrm{PE}[t^2+t^{N_f-2}+t^{N_f-1} -t^{2N_f-2}]$, which identifies the generators as Casimir invariant, bare monopole operator, and dressed monopole operator of charge $2$, $N_f -2$, and $N_f-1$, respectively,  subject to one relation at order $2N_f-2$. For $N_f=3$, the bare monopole operator is part of the global symmetry multiplet of R-charge $1$, hence the IR Coulomb branch global symmetry emerges due to monopole operators. The same argument holds for $\sprm(k)$ with $N_f=2k+1$, see for instance \cite[eq.\ (5.14)]{Cremonesi:2013lqa} for the explicit monopole formula.

For $N_f=2k$, the theory is conventionally labelled as \emph{bad}, because there exists at least one monopole operator that violates the unitarity bound. Hence, the IR theory needs to be carefully evaluated, see for instance \cite{Assel:2018exy} for a detailed discussion. Unless otherwise stated, bad theories are omitted in this paper because most of the partition functions are ill-defined.

Interestingly, two different realisations exist which give rise to the same effective theory, namely:
\begin{compactitem}
 \item A stack of $k$ full D3 branes on top of an O$3^+$ plane, intersected by $2N_f$ half D5 branes \cite{Feng:2000eq}.
\item A stack of $k$ full D3 branes crossing an O$5^-$ plane, together with a stack of $2N_f$ half D5 branes parallel to the orientifold \cite{Kapustin:1998fa,Hanany:1999sj}.
\end{compactitem}

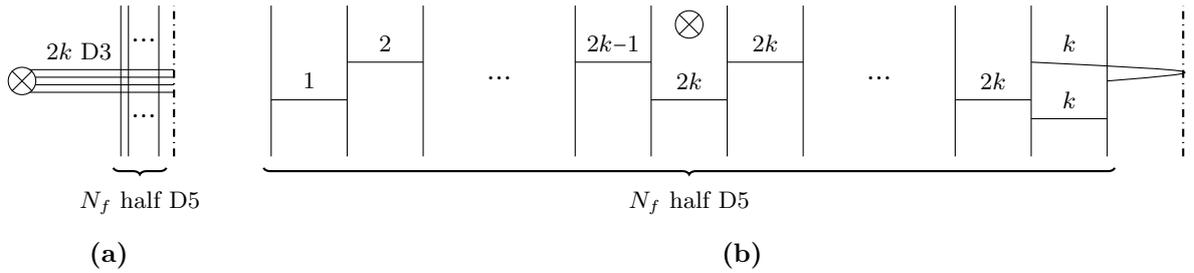
\begin{figure}[t]
    \centering
    \begin{subfigure}{0.2\textwidth}
    \centering
        \begin{tikzpicture}
\draw[dashdotted,thick] (0,-1)--(0,1);
\draw (-0.2,-1)--(-0.2,1) (-0.6,-1)--(-0.6,1) (-0.7,-1)--(-0.7,1);
\node at (-0.4,0.5) {$\cdots$};
\node at (-0.4,-0.5) {$\cdots$};
\draw (-2,0.15)--(0,0.15) (-2,0.05)--(0,0.05)  (-2,-0.05)--(0,-0.05)  (-2,-0.15)--(0,-0.15);
\ns{-2,0}
\node at (-1.25,0.4) {\footnotesize{$2k$ D3}};
\draw[decoration={brace,mirror,raise=10pt},decorate,thick]
  (-0.8,-0.8) -- node[below=15pt] {\footnotesize{$N_f$ half D5}} (-0.1,-0.8);
\end{tikzpicture}
\caption{}
\label{subfig:branes_O5_Sp_a}
    \end{subfigure}
\hfill
    \begin{subfigure}{0.79\textwidth}
    \centering
        \begin{tikzpicture}
\draw (0,-1)--(0,1) (1,-1)--(1,1) (2,-1)--(2,1);
\node at (3,0) {$\cdots$};
\draw (4,-1)--(4,1) (5,-1)--(5,1) (6,-1)--(6,1) (7,-1)--(7,1);
\node at (8,0) {$\cdots$};
\draw (9,-1)--(9,1) (10,-1)--(10,1) (11,-1)--(11,1);
\draw[dashdotted,thick] (12,-1)--(12,1);
\ns{5.5,0.75}
\draw (0,-0.25)--(1,-0.25) (1,0.25)--(2,0.25)
(4,0.25)--(5,0.25) (5,-0.25)--(6,-0.25) (6,0.25)--(7,0.25)
(9,-0.25)--(10,-0.25) (10,-0.5)--(11,-0.5);
\draw (10,0.25) .. controls (12.5,0.125) .. (11,0);
\node at (0.5,0) {\footnotesize{$1$}};
\node at (1.5,0.5) {\footnotesize{$2$}};
\node at (4.5,0.5) {\footnotesize{$2k{-1}$}};
\node at (5.5,0) {\footnotesize{$2k$}};
\node at (6.5,0.5) {\footnotesize{$2k$}};
\node at (9.5,0) {\footnotesize{$2k$}};
\node at (10.5,0.5) {\footnotesize{$k$}};
\node at (10.5,-0.25) {\footnotesize{$k$}};
\draw[decoration={brace,mirror,raise=10pt},decorate,thick]
  (-0.1,-0.8) -- node[below=15pt] {\footnotesize{$N_f$ half D5}} (11.1,-0.8);
\end{tikzpicture}
\caption{}
\label{subfig:branes_O5_Sp_b}
    \end{subfigure}
        \caption{Brane configuration for 3d $\sprm(k)$ with $N_f$ fundamental flavours. (See \eqref{notation} for notations.) In (\subref{subfig:branes_O5_Sp_a}), the brane configuration is in the electric phase or Coulomb branch phase, in which all D3s are suspended between NS5 branes. To transition to the Higgs branch phase, or magnetic phase, shown on (\subref{subfig:branes_O5_Sp_b}), $2k$ half D5s are moved through the half NS5 towards the left hand-side. The mirror theory is obtained from (\subref{subfig:branes_O5_Sp_b}) via S-duality, \textit{i.e.}\ exchanging D5 and NS5, and replacing the O$5^-$ with an ON${}^-$.}
    \label{fig:branes_O5_Sp}
\end{figure}

\paragraph{Unitary mirror quiver.}
To begin with, consider the first setup for which the brane realisation is provided in Figure \ref{fig:branes_O5_Sp}. The corresponding mirror quiver \cite{Hanany:1999sj} is given by
\begin{subequations}\label{eq:unitary-mirror}
\begin{alignat}{2}
N_f &\geq2k +2:\qquad  &
      &\raisebox{-.5\height}{
    \begin{tikzpicture}
	\node (g1) [gauge,label=below:{\footnotesize{$1$}}] {};
	\node (g2) [gauge,right of =g1,label=below:{\footnotesize{$2$}}] {};
	\node (g3) [right of =g2] {$\ldots$};
	\node (g4) [gauge,right of =g3,label=below:{\footnotesize{$2k{-}1$}}] {};
	\node (g5) [gauge,right of =g4,label=below:{\footnotesize{${2k}$}}] {};
	\node (g6) [gauge,right of =g5,label=below:{\footnotesize{${2k}$}}] {};
	\node (g7) [right of =g6] {$\ldots$};
	\node (g8) [gauge,right of =g7,label=below:{\footnotesize{$2k$}}] {};
	\node (g9) [gauge,below right of =g8,label=below:{\footnotesize{$k$}}] {};
	\node (g10) [gauge,above right of =g8,label=below:{\footnotesize{$k$}}] {};
	\node (f1) [flavour,above of=g5, label=left:{\footnotesize{$1$}}] {};
	\draw (g1)--(g2) (g2)--(g3) (g3)--(g4) (g4)--(g5) (g5)--(g6)
	(g6)--(g7) (g7)--(g8) (g8)--(g9) (g8)--(g10) (g5)--(f1);
\draw[decoration={brace,mirror,raise=10pt},decorate,thick]
  (3.75,-0.25) -- node[below=10pt] {\footnotesize{$N_f-2k-1$ nodes}} (7.25,-0.25);
	\end{tikzpicture}
    }
    \label{eq:U_mirror_Nf>2k+1}
    \\
N_f &= 2k +1:\qquad  &
      &\raisebox{-.5\height}{
    \begin{tikzpicture}
	\node (g1) [gauge,label=below:{\footnotesize{$1$}}] {};
	\node (g2) [gauge,right of =g1,label=below:{\footnotesize{$2$}}] {};
	\node (g3) [right of =g2] {$\ldots$};
	\node (g4) [gauge,right of =g3,label=below:{\footnotesize{$2k{-}2$}}] {};
	\node (g5) [gauge,right of =g4,label=below:{\footnotesize{$2k{-}1$}}] {};
	\node (g6) [gauge,below right of =g5,label=below:{\footnotesize{$k$}}] {};
	\node (g7) [gauge,above right of =g5,label=below:{\footnotesize{$k$}}] {};
	\node (f1) [flavour,right of=g6, label=below:{\footnotesize{$1$}}] {};
	\node (f2) [flavour,right of=g7, label=below:{\footnotesize{$1$}}] {};
	\draw (g1)--(g2) (g2)--(g3) (g3)--(g4) (g4)--(g5) (g5)--(g6)
	(g5)--(g7) (g6)--(f1) (g7)--(f2);
	\end{tikzpicture}
    }
    \label{eq:U_mirror_Nf=2k+1}\\
 N_f &=2k :\qquad  &
      &\raisebox{-.5\height}{
    \begin{tikzpicture}
	\node (g1) [gauge,label=below:{\footnotesize{$1$}}] {};
	\node (g2) [gauge,right of =g1,label=below:{\footnotesize{$2$}}] {};
	\node (g3) [right of =g2] {$\ldots$};
	\node (g4) [gauge,right of =g3,label=below:{\footnotesize{$2k{-}3$}}] {};
	\node (g5) [gauge,right of =g4,label=below:{\footnotesize{$2k{-}2$}}] {};
	\node (g6) [gauge,below right of =g5,label=below:{\footnotesize{$k$}}] {};
	\node (g7) [gauge,above right of =g5,label=below:{\footnotesize{$k{-}1$}}] {};
	\node (f1) [flavour,right of=g6, label=below:{\footnotesize{$2$}}] {};
	\draw (g1)--(g2) (g2)--(g3) (g3)--(g4) (g4)--(g5) (g5)--(g6)
	(g5)--(g7)  (g6)--(f1);
	\end{tikzpicture}
    }
    \label{eq:U_mirror_Nf=2k}
\end{alignat}
\end{subequations}
and, in each case, there are $N_f$ balanced nodes. The case $N_f=2k+1$ displays a $S(\urm(1)\times\urm(1))\cong \urm(1)$ flavour symmetry, which indicates the non-trivial $\urm(1)_J$ Coulomb branch symmetry for the balanced $\sprm(k)$ gauge theory on the mirror side.

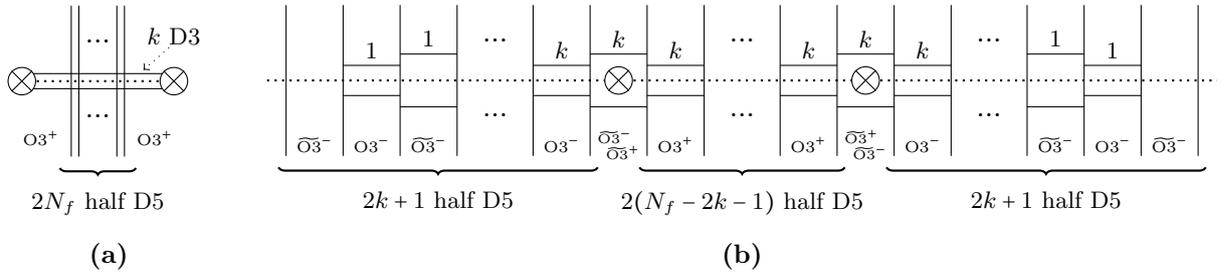
\begin{figure}[ht]
    \centering
    \begin{subfigure}{0.2\textwidth}
    \centering
    \begin{tikzpicture}
    \draw[dotted,thick] (0,0)--(2,0);
    \draw (1-0.35,1)--(1-0.35,-1) (1-0.25,1)--(1-0.25,-1)
    (1+0.35,1)--(1+0.35,-1) (1+0.25,1)--(1+0.25,-1);
    \node at (1,0.5) {$\cdots$};
    \node at (1,-0.5) {$\cdots$};
    \draw (0,0.1)--(2,0.1) (0,-0.1)--(2,-0.1);
    \ns{0,0}
    \ns{2,0}
    \draw[decoration={brace,mirror,raise=10pt},decorate,thick]
  (0.5,-0.8) -- node[below=15pt] {\footnotesize{$2N_f$ half D5}} (1.5,-0.8);
   \node at (1.75,-0.75) {\tiny{$\mathrm{O}3^+$}};
   \node at (0.25,-0.75) {\tiny{$\mathrm{O}3^+$}};
   \node at (2,0.6) {\footnotesize{$k$ D3}};
   \draw[dotted,->] (2,0.5)--(1.6,0.15);
    \end{tikzpicture}
     \caption{}
 \label{subfig:brane_O3_Sp_a}
    \end{subfigure}
\hfill
\begin{subfigure}{0.79\textwidth}
\centering
 \begin{tikzpicture}
 \draw[dotted,thick] (-1,0)--(11.5,0);
  \draw (-0.75,1)--(-0.75,-1) (0,1)--(0,-1) (0.75,1)--(0.75,-1) (1.5,1)--(1.5,-1)
  (2.5,1)--(2.5,-1) (3.25,1)--(3.25,-1) (4,1)--(4,-1)  (4.75,1)--(4.75,-1)
  (8,1)--(8,-1) (7.25,1)--(7.25,-1) (6.5,1)--(6.5,-1)  (5.75,1)--(5.75,-1)
  (11.25,1)--(11.25,-1) (10.5,1)--(10.5,-1) (9.75,1)--(9.75,-1) (9,1)--(9,-1);
  \node at (2,0.5) {$\cdots$};
  \node at (2,-0.5) {$\cdots$};
  \node at (5.25,0.5) {$\cdots$};
  \node at (5.25,-0.5) {$\cdots$};
  \node at (8.5,0.5) {$\cdots$};
  \node at (8.5,-0.5) {$\cdots$};
\draw (0,0.2)--(0.75,0.2) (0,-0.2)--(0.75,-0.2)
(0.75,0.35)--(1.5,0.35) (0.75,-0.35)--(1.5,-0.35)
(2.5,0.2)--(3.25,0.2) (2.5,-0.2)--(3.25,-0.2)
(3.25,0.35)--(4,0.35) (3.25,-0.35)--(4,-0.35)
(4,0.2)--(4.75,0.2) (4,-0.2)--(4.75,-0.2)
(7.25,0.2)--(8,0.2) (7.25,-0.2)--(8,-0.2)
(6.5,0.35)--(7.25,0.35) (6.5,-0.35)--(7.25,-0.35)
(5.75,0.2)--(6.5,0.2) (5.75,-0.2)--(6.5,-0.2)
(9,0.35)--(9.75,0.35) (9,-0.35)--(9.75,-0.35)
(9.75,0.2)--(10.5,0.2) (9.75,-0.2)--(10.5,-0.2);
\node at (0.75/2,0.4) {\footnotesize{$1$}};
\node at (0.75+0.75/2,0.55) {\footnotesize{$1$}};
\node at (2.5+0.75/2,0.4) {\footnotesize{$k$}};
\node at (3.25+0.75/2,0.55) {\footnotesize{$k$}};
\node at (4+0.75/2,0.4) {\footnotesize{$k$}};
\node at (5.75+0.75/2,0.4) {\footnotesize{$k$}};
\node at (6.5+0.75/2,0.55) {\footnotesize{$k$}};
\node at (7.25+0.75/2,0.4) {\footnotesize{$k$}};
\node at (9+0.75/2,0.55) {\footnotesize{$1$}};
\node at (9.75+0.75/2,0.4) {\footnotesize{$1$}};
\ns{3.25+0.75/2,0}
\ns{6.5+0.75/2,0}
 \node at (-0.75/2,-0.85) {\tiny{$\widetilde{\mathrm{O}3}^-$}};
 \node at (0.75/2,-0.85) {\tiny{$\mathrm{O}3^-$}};
 \node at (0.75+0.75/2,-0.85) {\tiny{$\widetilde{\mathrm{O}3}^-$}};
 \node at (2.5+0.75/2,-0.85) {\tiny{$\mathrm{O}3^-$}};
 \node at (3.25+0.75/2,-0.85) {\tiny{$\substack{\widetilde{\mathrm{O}3}^-  \; \; \\ \; \; \widetilde{\mathrm{O}3}^+ }$}};
 \node at (4+0.75/2,-0.85) {\tiny{$\mathrm{O}3^+$}};
 \node at (5.75+0.75/2,-0.85) {\tiny{$\mathrm{O}3^+$}};
 \node at (6.5+0.75/2,-0.85) {\tiny{$\substack{\widetilde{\mathrm{O}3}^+  \; \; \\ \; \; \widetilde{\mathrm{O}3}^- }$}};
 \node at (7.25+0.75/2,-0.85) {\tiny{$\mathrm{O}3^-$}};
 \node at (9+0.75/2,-0.85) {\tiny{$\widetilde{\mathrm{O}3}^-$}};
 \node at (9.75+0.75/2,-0.85) {\tiny{$\mathrm{O}3^-$}};
 \node at (10.5+0.75/2,-0.85) {\tiny{$\widetilde{\mathrm{O}3}^-$}};
 \draw[decoration={brace,mirror,raise=10pt},decorate,thick]
  (-0.85,-0.8) -- node[below=15pt] {\footnotesize{$2k+1$ half D5}} (3.35,-0.8);
  \draw[decoration={brace,mirror,raise=10pt},decorate,thick]
  (3.9,-0.8) -- node[below=15pt] {\footnotesize{$2(N_f-2k-1)$ half D5}} (6.6,-0.8);
  \draw[decoration={brace,mirror,raise=10pt},decorate,thick]
  (7.15,-0.8) -- node[below=15pt] {\footnotesize{$2k+1$ half D5}} (11.35,-0.8);
 \end{tikzpicture}
 \caption{}
 \label{subfig:brane_O3_Sp_b}
\end{subfigure}
    \caption{Brane configuration for 3d $\sprm(k)$ with $N_f$ fundamental flavours using O3 planes. (See \eqref{notation} for notations.) In (\subref{subfig:brane_O3_Sp_a}), the electric phase of the brane configuration is displayed, which gives rise to the $\sprm(k)$ gauge group due to the stack of k full D3s on top of an O$3^+$ plane. The magnetic phase displayed in (\subref{subfig:brane_O3_Sp_b}) is reached by moving 2k half D5 through each of the half NS5s. In the general case $N_f\geq 2k+2$, it is convenient to move an additional half D5 through each half NS5. The mirror configuration is then derived via S-duality, \textit{i.e.}\ exchanging D5 and NS5, and exchanging O$3^+$ and $\widetilde{\mathrm{O}3}^-$, while O$3^-$ and $\widetilde{\mathrm{O}3}^+$ are invariant.}
    \label{fig:brane_O3_Sp}
\end{figure}

\paragraph{Orthosymplectic mirror quiver.}
Next, consider the alternative setup which is given by the brane configuration in Figure \ref{fig:brane_O3_Sp}. The resulting mirror quiver \cite{Feng:2000eq} reads
\begin{subequations}
\begin{alignat}{2}
N_f &\geq 2k+2 :\quad  &
      &\raisebox{-.5\height}{
    \begin{tikzpicture}
	\node (g1) [gaugeSO,label=below:{\footnotesize{$2$}}] {};
	\node (g2) [gaugeSp,right of =g1,label=below:{\footnotesize{$2$}}] {};
	\node (g3) [right of =g2] {$\ldots$};
	\node (g4) [gaugeSO,right of =g3,label=below:{\footnotesize{$2k$}}] {};
	\node (g5) [gaugeSp,right of =g4,label=below:{\footnotesize{$2k$}}] {};
	\node (g6) [gaugeSO,right of =g5,label=below:{\footnotesize{$2k{+}1$}}] {};
	\node (g7) [right of =g6] {$\ldots$};
	\node (g8) [gaugeSO,right of =g7,label=below:{\footnotesize{$2k{+}1$}}] {};
	\node (g9) [gaugeSp,right of =g8,label=below:{\footnotesize{$2k$}}] {};
	\node (g10) [gaugeSO,right of =g9,label=below:{\footnotesize{$2k$}}] {};
	\node (g11) [right of =g10] {$\ldots$};
	\node (g12) [gaugeSp,right of =g11,label=below:{\footnotesize{$2$}}] {};
	\node (g13) [gaugeSO,right of =g12,label=below:{\footnotesize{$2$}}] {};
	\node (f1) [flavourSO,above of=g5, label=left:{\footnotesize{$1$}}] {};
	\node (f2) [flavourSO,above of=g9, label=left:{\footnotesize{$1$}}] {};
	\draw (g1)--(g2) (g2)--(g3) (g3)--(g4) (g4)--(g5) (g5)--(g6)
	(g6)--(g7) (g7)--(g8) (g8)--(g9) (g9)--(g10) (g10)--(g11) (g11)--(g12) (g12)--(g13) (g5)--(f1) (g9)--(f2);
\draw[decoration={brace,mirror,raise=10pt},decorate,thick]
  (3.75,-0.25) -- node[below=10pt] {\footnotesize{$\substack{
  \text{$( N_f{-}2k{-}1)\times  \sorm(2k+1)$ nodes}   \\ \text{$( N_f{-}2k)\times  \sprm(k)$ nodes} }  $ }} (8.25,-0.25);
	\end{tikzpicture}
    }
    \label{eq:OSp_mirror_Nf>2k+1}
    \\
    N_f &= 2k+1 :\quad  &
      &\raisebox{-.5\height}{
        \begin{tikzpicture}
	\node (g1) [gaugeSO,label=below:{\footnotesize{$2$}}] {};
	\node (g2) [gaugeSp,right of =g1,label=below:{\footnotesize{$2$}}] {};
	\node (g3) [right of =g2] {$\ldots$};
	\node (g4) [gaugeSO,right of =g3,label=below:{\footnotesize{$2k{-}2$}}] {};
	\node (g5) [gaugeSp,right of =g4,label=below:{\footnotesize{$2k{-}2$}}] {};
	\node (g6) [gaugeSO,right of =g5,label=below:{\footnotesize{$2k$}}] {};
	\node (g7) [gaugeSp,right of =g6,label=below:{\footnotesize{$2k$}}] {};
	\node (g8) [gaugeSO,right of =g7,label=below:{\footnotesize{$2k$}}] {};
	\node (g9) [gaugeSp,right of =g8,label=below:{\footnotesize{$2k{-}2$}}] {};
	\node (g10) [gaugeSO,right of=g9,label=below:{\footnotesize{$2k{-}2$}}] {};
	\node (g11) [right of =g10] {$\ldots$};
	\node (g12) [gaugeSp,right of =g11,label=below:{\footnotesize{$2$}}] {};
	\node (g13) [gaugeSO,right of =g12,label=below:{\footnotesize{$2$}}] {};
	\node (f1) [flavourSO,above of=g7, label=left:{\footnotesize{$2$}}] {};
	\draw (g1)--(g2) (g2)--(g3) (g3)--(g4) (g4)--(g5) (g5)--(g6)
	(g6)--(g7) (g7)--(g8) (g8)--(g9) (g9)--(g10) (g10)--(g11) (g11)--(g12) (g12)--(g13) (g7)--(f1);
	\end{tikzpicture}
    }
    \label{eq:OSp_mirror_Nf=2k+1}\\
    N_f &= 2k:\quad  &
      &\raisebox{-.5\height}{
    \begin{tikzpicture}
	\node (g1) [gaugeSO,label=below:{\footnotesize{$2$}}] {};
	\node (g2) [gaugeSp,right of =g1,label=below:{\footnotesize{$2$}}] {};
	\node (g3) [right of =g2] {$\ldots$};
	\node (g4) [gaugeSp,right of =g3,label=below:{\footnotesize{$2k{-}4$}}] {};
	\node (g5) [gaugeSO,right of =g4,label=below:{\footnotesize{$2k{-}2$}}] {};
	\node (g6) [gaugeSp,right of =g5,label=below:{\footnotesize{$2k{-}2$}}] {};
	\node (g7) [gaugeSO,right of =g6,label=below:{\footnotesize{$2k$}}] {};
	\node (g8) [gaugeSp,right of =g7,label=below:{\footnotesize{$2k{-}2$}}] {};
	\node (g9) [gaugeSO,right of =g8,label=below:{\footnotesize{$2k{-}2$}}] {};
	\node (g10) [gaugeSp,right of=g9,label=below:{\footnotesize{$2k{-}4$}}] {};
	\node (g11) [right of =g10] {$\ldots$};
	\node (g12) [gaugeSp,right of =g11,label=below:{\footnotesize{$2$}}] {};
	\node (g13) [gaugeSO,right of =g12,label=below:{\footnotesize{$2$}}] {};
	\node (f1) [flavourSp,above of=g7, label=left:{\footnotesize{$2$}}] {};
	\draw (g1)--(g2) (g2)--(g3) (g3)--(g4) (g4)--(g5) (g5)--(g6)
	(g6)--(g7) (g7)--(g8) (g8)--(g9) (g9)--(g10) (g10)--(g11) (g11)--(g12) (g12)--(g13) (g7)--(f1);
	\end{tikzpicture}
    }
    \label{eq:OSp_mirror_Nf=2k}
\end{alignat}
\end{subequations}
and the linear chain of $p=2N_f-3$ balanced nodes with $\sorm(2)$ at both ends gives rise to an enhanced $\sorm(p+3)=\sorm(2N_f)$ Coulomb branch symmetry, see \cite{Gaiotto:2008ak} for details and recall \eqref{eq:def_balance}. As in the unitary case, the $N_f=2k+1$ case displays a $\sorm(2)\cong \urm(1)$ flavour symmetry indicating the non-trivial $\urm(1)_J$ topological symmetry on the mirror side.

We obtain a pair of 3d $\Ncal=4$ unitary and orthosymplectic mirror quivers from the $\sprm(k)$ SQCD, and it is expected that they flow to the same IR fixed point \cite{Hanany:1999sj,Feng:2000eq}. To verify it,
we evaluate the superconformal index for both quivers. A straightforward but tedious computation shows that the indices are compatible with each other in the given order of perturbative evaluation. The computational details for the cases $k=1$ with $N_f=3,4,5,6$ and $k=2$ with $N_f=4,5,6$ are presented in Table \ref{tab:SCI_Sp_mirrors}. Moreover, the limits \eqref{HC-limit} of the computed superconformal indices reproduce the known Higgs and Coulomb branch Hilbert series \cite{Hanany:2016gbz,Cabrera:2017ucb,Cabrera:2018ldc}, which provides a consistency check of the results.

\subsection{Match of Wilson line defects in the two mirror quivers}
\label{sec:Wilson_mirrors}

It is natural to ask whether the unitary and orthosymplectic mirror quivers are endowed with the same set of line defects. If so, it is important to understand how they are mapped under the duality. Thereafter, we may proceed to include Wilson line defects in the two mirrors. By using the B-twisted index, we create a dictionary for the corresponding Wilson lines under the duality.
The guiding principle is the following:
Wilson lines defects in the unitary and orthosymplectic quivers can be compatible only if they originate from an equal number of brane configurations. That is, the F1 defining the Wilson line needs to have as many possibilities to end on a stack of D3 branes in one mirror configuration as in the other mirror configuration. This means that the total dimensions of representations for Wilson lines are equal under the duality.

After careful analysis of the B-twisted indices for the cases $k=1$ with $N_f=3,4,5,6$ -- see Tables \ref{Sp1Nf3WilsonHS}, \ref{Sp1Nf4WilsonHS}, \ref{Sp1Nf5WilsonHS}, \ref{Sp1Nf6WilsonHS}-- and $k=2$ with $N_f=4,5,6$ -- see Tables \ref{Sp2Nf4WilsonHS}, \ref{Sp2Nf5WilsonHS}, \ref{Sp2Nf6WilsonHS} -- an interesting pattern between Wilson lines in \eqref{eq:U_mirror_Nf>2k+1}  and in \eqref{eq:OSp_mirror_Nf>2k+1} arises.
\begin{itemize}
    \item \textbf{Observation 1:} For a fundamental Wilson line at the $\sprm(r)$ node, for $1\leq r \leq k$
    \begin{align}
    &\raisebox{-.5\height}{

    }
\end{align}
where the Wilson line of the $\urm(2r)$ gauge node is in the fundamental representation.
\item \textbf{Observation 2:} For a fundamental Wilson line at the $\sorm(2r)$ node, for $1\leq r \leq k$
    \begin{align}
    &\raisebox{-.5\height}{
    %
    } \notag
\end{align}
and the Wilson lines of the unitary gauge nodes transform in the fundamental representations. Here and similarly below, the equality is understood as a single Wilson line expectation value in the orthosymplectic quiver equals a linear combination of two Wilson line expectation values in the unitary quiver.
\item \textbf{Observation 3:} For $\sprm(k)$ node at position $l\geq 2k$, which is not the central node:
    \begin{align}
    &\raisebox{-.5\height}{
    %
    }
\end{align}
and the Wilson lines in unitary nodes transform in the fundamental representation.
\item \textbf{Observation 4:} For $\sorm(2k+1)$ node at position $l>2k$, which is not the central node:
    \begin{align}
    &\raisebox{-.5\height}{
    %
    } \notag
\end{align}
and the Wilson lines in the unitary nodes transform in the fundamental representation.
\item \textbf{Observation 5:} For a fundamental Wilson line at the central $\sprm(k)$ node
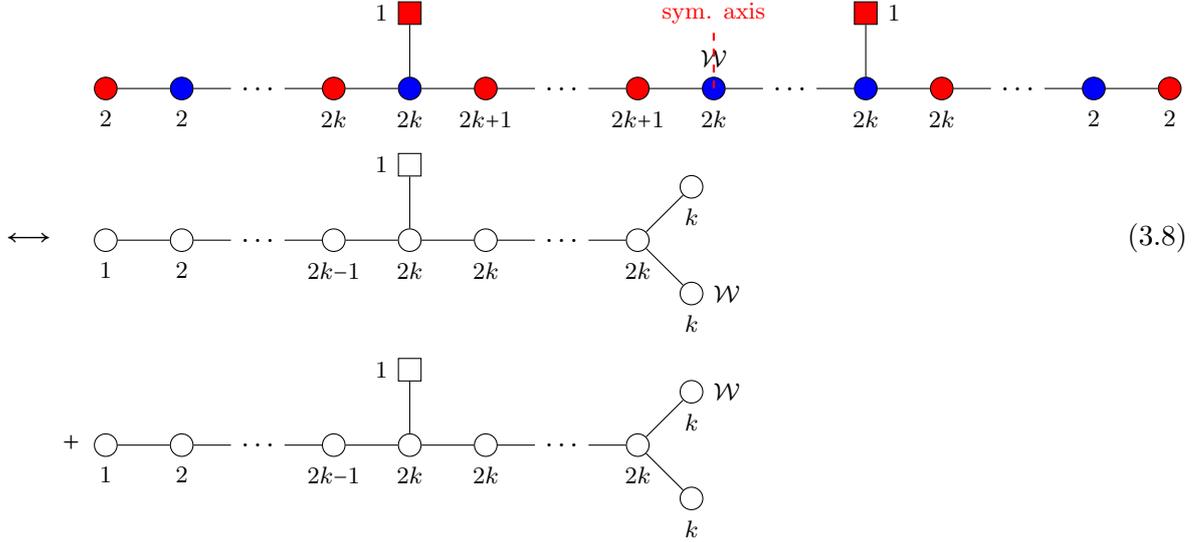
    \begin{align}
    &\raisebox{-.5\height}{
    %
    } \notag
\end{align}
and the Wilson lines in the unitary nodes transform in the fundamental representation.
\item \textbf{Observation 6:} For a fundamental Wilson line at the central $\sorm(2k+1)$ node
    \begin{align}
    &\raisebox{-.5\height}{
    %
    } \notag
\end{align}
and the Wilson lines in the unitary nodes transform in the fundamental representation.
\end{itemize}
In this dictionary, a Wilson line at an Sp gauge node is dual to a Wilson line at the corresponding unitary gauge node from the left in the quiver diagrams. (Observation 1, 3.) Note that a Wilson line at the middle Sp gauge node corresponds to a direct sum of those at the two spinor nodes. (Observation 5.) On the other hand, a Wilson line at a non-abelian SO gauge node is dual to a direct sum of one(s) at the corresponding unitary gauge node(s) and one at the left $\urm(1)$ node. (Observation 2, 4, 6.) As a result, it is consistent with the guiding principle so that the total dimensions of representations are equal.

The observed pattern straightforwardly applies to the balanced case $N_f=2k+1$. Since the flavour nodes are at the spinor nodes in \eqref{eq:U_mirror_Nf=2k+1} or at the central $\sprm$ node in \eqref{eq:OSp_mirror_Nf=2k+1}, Observations 1 and 2 apply for all $\sorm(2l)$ nodes with $1\leq l \leq k$ and all $\sprm(l)$ nodes with $1\leq l \leq k-1$, respectively. Similarly, Observation 5 applies for the Wilson line at the central $\sprm(k)$ node.

\subsection{Mirror symmetry for Wilson and vortex defect}
\label{sec:mirror_Wilson_vortex}
Now let us consider the 3d $\Ncal=4$ mirror symmetry with line defects. As studied in \cite{Assel:2015oxa,Dimofte:2019zzj,Dey:2021jbf}, Wilson and vortex defects are exchanged under the mirror symmetry. A vortex defect in the $\sprm(k)$ SQCD with $N_f$ flavours is characterised by an $\sprm(k)$ representation $R$ and a splitting $N_f=L+(N_f-L)$. Here, the aim is to derive the mirror Wilson line defects from the vortex defects of $\sprm(k)$ by brane dynamics and confirm that the exact partition functions match. The starting point is the brane configuration of $\sprm(k)$ SQCD in the presence of a vortex line defect:
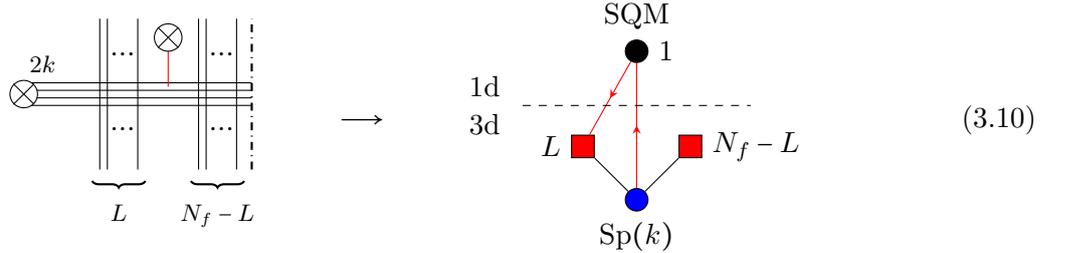
\begin{align}
  \raisebox{-.5\height}{
    \begin{tikzpicture}
\draw[dashdotted,thick] (0,-1)--(0,1);
\draw (-0.2,-1)--(-0.2,1) (-0.6,-1)--(-0.6,1) (-0.7,-1)--(-0.7,1);
\node at (-0.4,0.5) {$\cdots$};
\node at (-0.4,-0.5) {$\cdots$};
\draw (-1.5,-1)--(-1.5,1) (-1.9,-1)--(-1.9,1) (-2,-1)--(-2,1);
\node at (-1.7,0.5) {$\cdots$};
\node at (-1.7,-0.5) {$\cdots$};
\draw (-3,0.05)--(0,0.05) (-3,-0.05)--(0,-0.05)
(-3,0.15)--(0,0.15) (-3,-0.15)--(0,-0.15);
\ns{-3,0}
\draw[red] (-1.1,0.1)--(-1.1,0.75);
\ns{-1.1,0.75}
\node at (-2.75,0.4) {\footnotesize{$2k$}};
\draw[decoration={brace,mirror,raise=10pt},decorate,thick]
  (-0.8,-0.8) -- node[below=15pt] {\footnotesize{$N_f-L$}} (-0.1,-0.8);
  \draw[decoration={brace,mirror,raise=10pt},decorate,thick]
  (-2.1,-0.8) -- node[below=15pt] {\footnotesize{$L$}} (-1.4,-0.8);
\end{tikzpicture}
}
\qquad \longrightarrow \qquad
  \raisebox{-.5\height}{
    \begin{tikzpicture}
	\node (g1) [gauge,label=below:{$\sprm(k)$},fill=blue] {};
	\node (f1) [flavour,above left of=g1, label=left:{$L$},fill=red] {};
	\node (f2) [flavour,above right of=g1, label=right:{$N_f-L$},fill=red] {};
	\draw (g1)--(f1) (g1)--(f2);
  \node (SQM) [defect, above=10ex of g1,label=above:{SQM},label=right:{$1$}] {};
  \begin{scope}[decoration={markings,mark =at position 0.5 with {\arrow{stealth}}}]
\draw[postaction={decorate},color=red] (g1)--(SQM) ;
  \draw[postaction={decorate},color=red] (SQM)--(f1);
\end{scope}
    \draw[dashed] (-1.5,1.25)--(1.5,1.25);
    \node at (-2,1.5) {1d};
    \node at (-2,1) {3d};
	\end{tikzpicture}
    }\label{fig:vortex}
\end{align}
with the notations of \eqref{notation}. It is understood that the D1 can end on either of the $2k$ D3 branes; hence, the defect is characterised by the fundamental representation of $\sprm(k)$.

The $S^3$ partition function for $\sprm(k)$ SQCD with $N_f$ fundamental hypermultiplets of masses $m_i$ ($i=1,\ldots,N_f$) is given by
\begin{equation}\label{eq:S3-sp-SQCD}
Z^{S^3}_{\sprm(k),N_f}(m)=\sum_{I\in C_{k}^{N_f}}\prod_{j=1}^k\frac{ m_{I_j}\sh(2m_{I_j})}{\prod_{\ell\not \in I} \sh(m_{\ell}\pm m_{I_j})}
\end{equation}
where $I$ runs over all combinations $C_{k}^{N_{f}}$ of $k$ different integers in $\left\{1, \ldots, N_{f}\right\}$. Once we include the vortex of type $L+\left(N_{f}-L\right)$ in \eqref{fig:vortex}, the partition function is given by
\begin{align}\label{eq:S3-sp-SQCD-vortex}
Z^{S^3}_{(L,N_f-L),\mathbf{2k}}=&(2k-L)\cdot Z^{S^3}_{\sprm(k),N_f}(m)+\sum_{j=1}^L Z^{S^3}_{\sprm(k),N_f}(m_j\to m_j-i)~.
\end{align}
where $m_j$ ($j=1,\ldots,L$) are the masses of the hypermultiplets charged under the vortex defect.
In Appendix \ref{app:sphere}, we provide a brief derivation of the sphere partition functions \eqref{eq:S3-sp-SQCD}--\eqref{eq:S3-sp-SQCD-vortex} and an analytic proof of the equalities of sphere partition functions with line operators under the mirror symmetry for $k=1$. In fact, \eqref{eq:S3-sp-SQCD-vortex} encodes the information about $2k$ configurations of the D1 ending on the D3 brane as well as a splitting of the flavours. Taking the S-duality, $2k$ individual brane configurations give rise to a linear combination of Wilson line defects in the mirror theory. Hence, the sum of the dimensions of the Wilson line representations is equal to $2k$.
We see this below in various flavour splittings. To begin with, focus on the $N_f>2k+1$ case \eqref{eq:U_mirror_Nf>2k+1}.

\paragraph{Splitting $L + (N_f-L)$, $1\leq L \leq 2k-1$.}
Performing S-duality of \eqref{fig:vortex} yields brane configurations of two types: the first configuration has the F1 stretched between the defect D5 and one of the $L$ D3 branes in segment of the form
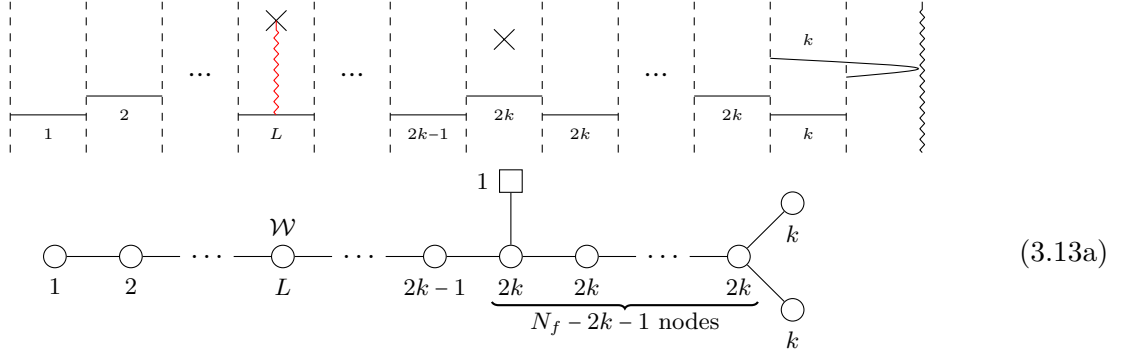
\begin{subequations}
\label{eq:split_L_Nf-L}
\begin{align}
&\raisebox{-.5\height}{
\begin{tikzpicture}
    \foreach \i in {-5,-4,...,6}
    \draw[dashed] (\i,-1)--(\i,1);
    \draw[line join=round,decorate, decoration={zigzag, segment length=4,amplitude=.9,post=lineto,post length=2pt}] (7,-1)--(7,1);
    \draw (-5,-0.5)--(-4,-0.5) (-4,-0.25)--(-3,-0.25) (-2,-0.5)--(-1,-0.5)
    (0,-0.5)--(1,-0.5) (1,-0.25)--(2,-0.25) (2,-0.5)--(3,-0.5) (4,-0.25)--(5,-0.25) (5,-0.5)--(6,-0.5);
    \draw (5,0.25) .. controls (7.4,0.125) .. (6,0);
    \dfive{1.5,0.5}
    \node at (3.5,0) {$\cdots$};
    \node at (-0.5,0) {$\cdots$};
    \node at (-2.5,0) {$\cdots$};
    \dfive{-1.5,0.75}
    \draw [red,line join=round,decorate, decoration={zigzag, segment length=4,amplitude=.9,post=lineto,post length=2pt}] (-1.5,-0.5) -- (-1.5,0.75);
    \node at (-4.5,-0.75) {\tiny{$1$}};
    \node at (-3.5,-0.5) {\tiny{$2$}};
    \node at (-1.5,-0.75) {\tiny{$L$}};
    \node at (0.5,-0.75) {\tiny{$2k{-}1$}};
    \node at (1.5,-0.5) {\tiny{$2k$}};
    \node at (2.5,-0.75) {\tiny{$2k$}};
    \node at (4.5,-0.5) {\tiny{$2k$}};
    \node at (5.5,-0.75) {\tiny{$k$}};
    \node at (5.5,0.5) {\tiny{$k$}};
\end{tikzpicture}
}\notag \\
&\quad \raisebox{-.5\height}{
    \begin{tikzpicture}
    \node (g1) [gauge,label=below:{\footnotesize{${1}$}}] {};
	\node (g2) [gauge,right of =g1,label=below:{\footnotesize{${2}$}}] {};
	\node (g3) [right of =g2] {$\ldots$};
	\node (g4) [gauge,right of =g3,label=below:{\footnotesize{${L}$}},label=above:{\footnotesize{$\mathcal{W}$}}] {};
	\node (g5) [right of =g4] {$\ldots$};
	\node (g6) [gauge,right of =g5,label=below:{\footnotesize{$2k-1$}}] {};
	\node (g7) [gauge,right of =g6,label=below:{\footnotesize{${2k}$}}] {};
	\node (g8) [gauge,right of =g7,label=below:{\footnotesize{${2k}$}}] {};
	\node (g9) [right of =g8] {$\ldots$};
	\node (g10) [gauge,right of =g9,label=below:{\footnotesize{$2k$}}] {};
	\node (g11) [gauge,below right of =g10,label=below:{\footnotesize{$k$}}] {};
	\node (g12) [gauge,above right of =g10,label=below:{\footnotesize{$k$}}] {};
	\node (f1) [flavour,above of=g7, label=left:{\footnotesize{$1$}}] {};
	\draw (g1)--(g2) (g2)--(g3) (g3)--(g4) (g4)--(g5) (g5)--(g6) (g6)--(g7)
	(g7)--(g8) (g8)--(g9) (g9)--(g10) (g10)--(g11) (g10)--(g12) (g7)--(f1);
\draw[decoration={brace,mirror,raise=10pt},decorate,thick]
  (5.75,-0.25) -- node[below=10pt] {\footnotesize{$N_f-2k-1$ nodes}} (9.25,-0.25);
	\end{tikzpicture}
    }
    \label{eq:Sp1_split_1_n-1_a}
    \end{align}
which gives rise to a fundamental Wilson line in the $\urm(L)$ gauge group. Again the notations for branes are summarised in \eqref{notation}.
The other configuration has the F1 stretched between the defect D5 and flavour D5. This setup is the result of the F1 sliding off the D5 due to the unequal number of connecting D3s from the left and right -- this has been referred to as D3 brane spike \cite{Callan:1997kz}. There are $(2k -L)$ configurations that give rise to the same brane system
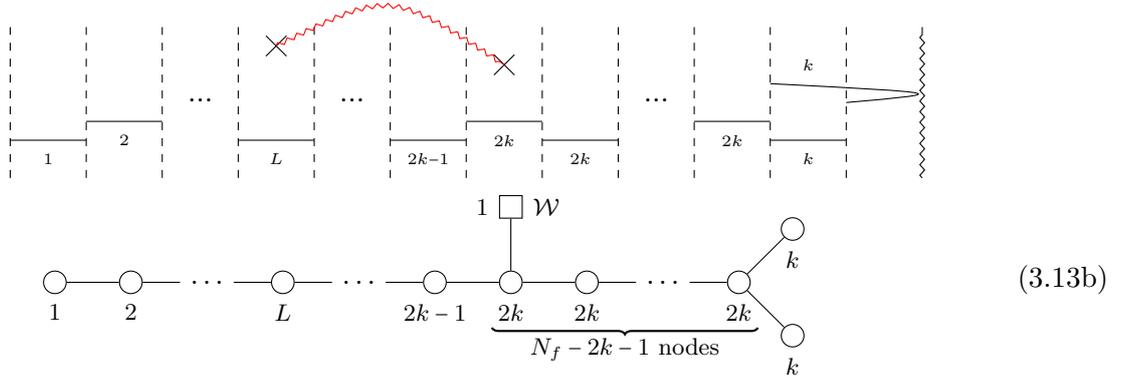
\begin{align}
&\raisebox{-.5\height}{
\begin{tikzpicture}
    \foreach \i in {-5,-4,...,6}
    \draw[dashed] (\i,-1)--(\i,1);
    \draw[line join=round,decorate, decoration={zigzag, segment length=4,amplitude=.9,post=lineto,post length=2pt}] (7,-1)--(7,1);
    \draw (-5,-0.5)--(-4,-0.5) (-4,-0.25)--(-3,-0.25) (-2,-0.5)--(-1,-0.5)
    (0,-0.5)--(1,-0.5) (1,-0.25)--(2,-0.25) (2,-0.5)--(3,-0.5) (4,-0.25)--(5,-0.25) (5,-0.5)--(6,-0.5);
    \draw (5,0.25) .. controls (7.4,0.125) .. (6,0);
    \dfive{1.5,0.5}
    \node at (3.5,0) {$\cdots$};
    \node at (-0.5,0) {$\cdots$};
    \node at (-2.5,0) {$\cdots$};
    \dfive{-1.5,0.75}
    \draw [red,line join=round,decorate, decoration={zigzag, segment length=4,amplitude=.9,post=lineto,post length=2pt}] (-1.5,0.75) .. controls (0,1.5) .. (1.5,0.5);
    \node at (-4.5,-0.75) {\tiny{$1$}};
    \node at (-3.5,-0.5) {\tiny{$2$}};
    \node at (-1.5,-0.75) {\tiny{$L$}};
    \node at (0.5,-0.75) {\tiny{$2k{-}1$}};
    \node at (1.5,-0.5) {\tiny{$2k$}};
    \node at (2.5,-0.75) {\tiny{$2k$}};
    \node at (4.5,-0.5) {\tiny{$2k$}};
    \node at (5.5,-0.75) {\tiny{$k$}};
    \node at (5.5,0.5) {\tiny{$k$}};
\end{tikzpicture}
} \notag \\
&\quad \raisebox{-.5\height}{
    \begin{tikzpicture}
    \node (g1) [gauge,label=below:{\footnotesize{${1}$}}] {};
	\node (g2) [gauge,right of =g1,label=below:{\footnotesize{${2}$}}] {};
	\node (g3) [right of =g2] {$\ldots$};
	\node (g4) [gauge,right of =g3,label=below:{\footnotesize{${L}$}}] {};
	\node (g5) [right of =g4] {$\ldots$};
	\node (g6) [gauge,right of =g5,label=below:{\footnotesize{$2k-1$}}] {};
	\node (g7) [gauge,right of =g6,label=below:{\footnotesize{${2k}$}}] {};
	\node (g8) [gauge,right of =g7,label=below:{\footnotesize{${2k}$}}] {};
	\node (g9) [right of =g8] {$\ldots$};
	\node (g10) [gauge,right of =g9,label=below:{\footnotesize{$2k$}}] {};
	\node (g11) [gauge,below right of =g10,label=below:{\footnotesize{$k$}}] {};
	\node (g12) [gauge,above right of =g10,label=below:{\footnotesize{$k$}}] {};
	\node (f1) [flavour,above of=g7, label=left:{\footnotesize{$1$}},label=right:{\footnotesize{$\mathcal{W}$}}] {};
	\draw (g1)--(g2) (g2)--(g3) (g3)--(g4) (g4)--(g5) (g5)--(g6) (g6)--(g7)
	(g7)--(g8) (g8)--(g9) (g9)--(g10) (g10)--(g11) (g10)--(g12) (g7)--(f1);
\draw[decoration={brace,mirror,raise=10pt},decorate,thick]
  (5.75,-0.25) -- node[below=10pt] {\footnotesize{$N_f-2k-1$ nodes}} (9.25,-0.25);
	\end{tikzpicture}
    }
    \label{eq:Sp1_split_1_n-1_b}
\end{align}
\end{subequations}
such that the resulting flavour Wilson line comes with multiplicity $2k-L$.

In the mirror unitary quiver \eqref{eq:unitary-mirror}, we assign the FI parameter $\xi_{l}-\xi_{l+1}$ to  the $l$-th gauge node from the left whereas we associate $\xi_{N_f-1}\pm\xi_{N_f}$ as the FI parameters to the two spinor nodes of the balanced $D$-type diagram. The $S^3$ partition function of the mirror unitary quiver is equal to \eqref{eq:S3-sp-SQCD} by mapping the FI parameters to the mass parameters  $\xi_i\longleftrightarrow m_i$. Since the mirror unitary quiver \eqref{eq:Sp1_split_1_n-1_a} contains the $T[\surm(2k)]$ tail \cite{Gaiotto:2008ak}, the fundamental Wilson loop at the $L$-th node ($L\le 2k-1$) gives rise to the shifts of the FI parameters \cite[\S5.4.2]{Assel:2015oxa} as
$$
Z^{S^3}_{\mathcal{W} \text{ at } \urm(L)}=\sum_{j=1}^L Z^{S^3}(\xi_j\to \xi_j-i)~.
$$
in the $S^3$ partition function. As a result, \eqref{eq:S3-sp-SQCD-vortex} can be recast into
\begin{align}
Z^{S^3}_{(L,N_f-L),\mathbf{2k}}\big|_{m_i \rightarrow \xi_i}=
Z^{S^3}_{\mathcal{W} \text{ at } \urm(L)}
+(2k-L)\cdot Z^{S^3}(\xi)
\end{align}
\textit{i.e.} the sum of a fundamental Wilson loop at the $L$-th node and $(2k-L)$ flavour Wilson loops in the unitary mirror quiver under the exchange  $\xi_i\longleftrightarrow m_i$.

In terms of twisted indices, an analogous relation is implied
\begin{align}
    \mathbb{I}^A_{(L,N_f-L),\mathbf{2k}} = \mathbb{I}^B_{\mathcal{W} \text{ at } \urm(L)} + (2k-L) \cdot \mathbb{I}^B
\end{align}
where the index $\mathbb{I}^B$ without defect reflects the contribution of the flavour Wilson line, due to the unrefined computation.
For the case $k=1$, the explicit results for the B-twisted index are provided in Tables \ref{Sp1Nf4WilsonHS}, \ref{Sp1Nf5WilsonHS}, \ref{Sp1Nf6WilsonHS} and for the A-twisted index in Table \ref{tab:vortx_SU2_results}.

\paragraph{Splitting $L+(N_f-L)$ for $2k\leq L\leq N_f-1$.}
Performing S-duality of \eqref{fig:vortex} in this range of $L$ yields only one brane configuration
\begin{align}
&\raisebox{-.5\height}{
\begin{tikzpicture}
    \foreach \i in {-3,-2,...,10}
    \draw[dashed] (\i,-1)--(\i,1);
    \draw[line join=round,decorate, decoration={zigzag, segment length=4,amplitude=.9,post=lineto,post length=2pt}] (11,-1)--(11,1);
    \draw (-3,-0.5)--(-2,-0.5) (-2,-0.25)--(-1,-0.25)
    (0,-0.5)--(1,-0.5) (1,-0.25)--(2,-0.25) (2,-0.5)--(3,-0.5)
    (4,-0.25)--(5,-0.25) (5,-0.5)--(6,-0.5)
    (6,-0.25)--(7,-0.25)
    (8,-0.25)--(9,-0.25) (9,-0.5)--(10,-0.5);
    \draw (9,0.25) .. controls (11.5,0.125) .. (10,0);
    \dfive{1.5,0.5}
    \node at (7.5,0) {$\cdots$};
    \node at (3.5,0) {$\cdots$};
    \node at (-0.5,0) {$\cdots$};
    \dfive{5.5,0.75}
    \draw [red,line join=round,decorate, decoration={zigzag, segment length=4,amplitude=.9,post=lineto,post length=2pt}] (5.5,-0.5) -- (5.5,0.75);
    \node at (-2.5,-0.75) {\tiny{$1$}};
    \node at (-1.5,-0.5) {\tiny{$2$}};
    \node at (0.5,-0.75) {\tiny{$2k{-}1$}};
    \node at (1.5,-0.5) {\tiny{$2k$}};
    \node at (2.5,-0.75) {\tiny{$2k$}};
    \node at (4.5,-0.5) {\tiny{$2k$}};
    \node at (5.5,-0.75) {\tiny{$2k$}};
    \node at (6.5,-0.5) {\tiny{$2k$}};
    \node at (8.5,-0.5) {\tiny{$2k$}};
    \node at (9.5,-0.75) {\tiny{$k$}};
    \node at (9.5,0.5) {\tiny{$k$}};
    \draw[decoration={brace,mirror,raise=10pt},decorate,thick]
  (-3.1,-0.85) -- node[below=10pt] {\footnotesize{$L$ NS5}} (5.1,-0.85);
  \draw[decoration={brace,mirror,raise=10pt},decorate,thick]
  (5.9,-0.85) -- node[below=10pt] {\footnotesize{$N_f{-}L$ NS5}} (10.1,-0.85);
\end{tikzpicture}
}
\notag \\
&\quad \raisebox{-.5\height}{
    \begin{tikzpicture}
    \node (h1) [gauge,label=below:{\footnotesize{${1}$}}] {};
	\node (h2) [gauge,right of =h1,label=below:{\footnotesize{${2}$}}] {};
	\node (h3) [right of =h2] {$\cdots$};
	\node (g1) [gauge,right of =h3,label=below:{\footnotesize{$2k{-}1$}}] {};
	\node (g2) [gauge,right of =g1,label=below:{\footnotesize{${2k}$}}] {};
	\node (g3) [gauge,right of =g2,label=below:{\footnotesize{${2k}$}}] {};
	\node (g4) [right of =g3] {$\cdots$};
	\node (g5) [gauge,right of =g4,label=below:{\footnotesize{${2k}$}}] {};
	\node (g6) [gauge,right of =g5,label=below:{\footnotesize{${2k}$}},label=above:{\footnotesize{$\mathcal{W}$}}] {};
	\node (g7) [gauge,right of =g6,label=below:{\footnotesize{${2k}$}}] {};
	\node (g8) [right of =g7] {$\ldots$};
	\node (g9) [gauge,right of =g8,label=below:{\footnotesize{$2k$}}] {};
	\node (g10) [gauge,below right of =g9,label=below:{\footnotesize{$k$}}] {};
	\node (g11) [gauge,above right of =g9,label=below:{\footnotesize{$k$}}] {};
	\node (f1) [flavour,above of=g2, label=left:{\footnotesize{$1$}}] {};
	\draw (h1)--(h2) (h2)--(h3) (h3)--(g1) (g1)--(g2) (g2)--(g3) (g3)--(g4) (g4)--(g5) (g5)--(g6) (g6)--(g7) (g7)--(g8) (g8)--(g9) (g9)--(g10) (g9)--(g11) (g2)--(f1);
\draw[decoration={brace,mirror,raise=10pt},decorate,thick]
  (-0.25,-0.25) -- node[below=10pt] {\footnotesize{$L-1$ nodes}} (7.25,-0.25);
	\end{tikzpicture}
    }
    \label{eq:Sp1_split_L_n-L}
\end{align}
which results in a fundamental Wilson line of an $\urm(2k)$ node in the linear part of the quiver theory. We compute the $S^3$ partition function with the Wilson loop for $k=1$ in \eqref{eq:Wilson2}, which agrees with the mirror vortex.
Likewise, in terms of twisted indices,  \eqref{eq:Sp1_split_L_n-L} implies the following relation:
\begin{align}
    \mathbb{I}^A_{(L,N_f-L),\mathbf{2}} = \mathbb{I}^B_{\mathcal{W} \text{ at } [0,\ldots,0,1,0,\ldots,0]}
\end{align}
where $[0,\ldots,0,1,0,\ldots,0]$ denotes the $L$-th node in the $D_{N_f}$-type Dynkin quiver.
For the $k=1$ case, the explicit results for the B-twisted index are provided in Tables \ref{Sp1Nf4WilsonHS}, \ref{Sp1Nf5WilsonHS}, \ref{Sp1Nf6WilsonHS} and for the A-twisted index in Table \ref{tab:vortx_SU2_results}.

\paragraph{Splitting $(N_f-1)+1$.}
Performing S-duality of \eqref{fig:vortex} yields two different brane configurations, depending on which D3 branes the F1 ends. To begin with, we consider that the F1 stretches between the defect D5 and the stack of $k$ D3s that does not pass through the ON plane. We find
\begin{align}
&\raisebox{-.5\height}{
\begin{tikzpicture}
    \foreach \i in {-3,-2,...,6}
    \draw[dashed] (\i,-1)--(\i,1);
    \draw[line join=round,decorate, decoration={zigzag, segment length=4,amplitude=.9,post=lineto,post length=2pt}] (7,-1)--(7,1);
    \draw (-3,-0.5)--(-2,-0.5) (-2,-0.25)--(-1,-0.25)
    (0,-0.5)--(1,-0.5) (1,-0.25)--(2,-0.25) (2,-0.5)--(3,-0.5) (4,-0.25)--(5,-0.25) (5,-0.5)--(6,-0.5);
    \draw (5,0.25) .. controls (7.4,0.125) .. (6,0);
    \dfive{1.5,0.5}
    \node at (3.5,0) {$\cdots$};
    \node at (-0.5,0) {$\cdots$};
    \dfive{5.5,0.75}
    \draw [red,line join=round,decorate, decoration={zigzag, segment length=4,amplitude=.9,post=lineto,post length=2pt}] (5.5,0.75) -- (5.5,-0.5);
    \node at (-2.5,-0.75) {\tiny{$1$}};
    \node at (-1.5,-0.5) {\tiny{$2$}};
    \node at (0.5,-0.75) {\tiny{$2k{-}1$}};
    \node at (1.5,-0.5) {\tiny{$2k$}};
    \node at (2.5,-0.75) {\tiny{$2k$}};
    \node at (4.5,-0.5) {\tiny{$2k$}};
    \node at (5.5,-0.75) {\tiny{$k$}};
    \node at (5.75,0.5) {\tiny{$k$}};
\end{tikzpicture}
} \notag \\
&\quad \raisebox{-.5\height}{
    \begin{tikzpicture}
	\node (g1) [gauge,label=below:{\footnotesize{${1}$}}] {};
	\node (g2) [gauge,right of =g1,label=below:{\footnotesize{${2}$}}] {};
	\node (g3) [right of =g2] {$\ldots$};
	\node (g4) [gauge,right of =g3,label=below:{\footnotesize{$2k{-}1$}}] {};
	\node (g5) [gauge,right of =g4,label=below:{\footnotesize{${2k}$}}] {};
	\node (g6) [gauge,right of =g5,label=below:{\footnotesize{${2k}$}}] {};
	\node (g7) [right of =g6] {$\ldots$};
	\node (g8) [gauge,right of =g7,label=below:{\footnotesize{$2k$}}] {};
	\node (g9) [gauge,below right of =g8,label=below:{\footnotesize{$k$}},label=right:{\footnotesize{$\mathcal{W}$}}] {};
	\node (g10) [gauge,above right of =g8,label=below:{\footnotesize{$k$}}] {};
	\node (f1) [flavour,above of=g5, label=left:{\footnotesize{$1$}}] {};
	\draw (g1)--(g2) (g2)--(g3) (g3)--(g4) (g4)--(g5) (g5)--(g6) (g6)--(g7) (g7)--(g8) (g8)--(g9) (g8)--(g10) (g5)--(f1);
\draw[decoration={brace,mirror,raise=10pt},decorate,thick]
  (3.75,-0.25) -- node[below=10pt] {\footnotesize{$N_f-2k-1$ nodes}} (7.25,-0.25);
	\end{tikzpicture}
    }
    \label{eq:Sp1_split_n-1_1_a}
\end{align}
which induces a fundamental Wilson line in one of the $\urm(k)$ nodes, \textit{i.e.} a spinor node of the balanced $D$-type diagram. It is straightforward to read off from the brane configuration that the corresponding FI parameter is $\xi_{N_f-1}-\xi_{N_f}$.
Next, the F1 could also be stretched between a defect D5 and the stack of $k$ D3s that goes through the ON plane. This yields
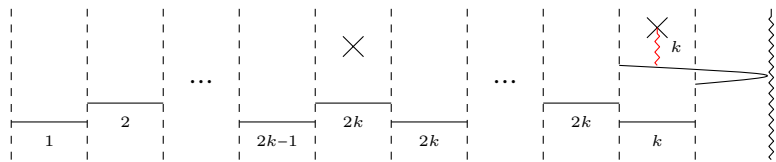
\begin{align}
&\raisebox{-.5\height}{
\begin{tikzpicture}
    \foreach \i in {-3,-2,...,6}
    \draw[dashed] (\i,-1)--(\i,1);
    \draw[line join=round,decorate, decoration={zigzag, segment length=4,amplitude=.9,post=lineto,post length=2pt}] (7,-1)--(7,1);
    \draw (-3,-0.5)--(-2,-0.5) (-2,-0.25)--(-1,-0.25)
    (0,-0.5)--(1,-0.5) (1,-0.25)--(2,-0.25) (2,-0.5)--(3,-0.5) (4,-0.25)--(5,-0.25) (5,-0.5)--(6,-0.5);
    \draw (5,0.25) .. controls (7.4,0.125) .. (6,0);
    \dfive{1.5,0.5}
    \node at (3.5,0) {$\cdots$};
    \node at (-0.5,0) {$\cdots$};
    \dfive{5.5,0.75}
    \draw [red,line join=round,decorate, decoration={zigzag, segment length=4,amplitude=.9,post=lineto,post length=2pt}] (5.5,0.25) -- (5.5,0.75);
    \node at (-2.5,-0.75) {\tiny{$1$}};
    \node at (-1.5,-0.5) {\tiny{$2$}};
    \node at (0.5,-0.75) {\tiny{$2k{-}1$}};
    \node at (1.5,-0.5) {\tiny{$2k$}};
    \node at (2.5,-0.75) {\tiny{$2k$}};
    \node at (4.5,-0.5) {\tiny{$2k$}};
    \node at (5.5,-0.75) {\tiny{$k$}};
    \node at (5.75,0.5) {\tiny{$k$}};
\end{tikzpicture}
} \notag \\
 &\quad \raisebox{-.5\height}{
    \begin{tikzpicture}
    \node (g1) [gauge,label=below:{\footnotesize{${1}$}}] {};
	\node (g2) [gauge,right of =g1,label=below:{\footnotesize{${2}$}}] {};
	\node (g3) [right of =g2] {$\ldots$};
	\node (g4) [gauge,right of=g3,label=below:{\footnotesize{$2k{-}1$}}] {};
	\node (g5) [gauge,right of =g4,label=below:{\footnotesize{${2k}$}}] {};
	\node (g6) [gauge,right of =g5,label=below:{\footnotesize{${2k}$}}] {};
	\node (g7) [right of =g6] {$\ldots$};
	\node (g8) [gauge,right of =g7,label=below:{\footnotesize{$2k$}}] {};
	\node (g9) [gauge,below right of =g8,label=below:{\footnotesize{$k$}}] {};
	\node (g10) [gauge,above right of =g8,label=below:{\footnotesize{$k$}},label=right:{\footnotesize{$\mathcal{W}$}}] {};
	\node (f1) [flavour,above of=g5, label=left:{\footnotesize{$1$}}] {};
	\draw (g1)--(g2) (g2)--(g3) (g3)--(g4) (g4)--(g5) (g5)--(g6) (g6)--(g7) (g7)--(g8) (g8)--(g9) (g8)--(g10) (g5)--(f1);
\draw[decoration={brace,mirror,raise=10pt},decorate,thick]
  (3.75,-0.25) -- node[below=10pt] {\footnotesize{$N_f-2k-1$ nodes}} (7.25,-0.25);
	\end{tikzpicture}
    }
    \label{eq:Sp1_split_n-1_1_b}
    \end{align}
which induces a fundamental Wilson line on the other $\urm(k)$ node. Note that the corresponding FI parameter of this gauge node is $\xi_{N_f-1}+\xi_{N_f}$.

In summary, the configuration \eqref{eq:Sp1_split_n-1_1_a} leads to a Wilson line in a $\urm(k)$ gauge group at one of the spinor nodes of the $D$-type quiver, while the configuration \eqref{eq:Sp1_split_n-1_1_b} induces a Wilson line in the other spinor node. We compute the sum of the $S^3$ partition functions with the Wilson loops for $k=1$ in \eqref{eq:Wilson3}, which agrees with the mirror vortex. In terms of twisted indices, this implies
\begin{align}
    \mathbb{I}^A_{(N_f-1,1),\mathbf{2k}} = \mathbb{I}^B_{\mathcal{W} \text{ at } [0,\ldots,0,1,0]} + \mathbb{I}^B_{\mathcal{W} \text{ at } [0,\ldots,0,0,1]}
\end{align}
where $[0,\ldots,0,1,0]$ and $[0,\ldots,0,0,1]$ denote the two spinor nodes of the $D_{N_f}$-type Dynkin quiver.
For the $k=1$ case, the explicit results for the B-twisted index are provided in Tables \ref{Sp1Nf4WilsonHS}, \ref{Sp1Nf5WilsonHS}, \ref{Sp1Nf6WilsonHS} and for the A-twisted index in Table \ref{tab:vortx_SU2_results}.
\paragraph{Splitting $N_f+0$.}
The mirror symmetry of the vortex loop of splitting $N_f+0$ is more subtle. When the defect D5 brane is present between the NS5 brane and the ON plane,
there are two ways for the F1 to end to the stack of $k$ D3s: before and after the D3s goes through the ON plane. This gives
\begin{align}
&\raisebox{-.5\height}{
\begin{tikzpicture}
    \foreach \i in {-3,-2,...,6}
    \draw[dashed] (\i,-1)--(\i,1);
    \draw[line join=round,decorate, decoration={zigzag, segment length=4,amplitude=.9,post=lineto,post length=2pt}] (7,-1)--(7,1);
    \draw (-3,-0.5)--(-2,-0.5) (-2,-0.25)--(-1,-0.25)
    (0,-0.5)--(1,-0.5) (1,-0.25)--(2,-0.25) (2,-0.5)--(3,-0.5) (4,-0.25)--(5,-0.25) (5,-0.5)--(6,-0.5);
    \draw (5,0.25) .. controls (7.4,0.125) .. (6,0);
    \dfive{1.5,0.5}
    \node at (3.5,0) {$\cdots$};
    \node at (-0.5,0) {$\cdots$};
    \dfive{6.3,0.75}
    \draw [red,line join=round,decorate, decoration={zigzag, segment length=4,amplitude=.9,post=lineto,post length=2pt}] (6.3,0.25) -- (6.3,0.75);
    \node at (-2.5,-0.75) {\tiny{$1$}};
    \node at (-1.5,-0.5) {\tiny{$2$}};
    \node at (0.5,-0.75) {\tiny{$2k{-}1$}};
    \node at (1.5,-0.5) {\tiny{$2k$}};
    \node at (2.5,-0.75) {\tiny{$2k$}};
    \node at (4.5,-0.5) {\tiny{$2k$}};
    \node at (5.5,-0.75) {\tiny{$k$}};
    \node at (5.75,0.5) {\tiny{$k$}};
\end{tikzpicture}
} \notag \\
\notag \\
&\raisebox{-.5\height}{
\begin{tikzpicture}
    \foreach \i in {-3,-2,...,6}
    \draw[dashed] (\i,-1)--(\i,1);
    \draw[line join=round,decorate, decoration={zigzag, segment length=4,amplitude=.9,post=lineto,post length=2pt}] (7,-1)--(7,1);
    \draw (-3,-0.5)--(-2,-0.5) (-2,-0.25)--(-1,-0.25)
    (0,-0.5)--(1,-0.5) (1,-0.25)--(2,-0.25) (2,-0.5)--(3,-0.5) (4,-0.25)--(5,-0.25) (5,-0.5)--(6,-0.5);
    \draw (5,0.25) .. controls (7.4,0.125) .. (6,0);
    \dfive{1.5,0.5}
    \node at (3.5,0) {$\cdots$};
    \node at (-0.5,0) {$\cdots$};
    \dfive{6.3,0.75}
    \draw [red,line join=round,decorate, decoration={zigzag, segment length=4,amplitude=.9,post=lineto,post length=2pt}] (6.3,-0.05) -- (6.3,0.75);
    \node at (-2.5,-0.75) {\tiny{$1$}};
    \node at (-1.5,-0.5) {\tiny{$2$}};
    \node at (0.5,-0.75) {\tiny{$2k{-}1$}};
    \node at (1.5,-0.5) {\tiny{$2k$}};
    \node at (2.5,-0.75) {\tiny{$2k$}};
    \node at (4.5,-0.5) {\tiny{$2k$}};
    \node at (5.5,-0.75) {\tiny{$k$}};
    \node at (5.75,0.5) {\tiny{$k$}};
\end{tikzpicture}
}\notag \\
 &\quad 2\times \raisebox{-.5\height}{
    \begin{tikzpicture}
    \node (g1) [gauge,label=below:{\footnotesize{${1}$}}] {};
  \node (g2) [gauge,right of =g1,label=below:{\footnotesize{${2}$}}] {};
  \node (g3) [right of =g2] {$\ldots$};
  \node (g4) [gauge,right of=g3,label=below:{\footnotesize{$2k{-}1$}}] {};
  \node (g5) [gauge,right of =g4,label=below:{\footnotesize{${2k}$}}] {};
  \node (g6) [gauge,right of =g5,label=below:{\footnotesize{${2k}$}}] {};
  \node (g7) [right of =g6] {$\ldots$};
  \node (g8) [gauge,right of =g7,label=below:{\footnotesize{$2k$}}] {};
  \node (g9) [gauge,below right of =g8,label=below:{\footnotesize{$k$}}] {};
  \node (g10) [gauge,above right of =g8,label=below:{\footnotesize{$k$}},label=right:{\footnotesize{$\mathcal{W}$}}] {};
  \node (f1) [flavour,above of=g5, label=left:{\footnotesize{$1$}}] {};
  \draw (g1)--(g2) (g2)--(g3) (g3)--(g4) (g4)--(g5) (g5)--(g6) (g6)--(g7) (g7)--(g8) (g8)--(g9) (g8)--(g10) (g5)--(f1);
\draw[decoration={brace,mirror,raise=10pt},decorate,thick]
  (3.75,-0.25) -- node[below=10pt] {\footnotesize{$N_f-2k-1$ nodes}} (7.25,-0.25);
  \end{tikzpicture}
    }
    \label{eq:Sp1_split_n_0_a}
    \end{align}
which results in the Wilson line at the spinor node of type $[0,\ldots,0,1]$ with multiplicity two. The difference from the $(N_f-1)+1$ splitting can be seen explicitly in the $S^3$ partition function. A fundamental Wilson line is charged under the gauge group with the FI parameter $\xi_{N_f-1}+\xi_{N_f}$.

In terms of $S^3$ partition functions, we have
\begin{equation}
  Z^{S^3}_{(N_f,0),\mathbf{2k}} =2 \cdot Z^{S^3}_{\mathcal{W} \text{ at } [0,\ldots,0,1]}~.
\end{equation}
We check this for $k=1$ in \eqref{eq:Wilson4}, which agrees with the mirror vortex.
We observe that the relation of twisted indices under the mirror symmetry is given by
\begin{align}
    \mathbb{I}^A_{(N_f,0),\mathbf{2k}} =(t+t^{-1}) \cdot \mathbb{I}^B_{\mathcal{W} \text{ at } [0,\ldots,0,1]}~.
\end{align}
The appearing prefactor $t+t^{-1}$ is computationally verified, but the reason behind is unclear at this point.
For the $k=1$ case, the explicit results for the B-twisted index are provided in Tables \ref{Sp1Nf4WilsonHS}, \ref{Sp1Nf5WilsonHS}, \ref{Sp1Nf6WilsonHS} and for the A-twisted index in Table \ref{tab:vortx_SU2_results}.

\paragraph{Balanced case.}
The analysis can be readily extended to the balanced case $N_f=2k+1$. Since the flavour nodes in \eqref{eq:U_mirror_Nf=2k+1} reside at the two spinor nodes, the pattern presented in \eqref{eq:split_L_Nf-L} applies for all splittings $L+(N_f-L)$, $1\leq L \leq 2k-1=N_f-2$. The remaining two splitting behave as follows: $L=2k=N_f-1$ follows the logic of \eqref{eq:Sp1_split_n-1_1_a}--\eqref{eq:Sp1_split_n-1_1_b}, i.e.\ the vortex line is dual to the sum of  Wilson line expectation values at the two spinor nodes. Lastly, the splitting $L=2k+1 =N_f$ corresponds to \eqref{eq:Sp1_split_n_0_a}, meaning a Wilson line on a spinor node with a multiplicity of 2.

\paragraph{Comment on orthosymplectic mirror quiver.}
One may ask why the mirror symmetry considerations are relevant for the objective of studying a duality between unitary and orthosymplectic quivers. In principle, one possible way to derive a pattern between Wilson lines as in \S \ref{sec:Wilson_mirrors} is to trace the Wilson lines back to vortex line defects in the $\sprm(k)$ SQCD. This, however, is not straightforward because they are realised by two different brane configurations. To exemplify, for an O3 configuration one can place a defect D1 between any two half D5 branes as follows
\begin{align}
\raisebox{-.5\height}{
    \begin{tikzpicture}
    \draw[dotted,thick] (0,0)--(2,0);
    \draw (1-0.45,1)--(1-0.45,-1) (1-0.35,1)--(1-0.35,-1)
    (1+0.45,1)--(1+0.45,-1) (1+0.35,1)--(1+0.35,-1);
    \draw[red] (1,0.1)--(1,0.75) (1,-0.1)--(1,-0.75);
    \ns{1,0.75}
    \ns{1,-0.75}
    \draw (0,0.1)--(2,0.1) (0,-0.1)--(2,-0.1);
    \ns{0,0}
    \ns{2,0}
    \draw[decoration={brace,mirror,raise=10pt},decorate,thick]
  (0.45,-0.8) -- node[below=15pt] {\footnotesize{$L$}} (0.75,-0.8);
    \draw[decoration={brace,mirror,raise=10pt},decorate,thick]
  (1.25,-0.8) -- node[below=15pt] {$\quad$\footnotesize{$2N_f-L$}} (1.55,-0.8);
   \node at (1.8,-0.75) {\tiny{$\mathrm{O}3^+$}};
   \node at (0.2,-0.75) {\tiny{$\mathrm{O}3^+$}};
   \node at (2,0.6) {\footnotesize{$k$ D3}};
   \draw[dotted,->] (2,0.5)--(1.6,0.15);
    \end{tikzpicture}
    }
\end{align}
which should induce a vortex defect of $L+(2N_f-L)$ splitting. Therefore, unlike \eqref{fig:vortex}, the $2N_f$ half-hypermultiplets admit $2N_f+2$ different ways of splittings. While one can argue that half of these are redundant due to the symmetry along the $x^6$ direction, the introduced 1d SQM and its coupling to the 3d bulk theory are rather delicate. If $L$ is odd, then the vortex defect is described by the SQM coupled to the $\sprm(k)$ bulk gauge symmetry and an $\sorm(L)$ flavour symmetry, which does not appear in \eqref{fig:vortex}. Moreover, evaluating the expectation value of such a vortex defect remains a challenging task, both in the A-twisted index and in the sphere partition function. It requires a precise microscopic description of the 1d/3d coupled system, and we leave this problem to future research.

%
\subsection{A comment on vortex line defects in the two mirrors}
\label{sec:vortex_in_mirrors}
In view of \S\ref{sec:Wilson_mirrors}, it is tempting to ask whether vortex line defects can be matched between the two different mirrors of $\sprm(k)$ gauge theory.
An interesting class of vortex lines is realised as 3d mirrors of a fundamental Wilson line in the $\sprm(k)$ gauge symmetry.
\paragraph{Unitary mirror quiver.}
The brane construction for a fundamental Wilson line is given by
\begin{align}
\raisebox{-.5\height}{
\begin{tikzpicture}
    \draw[dashed] (0,-1)--(0,1);
    \draw (0,0.05)--(4,0.05) (0,-0.05)--(4,-0.05);
    \dfive{0.5,0}
    \dfive{1,0}
    \dfive{1.5,0}
    \dfive{2,0}
    \dfive{2.5,0}
    \dfive{3,0}
    \dfive{3.5,0}
    \node[gauge] at (4,0) {};
    %
    \dfive{1.75,1}
    \draw [red,line join=round,decorate, decoration={zigzag, segment length=4,amplitude=.9,post=lineto,post length=2pt}] (1.75,1) -- (1.75,0);
    \draw[decoration={brace,mirror,raise=10pt},decorate,thick]
  (0.25,-0.25) -- node[below=10pt] {\footnotesize{$N_f$ D5}} (3.75,-0.25);
  \node at (1,0.5) {\footnotesize{$2k$ D3}};
  \node at (4.2,0.35) {\footnotesize{O$5^-$}};
\end{tikzpicture}
}
\qquad \longrightarrow \qquad
\raisebox{-.5\height}{
    \begin{tikzpicture}
	\node (g1) [gaugeSp,label=right:{$\sprm(k)$},label=left:{\footnotesize{$\mathcal{W}$}}] {};
	\node (f1) [flavourSO,above of=g1,label=right:{$\sorm(2N_f)$}] {};
	\draw (g1)--(f1);
	\end{tikzpicture}
    }
    \label{eq:Wilson_O5_branes}
\end{align}
and the position of the F1 with respect to the flavour D5s is not physical. This is because the F1 can move across any of the D5 branes since the number of D3 branes on the left and right agree.

Based on the brane configuration \eqref{eq:Wilson_O5_branes} and assuming $N_f>2k+1$, the mirror vortex defect is realised by
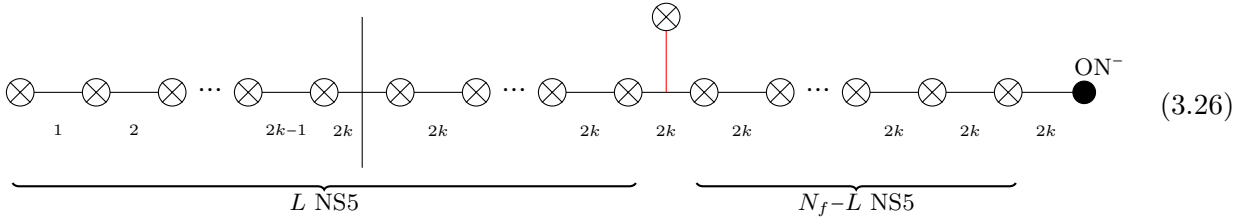
\begin{align}
    &\raisebox{-.5\height}{
    \begin{tikzpicture}
    \draw (-3,0)--(-1,0)
    (0,0)--(3,0)
    (4,0)--(7,0)
    (8,0)--(11,0);
    \foreach \i in {-3,-2,...,10}
    \ns{\i,0};
    \node[defect] at (11,0) {};
    \node at (11.2,0.35) {\footnotesize{ON${}^-$}};
    \draw (1.5,-1)--(1.5,1);
    \node at (7.5,0) {$\cdots$};
    \node at (3.5,0) {$\cdots$};
    \node at (-0.5,0) {$\cdots$};
    \draw[red] (5.5,0) -- (5.5,1);
    \ns{5.5,1}
    \node at (-2.5,-0.5) {\tiny{$1$}};
    \node at (-1.5,-0.5) {\tiny{$2$}};
    \node at (0.5,-0.5) {\tiny{$2k{-}1$}};
    \node at (1.25,-0.5) {\tiny{$2k$}};
    \node at (2.5,-0.5) {\tiny{$2k$}};
    \node at (4.5,-0.5) {\tiny{$2k$}};
    \node at (5.5,-0.5) {\tiny{$2k$}};
    \node at (6.5,-0.5) {\tiny{$2k$}};
    \node at (8.5,-0.5) {\tiny{$2k$}};
    \node at (9.5,-0.5) {\tiny{$2k$}};
    \node at (10.5,-0.5) {\tiny{$2k$}};
    \draw[decoration={brace,mirror,raise=10pt},decorate,thick]
  (-3.1,-0.85) -- node[below=10pt] {\footnotesize{$L$ NS5}} (5.1,-0.85);
  \draw[decoration={brace,mirror,raise=10pt},decorate,thick]
  (5.9,-0.85) -- node[below=10pt] {\footnotesize{$N_f{-}L$ NS5}} (10.1,-0.85);
\end{tikzpicture}
}
\label{eq:branes_vortex_U_mirror}
\end{align}
Moving the D1 to the nearest NS5 on the left-hand side, yields the following defect insertion:
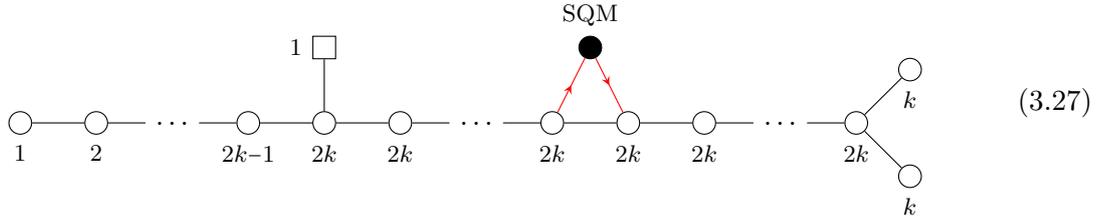
\begin{align}\label{vortex1}
    \raisebox{-.5\height}{
    \begin{tikzpicture}
	\node (g1) [gauge,label=below:{\footnotesize{$1$}}] {};
	\node (g2) [gauge,right of =g1,label=below:{\footnotesize{$2$}}] {};
	\node (g3) [right of =g2] {$\ldots$};
	\node (g4) [gauge,right of =g3,label=below:{\footnotesize{$2k{-}1$}}] {};
	\node (g5) [gauge,right of =g4,label=below:{\footnotesize{${2k}$}}] {};
	\node (g6) [gauge,right of =g5,label=below:{\footnotesize{${2k}$}}] {};
	\node (h0) [right of =g6] {$\ldots$};
	\node (h1) [gauge,right of =h0,label=below:{\footnotesize{${2k}$}}] {};
	\node (h2) [gauge,right of =h1,label=below:{\footnotesize{${2k}$}}] {};
	\node (h3) [gauge,right of =h2,label=below:{\footnotesize{${2k}$}}] {};
	\node (g7) [right of =h3] {$\ldots$};
	\node (g8) [gauge,right of =g7,label=below:{\footnotesize{$2k$}}] {};
	\node (g9) [gauge,below right of =g8,label=below:{\footnotesize{$k$}}] {};
	\node (g10) [gauge,above right of =g8,label=below:{\footnotesize{$k$}}] {};
	\node (f1) [flavour,above of=g5, label=left:{\footnotesize{$1$}}] {};
	\draw (g1)--(g2) (g2)--(g3) (g3)--(g4) (g4)--(g5) (g5)--(g6)
	(g6)--(h0) (h0)--(h1) (h1)--(h2) (h2)--(h3)
	(h3)--(g7) (g7)--(g8) (g8)--(g9) (g8)--(g10) (g5)--(f1);
%
    \node (SQM) at (7.5,1) [defect,label=above:{\footnotesize{SQM}}] {};
    \begin{scope}[decoration={markings,mark =at position 0.5 with {\arrow{stealth}}}]
    \draw[postaction={decorate},color=red] (h1) -- (SQM);
    \draw[postaction={decorate},color=red] (SQM) -- (h2);
     \end{scope}
	\end{tikzpicture}
    }
    \end{align}
whereas moving the D1 to the nearest NS5 on the right-hand side yields the following:
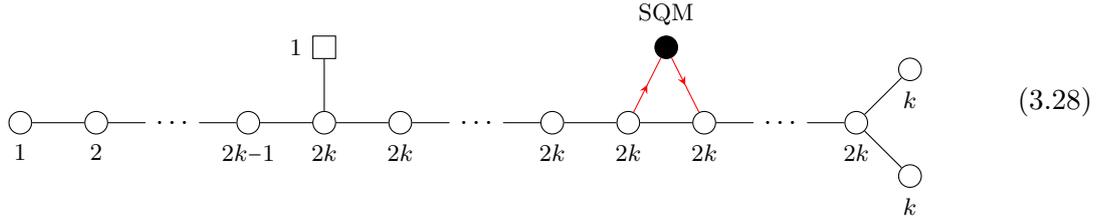
\begin{align}\label{vortex2}
    \raisebox{-.5\height}{
    \begin{tikzpicture}
	\node (g1) [gauge,label=below:{\footnotesize{$1$}}] {};
	\node (g2) [gauge,right of =g1,label=below:{\footnotesize{$2$}}] {};
	\node (g3) [right of =g2] {$\ldots$};
	\node (g4) [gauge,right of =g3,label=below:{\footnotesize{$2k{-}1$}}] {};
	\node (g5) [gauge,right of =g4,label=below:{\footnotesize{${2k}$}}] {};
	\node (g6) [gauge,right of =g5,label=below:{\footnotesize{${2k}$}}] {};
	\node (h0) [right of =g6] {$\ldots$};
	\node (h1) [gauge,right of =h0,label=below:{\footnotesize{${2k}$}}] {};
	\node (h2) [gauge,right of =h1,label=below:{\footnotesize{${2k}$}}] {};
	\node (h3) [gauge,right of =h2,label=below:{\footnotesize{${2k}$}}] {};
	\node (g7) [right of =h3] {$\ldots$};
	\node (g8) [gauge,right of =g7,label=below:{\footnotesize{$2k$}}] {};
	\node (g9) [gauge,below right of =g8,label=below:{\footnotesize{$k$}}] {};
	\node (g10) [gauge,above right of =g8,label=below:{\footnotesize{$k$}}] {};
	\node (f1) [flavour,above of=g5, label=left:{\footnotesize{$1$}}] {};
	\draw (g1)--(g2) (g2)--(g3) (g3)--(g4) (g4)--(g5) (g5)--(g6)
	(g6)--(h0) (h0)--(h1) (h1)--(h2) (h2)--(h3)
	(h3)--(g7) (g7)--(g8) (g8)--(g9) (g8)--(g10) (g5)--(f1);
%
    \node (SQM) at (8.5,1) [defect,label=above:{\footnotesize{SQM}}] {};
    \begin{scope}[decoration={markings,mark =at position 0.5 with {\arrow{stealth}}}]
    \draw[postaction={decorate},color=red] (h2) -- (SQM);
    \draw[postaction={decorate},color=red] (SQM) -- (h3);
     \end{scope}
	\end{tikzpicture}
    }
\end{align}
In terms of the brane configuration \eqref{eq:branes_vortex_U_mirror}, the D-string can cross any of the NS5 as long as the numbers of D3 branes on the left and right are equal. Hence, the two defects should be equivalent. This is indeed true due to the hopping duality discussed in \cite{Assel:2015oxa}. In other words, the vortex defects defined by two adjacent $\urm(2k)$ nodes are equivalent. This is evident from the brane configuration and has been proven on the level of the $S^3$ partition function in \cite{Assel:2015oxa}.

By the same logic, we find two more configurations describing dual defects by utilising hopping duality. These are given by
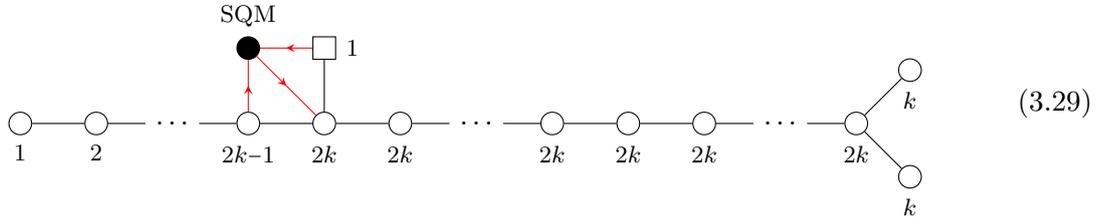
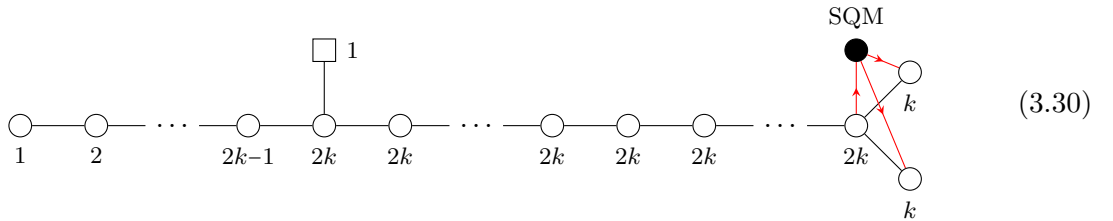
\begin{align}\label{vortex3}
    \raisebox{-.5\height}{
    \begin{tikzpicture}
	\node (g1) [gauge,label=below:{\footnotesize{$1$}}] {};
	\node (g2) [gauge,right of =g1,label=below:{\footnotesize{$2$}}] {};
	\node (g3) [right of =g2] {$\ldots$};
	\node (g4) [gauge,right of =g3,label=below:{\footnotesize{$2k{-}1$}}] {};
	\node (g5) [gauge,right of =g4,label=below:{\footnotesize{${2k}$}}] {};
	\node (g6) [gauge,right of =g5,label=below:{\footnotesize{${2k}$}}] {};
	\node (h0) [right of =g6] {$\ldots$};
	\node (h1) [gauge,right of =h0,label=below:{\footnotesize{${2k}$}}] {};
	\node (h2) [gauge,right of =h1,label=below:{\footnotesize{${2k}$}}] {};
	\node (h3) [gauge,right of =h2,label=below:{\footnotesize{${2k}$}}] {};
	\node (g7) [right of =h3] {$\ldots$};
	\node (g8) [gauge,right of =g7,label=below:{\footnotesize{$2k$}}] {};
	\node (g9) [gauge,below right of =g8,label=below:{\footnotesize{$k$}}] {};
	\node (g10) [gauge,above right of =g8,label=below:{\footnotesize{$k$}}] {};
	\node (f1) [flavour,above of=g5, label=right:{\footnotesize{$1$}}] {};
	\draw (g1)--(g2) (g2)--(g3) (g3)--(g4) (g4)--(g5) (g5)--(g6)
	(g6)--(h0) (h0)--(h1) (h1)--(h2) (h2)--(h3)
	(h3)--(g7) (g7)--(g8) (g8)--(g9) (g8)--(g10) (g5)--(f1);
%
    \node (SQM) at (3,1) [defect,label=above:{\footnotesize{SQM}}] {};
    \begin{scope}[decoration={markings,mark =at position 0.5 with {\arrow{stealth}}}]
    \draw[postaction={decorate},color=red] (g4) -- (SQM);
    \draw[postaction={decorate},color=red] (f1) -- (SQM);
    \draw[postaction={decorate},color=red] (SQM) -- (g5);
     \end{scope}
	\end{tikzpicture}
    } \\
\raisebox{-.5\height}{
    \begin{tikzpicture}
	\node (g1) [gauge,label=below:{\footnotesize{$1$}}] {};
	\node (g2) [gauge,right of =g1,label=below:{\footnotesize{$2$}}] {};
	\node (g3) [right of =g2] {$\ldots$};
	\node (g4) [gauge,right of =g3,label=below:{\footnotesize{$2k{-}1$}}] {};
	\node (g5) [gauge,right of =g4,label=below:{\footnotesize{${2k}$}}] {};
	\node (g6) [gauge,right of =g5,label=below:{\footnotesize{${2k}$}}] {};
	\node (h0) [right of =g6] {$\ldots$};
	\node (h1) [gauge,right of =h0,label=below:{\footnotesize{${2k}$}}] {};
	\node (h2) [gauge,right of =h1,label=below:{\footnotesize{${2k}$}}] {};
	\node (h3) [gauge,right of =h2,label=below:{\footnotesize{${2k}$}}] {};
	\node (g7) [right of =h3] {$\ldots$};
	\node (g8) [gauge,right of =g7,label=below:{\footnotesize{$2k$}}] {};
	\node (g9) [gauge,below right of =g8,label=below:{\footnotesize{$k$}}] {};
	\node (g10) [gauge,above right of =g8,label=below:{\footnotesize{$k$}}] {};
	\node (f1) [flavour,above of=g5, label=right:{\footnotesize{$1$}}] {};
	\draw (g1)--(g2) (g2)--(g3) (g3)--(g4) (g4)--(g5) (g5)--(g6)
	(g6)--(h0) (h0)--(h1) (h1)--(h2) (h2)--(h3)
	(h3)--(g7) (g7)--(g8) (g8)--(g9) (g8)--(g10) (g5)--(f1);
%
    \node (SQM) at (11,1) [defect,label=above:{\footnotesize{SQM}}] {};
    \begin{scope}[decoration={markings,mark =at position 0.5 with {\arrow{stealth}}}]
    \draw[postaction={decorate},color=red] (g8) -- (SQM);
    \draw[postaction={decorate},color=red] (SQM) -- (g9);
    \draw[postaction={decorate},color=red] (SQM) -- (g10);
     \end{scope}
	\end{tikzpicture}
    }
    \label{eq:vortex_D-type_Nf>2k+1}
\end{align}
and the equivalence follows straightforwardly from the arguments in \cite{Assel:2015oxa}.

For $k=1$ $N_f>3$, we show explicitly by $S^3$ partition functions \eqref{k1-vortex} that the vortex loop in the $D$-type quiver is mirror dual to the Wilson loop in the $\sprm(1)$ SQCD.

\paragraph{Orthosymplectic mirror quiver.}
Turning to the brane configuration of a $\sprm(k)$ gauge theory with a fundamental Wilson line via an O3 plane
\begin{align}
\raisebox{-.5\height}{
\begin{tikzpicture}
    \draw[dashed] (0,-1)--(0,1) (6,-1)--(6,1);
    \draw (0,0.075)--(6,0.075) (0,-0.075)--(6,-0.075);
    \draw[dotted,thick] (-1,0)--(7,0);
    \dfive{1,0}
    \dfive{1.5,0}
    \dfive{2,0}
    \dfive{2.5,0}
    \dfive{3,0}
    \dfive{3.5,0}
    \dfive{4,0}
    \dfive{4.5,0}
    \dfive{5,0}
    %
    \dfive{1.75,1}
    \draw [red,line join=round,decorate, decoration={zigzag, segment length=4,amplitude=.9,post=lineto,post length=2pt}] (1.75,1) -- (1.75,0);
    \draw[decoration={brace,mirror,raise=10pt},decorate,thick]
  (0.75,-0.25) -- node[below=10pt] {\footnotesize{$2N_f$ half D5}} (5.25,-0.25);
  \node at (1,0.5) {\footnotesize{$2k$ D3}};
  \node at (-0.35,-0.75) {\tiny{O$3^-$}};
  \node at (0.35,-0.75) {\tiny{O$3^+$}};
  \node at (6.35,-0.75) {\tiny{O$3^-$}};
  \node at (5.65,-0.75) {\tiny{O$3^+$}};
\end{tikzpicture}
}
\qquad \longrightarrow \qquad
\raisebox{-.5\height}{
    \begin{tikzpicture}
	\node (g1) [gaugeSp,label=right:{$\sprm(k)$},label=left:{\footnotesize{$\mathcal{W}$}}] {};
	\node (f1) [flavourSO,above of=g1,label=right:{$\sorm(2N_f)$}] {};
	\draw (g1)--(f1);
	\end{tikzpicture}
    }
    \label{eq:Wilson_O3_branes}
\end{align}
and the position of the F1 string with respect to the flavour D5s is not physical. As above, this is because the F1 can move across any of the half D5 branes since the number of D3 branes on the left and right are the same.

Based on the brane configuration \eqref{eq:Wilson_O3_branes}  and assuming $N_f>2k+1$, the mirror configuration is given by
\begin{align}
    \raisebox{-0.5\height}{
    \begin{tikzpicture}
 \draw[dotted,thick] (-0.5,0)--(3*0.75,0) (4*0.75,0)--(7*0.75,0) (8*0.75,0)--(11*0.75,0) (12*0.75,0)--(15*0.75,0) (16*0.75,0)--(19*0.75+0.5,0);
  \node at (3.5*0.75,0) {$\cdots$};
  \node at (7.5*0.75,0) {$\cdots$};
  \node at (11.5*0.75,0) {$\cdots$};
  \node at (15.5*0.75,0) {$\cdots$};
\draw (5.5*0.75,-1)--(5.5*0.75,1) (13.5*0.75,-1)--(13.5*0.75,1);
\foreach \i in {1,-1}
\draw (0.75,0.1*\i)--(3*0.75,0.1*\i) (4*0.75,0.1*\i)--(7*0.75,0.1*\i) (8*0.75,0.1*\i)--(11*0.75,0.1*\i) (12*0.75,0.1*\i)--(15*0.75,0.1*\i) (16*0.75,0.1*\i)--(18*0.75,0.1*\i);
\node at (1.5*0.75,-0.4) {\footnotesize{$1$}};
\node at (2.5*0.75,-0.4) {\footnotesize{$1$}};
\node at (4.5*0.75,-0.4) {\footnotesize{$k$}};
\node at (5.5*0.75-0.2,-0.4) {\footnotesize{$k$}};
\node at (6.5*0.75,-0.4) {\footnotesize{$k$}};
\node at (8.5*0.75,-0.4) {\footnotesize{$k$}};
\node at (9.5*0.75-0.15,-0.4) {\footnotesize{$k$}};
\node at (10.5*0.75,-0.4) {\footnotesize{$k$}};
\node at (12.5*0.75,-0.4) {\footnotesize{$k$}};
\node at (13.5*0.75+0.2,-0.4) {\footnotesize{$k$}};
\node at (14.5*0.75,-0.4) {\footnotesize{$k$}};
\node at (16.5*0.75,-0.4) {\footnotesize{$1$}};
\node at (17.5*0.75,-0.4) {\footnotesize{$1$}};
  \foreach \i in {0,1,...,19}
    \ns{0.75*\i,0};
\draw[red] (9.5*0.75-0.05,0.1)--(9.5*0.75-0.05,1)
 (9.5*0.75+0.05,-0.1)--(9.5*0.75+0.05,-1);
\ns{9.5*0.75,1}
\ns{9.5*0.75,-1}
 \node at (0.5*0.75,-0.85) {\tiny{$\mathrm{O}3^+$}};
 \node at (1.5*0.75,-0.85) {\tiny{$\mathrm{O}3^-$}};
 \node at (2.5*0.75,-0.85) {\tiny{$\mathrm{O}3^+$}};
 \node at (4.5*0.75,-0.85) {\tiny{$\mathrm{O}3^-$}};
 \node at (6.5*0.75,-0.85) {\tiny{$\widetilde{\mathrm{O}3}^-$}};
\node at (8.5*0.75,-0.85) {\tiny{$\widetilde{\mathrm{O}3}^+$}};
\node at (10.5*0.75,-0.85) {\tiny{$\widetilde{\mathrm{O}3}^+$}};
\node at (12.5*0.75,-0.85) {\tiny{$\widetilde{\mathrm{O}3}^-$}};
\node at (14.5*0.75,-0.85) {\tiny{$\mathrm{O}3^-$}};
 \node at (16.5*0.75,-0.85) {\tiny{$\mathrm{O}3^+$}};
 \node at (17.5*0.75,-0.85) {\tiny{$\mathrm{O}3^-$}};
 \node at (18.5*0.75,-0.85) {\tiny{$\mathrm{O}3^+$}};
 \end{tikzpicture}
    }
    \label{eq:branes_vortex_OSp_mirror}
\end{align}
and the D1 is free to move across any of the central half NS5 branes because the numbers of D3 branes on the left and right are equal.
To derive the defect description, the D1 brane in \eqref{eq:branes_vortex_OSp_mirror} can be moved to the nearest NS5 brane on the left such that the mirror theory with vortex defect is realised by
\begin{align}
    \raisebox{-.5\height}{
    \begin{tikzpicture}
	\node (g1) [gaugeSO,label=below:{\footnotesize{$2$}}] {};
	\node (g2) [gaugeSp,right of =g1,label=below:{\footnotesize{$2$}}] {};
	\node (g3) [right of =g2] {$\ldots$};
	\node (g4) [gaugeSO,right of =g3,label=below:{\footnotesize{$2k$}}] {};
	\node (g5) [gaugeSp,right of =g4,label=below:{\footnotesize{$2k$}}] {};
	\node (g6) [right of =g5] {$\ldots$};
	\node (g7) [gaugeSp,right of =g6,label=below:{\footnotesize{$2k$}}] {};
	\node (g8) [gaugeSO,right of =g7,label=below:{\footnotesize{$2k{+}1$}}] {};
	\node (g9) [gaugeSp,right of =g8,label=below:{\footnotesize{$2k$}}] {};
	\node (g10) [right of =g9] {$\ldots$};
	\node (f1) [flavourSO,above of=g5, label=left:{\footnotesize{$1$}}] {};
	\draw (g1)--(g2) (g2)--(g3) (g3)--(g4) (g4)--(g5) (g5)--(g6)
	(g6)--(g7) (g7)--(g8) (g8)--(g9) (g9)--(g10) (g5)--(f1);
	%
    \node (SQM) at (6.5,1) [defect,label=above:{\footnotesize{SQM}}] {};
    \begin{scope}[decoration={markings,mark =at position 0.5 with {\arrow{stealth}}}]
    \draw[postaction={decorate},color=red] (g7) -- (SQM);
    \draw[postaction={decorate},color=red] (SQM) -- (g8);
     \end{scope}
    %
    \node (b2) [gaugeSp,right of =g10,label=below:{\footnotesize{$2k$}}] {};
	\node (b3) [gaugeSO,right of =b2,label=below:{\footnotesize{$2k$}}] {};
	\node (b4) [right of =b3] {$\ldots$};
	\node (b5) [gaugeSp,right of =b4,label=below:{\footnotesize{$2$}}] {};
	\node (b6) [gaugeSO,right of =b5,label=below:{\footnotesize{$2$}}] {};
	\node (f2) [flavourSO,above of=b2, label=right:{\footnotesize{$1$}}] {};
	\draw (g10)--(b2) (b2)--(b3) (b3)--(b4) (b4)--(b5) (b5)--(b6) (b2)--(f2);
	\end{tikzpicture}
    }
    \end{align}
Alternatively, the D1 brane in \eqref{eq:branes_vortex_OSp_mirror} can be moved the nearest NS5 brane on the right-hand side, which implies the following realisation:
\begin{align}
\raisebox{-.5\height}{
    \begin{tikzpicture}
	\node (g1) [gaugeSO,label=below:{\footnotesize{$2$}}] {};
	\node (g2) [gaugeSp,right of =g1,label=below:{\footnotesize{$2$}}] {};
	\node (g3) [right of =g2] {$\ldots$};
	\node (g4) [gaugeSO,right of =g3,label=below:{\footnotesize{$2k$}}] {};
	\node (g5) [gaugeSp,right of =g4,label=below:{\footnotesize{$2k$}}] {};
	\node (g6) [right of =g5] {$\ldots$};
	\node (g7) [gaugeSp,right of =g6,label=below:{\footnotesize{$2k$}}] {};
	\node (g8) [gaugeSO,right of =g7,label=below:{\footnotesize{$2k{+}1$}}] {};
	\node (g9) [gaugeSp,right of =g8,label=below:{\footnotesize{$2k$}}] {};
	\node (g10) [right of =g9] {$\ldots$};
	\node (f1) [flavourSO,above of=g5, label=left:{\footnotesize{$1$}}] {};
	\draw (g1)--(g2) (g2)--(g3) (g3)--(g4) (g4)--(g5) (g5)--(g6)
	(g6)--(g7) (g7)--(g8) (g8)--(g9) (g9)--(g10) (g5)--(f1);
	%
    \node (SQM) at (7.5,1) [defect,label=above:{\footnotesize{SQM}}] {};
    \begin{scope}[decoration={markings,mark =at position 0.5 with {\arrow{stealth}}}]
    \draw[postaction={decorate},color=red] (g8) -- (SQM);
    \draw[postaction={decorate},color=red] (SQM) -- (g9);
     \end{scope}
    %
    \node (b2) [gaugeSp,right of =g10,label=below:{\footnotesize{$2k$}}] {};
	\node (b3) [gaugeSO,right of =b2,label=below:{\footnotesize{$2k$}}] {};
	\node (b4) [right of =b3] {$\ldots$};
	\node (b5) [gaugeSp,right of =b4,label=below:{\footnotesize{$2$}}] {};
	\node (b6) [gaugeSO,right of =b5,label=below:{\footnotesize{$2$}}] {};
	\node (f2) [flavourSO,above of=b2, label=right:{\footnotesize{$1$}}] {};
	\draw (g10)--(b2) (b2)--(b3) (b3)--(b4) (b4)--(b5) (b5)--(b6) (b2)--(f2);
	\end{tikzpicture}
    }
\end{align}
and again, the equivalence of these two defects follows from generalising the hopping duality to brane configurations with O3 planes. Consequently, any vortex defect defined by two consecutive $\sprm(k)$ and $\sorm(2k+1)$ gauge nodes is expected to be the same defect, which is mirror dual to a fundamental Wilson line defect.

Moreover, we find another mirror realisation of the fundamental Wilson line in the $\sprm(k)$ theories, which is given by
\begin{align}
    \raisebox{-.5\height}{
    \begin{tikzpicture}
	\node (g1) [gaugeSO,label=below:{\footnotesize{$2$}}] {};
	\node (g2) [gaugeSp,right of =g1,label=below:{\footnotesize{$2$}}] {};
	\node (g3) [right of =g2] {$\ldots$};
	\node (g4) [gaugeSO,right of =g3,label=below:{\footnotesize{$2k$}}] {};
	\node (g5) [gaugeSp,right of =g4,label=below:{\footnotesize{$2k$}}] {};
	\node (g6) [right of =g5] {$\ldots$};
	\node (g7) [gaugeSp,right of =g6,label=below:{\footnotesize{$2k$}}] {};
	\node (g8) [gaugeSO,right of =g7,label=below:{\footnotesize{$2k{+}1$}}] {};
	\node (g9) [gaugeSp,right of =g8,label=below:{\footnotesize{$2k$}}] {};
	\node (g10) [right of =g9] {$\ldots$};
	\node (f1) [flavourSO,above of=g5, label=right:{\footnotesize{$1$}}] {};
	\draw (g1)--(g2) (g2)--(g3) (g3)--(g4) (g4)--(g5) (g5)--(g6)
	(g6)--(g7) (g7)--(g8) (g8)--(g9) (g9)--(g10) (g5)--(f1);
	%
    \node (SQM) at (3,1) [defect,label=above:{\footnotesize{SQM}}] {};
    \begin{scope}[decoration={markings,mark =at position 0.5 with {\arrow{stealth}}}]
    \draw[postaction={decorate},color=red] (g4) -- (SQM);
    \draw[postaction={decorate},color=red] (f1) -- (SQM);
    \draw[postaction={decorate},color=red] (SQM) -- (g5);
     \end{scope}
    %
    \node (b2) [gaugeSp,right of =g10,label=below:{\footnotesize{$2k$}}] {};
	\node (b3) [gaugeSO,right of =b2,label=below:{\footnotesize{$2k$}}] {};
	\node (b4) [right of =b3] {$\ldots$};
	\node (b5) [gaugeSp,right of =b4,label=below:{\footnotesize{$2$}}] {};
	\node (b6) [gaugeSO,right of =b5,label=below:{\footnotesize{$2$}}] {};
	\node (f2) [flavourSO,above of=b2, label=right:{\footnotesize{$1$}}] {};
	\draw (g10)--(b2) (b2)--(b3) (b3)--(b4) (b4)--(b5) (b5)--(b6) (b2)--(f2);
	\end{tikzpicture}
    }
\end{align}
which follows from moving the D1 in \eqref{eq:branes_vortex_OSp_mirror} to the left-hand side of the interval containing the half D5 flavour brane.

Unfortunately, it is not clear how to verify these predictions by A-twisted index or sphere partition function. A promising tool could be the squashed sphere partition function \cite{Hama:2010av,Hama:2011ea}.

\paragraph{Comments on the balanced case $N_f=2k+1$.}
By an analogous line of reasoning, one arrives at the following conjectural statements: a fundamental Wilson line in $\sprm(k)$ SQCD with $N_f=2k+1$ should be mirror dual to
\begin{align}
    \raisebox{-.5\height}{
    \begin{tikzpicture}
	\node (g1) [gauge,label=below:{\footnotesize{$1$}}] {};
	\node (g2) [gauge,right of =g1,label=below:{\footnotesize{$2$}}] {};
	\node (g3) [right of =g2] {$\ldots$};
	\node (g7) [gauge,right of =g3,label=below:{\footnotesize{${2k{-}2}$}}] {};
	\node (g8) [gauge,right of =g7,label=below:{\footnotesize{$2k{-}1$}}] {};
	\node (g9) [gauge,below right of =g8,label=below:{\footnotesize{$k$}}] {};
	\node (g10) [gauge,above right of =g8,label=above:{\footnotesize{$k$}}] {};
	\node (f1) [flavour,right of=g9, label=below:{\footnotesize{$1$}}] {};
	\node (f2) [flavour,right of=g10, label=above:{\footnotesize{$1$}}] {};
	\draw (g1)--(g2) (g2)--(g3) (g3)--(g7) (g7)--(g8) (g8)--(g9) (g8)--(g10) (g9)--(f1)  (g10)--(f2);
%
    \node (SQM) at (6,0) [defect,label=right:{\footnotesize{SQM}}] {};
    \begin{scope}[decoration={markings,mark =at position 0.5 with {\arrow{stealth}}}]
    \draw[postaction={decorate},color=red] (SQM) -- (g9);
    \draw[postaction={decorate},color=red] (SQM) -- (g10);
     \end{scope}
	\end{tikzpicture}
    }
\end{align}
which is similar in spirit to \eqref{eq:vortex_D-type_Nf>2k+1}. Moreover, the Wilson line expectation value should also be mirror dual to the follow vortex line in the orthosymplectic mirror:
\begin{align}
    \raisebox{-.5\height}{
        \begin{tikzpicture}
	\node (g1) [gaugeSO,label=below:{\footnotesize{$2$}}] {};
	\node (g2) [gaugeSp,right of =g1,label=below:{\footnotesize{$2$}}] {};
	\node (g3) [right of =g2] {$\ldots$};
	\node (g4) [gaugeSO,right of =g3,label=below:{\footnotesize{$2k{-}2$}}] {};
	\node (g5) [gaugeSp,right of =g4,label=below:{\footnotesize{$2k{-}2$}}] {};
	\node (g6) [gaugeSO,right of =g5,label=below:{\footnotesize{$2k$}}] {};
	\node (g7) [gaugeSp,right of =g6,label=below:{\footnotesize{$2k$}}] {};
	\node (g8) [gaugeSO,right of =g7,label=below:{\footnotesize{$2k$}}] {};
	\node (g9) [gaugeSp,right of =g8,label=below:{\footnotesize{$2k{-}2$}}] {};
	\node (g10) [gaugeSO,right of=g9,label=below:{\footnotesize{$2k{-}2$}}] {};
	\node (g11) [right of =g10] {$\ldots$};
	\node (g12) [gaugeSp,right of =g11,label=below:{\footnotesize{$2$}}] {};
	\node (g13) [gaugeSO,right of =g12,label=below:{\footnotesize{$2$}}] {};
	\node (f1) [flavourSO,above left of=g7, label=below:{\footnotesize{$1$}}] {};
	\node (f2) [flavourSO,above right of=g7, label=above:{\footnotesize{$1$}}] {};
	\draw (g1)--(g2) (g2)--(g3) (g3)--(g4) (g4)--(g5) (g5)--(g6)
	(g6)--(g7) (g7)--(g8) (g8)--(g9) (g9)--(g10) (g10)--(g11) (g11)--(g12) (g12)--(g13) (g7)--(f1) (g7)--(f2);
%
    \node (SQM) at (5,2) [defect,label=right:{\footnotesize{SQM}}] {};
    \begin{scope}[decoration={markings,mark =at position 0.5 with {\arrow{stealth}}}]
    \draw[postaction={decorate},color=red] (g6) -- (SQM);
    \draw[postaction={decorate},color=red] (SQM) -- (g7);
    \draw[postaction={decorate},color=red] (f1) -- (SQM);
     \end{scope}
	\end{tikzpicture}
    }
\end{align}
As we explicitly describe the case of $k=1$ in \S\ref{sec:summary-open}, we do not have the precise description of the coupling of the SQM to the 3d theory in both unitary and orthosymplectic cases. This is left for future work.

\section{\texorpdfstring{5d $\sprm(k)$ SQCD: unitary and orthosymplectic magnetic quivers}{5d Sp(k) SQCD: unitary and orthosymplectic magnetic quivers}}
\label{sec:5d_Sp}

Consider 5d $\Ncal=1$ theories which have a low-energy effective description as a 5d $\Ncal=1$ gauge theory with $\sprm(k)$ gauge group and $N_f$ fundamental flavours. However, the number of flavours is constrained by $N_f \leq 2k+5$, which ensures that the low-energy theory admits a UV completion into a 5d $\Ncal=1$ SCFT. Field theory \cite{Intriligator:1997pq} and 5-brane web \cite{Brunner:1997gk} arguments restraint $N_f \leq 2k+4$, while the case $N_f=2k+5$ has been shown to be admissible in \cite{Bergman:2015dpa}.

The Higgs branch of a 5d $\Ncal=1$ theory changes from the IR theory, referred to as finite (gauge) coupling, to the UV fixed point known as infinite coupling since new massless instantons appear. The \emph{magnetic quiver} technique, introduced in \cite{Cremonesi:2015lsa,Ferlito:2017xdq,Cabrera:2018jxt,Bourget:2020gzi}, allows us to analyse the Higgs branches at both finite and infinite coupling.

Focusing in $\sprm(k)$ gauge groups, the same logic as in \S\ref{sec:3d_Sp} applies. Due to two different string theory realisations, we have two distinct types of magnetic quivers:
\begin{compactitem}
\item 5-brane webs in the presence of an O$7^-$ plane yield unitary magnetic quivers \cite{Bourget:2020gzi}.
\item 5-brane webs with an O$5^+$ plane yield (unitary-)orthosymplectic magnetic quivers \cite{Bourget:2020gzi,Akhond:2020vhc}.
\end{compactitem}
The magnetic quivers for finite coupling are the same as the mirror quivers for $\sprm(k)$ in 3d, which have been explained in \S\ref{sec:3d_Sp}. Now, the focus is placed on the magnetic quivers for the Higgs branches at infinite coupling. The relevant unitary and orthosymplectic quivers are given in \cite[Tab.\ 1]{Bourget:2020gzi}.

The unitary magnetic quivers derived from 6d or 5d brane configurations at the strong coupling fixed point are conventionally written as \emph{unframed quivers}, meaning that no explicit flavour node exists. As detailed in \S\ref{sec:lattice}, one overall $\urm(1)$ needs to be removed from the appearing product gauge group in order to render the theory well-defined. Starting from the unitary magnetic quivers in \cite[Tab.\ 1]{Bourget:2020gzi}, the diagonal $\urm(1)$ is removed as follows: firstly, it is convenient to ungauge a $\urm(1)$ gauge node, as it is the simplest ungauging procedure. Secondly, it is advisable to choose a $\urm(1)$ node that is not balanced (if such a node exists). Since the magnetic quivers are composed of good nodes only, the suitable $\urm(1)$ is always over-balanced. The reason for this choice is that the set of balanced nodes provides a Dynkin diagram of the enhanced Coulomb branch symmetry. Hence, it seems ill-advised to sacrifice this pattern. See also Appendix \ref{choiceofU1}.

Turning to orthosymplectic magnetic quivers, these are also unframed if they correspond to Higgs branches of 5d or 6d theories at the strong coupling fixed point. As discussed in \S\ref{sec:lattice}, only a diagonal $\Z_2$ needs to be taken care of, which results in extending the magnetic lattice from a pure integer lattice to \emph{integer lattice + half-integer lattice}. This is particularly relevant for the superconformal index, the A-twisted index, and the Coulomb branch Hilbert series.

\subsection{Duality of unitary and orthosymplectic magnetic quivers}
\label{sec:UOSp-5d}
It is observed in \cite{Bourget:2020xdz} that the 3d $\Ncal=4$ unitary and orthosymplectic magnetic quivers have not only the same Coulomb branch, but also the same Higgs branch. Namely, the Hilbert series of the Higgs and Coulomb branches agree for a pair of the magnetic quivers. Nevertheless, it remains unsettled to see whether they are dual as quantum field theory. In this section, we show solid evidence by exact results that they flow to the same IR fixed point.

\paragraph{$E_4$ quiver.}
To begin with, consider $k=1$ and $N_f=3$. The $E_4\cong \mathfrak{su}(5)$ quiver is given by the Dynkin diagram of $\mathfrak{su}(5)$:
\begin{equation}\label{A4}
\raisebox{-.5\height}{
    \begin{tikzpicture}
	\begin{pgfonlayer}{nodelayer}
		\node [style=gauge3] (0) at (0, 0) {};
		\node [style=gauge3] (1) at (-2, 0) {};
		\node [style=gauge3] (3) at (1, 0) {};
		\node [style=gauge3] (4) at (-1, 0) {};
		\node [style=flavour2] (5) at (-2, 1.25) {};
		\node [style=flavour2] (6) at (1, 1.25) {};
		\node [style=none] (7) at (-2, 1.75) {1};
		\node [style=none] (8) at (1, 1.75) {1};
		\node [style=none] (9) at (-2, -0.5) {1};
		\node [style=none] (10) at (-1, -0.5) {1};
		\node [style=none] (11) at (0, -0.5) {1};
		\node [style=none] (12) at (1, -0.5) {1};
	\end{pgfonlayer}
	\begin{pgfonlayer}{edgelayer}
		\draw (5) to (1);
		\draw (1) to (4);
		\draw (4) to (0);
		\draw (0) to (3);
		\draw (3) to (6);
	\end{pgfonlayer}
\end{tikzpicture}
\begin{tikzpicture}
\begin{pgfonlayer}{nodelayer}
		\node [style=redgauge] (3) at (-2, 0) {};
		\node [style=bluegauge] (9) at (-1, 0) {};
		\node [style=redgauge] (16) at (0, 0) {};
		\node [style=none] (29) at (-1, -0.5) {2};
		\node [style=none] (35) at (0, -0.5) {2};
		\node [style=none] (39) at (-2, -0.5) {2};
		\node [style=gauge3] (41) at (-1, 1) {};
 		\node [style=none] (43) at (-0.5, 1) {1};
		\node [style=none] (44) at (-0.5, 2.3) {$1$};
		\node [style=flavor2] (48) at (-1, 2.3) {};
	\end{pgfonlayer}
	\begin{pgfonlayer}{edgelayer}
		\draw [<->] (-4.5,.5) -- (-2.5,0.5);
    \node at (-4.5,.5) {$\ $};
		\draw (3) to (9);
		\draw (9) to (41);
		\draw (9) to (16);
		\draw [line join=round,decorate, decoration={zigzag, segment length=4,amplitude=.9,post=lineto,post length=2pt}]  (48) -- (41);
	\end{pgfonlayer}
	\end{tikzpicture}
	}
\end{equation}
The wiggly line denotes a charge 2 hypermultiplet (under the $\urm(1)$ gauge node). The definition of the index for the framed unitary quiver is standard; in contrast, the index for the orthosymplectic quiver requires a careful consideration of the magnetic lattice as emphasised in \S\ref{sec:lattice}. After these preliminary remarks, a straightforward perturbative computation shows that both quivers have the same  superconformal indices
\begin{align}
    \mathbb{I}=1&+\sqrt{\mathfrak{q}}\left(\frac{24}{\mathfrak{t}^{2}}+\mathfrak{t}^{2}\right)+\mathfrak{q}\left(-26+\frac{200}{\mathfrak{t}^{4}}+\mathfrak{t}^{4}\right)+2 \mathfrak{q}^{\frac54} \mathfrak{t}^{5}+\mathfrak{q}^{\frac32}\left(\frac{1000}{\mathfrak{t}^{6}}-\frac{451}{\mathfrak{t}^{2}}+\mathfrak{t}^{6}\right)+\mathfrak{q}^{\frac74}\left(-2 \mathfrak{t}^{3}+2 \mathfrak{t}^{7}\right) \notag \\
    &+\mathfrak{q}^{2}\left(373+\frac{3675}{\mathfrak{t}^{8}}-\frac{2824}{\mathfrak{t}^{4}}+\mathfrak{t}^{8}\right)+\ldots
\end{align}
up to order $\mathfrak{q}^2$.
\paragraph{$E_5$ quiver.}
In the case of $N_f=4$, the unitary quiver is the same as the $k=1$, $N_f=5$ case discussed in \S\ref{sec:3d_Sp}, whereas the orthosymplectic quiver is different. Again, to evaluate the superconformal index of the unframed orthosymplectic, we need to adjust the magnetic lattice.
It is then straightforward but tedious to verify that the superconformal index of the orthosymplectic quiver below agrees with the $k=1$, $N_f=5$ case of Table \ref{tab:SCI_Sp_mirrors} up to $\Ocal(\mathfrak{q}^{3/2})$.
\begin{equation}
\raisebox{-.5\height}{
\begin{tikzpicture}
    \node[flavor2,label=left:{$1$}] (1) at (-1,1) {};
    \node[gauge3,label=left:{$1$}] (2) at (-1,-1) {};
    \node[gauge3,label=right:{$1$}] (3) at (2,1) {};
    \node[gauge3,label=right:{$1$}] (4) at (2,-1) {};
    \node[gauge3,label=below:{$2$}] (5) at (0,0) {};
        \node[gauge3,label=below:{$2$}] (6) at (1,0) {};
    \draw (1)--(5)--(2) (3)--(6)--(4) (5)--(6);
\end{tikzpicture}
\begin{tikzpicture}
	\begin{pgfonlayer}{nodelayer}
		\node [style=bluegauge] (0) at (0, 0) {};
		\node [style=redgauge] (1) at (1.25, 0) {};
		\node [style=bluegauge] (2) at (2.5, 0) {};
		\node [style=redgauge] (3) at (3.75, 0) {};
		\node [style=redgauge] (4) at (-1.25, 0) {};
		\node [style=none] (8) at (0, -0.5) {2};
		\node [style=none] (9) at (1.25, -0.5) {4};
		\node [style=none] (10) at (2.5, -0.5) {2};
		\node [style=none] (11) at (3.75, -0.5) {2};
		\node [style=none] (12) at (-1.25, -0.5) {2};
		\node [style=none] (29) at (1.75, 1.25) {1};
		\node [style=gauge3] (30) at (1.25, 1.25) {};
	\end{pgfonlayer}
	\begin{pgfonlayer}{edgelayer}
	\draw [<->] (-5,.5) -- (-2,0.5);
		\draw (4) to (0);
		\draw (0) to (1);
		\draw (1) to (2);
		\draw (2) to (3);
		\draw (30) to (1);
	\end{pgfonlayer}
\end{tikzpicture}
}
\end{equation}
\paragraph{$E_6$ quiver.}
Consider the unitary quiver whose Coulomb branch is $\overline{\mathcal{O}}^{\mathfrak{e}_6}_{\text{min}}$:
\begin{equation}
 \raisebox{-0.5\height}{
    \begin{tikzpicture}
	\begin{pgfonlayer}{nodelayer}
		\node [style=gauge3] (0) at (0, 0) {};
		\node [style=gauge3] (1) at (-1, 0) {};
		\node [style=gauge3] (2) at (-2, 0) {};
		\node [style=gauge3] (3) at (0, 1) {};
		\node [style=flavour2] (4) at (0, 2) {};
		\node [style=gauge3] (5) at (1, 0) {};
		\node [style=gauge3] (6) at (2, 0) {};
		\node [style=none] (7) at (0, -0.5) {3};
		\node [style=none] (8) at (-1, -0.5) {2};
		\node [style=none] (9) at (-2, -0.5) {1};
		\node [style=none] (10) at (1, -0.5) {2};
		\node [style=none] (11) at (2, -0.5) {1};
		\node [style=none] (12) at (0.5, 1) {2};
		\node [style=none] (13) at (0.5, 2) {1};
	\end{pgfonlayer}
	\begin{pgfonlayer}{edgelayer}
		\draw (4) to (3);
		\draw (3) to (0);
		\draw (0) to (5);
		\draw (5) to (6);
		\draw (2) to (1);
		\draw (1) to (0);
	\end{pgfonlayer}
\end{tikzpicture}
}
\label{E6_simply_laced}
\end{equation}
The unitary-orthosymplectic quiver whose Coulomb branch is the closure of the $E_6$ minimal nilpotent orbit $\overline{\mathcal{O}}^{\mathfrak{e}_6}_{\text{min}}$ takes the following form \cite{Bourget:2020gzi} (see also \cite[\S  A.1.5]{Chacaltana:2012ch} for class $\mathcal{S}$ description):
\begin{eqnarray}
\scalebox{.8}{\raisebox{-.5\height}{
\begin{tikzpicture}
	\begin{pgfonlayer}{nodelayer}
		\node [style=bluegauge] (0) at (0, 0) {};
		\node [style=redgauge] (1) at (1.25, 0) {};
		\node [style=bluegauge] (2) at (2.5, 0) {};
		\node [style=redgauge] (3) at (3.75, 0) {};
		\node [style=redgauge] (4) at (-1.25, 0) {};
		\node [style=redgauge] (5) at (-3.75, 0) {};
		\node [style=bluegauge] (6) at (-2.5, 0) {};
		\node [style=gauge3] (7) at (0, 1) {};
		\node [style=none] (8) at (0, -0.5) {4};
		\node [style=none] (9) at (1.25, -0.5) {4};
		\node [style=none] (10) at (2.5, -0.5) {2};
		\node [style=none] (11) at (3.75, -0.5) {2};
		\node [style=none] (12) at (-1.25, -0.5) {4};
		\node [style=none] (13) at (-2.5, -0.5) {2};
		\node [style=none] (14) at (-3.75, -0.5) {2};
		\node [style=none] (15) at (0, 1.5) {1};
	\end{pgfonlayer}
	\begin{pgfonlayer}{edgelayer}
		\draw (5) to (6);
		\draw (6) to (4);
		\draw (4) to (0);
		\draw (0) to (7);
		\draw (0) to (1);
		\draw (1) to (2);
		\draw (2) to (3);
	\end{pgfonlayer}
\end{tikzpicture}}
} {~}={~}
\scalebox{.8}{\raisebox{-.5\height}{
\begin{tikzpicture}
	\begin{pgfonlayer}{nodelayer}
		\node [style=bluegauge] (0) at (0, 0) {};
		\node [style=redgauge] (1) at (1.25, 0) {};
		\node [style=bluegauge] (2) at (2.5, 0) {};
		\node [style=redgauge] (3) at (3.75, 0) {};
		\node [style=redgauge] (4) at (-1.25, 0) {};
		\node [style=redgauge] (5) at (-3.75, 0) {};
		\node [style=bluegauge] (6) at (-2.5, 0) {};
		\node [style=redgauge] (7) at (0, 1) {};
		\node [style=none] (8) at (0, -0.5) {4};
		\node [style=none] (9) at (1.25, -0.5) {4};
		\node [style=none] (10) at (2.5, -0.5) {2};
		\node [style=none] (11) at (3.75, -0.5) {2};
		\node [style=none] (12) at (-1.25, -0.5) {4};
		\node [style=none] (13) at (-2.5, -0.5) {2};
		\node [style=none] (14) at (-3.75, -0.5) {2};
		\node [style=none] (15) at (0, 1.5) {2};
	\end{pgfonlayer}
	\begin{pgfonlayer}{edgelayer}
		\draw (5) to (6);
		\draw (6) to (4);
		\draw (4) to (0);
		\draw (0) to (7);
		\draw (0) to (1);
		\draw (1) to (2);
		\draw (2) to (3);
	\end{pgfonlayer}
\end{tikzpicture}}
}
\label{quiverE6}
\end{eqnarray}
Since the orthosymplectic quiver is rather large, the perturbative calculation of the superconformal indices of these theories is limited to order $\Ocal(\mathfrak{q})$. Nonetheless, both computations yield the same result
\begin{equation}
\mathbb{I}=1+\frac{78 \sqrt{\mathfrak{q}}}{\mathfrak{t}^{2}}+\mathfrak{q}\left(-79+\frac{2430}{\mathfrak{t}^{4}}\right)+\ldots~.
\end{equation}
\paragraph{$E_7$ quiver.}
We move on to the unitary-orthosymplectic quiver whose Coulomb branch is the closure of the $E_7$ minimal nilpotent orbit $\overline{\mathcal{O}}^{\mathfrak{e}_7}_{\text{min}}$.
\begin{equation}
\raisebox{-.5\height}{
 \begin{tikzpicture}
	\begin{pgfonlayer}{nodelayer}
 \node (g1) [gauge,label=below:{1}] {};
 \node (g2) [gauge,right of=g1,label=below:{2}] {};
 \node (g3) [gauge,right of=g2,label=below:{3}] {};
 \node (g4) [gauge,right of=g3,label=below:{4}] {};
 \node (g5) [gauge,right of=g4,label=below:{3}] {};
 \node (g6) [gauge,right of=g5,label=below:{2}] {};
 \node (g7) [gauge,right of=g6,label=below:{1}] {};
 \node (g8) [gauge,above of=g4,label=above:{2}] {};
 \draw (g1)--(g2) (g2)--(g3) (g3)--(g4) (g4)--(g5) (g5)--(g6) (g6)--(g7) (g4)--(g8);
 	\end{pgfonlayer}
\end{tikzpicture}
 }
\scalebox{.8}{\raisebox{-.5\height}{
\begin{tikzpicture}
	\begin{pgfonlayer}{nodelayer}
		\node [style=bluegauge] (0) at (0, 0) {};
		\node [style=redgauge] (1) at (1.25, 0) {};
		\node [style=bluegauge] (2) at (2.5, 0) {};
		\node [style=redgauge] (3) at (3.75, 0) {};
		\node [style=redgauge] (4) at (-1.25, 0) {};
		\node [style=redgauge] (5) at (-3.75, 0) {};
		\node [style=bluegauge] (6) at (-2.5, 0) {};
		\node [style=none] (8) at (0, -0.5) {4};
		\node [style=none] (9) at (1.25, -0.5) {6};
		\node [style=none] (10) at (2.5, -0.5) {4};
		\node [style=none] (11) at (3.75, -0.5) {4};
		\node [style=none] (12) at (-1.25, -0.5) {4};
		\node [style=none] (13) at (-2.5, -0.5) {2};
		\node [style=none] (14) at (-3.75, -0.5) {2};
		\node [style=bluegauge] (23) at (5, 0) {};
		\node [style=redgauge] (24) at (6.25, 0) {};
		\node [style=none] (25) at (5, -0.5) {2};
		\node [style=none] (26) at (6.25, -0.5) {2};
		\node [style=bluegauge,label=right:{2}] (27) at (1.25, 1.25) {};
		\node [style=gauge3,label=right:{1}] (28) at (1.25, 2.5) {};
	\end{pgfonlayer}
	\begin{pgfonlayer}{edgelayer}
		\draw (5) to (6);
		\draw (6) to (4);
		\draw (4) to (0);
		\draw (0) to (1);
		\draw (1) to (2);
		\draw (2) to (3);
		\draw (3) to (23);
		\draw (23) to (24);
		\draw (28) to (27);
		\draw (27) to (1);
	\end{pgfonlayer}
  	\draw [<->] (-5.2,.5) -- (-4.2,0.5);
        \node at (-4,.5) {$\ $};
\end{tikzpicture}
}}
\label{quiverE7}
\end{equation}
The unitary quiver is of Dynkin type, while the orthosymplectic quiver has been derived in \cite{Bourget:2020gzi} (see also \cite[\S 3.2.2]{Chacaltana:2011ze} for the class $\mathcal{S}$ viewpoint).
Here we present the unframed unitary quiver. Due to the large rank of the gauge group, it is computationally challenging to evaluate the superconformal indices of these theories. Despite the obstacles, the sphere partition function is insightful as we can utilise the equivalence of the $S^3$ partition function on the level of the different legs in the two star-shaped quivers \eqref{quiverE7}.
For this computation, we ungauge $\urm(1)$ from the middle gauge group $\urm(4)$ in the unitary quiver. Since the $S^3$ partition function is insensitive to the magnetic lattice, we can compute it by naively decoupling the $\urm(1)$ factor from the middle gauge node to $\surm(4)$. (See \S\ref{sec:lattice}.)

Both the quivers in \eqref{quiverE7} have three tails, and two of them are $T[\surm(4)] \simeq T[\sorm(6)]$  \cite{Gaiotto:2008ak}. The $S^3$ partition functions of the $T[G]$ theory is conjectured in \eqref{eq:TG-S3}, and in particular, for $T[\surm(4)] \simeq T[\sorm(6)]$, explicit computation yields
\begin{align}\label{eq:tail1}
  Z^{S^3}_{T[\surm(4)]}=Z^{S^3}_{T[\sorm(6)]}= \frac{1}{12} \prod_{1\le i<j\le 3}\frac{(m_i\pm m_j)}{\sh(m_i\pm m_j)}~.
\end{align}
Taking residues, we calculate the $S^3$ partition function of the other tail as
\begin{equation}\label{eq:tail2}
  Z^{S^3}\left[ \scalebox{.8}{\raisebox{-.2\height}{
\begin{tikzpicture}
	\begin{pgfonlayer}{nodelayer}
		\node [style=flavour,label=above:{$\surm(4)$}] (2) at (1.25,0) {};
		\node [style=gauge3,label=above:{2}] (3) at (0,0) {};
	\end{pgfonlayer}
	\begin{pgfonlayer}{edgelayer}
		\draw (3) to (2);
	\end{pgfonlayer}
\end{tikzpicture}}}\ \right]=  Z^{S^3}\left[ \scalebox{.8}{\raisebox{-.2\height}{
\begin{tikzpicture}
	\begin{pgfonlayer}{nodelayer}
		\node [style=flavorRed,label=above:{6}] (1) at (2.5,0) {};
		\node [style=bluegauge,label=above:{2}] (2) at (1.25,0) {};
		\node [style=gauge3,label=above:{1}] (3) at (0,0) {};
	\end{pgfonlayer}
	\begin{pgfonlayer}{edgelayer}
		\draw (3) to (2);
		\draw (2) to (1);
	\end{pgfonlayer}
\end{tikzpicture}}}\ \right]=\sum_{i=1}^3\frac{m_i^2}{\prod_{j\neq i}\sh(m_j\pm m_i)}~.
\end{equation}
Note that the $S^3$ partition function of $\urm(2)$ SQCD with $N_f=4$ flavours can be read off from \eqref{eq:U-SQCD}. To obtain \eqref{eq:tail2}, we impose $m_4=m_1+m_2+m_3$ on \eqref{eq:U-SQCD} for the $\surm(4)$ flavour symmetry.
 Therefore, the equalities of \eqref{eq:tail1} and \eqref{eq:tail2} imply that the $S^3$ partition functions of the two quivers \eqref{quiverE7} agree.

\subsection{Wilson lines and unframed quivers}
\label{sec:Wilson_unframed}
As a next step, we include Wilson lines defects into the unitary and orthosymplectic magnetic quivers.  As found in \cite{Bourget:2020xdz,Closset:2020afy,Closset:2020scj}, some unframed magnetic quivers are endowed with non-trivial 1-form symmetries which dictate the spectra of admissible line defects. Here, we therefore study the spectra of admissible Wilson lines in the unitary and orthosymplectic magnetic quivers from the viewpoint of 1-form symmetries and ungauging schemes.

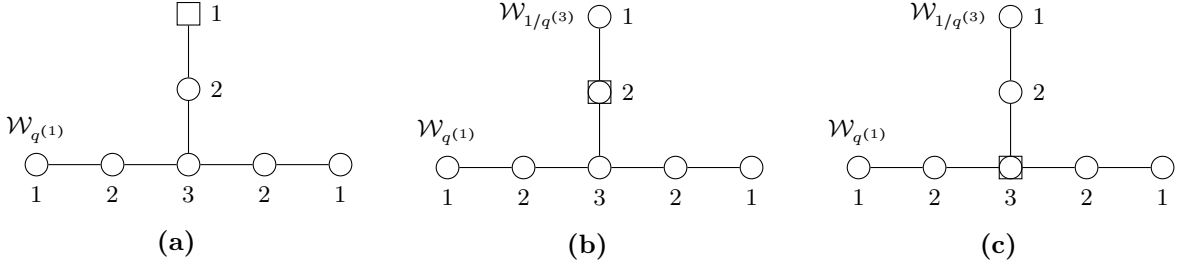
\begin{figure}[t]
    \centering
    \begin{subfigure}{0.32\textwidth}
    \centering
    \raisebox{-.5\height}{
     \begin{tikzpicture}
        \node (g1) [gauge,label=below:{\footnotesize{$1$}},label=above:{\footnotesize{$\mathcal{W}_{q^{(1)}}$}}] {};
        \node (g2) [gauge, right of =g1,label=below:{\footnotesize{$2$}}] {};
        \node (g3) [gauge, right of =g2,label=below:{\footnotesize{$3$}}] {};
        \node (g4) [gauge, right of =g3,label=below:{\footnotesize{$2$}}] {};
        \node (g5) [gauge, right of =g4,label=below:{\footnotesize{$1$}}] {};
        \node (h2) [gauge, above of=g3, label=right:{\footnotesize{$2$}}] {};
        \node (h1) [flavour, above of=h2, label=right:{\footnotesize{$1$}}] {};
        \draw (g1)--(g2) (g2)--(g3) (g3)--(g4) (g4)--(g5) (g3)--(h2) (h2)--(h1);
    \end{tikzpicture}
    }
    \caption{}
    \label{fig:Wilson_ex_Z1}
    \end{subfigure}
    \begin{subfigure}{0.32\textwidth}
    \centering
    \raisebox{-.5\height}{
     \begin{tikzpicture}
        \node (g1) [gauge,label=below:{\footnotesize{$1$}},label=above:{\footnotesize{$\mathcal{W}_{q^{(1)}}$}}] {};
        \node (g2) [gauge, right of =g1,label=below:{\footnotesize{$2$}}] {};
        \node (g3) [gauge, right of =g2,label=below:{\footnotesize{$3$}}] {};
        \node (g4) [gauge, right of =g3,label=below:{\footnotesize{$2$}}] {};
        \node (g5) [gauge, right of =g4,label=below:{\footnotesize{$1$}}] {};
        \node [flavour, above of =g3] {};
        \node (h2) [gauge, above of=g3, label=right:{\footnotesize{$2$}}] {};
        \node (h1) [gauge, above of=h2, label=right:{\footnotesize{$1$}},label=left:{\footnotesize{$\mathcal{W}_{1/q^{(3)}}$}}] {};
        \draw (g1)--(g2) (g2)--(g3) (g3)--(g4) (g4)--(g5) (g3)--(h2) (h2)--(h1);
    \end{tikzpicture}
    }
    \caption{}
    \label{fig:Wilson_ex_Z2}
    \end{subfigure}
    \begin{subfigure}{0.32\textwidth}
    \centering
    \raisebox{-.5\height}{
     \begin{tikzpicture}
        \node (g1) [gauge,label=below:{\footnotesize{$1$}},label=above:{\footnotesize{$\mathcal{W}_{q^{(1)}}$}}] {};
        \node (g2) [gauge, right of =g1,label=below:{\footnotesize{$2$}}] {};
        \node [flavour, right of =g2] {};
        \node (g3) [gauge, right of =g2,label=below:{\footnotesize{$3$}}] {};
        \node (g4) [gauge, right of =g3,label=below:{\footnotesize{$2$}}] {};
        \node (g5) [gauge, right of =g4,label=below:{\footnotesize{$1$}}] {};
        \node (h2) [gauge, above of=g3, label=right:{\footnotesize{$2$}}] {};
        \node (h1) [gauge, above of=h2, label=right:{\footnotesize{$1$}},label=left:{\footnotesize{$\mathcal{W}_{1/q^{(3)}}$}}] {};
        \draw (g1)--(g2) (g2)--(g3) (g3)--(g4) (g4)--(g5) (g3)--(h2) (h2)--(h1);
    \end{tikzpicture}
    }
    \caption{}
    \label{fig:Wilson_ex_Z3}
    \end{subfigure}
    \caption{Affine $E_6$ Dynkin quiver and different choices of ungauging. Here the superposition of a circle and a  square, called squircle, represents the node where the diagonal U(1) is ungauged. The Wilson lines defined in each theory have identical B-twisted indices, see Table \ref{E6WilsonHSk1}.}
    \label{fig:Wilson_ex}
\end{figure}

\paragraph{Admissible Wilson lines for unframed unitary quivers.}
For unframed unitary magnetic quivers, the quotient of the product gauge group $\prod_i \urm(n_i)$ by the continuous $\mathrm{ker}\ \phi \cong \urm(1)$ group can be realised in various ways, see \S\ref{sec:lattice}. Although the resulting theories have been argued to give rise to the same moduli spaces, the allowed line defects require a careful treatment:
\begin{compactitem}
\item If $n_i =1$ holds for some $i$, and the gauge node is over-balanced, we can simply ungauge the group and obtain $G\slash \mathrm{ker}\ \phi \cong \prod_{j\neq i} \urm(n_j)$. Notably, $G^\prime \equiv G\slash \mathrm{ker}\ \phi$ has no trivially acting subgroup anymore, due to the appearance of the $\urm(n_i) =\urm(1)$ flavour node. The allowed representations for Wilson lines lie in the weight lattice of the gauge group, \textit{i.e.}
\begin{align}
    \mathcal{W} \in \Lambda_w^{G^\prime} = \bigoplus_{j\neq i} \Z^{n_j} \,.
\end{align}
In other words, the Wilson line can transform in any product representation composed of the $\urm(n_j)|_{j\neq i}$ representations.
\item We could also choose to remove $\ker \ \phi$ from any $\urm(n_i)$ node, with $n_i \geq 2$. Naively, the gauge group becomes $G^{\mathrm{naive}} = \surm(n_i) \times \prod_{j\neq i} \urm(n_j)$. In contrast to the case above, there still exists a non-trivial $\Z_{n_i} = Z(\surm(n_i))$, which acts trivial on the matter content and is embedded via $\Z_{n_i} \subset \urm(1) = Z(\urm(n_j))$ for $j\neq i$. This $\Z_{n_i}^{\mathrm{diag}}$ groups defines an electric 1-form symmetry for the $G^{\mathrm{naive}}$ gauge theory \cite{Gaiotto:2014kfa}. However, as argued in \cite{Bourget:2020xdz}, the gauge group, which seems to be preferred from the brane system, is given by the quotient $G^{\prime} = G^{\mathrm{naive}} \slash \Z_{n_i}^{\mathrm{diag}}$. As a result, the allowed Wilson line representations $\mathcal{R}$ are any product representation of the gauge group factors $\surm(n_i) \times \prod_{j\neq i} \urm(n_j)$, such that the $\mathcal{R}$ is $\Z_{n_i}^{\mathrm{diag}}$ invariant, \textit{i.e.} the charge under the diagonal $\urm(1)$ has to be $0$ mod $n_i$.
\end{compactitem}
To demonstrate, consider the affine $E_6$ unitary quiver
\begin{align}
\raisebox{-.5\height}{
    \begin{tikzpicture}
        \node (g1) [gauge,label=below:{\footnotesize{$1$}},label=above:{\footnotesize{$q^{(1)}$}}] {};
        \node (g2) [gauge, right of =g1,label=below:{\footnotesize{$2$}},label=above:{\footnotesize{$a_i^{(1)}$}}] {};
        \node (g3) [gauge, right of =g2,label=below:{\footnotesize{$3$}}] {};
        \node (g4) [gauge, right of =g3,label=below:{\footnotesize{$2$}},label=above:{\footnotesize{$a_i^{(2)}$}}] {};
        \node (g5) [gauge, right of =g4,label=below:{\footnotesize{$1$}},label=above:{\footnotesize{$q^{(2)}$}}] {};
        \node (h2) [gauge, above of=g3, label=right:{\footnotesize{$2$}},label=left:{\footnotesize{$a_i^{(3)}$}}] {};
        \node (h1) [gauge, above of=h2, label=right:{\footnotesize{$1$}},label=left:{\footnotesize{$q^{(3)}$}}] {};
        \draw (g1)--(g2) (g2)--(g3) (g3)--(g4) (g4)--(g5) (g3)--(h2) (h2)--(h1);
    \end{tikzpicture}
    }
\end{align}
such that the $\urm(1)_I$ charges are $q^{(I)}$, the $\urm(2)_I$ fugacities are $\{a_i^{(I)}\}_{i=1,2}$, and the $\urm(3)$ fugacities are $\{b_j\}_{j=1,2,3}$.
Starting from ungauging the diagonal $\urm(1)$ at, say, the $\urm(1)_3$ node. Then we can without doubt consider the following Wilson line:
\begin{align}
\mathcal{R} = q^{(1)} \otimes \prod_I [0,0]_{a_i^{(I)}} \otimes  [0,0,0]_{b_j}
\label{eq:Wilson_ex_Z1}
\end{align}
see also Figure \ref{fig:Wilson_ex_Z1}.
This Wilson line is non-trivial.

Next, we ungauge the diagonal $\urm(1)$ at, say, the $\urm(2)_3$ node. Hence, the admissible Wilson lines need to be invariant under $\Z_2^{\mathrm{diag}}$. We may ask what happens to the Wilson line \eqref{eq:Wilson_ex_Z1} in this frame. Given the representation $\mathcal{R}$, we need to implement the $\Z_2$ invariance.  Using the additional $q^{(3)}$ charge, the candidate Wilson line is given by
\begin{align}
\begin{aligned}
\mathcal{R}^{\prime} &= q^{(1)} \otimes \left(q^{(3)} \right)^x \otimes \prod_{I\neq 3} [0,0]_{a_i^{(I)}} \otimes [0]_{\surm(2)_{(3)}} \otimes  [0,0,0]_{b_j}   \\
&\Z_2-\text{invariance} \qquad \Rightarrow \qquad 1+x = 0 \; \mathrm{mod} \; 2
    \end{aligned}
    \label{eq:Wilson_ex_Z2}
\end{align}
see Figure \ref{fig:Wilson_ex_Z2}.

Likewise, ungauging at the central $\urm(3)$ node leads to Wilson line configuration that needs to be $\Z_3$ invariant. Utilising the $q^{(3)}$ charge, the candidate Wilson line is given by
\begin{align}
\begin{aligned}
\mathcal{R}^{\prime \prime} &= q^{(1)} \otimes \left(q^{(3)} \right)^x \otimes \prod_I [0,0]_{a_i^{(I)}} \otimes  [0,0]_{\surm(3)}  \\
    &\Z_3-\text{invariance} \qquad \Rightarrow \qquad 1+x = 0 \; \mathrm{mod} \; 3
\end{aligned}
    \label{eq:Wilson_ex_Z3}
\end{align}
see Figure \ref{fig:Wilson_ex_Z3}.

Demanding that the simplest solution satisfies both \eqref{eq:Wilson_ex_Z2} and \eqref{eq:Wilson_ex_Z3}, and is moreover compatible with \eqref{eq:Wilson_ex_Z1}, we find $x=-1$. Explicit computations for all Wilson lines in Figure \ref{fig:Wilson_ex} show that the B-twisted indices agree, see Table \ref{E6WilsonHSk1} for the result.

\paragraph{Admissible Wilson lines for unframed orthosymplectic quivers.}
For unframed orthosymplectic magnetic quivers, the removal of the diagonal $\mathrm{ker}\ \phi =\Z_2$ modifies not only the magnetic lattice, but also the weight lattice of the gauge group, see \eqref{eq:lattice_example} for an explicit example. An immediate consequence is that a Wilson line transforming in the fundamental representation of either a single $\sorm(2n)$ node or single $\sprm(k)$ node is not an admissible line operator, as these representations are outside the weight lattice. There are two possible ways out if we aim for Wilson lines in ``simple" representations:
\begin{compactitem}
\item A Wilson line transforming in the \emph{product representation} of two fundamental representations of two distinct gauge nodes. This corresponds to the $ \Z_{\sum|\mathrm{odd}}^{n} \oplus \Z_{\sum|\mathrm{odd}}^{k}$ factor in \eqref{eq:lattice_example}.
\item A Wilson line that transforms in a \emph{representation on the root lattice} of a single gauge node. This corresponds to the $ \Z_{\sum|\mathrm{even}}^{n} \oplus \Z_{\sum|\mathrm{even}}^{k} $ factor in \eqref{eq:lattice_example}, where each lattice is a root lattice of either $\sorm(2n)$ or $\sprm(k)$.
\end{compactitem}
\subsection{General observations and conjectures}
\label{sec:Wilson_unframed_MagQuiv}

Given the spectra of admissible Wilson lines analysed above, our next goal is to establish a pattern between the admissible Wilson lines of the different magnetic quiver constructions. The motivation comes from the pattern observed in the two different mirror quivers of 3d $\sprm(k)$ of \S\ref{sec:3d_Sp}.

The exceptional families \cite{Bourget:2020gzi} fall into two groups:
\begin{itemize}
    \item The $E_{8,7,6,5}$ families are labelled by the rank $k$ of the 5d $\sprm(k)$ gauge group.
    \item The $E_{4-2l}$ and $E_{3-2l}$ families depend on the rank $k$ as well as another integer $0\leq l\leq k$.
\end{itemize}
Almost all orthosymplectic magnetic quivers for the exceptional $E_n$ families contain a single $\urm(1)$, originating from 5-branes unaffected by the orientifold projection. The only exception is the $E_8$ family. Given this $\urm(1)$, it is particularly convenient to study Wilson lines in the product representation $q\otimes [1,0,\ldots,0]_{C/D}$ of the charge $1$ $\urm(1)$ representation and a fundamental representation of a single $\sorm(2n)$ or $\sprm(k)$ gauge node. This is because the $\urm(1)$ representation does not increase the dimensionality of the representation.

Explicit computations of B-twisted indices with Wilson loops have been performed for $E_{7,6,5}$ with $k=1$ -- see Tables \ref{Sp1Nf5WilsonHS}, \ref{E6WilsonHSk1}, \ref{E7WilsonHSk1} -- as well as  $E_4$ with $k=1,2$ and $E_3$ with $k=1,2,3$ -- see Tables \ref{E3WilsonHSk1}, \ref{E3WilsonHSk2}, \ref{E3WilsonHSk3}, \ref{E4WilsonHSk1}. The results obtained indicate a systematic pattern between the Wilson lines in the orthosymplectic quiver and Wilson lines in the unitary quiver\footnote{The unitary magnetic quivers of $E_n$ families obtained from brane webs are unframed. Therefore, an overall $\urm(1)$ needs to be ungauged which corresponds to fixing the centre of mass of the brane system. The choice of ungauging in this section is explained in Appendix \ref{choiceofU1}. }. Based on the computations and \emph{assuming regularity of the behaviour}, the pattern can be turned into the following conjectures:

\paragraph{$E_7$ family.}
\begin{itemize}
    \item For a Wilson line transforming in the product representation $q\otimes [1,0,\ldots,0]_{C_r}$ of the $\urm(1)$ node and an $\sprm(r)$ node for $1\leq r \leq k+1$
\begin{align}
    &\raisebox{-.5\height}{

    }
\end{align}
\item For a Wilson line transforming in the product representation $q\otimes [1,0,\ldots,0]_{D_r}$ of the $\urm(1)$ node and an $\sorm(2r)$ node for $1\leq r \leq k+2$
\begin{align}
    &\raisebox{-.5\height}{
    %
    } \notag
    \end{align}
    if $r=1$, there is only one Wilson line in the unitary quiver.
    \item For a Wilson line transforming in the product representation $q\otimes [1,0,\ldots,0]_{D_{k+2}}$ of the $\urm(1)$ node and the central $\sorm(2k+4)$ node
    \begin{align}
    &\raisebox{-.5\height}{
    %
    } \notag
    \end{align}
\end{itemize}
\paragraph{$E_6$ family.}
\begin{itemize}
    \item For a Wilson line transforming in the product representation $q\otimes [1,0,\ldots,0]_{C_r}$ of the $\urm(1)$ node and an $\sprm(r)$ node for $1\leq r \leq k+1$
\begin{align}
    &\raisebox{-.5\height}{
    %
    }
\end{align}
\item For a Wilson line transforming in the product representation $q\otimes [1,0,\ldots,0]_{D_r}$ of the $\urm(1)$ node and an $\sorm(2r)$ node for $1\leq r \leq k+3$
\begin{align}
    &\raisebox{-.5\height}{
    %
    } \notag
    \end{align}
    if $r=1$, there is only one Wilson line in the unitary quiver.
    \item For a Wilson line transforming in the product representation $q\otimes [1,0,\ldots,0]_{C_{k+1}}$ of the $\urm(1)$ node and the central $\sprm(k+1)$ node
\begin{align}
    &\raisebox{-.5\height}{
    %
    } \notag
    \end{align}
\end{itemize}

\paragraph{$E_5$ family.}
\begin{itemize}
    \item For a Wilson line transforming in the product representation $q\otimes [1,0,\ldots,0]_{C_r}$ of the $\urm(1)$ node and an $\sprm(r)$ node for $1\leq r \leq k$
\begin{align}
    &\raisebox{-.5\height}{
    %
    }
\end{align}
\item For a Wilson line transforming in the product representation $q\otimes [1,0,\ldots,0]_{D_r}$ of the $\urm(1)$ node and an $\sorm(2r)$ node for $1\leq r \leq k$
\begin{align}
    &\raisebox{-.5\height}{
    %
    } \notag
\end{align}
\item For a Wilson line transforming in the product representation $q\otimes [1,0,\ldots,0]_{D_{k+1}}$ of the $\urm(1)$ node and the central $\sorm(2k+2)$ node
\begin{align}
    &\raisebox{-.5\height}{
    %
    } \notag
\end{align}
\end{itemize}

\paragraph{$E_4$ family.}
\begin{itemize}
    \item For a Wilson line transforming in the product representation $q\otimes [1,0,\ldots,0]_{C_r}$ of the $\urm(1)$ node and an $\sprm(r)$ node for $1\leq r \leq k-l$
    \begin{align}
    &\raisebox{-.5\height}{
    %
    }
\end{align}
\item For a Wilson line transforming in the product representation $q\otimes [1,0,\ldots,0]_{D_r}$ of the $\urm(1)$ node and an $\sorm(2r)$ node for $1\leq r \leq k-l$
\begin{align}
    &\raisebox{-.5\height}{
    %
    } \notag
\end{align}
\item For a Wilson line transforming in the product representation $q\otimes [1,0,\ldots,0]_{C_{k-l}}$ of the $\urm(1)$ node and the central $\sprm(k-l)$ node
\begin{align}
    &\raisebox{-.5\height}{
    %
    } \notag
\end{align}
\end{itemize}

\paragraph{$E_3$ family.}
\begin{itemize}
    \item For a Wilson line transforming in the product representation $q\otimes [1,0,\ldots,0]_{C_r}$ of the $\urm(1)$ node and an $\sprm(r)$ node, for $1\leq r \leq k-l-1$
    \begin{align}
    &\raisebox{-.5\height}{
    %
    }
\end{align}
    \item For a Wilson line transforming in the product representation $q\otimes [1,0,\ldots,0]_{D_r}$ of the $\urm(1)$ node and an $\sorm(r)$ node, for $1\leq r \leq k-l-1$
    \begin{align}
    &\raisebox{-.5\height}{
    %
    } \notag
\end{align}
\item For a Wilson line transforming in the product representation $q\otimes [1,0,\ldots,0]_{D_{k-l}}$ of the $\urm(1)$ node and the central $\sorm(2k-2l)$ node
\begin{align}
    &\raisebox{-.5\height}{
    %
    } \notag
\end{align}
\end{itemize}

\paragraph{$E_8$ family and 6d $\sprm(k)$ SQCD.}
One prominent family has not been presented yet: the $E_8$ family of \cite{Bourget:2020gzi}, \textit{i.e.}\ the infinite coupling limit of 5d $\Ncal=1$ $\sprm(k)$ with the maximally allowed number of flavours $N_f = 2k+5$. This family faces computational difficulties due to the large rank of the gauge group such that we are unable to evaluate the partition functions in the presence of defects. Moreover, the structure of the orthosymplectic quiver, with the absence of a $\urm(1)$ node, even prevents us from formulating a clear prediction for matching Wilson lines.

Similarly, one could consider 6d $\Ncal=(1,0)$ $\sprm(k)$ SQCD with $N_f=2k+8$ flavours, which has a unitary magnetic quiver \cite{Cabrera:2019izd} as well as an orthosymplectic magnetic quiver \cite{Cabrera:2019dob}. Again, the analysis of the partition functions and line defects is obstructed for the same reasons as the 5d $E_8$ family.

\paragraph{Wilson line transforming in other representations.}
For unframed orthosymplectic quivers, a Wilson line transforming in the fundamental representation of a single gauge group is not invariant under the $\mathbb{Z}_2$ 1-form symmetry. On the other hand, other representations such as charge-2 under the $\urm(1)$ or the symmetric representation $[2]_{C_1}$ of $\sprm(1)$ are $\mathbb{Z}_2$-invariant. Here, we demonstrate adding Wilson lines transforming under charge-2 and symmetric representations for the $E_4$ orthosymplectic quiver and match them with Wilson lines on the unitary counterpart in the $E_4$ quivers.
Firstly, we chose the charge-2 Wilson line at $\sorm(2)$ with character $q^2$:
\begin{subequations}
\begin{equation}
\scalebox{0.8}{   \raisebox{-.5\height}{ \begin{tikzpicture}
	\begin{pgfonlayer}{nodelayer}
		\node [style=gauge3] (0) at (0.5, 0) {};
		\node [style=gauge3] (1) at (1.75, 0) {};
		\node [style=gauge3] (2) at (3, 0) {};
		\node [style=none] (3) at (0.5, -0.5) {1};
		\node [style=none] (4) at (1.75, -0.5) {1};
		\node [style=none] (5) at (3, -0.5) {1};
		\node [style=gauge3] (6) at (4.25, 0) {};
		\node [style=none] (7) at (4.25, -0.5) {1};
		\node [style=flavour2] (8) at (0.5, 1.25) {};
		\node [style=flavour2] (9) at (4.25, 1.25) {};
		\node [style=none] (10) at (0.5, 1.75) {1};
		\node [style=none] (11) at (4.25, 1.75) {1};
		\node [style=redgauge] (12) at (10.25, 0) {};
		\node [style=redgauge] (13) at (12.25, 0) {};
		\node [style=bluegauge] (14) at (11.25, 0) {};
		\node [style=gauge3] (15) at (11.25, 1) {};
		\node [style=flavour2] (16) at (11.25, 2) {};
		\node [style=none] (17) at (10.25, -0.5) {2};
		\node [style=none] (18) at (11.25, -0.5) {2};
		\node [style=none] (19) at (12.25, -0.5) {2};
		\node [style=none] (20) at (12, 1) {1};
		\node [style=none] (21) at (11.25, 2.5) {1};
		\node [style=none] (22) at (10.25, -1) {$\mathcal{W}^2$};
		\node [style=none] (23) at (1.75, -1) {$\mathcal{W}$};
		\node [style=none] (24) at (3, -1) {$\mathcal{W}$};
		\node [style=none] (25) at (7.5, 0) {$  \longleftrightarrow$};
	\end{pgfonlayer}
	\begin{pgfonlayer}{edgelayer}
		\draw (0) to (2);
		\draw (6) to (2);
		\draw (8) to (0);
		\draw (9) to (6);
		\draw (12) to (14);
		\draw (14) to (13);
		\draw (15) to (14);
	\draw [line join=round,decorate, decoration={zigzag, segment length=4,amplitude=.9,post=lineto,post length=2pt}]  (16) -- (15);
	\end{pgfonlayer}
\end{tikzpicture}}}
\end{equation}
and the matching Wilson line in the unitary quiver transforms in two adjacent $\urm(1)$ nodes (in ($+1,+1$) charges).

Secondly, one may choose the second symmetric representation $[2]$ of $\sprm(1)$
\begin{equation}
 \scalebox{0.8}{ \raisebox{-.5\height}{\begin{tikzpicture}
	\begin{pgfonlayer}{nodelayer}
		\node [style=none] (25) at (7.5, 0) {$  \longleftrightarrow$};
		\node [style=gauge3] (26) at (-4.5, 1.75) {};
		\node [style=gauge3] (27) at (-3.25, 1.75) {};
		\node [style=gauge3] (28) at (-2, 1.75) {};
		\node [style=gauge3] (29) at (-0.75, 1.75) {};
		\node [style=flavour2] (30) at (-4.5, 3) {};
		\node [style=flavour2] (31) at (-0.75, 3) {};
		\node [style=none] (32) at (-4.5, 3.5) {1};
		\node [style=none] (33) at (-0.75, 3.5) {1};
		\node [style=redgauge] (34) at (11.25, 0) {};
		\node [style=redgauge] (35) at (13.25, 0) {};
		\node [style=bluegauge] (36) at (12.25, 0) {};
		\node [style=gauge3] (37) at (12.25, 1) {};
		\node [style=flavour2] (38) at (12.25, 2) {};
		\node [style=none] (39) at (11.25, -0.5) {2};
		\node [style=none] (40) at (12.25, -0.5) {2};
		\node [style=none] (41) at (13.25, -0.5) {2};
		\node [style=none] (42) at (13, 1) {1};
		\node [style=none] (43) at (12.25, 2.5) {1};
		\node [style=none] (44) at (12.25, -1) {$\mathcal{W}_{[2]}$};
		\node [style=gauge3] (45) at (1, 1.75) {};
		\node [style=gauge3] (46) at (2.25, 1.75) {};
		\node [style=gauge3] (47) at (3.5, 1.75) {};
		\node [style=gauge3] (48) at (4.75, 1.75) {};
		\node [style=flavour2] (49) at (1, 3) {};
		\node [style=flavour2] (50) at (4.75, 3) {};
		\node [style=none] (51) at (1, 3.5) {1};
		\node [style=none] (52) at (4.75, 3.5) {1};
		\node [style=gauge3] (53) at (-1.75, -2.25) {};
		\node [style=gauge3] (54) at (-0.5, -2.25) {};
		\node [style=gauge3] (55) at (0.75, -2.25) {};
		\node [style=gauge3] (56) at (2, -2.25) {};
		\node [style=flavour2] (57) at (-1.75, -1) {};
		\node [style=flavour2] (58) at (2, -1) {};
		\node [style=none] (59) at (-1.75, -0.5) {1};
		\node [style=none] (60) at (2, -0.5) {1};
		\node [style=none] (61) at (-1.75, -3.25) {$\mathcal{W}$};
		\node [style=none] (62) at (-3.25, 0.75) {$\mathcal{W}$};
		\node [style=none] (63) at (1, 0.75) {$\mathcal{W}$};
		\node [style=none] (64) at (2.25, 0.75) {$\mathcal{W}$};
		\node [style=none] (65) at (0, -0.25) {$+$};
		\node [style=none] (66) at (2, -3.25) {$\mathcal{W}$};
		\node [style=none] (67) at (-4.5, 1.25) {1};
		\node [style=none] (68) at (-3.25, 1.25) {1};
		\node [style=none] (69) at (-2, 1.25) {1};
		\node [style=none] (70) at (-0.75, 1.25) {1};
		\node [style=none] (71) at (1, 1.25) {1};
		\node [style=none] (72) at (2.25, 1.25) {1};
		\node [style=none] (73) at (3.5, 1.25) {1};
		\node [style=none] (74) at (4.75, 1.25) {1};
		\node [style=none] (75) at (-1.75, -2.75) {1};
		\node [style=none] (76) at (-0.5, -2.75) {1};
		\node [style=none] (77) at (0.75, -2.75) {1};
		\node [style=none] (78) at (2, -2.75) {1};
	\end{pgfonlayer}
	\begin{pgfonlayer}{edgelayer}
		\draw (26) to (28);
		\draw (29) to (28);
		\draw (30) to (26);
		\draw (31) to (29);
		\draw (34) to (36);
		\draw (36) to (35);
		\draw (37) to (36);
		\draw (45) to (47);
		\draw (48) to (47);
		\draw (49) to (45);
		\draw (50) to (48);
		\draw (53) to (55);
		\draw (56) to (55);
		\draw (57) to (53);
		\draw (58) to (56);
	\draw [line join=round,decorate, decoration={zigzag, segment length=4,amplitude=.9,post=lineto,post length=2pt}]  (38) -- (37);
	\end{pgfonlayer}
\end{tikzpicture}}}
\end{equation}
and the corresponding line defect in the unitary quiver appears to be a linear combination of three different Wilson lines. One Wilson line is charged under a single $\urm(1)$ and the two others are charged under two $\urm(1)$ nodes each. All U(1) charges are $+1$.

Lastly, we consider the charge-2 representation of the $\urm(1)$
\begin{equation}
\scalebox{0.8}{\raisebox{-.5\height}{\begin{tikzpicture}
	\begin{pgfonlayer}{nodelayer}
		\node [style=none] (25) at (7.5, 0) {$  \longleftrightarrow$};
		\node [style=redgauge] (34) at (11.25, 0) {};
		\node [style=redgauge] (35) at (13.25, 0) {};
		\node [style=bluegauge] (36) at (12.25, 0) {};
		\node [style=gauge3] (37) at (12.25, 1) {};
		\node [style=flavour2] (38) at (12.25, 2) {};
		\node [style=none] (39) at (11.25, -0.5) {2};
		\node [style=none] (40) at (12.25, -0.5) {2};
		\node [style=none] (41) at (13.25, -0.5) {2};
		\node [style=none] (42) at (13, 1) {1};
		\node [style=none] (43) at (12.25, 2.5) {1};
		\node [style=gauge3] (45) at (0.25, -0.25) {};
		\node [style=gauge3] (46) at (1.5, -0.25) {};
		\node [style=gauge3] (47) at (2.75, -0.25) {};
		\node [style=gauge3] (48) at (4, -0.25) {};
		\node [style=flavour2] (49) at (0.25, 1) {};
		\node [style=flavour2] (50) at (4, 1) {};
		\node [style=none] (51) at (0.25, 1.5) {1};
		\node [style=none] (52) at (4, 1.5) {1};
		\node [style=none] (53) at (0.25, -1.25) {$\mathcal{W}$};
		\node [style=none] (54) at (0.25, -0.75) {1};
		\node [style=none] (55) at (1.5, -0.75) {1};
		\node [style=none] (56) at (2.75, -0.75) {1};
		\node [style=none] (57) at (4, -0.75) {1};
		\node [style=none] (58) at (11.5, 1) {$\mathcal{W}^2$};
	\end{pgfonlayer}
	\begin{pgfonlayer}{edgelayer}
		\draw (34) to (36);
		\draw (36) to (35);
		\draw (37) to (36);
	\draw [line join=round,decorate, decoration={zigzag, segment length=4,amplitude=.9,post=lineto,post length=2pt}]  (38) -- (37);
		\draw (45) to (47);
		\draw (48) to (47);
		\draw (49) to (45);
		\draw (50) to (48);
	\end{pgfonlayer}
\end{tikzpicture}}}
\end{equation}
\label{single-Wilson}%
\end{subequations}
which corresponds to a Wilson line charged ($+1$) under just one $\urm(1)$ node in the unitary quiver.

In these three examples, we depict the Wilson lines in the unitary-orthosymplectic theory on the right side. While on the left side, we describe the corresponding Wilson lines in the unitary theory, consisting of charge-1 Wilson lines at one gauge group or a product of two gauge groups. As seen in the cases considered above, the dimension of the representations agrees on both sides. As the dimension of the representation of a Wilson line in the orthosymplectic theory becomes larger, it becomes more challenging to find the corresponding Wilson line in the unitary theory.

\subsection{Refining symmetries}\label{sec:refinement}
One may wonder how the global symmetries might be matched once we (partially) include their fugacities in the partition functions. In particular, it is intriguing to study the consequences of gauging the 1-form symmetry $\Z_2^{\mathrm{diag}}$ that is expected to introduce the corresponding 0-form symmetry \cite{Gaiotto:2014kfa}. In order to investigate this question, one approach can be the superconformal index refined by fugacities of the global symmetries. However, it is, by definition, clear that the $\Z_2^{\mathrm{diag}}$ has no effect on the Higgs branch operators; therefore, it is sufficient to focus on Coulomb branch operators. This then allows us to focus on Coulomb branch Hilbert series via the monopole formula \cite{Cremonesi:2013lqa} because it is much easier to evaluate it than the superconformal index.

For refinement, the following fugacities can be turned on
\begin{subequations}
\begin{alignat}{4}
	&\urm(k) \text{ gauge group } & \qquad  &\rightarrow  & \qquad &\urm(1)_{\mathrm{top}}  & \quad &\text{ with } z^{\sum_{i=1}^k m_i} \\
	&\sorm(2N) \text{ gauge group } & \qquad &\rightarrow & \qquad &\Z_2^{\mathrm{centre}} & \quad &\text{ with } y^{\sum_{i=1}^{N} m_i} \;, \quad y^2=1
\end{alignat}
\end{subequations}
for $\sorm(2)\cong \urm(1)$ the centre symmetry is continuous. The 0-form symmetry $\Z_{\ell}$ resulting from gauging the 1-form symmetry $\Z_{\ell}$ is taken into account as follows:
\begin{align}
\HS = \sum_{\kappa=0}^{\ell-1} q^\kappa \sum_{\mathbf{m} \in \left(\Z+\tfrac{\kappa}{\ell}\right)^{\mathrm{rk}} } P(t,\mathbf{m})\ \mathbf{z}^{\mathbf{m}}\ t^{\Delta(\mathbf{m})}
\end{align}
where $\left(\Z+\tfrac{\kappa}{\ell}\right)^{\mathrm{rk}}$ denotes a suitable lattice shifted by $+\tfrac{\kappa}{\ell}$ in every component. See \S \ref{sec:lattice} and \cite{Bourget:2020xdz}.

Here we outline a strategy that identifies the global symmetries of the orthosymplectic quivers apparent at UV as subgroups of the (much larger) symmetries of the unitary counterparts.  The hidden symmetries, i.e.\ the part that only emerges in the IR, cannot be addressed.

\paragraph{$E_4$ quiver.}
For the $E_4$ quiver, one may turn on the following symmetry fugacities:
 \begin{align}
	&\left[ \raisebox{-.5\height}{
	\begin{tikzpicture}
	\begin{pgfonlayer}{nodelayer}
		\node [style=redgauge] (12) at (10.25, 0) {};
		\node [style=redgauge] (13) at (12.25, 0) {};
		\node [style=bluegauge] (14) at (11.25, 0) {};
		\node [style=gauge3] (15) at (11.25, 1) {};
		\node [style=flavour2] (16) at (11.25, 2) {};
		\node [style=none] (17) at (10.25, -0.5) {\footnotesize{$2$}};
		\node [style=none] (18) at (11.25, -0.5) {\footnotesize{$2$}};
		\node [style=none] (19) at (12.25, -0.5) {\footnotesize{$2$}};
		\node [style=none] (20) at (12, 1) {\footnotesize{$1$}};
		\node [style=none] (21) at (11.25, 2.5) {\footnotesize{$1$}};
		\node [style=none] (22) at (10.25, -1) {\footnotesize{$z_1$}};
		\node [style=none] (23) at (10.5, 1) {\footnotesize{$x$}};
		\node [style=none] (24) at (12.25, -1) {\footnotesize{$z_2$}};
	\end{pgfonlayer}
	\begin{pgfonlayer}{edgelayer}
		   \draw [line join=round,decorate, decoration={zigzag, segment length=4,amplitude=.9,post=lineto,post length=2pt}]  (15) -- (16);
		\draw (12) to (14);
		\draw (14) to (13);
		\draw (15) to (14);
	\end{pgfonlayer}
\end{tikzpicture}
	}\right]_{\slash \mathbb{Z}_2^{\mathrm{diag}} \ni q}
	\longleftrightarrow \;\text{(A)}\quad   \raisebox{-.5\height}{
\begin{tikzpicture}
	\begin{pgfonlayer}{nodelayer}
		\node [style=gauge3] (0) at (0.5, 0) {};
		\node [style=gauge3] (1) at (1.75, 0) {};
		\node [style=gauge3] (2) at (3, 0) {};
		\node [style=none] (3) at (0.5, -0.5) {\footnotesize{$1$}};
		\node [style=none] (4) at (1.75, -0.5) {\footnotesize{$1$}};
		\node [style=none] (5) at (3, -0.5) {\footnotesize{$1$}};
		\node [style=gauge3] (6) at (4.25, 0) {};
		\node [style=none] (7) at (4.25, -0.5) {\footnotesize{$1$}};
		\node [style=flavour2] (8) at (0.5, 1.25) {};
		\node [style=flavour2] (9) at (4.25, 1.25) {};
		\node [style=none] (10) at (0.5, 1.75) {1};
		\node [style=none] (11) at (4.25, 1.75) {\footnotesize{$1$}};
		\node [style=none] (13) at (0.5, -1) {\footnotesize{$w_1$}};
		\node [style=none] (14) at (1.75, -1) {\footnotesize{$w_2$}};
		\node [style=none] (15) at (3, -1) {\footnotesize{$w_3$}};
		\node [style=none] (16) at (4.25, -1) {\footnotesize{$w_4$}};
	\end{pgfonlayer}
	\begin{pgfonlayer}{edgelayer}
		\draw (0) to (2);
		\draw (6) to (2);
		\draw (8) to (0);
		\draw (9) to (6);
	\end{pgfonlayer}
\end{tikzpicture}
	}
\end{align}
Evaluating the monopole formula, we see the following match of partially refined Hilbert series
\begin{align}
	\HS^{\mathrm{OSp}}&(z_i=z,x=w^2,q)
	= \HF_{\Z}(z_i=z,x=w^2) + q\cdot  \HF_{\Z+\tfrac{1}{2}}(z_i=z,x=w^2) \notag \\
	&=\HS^{(\mathrm{A})}\left(w_1=w_4=z,w_2=\tfrac{q}{w},w_3=qw\right)|_{q^2=1}  \\
&=\scriptstyle{1
+t \left(6+\frac{1}{z^2}+\frac{4}{z}+4 z+z^2
+q \left(\frac{2}{w}+2 w+\frac{1}{w z}+\frac{w}{z}+\frac{z}{w}+w z \right)
\right)
} \notag\\
&\scriptstyle{
\quad +t^2 \Big(26+\frac{4}{w^2}+4 w^2+\frac{1}{z^4}+\frac{4}{z^3}+\frac{12}{z^2}+\frac{1}{w^2 z^2}+\frac{w^2}{z^2}+\frac{20}{z}+\frac{2}{w^2 z}+\frac{2 w^2}{z}+20 z+\frac{2 z}{w^2}+2 w^2 z+12 z^2+\frac{z^2}{w^2}+w^2 z^2+4 z^3+z^4
}\notag \\
&\scriptstyle{
\qquad \qquad
+q \Big(\frac{12}{w}+12 w+\frac{1}{w z^3}+\frac{w}{z^3}+\frac{4}{w z^2}+\frac{4 w}{z^2}+\frac{9}{w z}+\frac{9 w}{z}+\frac{9 z}{w}+9 w z+\frac{4 z^2}{w}+4 w z^2+\frac{z^3}{w}+w z^3 \Big)
\Big)
} \notag\\
&\scriptstyle{
\quad +t^3 \Big(78+\frac{20}{w^2}+20 w^2+\frac{1}{z^6}+\frac{4}{z^5}+\frac{12}{z^4}+\frac{1}{w^2 z^4}+\frac{w^2}{z^4}+\frac{28}{z^3}+\frac{4}{w^2 z^3}+\frac{4 w^2}{z^3}+\frac{48}{z^2}+\frac{9}{w^2 z^2}+\frac{9 w^2}{z^2}+\frac{68}{z}
+\frac{16}{w^2 z}+\frac{16 w^2}{z}+68 z+\frac{16 z}{w^2}
}
\notag\\
&\scriptstyle{
\qquad \qquad
+16 w^2 z+48 z^2+\frac{9 z^2}{w^2}+9 w^2 z^2+28 z^3+\frac{4 z^3}{w^2}+4 w^2 z^3+12 z^4+\frac{z^4}{w^2}+w^2 z^4+4 z^5+z^6
}
\notag\\
&\scriptstyle{
\qquad \qquad
+q \Big( \frac{6}{w^3}+\frac{44}{w}+44 w+6 w^3+\frac{1}{w z^5}+\frac{w}{z^5}+\frac{4}{w z^4}+\frac{4 w}{z^4}+\frac{1}{w^3 z^3}+\frac{12}{w z^3}+\frac{12 w}{z^3}+\frac{w^3}{z^3}+\frac{2}{w^3 z^2}+\frac{24}{w z^2}+\frac{24 w}{z^2}+\frac{2 w^3}{z^2}+\frac{4}{w^3 z}
}
\notag\\
&\scriptstyle{
\qquad \qquad +\frac{37}{w z}+\frac{37 w}{z}+\frac{4 w^3}{z}+\frac{4 z}{w^3}+\frac{37 z}{w}+37 w z+4 w^3 z+\frac{2 z^2}{w^3}+\frac{24 z^2}{w}+24 w z^2+2 w^3 z^2+\frac{z^3}{w^3}+\frac{12 z^3}{w}+12 w z^3+w^3 z^3+\frac{4 z^4}{w}
}
\notag\\
&\scriptstyle{
\qquad \qquad+4 w z^4+\frac{z^5}{w}+w z^5\Big)
\Big)
} \notag\\
&\scriptstyle{\quad+ \mathcal{O}(t^4)} \notag
\end{align}
and the agreement has been verified by perturbative evaluation up to order $t^{10}$. This identifies two continuous global symmetries besides $\Z_2^{\mathrm{diag}}$.
\paragraph{$E_5$ quiver.} It turns out that the identification of global symmetries in the $E_5$ magnetic quivers is more interesting since it depends on the $\urm(1)$ ungauging scheme of the unitary counterparts, namely (A) and (B) below.  For the $E_5$ magnetic quivers, one may turn on the following symmetry fugacities:
    \begin{align}
	\left[ \raisebox{-.5\height}{
		\begin{tikzpicture}
			\node (g1) [gaugeSO,label=below:{\footnotesize{$2$}}] {};
			\node (g2) [gaugeSp,right of =g1,label=below:{\footnotesize{$2$}}] {};
			\node (g3) [gaugeSO,right of =g2,label=below:{\footnotesize{$4$}}] {};
			\node (g4) [gaugeSp,right of =g3,label=below:{\footnotesize{$2$}}] {};
			\node (g5) [gaugeSO,right of =g4,label=below:{\footnotesize{$2$}}] {};
			\node (b1) [gauge,above of=g3, label=right:{\footnotesize{$1$}}, label=left:{\footnotesize{$x$}}] {};
			\draw (g1)--(g2) (g2)--(g3) (g3)--(g4) (g4)--(g5) (g3)--(b1);
			\node [below of=g1 ] {\footnotesize{$z_1$}};
			\node [below of=g5 ] {\footnotesize{$z_2$}};
			\node [below of=g3 ] {\footnotesize{$y\in \mathbb{Z}_2$}};
		\end{tikzpicture}
	}\right]_{\slash \mathbb{Z}_2^{\mathrm{diag}} \ni q}
	%
	&\longleftrightarrow \; (\mathrm{A}) \raisebox{-.5\height}{
		\begin{tikzpicture}
			\node (g1) [gauge,label=below:{\footnotesize{$1$}}] {};
			\node (g2) [gauge,right of =g1,label=below:{\footnotesize{$2$}}] {};
			\node (g3) [gauge,right of =g2,label=below:{\footnotesize{$2$}}] {};
			\node (g4) [gauge,below right of =g3,label=right:{\footnotesize{$1$}}] {};
			\node (g5) [gauge,above right of =g3,label=right:{\footnotesize{$1$}}] {};
			\node (f1) [flavour,above of=g2, label=left:{\footnotesize{$1$}}] {};
			\draw (g1)--(g2) (g2)--(g3) (g3)--(g4) (g3)--(g5) (g2)--(f1);
			\node [below of=g1 ] {\footnotesize{$w_1$}};
			\node [below of=g2 ] {\footnotesize{$w_2$}};
			\node [below of=g3 ] {\footnotesize{$w_3$}};
			\node [right of=g4 ] {\footnotesize{$w_4$}};
			\node [right of=g5 ] {\footnotesize{$w_5$}};
		\end{tikzpicture}
	}
 \\
	&\longleftrightarrow \;(\mathrm{B}) \left[ \raisebox{-.5\height}{
	\begin{tikzpicture}
		\node (g2) [gauge,label=below:{\footnotesize{$\; \; \; \surm(2)$}}] {};
		\node (g1) [gauge,below left of =g2,label=left:{\footnotesize{$1$}}] {};
		\node (g3) [gauge,right of =g2,label=below:{\footnotesize{$2$}}] {};
		\node (g4) [gauge,below right of =g3,label=right:{\footnotesize{$1$}}] {};
		\node (g5) [gauge,above right of =g3,label=right:{\footnotesize{$1$}}] {};
		\node (g0) [gauge,above left of=g2, label=left:{\footnotesize{$1$}}] {};
		\draw (g1)--(g2) (g2)--(g3) (g3)--(g4) (g3)--(g5) (g2)--(g0);
		\node [left of=g0 ] {\footnotesize{$v_1$}};
		\node [left of=g1 ] {\footnotesize{$v_2$}};
		\node [below of=g3 ] {\footnotesize{$u_2$}};
		\node [right of=g4 ] {\footnotesize{$u_1$}};
		\node [right of=g5 ] {\footnotesize{$u_3$}};
	\end{tikzpicture}
} \right]_{\slash \mathbb{Z}_2^{\mathrm{diag}}\ni p}
\notag
\end{align}
The Hilbert series evaluation yields the following relations
\begin{align}
\HS^{\mathrm{OSp}}&(z_i=z,y=1,x=1,q)
= \HF_{\Z}(z_i=z,y=1,x=1) + q\cdot  \HF_{\Z+\tfrac{1}{2}}(z_i=z,y=1,x=1) \notag \\
&=\HS^{(\mathrm{A})}(w_1=w_4=w_5^{-1}=z,w_2=w_3=q)|_{q^2=1}  \\
&=\HS^{(\mathrm{B})}(u_1=u_3^{-1}=z,u_2=q,v_1=v_2^{-1}=z,p=1)|_{q^2=1} \notag\\
&=
\scriptstyle{1
+t \left(11
+8 z+\frac{8}{z}
+z^2+\frac{1}{z^2}
+q \left(4 z+\frac{4}{z}+8\right)\right)
}\notag \\
&\scriptstyle{
\quad +t^2 \left(130
+96 z+\frac{96}{z}
+47 z^2+\frac{47}{z^2}
+8 z^3+\frac{8}{z^3}
+z^4+\frac{1}{z^4}
+q \left(4 z^3+\frac{4}{z^3}+32 z^2+\frac{32}{z^2}+80 z+\frac{80}{z}+104\right)
\right)
} \notag\\
&\scriptstyle{
	\quad
+t^3 \big(942 +808 z+\frac{808}{z}
+487 z^2+\frac{487}{z^2}
+208 z^3+\frac{208}{z^3}
+47 z^4+\frac{47}{z^4}
+8 z^5+\frac{8}{z^5}
+z^6+\frac{1}{z^6}
}
\notag\\
&\scriptstyle{
	\qquad \qquad
+q \left(4 z^5+\frac{4}{z^5}+32 z^4+\frac{32}{z^4}+168 z^3+\frac{168}{z^3}+432 z^2+\frac{432}{z^2}+724 z+\frac{724}{z}+864\right)
\big)
}
\notag\\
&\scriptstyle{
	\quad
+t^4 \big(
5350
+4744 z+\frac{4744}{z}
+3381 z^2+\frac{3381}{z^2}
+1856 z^3+\frac{1856}{z^3}
+772 z^4+\frac{772}{z^4}
+208 z^5+\frac{208}{z^5}
+47 z^6+\frac{47}{z^6}
+8 z^7+\frac{8}{z^7}
+z^8+\frac{1}{z^8}
}
\notag\\
&\scriptstyle{
\qquad \qquad
+q \left(4 z^7+\frac{4}{z^7}+32 z^6+\frac{32}{z^6}+168 z^5+\frac{168}{z^5}+672 z^4+\frac{672}{z^4}+1712 z^3+\frac{1712}{z^3}+3152 z^2+\frac{3152}{z^2}+4500 z+\frac{4500}{z}+5056\right)
\big)
} \notag\\
&\scriptstyle{
\quad
+t^5 \big(
24218
+22264 z+\frac{22264}{z}
+17199 z^2+\frac{17199}{z^2}
+11168 z^3+\frac{11168}{z^3}
+5900 z^4+\frac{5900}{z^4}
+2480 z^5+\frac{2480}{z^5}
}
\notag\\
&\scriptstyle{
\qquad \qquad \qquad
+772 z^6+\frac{772}{z^6}
+208 z^7+\frac{208}{z^7}
+47 z^8+\frac{47}{z^8}
+8 z^9+\frac{8}{z^9}
+z^{10}+\frac{1}{z^{10}}
}
\notag\\
&\scriptstyle{
\qquad \qquad
+q \big(
23488
+21516 z+\frac{21516}{z}
+16592 z^2+\frac{16592}{z^2}
+10656 z^3+\frac{10656}{z^3}
+5568 z^4+\frac{5568}{z^4}
+2264 z^5+\frac{2264}{z^5}
}
\notag\\
&\scriptstyle{
\qquad \qquad \qquad
+672 z^6+\frac{672}{z^6}
+168 z^7+\frac{168}{z^7}
+32 z^8+\frac{32}{z^8}
+4 z^9+\frac{4}{z^9}
\big)
\big)
} \notag \\
&\scriptstyle{\quad+ \mathcal{O}(t^6)} \notag
\end{align}
and  agreement has been verified by perturbative evaluation up to order $t^5$.
This implies in particularly that the separation into integer and half-integer lattice of the orthosymplectic quiver and the unitary quiver (B) \emph{does not} coincide. Moreover, the identification of the suitable $\Z_2^{\mathrm{diag}}$ in the unitary quivers is realised by embedding the $\Z_2$ into a single $\urm(1)_{\mathrm{top}}$ for (B) and into a diagonal $\urm(1)\subset \urm(1)_{\mathrm{top}}\times \urm(1)_{\mathrm{top}}$ for (A).

For the unitary quiver (A) one can, moreover, identify all continuous global symmetries of the orthosymplectic quiver as follows:
\begin{align}
    \HS^{\mathrm{OSp}}&(z_i,y=1,x,q)
=\HS^{(\mathrm{A})}(w_1=z_1^{-1},w_2=q\sqrt{z_1 z_2 x},w_3= \tfrac{q}{\sqrt{z_1 z_2 x}},w_4= z_1,w_5=z_2)|_{q^2=1}  \\
&\scriptstyle{= 1
+t \bigg(
9
+4 z_1+\frac{4}{z_1}
+4 z_2+\frac{4}{z_2}
+z_1 z_2+\frac{z_1}{z_2}
+\frac{1}{z_1 z_2}
+\frac{z_2}{z_1}
} \notag \\
&\qquad \qquad \scriptstyle{
+2q z_1^{-\frac12} z_2^{-\frac12}(\sqrt{x}+1/\sqrt{x})\left(1+z_1+ z_2+z_1 z_2\right)
\bigg)
+ \mathcal{O}(t^2) } \notag
\end{align}
which has been verified up to order $t^5$.

%
\paragraph{$E_6$ quiver.} For the $E_6$ magnetic quivers, we turn on the following fugacities
    \begin{align}
	&\left[ \raisebox{-.5\height}{
		\begin{tikzpicture}
			\node (g1) [gaugeSO,label=below:{\footnotesize{$2$}}] {};
			\node (g2) [gaugeSp,right of =g1,label=below:{\footnotesize{$2$}}] {};
			\node (g3) [gaugeSO,right of =g2,label=below:{\footnotesize{$4$}}] {};
			\node (g4) [gaugeSp,right of =g3,label=below:{\footnotesize{$4$}}] {};
			\node (g5) [gaugeSO,right of =g4,label=below:{\footnotesize{$4$}}] {};
			\node (g6) [gaugeSp,right of =g5,label=below:{\footnotesize{$2$}}] {};
			\node (g7) [gaugeSO,right of =g6,label=below:{\footnotesize{$2$}}] {};
			\node (b1) [gauge,above of=g4, label=right:{\footnotesize{$1$}}, label=left:{\footnotesize{$x$}}] {};
			\draw (g1)--(g2) (g2)--(g3) (g3)--(g4) (g4)--(g5) (g5)--(g6) (g6)--(g7) (g4)--(b1);
			\node [below of=g1] {\footnotesize{$z_1$}};
			\node [below of=g7] {\footnotesize{$z_2$}};
			\node [below of=g3] {\footnotesize{$y_1\in \mathbb{Z}_2$}};
			\node [below of=g5] {\footnotesize{$y_2\in \mathbb{Z}_2$}};
		\end{tikzpicture}
	}\right]_{\slash \mathbb{Z}_2^{\mathrm{diag}} \ni q}
	\notag \\
	&\longleftrightarrow \; (\mathrm{A}) \raisebox{-.5\height}{
		\begin{tikzpicture}
			\node (g1) [gauge,label=below:{\footnotesize{$1$}}] {};
			\node (g2) [gauge,right of =g1,label=below:{\footnotesize{$2$}}] {};
			\node (g3) [gauge,right of =g2,label=below:{\footnotesize{$3$}}] {};
			\node (g4) [gauge,right of =g3,label=below:{\footnotesize{$2$}}] {};
			\node (g5) [gauge,right of =g4,label=below:{\footnotesize{$1$}}] {};
			\node (b2) [gauge,above of =g3,label=right:{\footnotesize{$2$}},label=left:{\footnotesize{$w_6$}}] {};
			\node (b1) [flavour,above of=b2, label=right:{\footnotesize{$1$}}] {};
			\draw (g1)--(g2) (g2)--(g3) (g3)--(g4) (g4)--(g5) (g3)--(b2) (b2)--(b1);
			\node [below of=g1 ] {\footnotesize{$w_1$}};
			\node [below of=g2 ] {\footnotesize{$w_2$}};
			\node [below of=g3 ] {\footnotesize{$w_3$}};
			\node [below of=g4 ] {\footnotesize{$w_4$}};
			\node [below of=g5 ] {\footnotesize{$w_5$}};
		\end{tikzpicture}
	}
	\\
	&\longleftrightarrow \;(\mathrm{B}) \left[ \raisebox{-.5\height}{
		\begin{tikzpicture}
			\node (g1) [gauge,label=below:{\footnotesize{$1$}}] {};
			\node (g2) [gauge,right of =g1,label=below:{\footnotesize{$2$}}] {};
			\node (g3) [gauge,right of =g2,label=below:{\footnotesize{$3$}}] {};
			\node (g4) [gauge,right of =g3,label=below:{\footnotesize{$2$}}] {};
			\node (g5) [gauge,right of =g4,label=below:{\footnotesize{$1$}}] {};
			\node (b2) [gauge,above of =g3,label=right:{\footnotesize{$\surm(2)$}}] {};
			\node (b1) [gauge,above of=b2, label=right:{\footnotesize{$1$}}, label=left:{\footnotesize{$v$}}] {};
			\draw (g1)--(g2) (g2)--(g3) (g3)--(g4) (g4)--(g5) (g3)--(b2) (b2)--(b1);
			\node [below of=g1 ] {\footnotesize{$u_1$}};
			\node [below of=g2 ] {\footnotesize{$u_2$}};
			\node [below of=g3 ] {\footnotesize{$u_3$}};
			\node [below of=g4 ] {\footnotesize{$u_4$}};
			\node [below of=g5 ] {\footnotesize{$u_5$}};
		\end{tikzpicture}
	} \right]_{\slash \mathbb{Z}_2^{\mathrm{diag}}\ni p}
	\notag
\end{align}
The evaluation of the monopole formula suggests the following
\begin{align}
	\HS^{\mathrm{OSp}}&(z_i=z,y_i=x=1,q)
	= \HF_{\Z}(z_i=z,y_i=x=1) + q\cdot  \HF_{\Z+\tfrac{1}{2}}(z_i=z,y_i=x=1) \notag \\
	&=\HS^{(\mathrm{A})}(w_1=w_5=1,w_2=w_4=q,w_3=w_6^{-1}=z)|_{q^2=1}  \\
	&=\HS^{(\mathrm{B})}(u_1=u_5=z,u_2=u_4=q,u_3=v=p=1)|_{q^2=1} \notag\\
	&=\HS^{(\mathrm{B})}(u_1=u_5=1,u_2=u_4=q,u_3=v^{-1}=z,p=1)|_{q^2=1} \notag \\
	&=\scriptstyle{1
	+t \left(
	20
	+12 z+\frac{12}{z}
	+z^2+\frac{1}{z^2}
		+q \left(16 +8 z+\frac{8}{z}\right)
	 \right)
} \notag\\
&\scriptstyle{
\quad
	+t^2 \left(
	422
	+300 z+\frac{300}{z}
	+115 z^2+\frac{115}{z^2}
	+12 z^3+\frac{12}{z^3}
	+z^4+\frac{1}{z^4}
	+q \left(384
	+280 z+\frac{280}{z}
	+96 z^2+\frac{96}{z^2}
	+8 z^3+\frac{8}{z^3}
	\right)
	\right)
} \notag\\
&\scriptstyle{
\quad
	+t^3 \big(
	5834
	+4756 z+\frac{4756}{z}
	+2518 z^2+\frac{2518}{z^2}
	+808 z^3+\frac{808}{z^3}
	+115 z^4+\frac{115}{z^4}
	+12 z^5+\frac{12}{z^5}
	+z^6+\frac{1}{z^6}
}
\notag\\
&\scriptstyle{
\qquad \qquad
		+q \left(
		5696
		+4616 z+\frac{4616}{z}
		+2432 z^2+\frac{2432}{z^2}
		+752 z^3+\frac{752}{z^3}
		+96 z^4+\frac{96}{z^4}
		+8 z^5+\frac{8}{z^5}
		\right)
	\big)
} \notag\\
&\scriptstyle{
\quad
+t^4 \big(
60006
+51688 z+\frac{51688}{z}
+33050 z^2+\frac{33050}{z^2}
+15104 z^3+\frac{15104}{z^3}
+4618 z^4+\frac{4618}{z^4}
}
\notag\\
&\scriptstyle{
\qquad \qquad \qquad
+808 z^5+\frac{808}{z^5}
+115 z^6+\frac{115}{z^6}
+12 z^7+\frac{12}{z^7}
+z^8+\frac{1}{z^8}
}
\notag\\
&\scriptstyle{
\qquad \qquad
+q \big(
59344
+51184 z+\frac{51184}{z}
+32576 z^2+\frac{32576}{z^2}
+14848 z^3+\frac{14848}{z^3}
}
\notag\\
&\scriptstyle{
\qquad \qquad \qquad
+4448 z^4+\frac{4448}{z^4}
+752 z^5+\frac{752}{z^5}
+96 z^6+\frac{96}{z^6}
+8 z^7+\frac{8}{z^7}
\big)
\big)
} \notag\\
&\scriptstyle{
\quad
+t^5 \big(
479893
+429036 z+\frac{429036}{z}
+305471 z^2+\frac{305471}{z^2}
+170968 z^3+\frac{170968}{z^3}
+72650 z^4+\frac{72650}{z^4}
}
\notag\\
&\scriptstyle{
\qquad \qquad \qquad
+22240 z^5+\frac{22240}{z^5}
+4618 z^6+\frac{4618}{z^6}
+808 z^7+\frac{808}{z^7}
+115 z^8+\frac{115}{z^8}
+12 z^9+\frac{12}{z^9}
+z^{10}+\frac{1}{z^{10}}
}
\notag\\
&\scriptstyle{
\qquad \qquad
+q \big(
477888
+427016 z+\frac{427016}{z}
+303920 z^2+\frac{303920}{z^2}
+169744 z^3+\frac{169744}{z^3}
+71936 z^4+\frac{71936}{z^4}
}
\notag\\
&\scriptstyle{
\qquad \qquad \qquad
+21824 z^5+\frac{21824}{z^5}
+4448 z^6+\frac{4448}{z^6}
+752 z^7+\frac{752}{z^7}
+96 z^8+\frac{96}{z^8}
+8 z^9+\frac{8}{z^9}
\big)
\big)
}
\notag \\
&\scriptstyle{
\quad +\mathcal{O}(t^6) } \notag
\end{align}
and the agreement has been verified up to order $t^5$.
Again, the GNO lattice split into integer and half-integer lattice for the orthosymplectic quiver is different compared to the split in the unitary quiver (B).
\paragraph{$E_7$ quiver.} For the $E_7$ magnetic quivers, consider the following theories and their symmetry refinement
    \begin{align}
	&\left[ \raisebox{-.5\height}{
		\begin{tikzpicture}
			\node (g1) [gaugeSO,label=below:{\footnotesize{$2$}}] {};
			\node (g2) [gaugeSp,right of =g1,label=below:{\footnotesize{$2$}}] {};
			\node (g3) [gaugeSO,right of =g2,label=below:{\footnotesize{$4$}}] {};
			\node (g4) [gaugeSp,right of =g3,label=below:{\footnotesize{$4$}}] {};
			\node (g5) [gaugeSO,right of =g4,label=below:{\footnotesize{$6$}}] {};
			\node (g6) [gaugeSp,right of =g5,label=below:{\footnotesize{$4$}}] {};
			\node (g7) [gaugeSO,right of =g6,label=below:{\footnotesize{$4$}}] {};
			\node (g8) [gaugeSp,right of =g7,label=below:{\footnotesize{$2$}}] {};
			\node (g9) [gaugeSO,right of =g8,label=below:{\footnotesize{$2$}}] {};
			\node (b1) [gaugeSp,above of=g5, label=right:{\footnotesize{$2$}}] {};
			\node (b0) [gauge,above of=b1,label=right:{\footnotesize{$1$}},label=left:{\footnotesize{$x$}}] {};
			\draw (g1)--(g2) (g2)--(g3) (g3)--(g4) (g4)--(g5) (g5)--(g6) (g6)--(g7) (g7)--(g8) (g8)--(g9) (g5)--(b1) (b0)--(b1);
			\node [below of=g1] {\footnotesize{$z_1$}};
			\node [below of=g9] {\footnotesize{$z_2$}};
			\node [below of=g3] {\footnotesize{$y_1\in \mathbb{Z}_2$}};
			\node [below of=g5] {\footnotesize{$y_2\in \mathbb{Z}_2$}};
			\node [below of=g7] {\footnotesize{$y_3\in \mathbb{Z}_2$}};
		\end{tikzpicture}
	}\right]_{\slash \mathbb{Z}_2^{\mathrm{diag}} \ni q}
	\notag \\
	&\longleftrightarrow \; (\mathrm{A}) \raisebox{-.5\height}{
		\begin{tikzpicture}
			\node (g1) [gauge,label=below:{\footnotesize{$1$}}] {};
			\node (g2) [gauge,right of =g1,label=below:{\footnotesize{$2$}}] {};
			\node (g3) [gauge,right of =g2,label=below:{\footnotesize{$3$}}] {};
			\node (g4) [gauge,right of =g3,label=below:{\footnotesize{$4$}}] {};
			\node (g5) [gauge,right of =g4,label=below:{\footnotesize{$3$}}] {};
			\node (g6) [gauge,right of =g5,label=below:{\footnotesize{$2$}}] {};
			\node (b2) [gauge,above of =g4,label=right:{\footnotesize{$2$}},label=left:{\footnotesize{$w_7$}}] {};
			\node (b1) [flavour,above of=g6, label=right:{\footnotesize{$1$}}] {};
			\draw (g1)--(g2) (g2)--(g3) (g3)--(g4) (g4)--(g5) (g5)--(g6) (g4)--(b2) (g6)--(b1);
			\node [below of=g1 ] {\footnotesize{$w_1$}};
			\node [below of=g2 ] {\footnotesize{$w_2$}};
			\node [below of=g3 ] {\footnotesize{$w_3$}};
			\node [below of=g4 ] {\footnotesize{$w_4$}};
			\node [below of=g5 ] {\footnotesize{$w_5$}};
			\node [below of=g6 ] {\footnotesize{$w_6$}};
		\end{tikzpicture}
	}
	\\
	&\longleftrightarrow \;(\mathrm{B}) \left[ \raisebox{-.5\height}{
		\begin{tikzpicture}
			\node (g1) [gauge,label=below:{\footnotesize{$1$}}] {};
			\node (g2) [gauge,right of =g1,label=below:{\footnotesize{$2$}}] {};
			\node (g3) [gauge,right of =g2,label=below:{\footnotesize{$3$}}] {};
			\node (g4) [gauge,right of =g3,label=below:{\footnotesize{$4$}}] {};
			\node (g5) [gauge,right of =g4,label=below:{\footnotesize{$3$}}] {};
			\node (g6) [gauge,right of =g5,label=below:{\footnotesize{$2$}}] {};
			\node (g7) [gauge,right of=g6, label=below:{\footnotesize{$1$}}] {};
			\node (b2) [gauge,above of =g4,label=right:{\footnotesize{$\surm(2)$}}] {};
			\draw (g1)--(g2) (g2)--(g3) (g3)--(g4) (g4)--(g5) (g5)--(g6) (g4)--(b2) (g6)--(g7);
			\node [below of=g1 ] {\footnotesize{$u_1$}};
			\node [below of=g2 ] {\footnotesize{$u_2$}};
			\node [below of=g3 ] {\footnotesize{$u_3$}};
			\node [below of=g4 ] {\footnotesize{$u_4$}};
			\node [below of=g5 ] {\footnotesize{$u_5$}};
			\node [below of=g6 ] {\footnotesize{$u_6$}};
			\node [below of=g7 ] {\footnotesize{$u_7$}};
		\end{tikzpicture}
	} \right]_{\slash \mathbb{Z}_2^{\mathrm{diag}}\ni p}
	\notag \\
		&\longleftrightarrow \;(\mathrm{C}) \left[ \raisebox{-.5\height}{
		\begin{tikzpicture}
			\node (g1) [gauge,label=below:{\footnotesize{$1$}}] {};
			\node (g2) [gauge,right of =g1,label=below:{\footnotesize{$2$}}] {};
			\node (g3) [gauge,right of =g2,label=below:{\footnotesize{$3$}}] {};
			\node (g4) [gauge,right of =g3,label=below:{\footnotesize{$4$}}] {};
			\node (g5) [gauge,right of =g4,label=below:{\footnotesize{$3$}}] {};
			\node (g6) [gauge,right of =g5,label=below:{\footnotesize{$\surm(2)$}}] {};
			\node (g7) [gauge,right of=g6, label=below:{\footnotesize{$1$}}] {};
			\node (b2) [gauge,above of =g4,label=right:{\footnotesize{$2$}},label=left:{\footnotesize{$v_6$}}] {};
			%
			\draw (g1)--(g2) (g2)--(g3) (g3)--(g4) (g4)--(g5) (g5)--(g6) (g4)--(b2) (g6)--(g7);
			\node [below of=g1 ] {\footnotesize{$v_1$}};
			\node [below of=g2 ] {\footnotesize{$v_2$}};
			\node [below of=g3 ] {\footnotesize{$v_3$}};
			\node [below of=g4 ] {\footnotesize{$v_4$}};
			\node [below of=g5 ] {\footnotesize{$v_5$}};
			\node [below of=g7 ] {\footnotesize{$w$}};
		\end{tikzpicture}
	} \right]_{\slash \mathbb{Z}_2^{\mathrm{diag}}\ni r}
	\notag
\end{align}
The Coulomb branch Hilbert series are related as follows:
\begin{align}
	\HS^{\mathrm{OSp}}&(z_i=z,y_i=x=1,q)
	= \HF_{\Z}(z_i=z,y_i=x=1) + q\cdot  \HF_{\Z+\tfrac{1}{2}}(z_i=z,y_i=x=1)\notag \\
	&=\HS^{(\mathrm{A})}(w_1=w_7=w_4^{-1}=z,w_2=w_6=q,w_3=w_5=1)|_{q^2=1}   \\
	&=\HS^{(\mathrm{B})}(u_1=u_7=z,u_2=u_6=q,u_{3,4,5}=p=1)|_{q^2=1} \notag\\
	&=\HS^{(\mathrm{C})}(v_1=v_5=v_6^{-1}=z,v_{2,3,4,5}=w=1,r=q) \notag \\
	&= \HF^{(\mathrm{C})}_{\Z }(v_1=v_5=v_6^{-1}=z,v_{2,3,4,5}=w=1)  \notag \\
	&\qquad + r|_{r=q} \cdot\HF^{(\mathrm{C})}_{\Z +\tfrac{1}{2}} (v_1=v_5=v_6^{-1}=z,v_{2,3,4,5}=w=1)\notag \\
	&\scriptstyle{
	=1
	+t \left(
	35
	+16 z+\frac{16}{z}
	+z^2+\frac{1}{z^2}
	+q \left(32+16 z+\frac{16}{z}\right)
	\right)
} \notag\\
&\scriptstyle{
\quad
	+t^2 \left(
	1351
	+896 z+\frac{896}{z}
	+273 z^2+\frac{273}{z^2}
	+16 z^3+\frac{16}{z^3}
	+z^4+\frac{1}{z^4}
	+q \left(
	1312
	+896 z+\frac{896}{z}
	+256 z^2+\frac{256}{z^2}
	+16 z^3+\frac{16}{z^3}
	\right)
	\right)
} \notag
\notag\\
&\scriptstyle{
	\quad +\mathcal{O}(t^3) } \notag
\end{align}
and the agreement has been verified up to order $t^3$.
We observe that the integer half-integer split of the orthosymplectic quiver \emph{does} coincide with the split in the unitary quiver (C), while it \emph{does not}  agree with the split in (B).

Interestingly, one can also verify the following relation
\begin{align}
    \HS^{\mathrm{OSp}}(z_i,y_i=x=1,q)
	&=\HS^{(\mathrm{C})}(v_1=z_1,v_5=v_6^{-1}=z_2,v_{2,3,4,5}=w=1,r=q)
	\notag \\
	&=\scriptstyle{
	1+
	t \big(
	+33
	+8 z_1+\frac{8}{z_1}+8 z_2+\frac{8}{z_2}
	+z_1 z_2+\frac{z_1}{z_2}+\frac{1}{z_1 z_2}+\frac{z_2}{z_1}
	} \notag\\
&\scriptstyle{
\quad
	+16q \sqrt{z_1z_2}\left(1+ z_1+ z_2+ z_1 z_2\right)
	\big)
	+\mathcal{O}(t^2)
	}
\end{align}
i.e.\ two global $\urm(1)_{z_i}$ symmetries and the $\Z_2^{\mathrm{diag}}$ can be identified.
\paragraph{Remarks.}
The results of this section demonstrate that the discrete 0-form symmetry $\Z_2^{\mathrm{diag}}$ resulting from gauging the 1-form $\Z_2^\mathrm{diag}$ in the orthosymplectic quivers, see \S \ref{sec:lattice}, can be recovered in the unitary counterparts $T^{(\alpha)}$. We also observe that the way the discrete 0-form symmetry $\Z_2^{\mathrm{diag}}$ is embedded depends on a $\urm(1)$ ungauging scheme $(\alpha)$. Of course, the orthosymplectic quiver offers less possibility of refining global symmetries compared to the unitary counterparts; however, it is a necessary and so far less studied question on how to embed the symmetries visible in the orthosymplectic quiver into the symmetries of the unitary quiver. A detailed study of matching the refined indices is left for future work.

One interesting outcome is the following: The identification of the discrete $\Z_2^{\mathrm{diag}}$ global symmetry in the unitary counterparts enables unitary magnetic quiver constructions for the orthosymplectic quivers where the $\Z_2^{\mathrm{diag}}$ 1-form symmetry has not been gauged. Schematically, one finds:
\begin{alignat}{2}\nonumber
\text{OSp quiver with }G^\prime &= \prod_i \left[\sorm(2n_i) \times \sprm(k_i)\right] \slash \Z_2^{\mathrm{diag}} &
&\longleftrightarrow \text{ known unitary counterpart(s) } T^{(\alpha)} \cr
\text{OSp quiver with }G &= \prod_i \left[\sorm(2n_i) \times \sprm(k_i)\right] &
&\longleftrightarrow \text{ to be explored counterpart(s) } \widetilde{T}^{(\alpha)}
\end{alignat}
where the quiver theories $\widetilde{T}^{(\alpha)}$ are derived from the unitary quivers $T^{(\alpha)}$ by gauging the corresponding $\Z_2^{\mathrm{diag}}$ global symmetry. As demonstrated in the examples of this section, the 0-form $\Z_2^{\mathrm{diag}}$ symmetry can be identified in the $T^{(\alpha)}$s. The implications are as follows:
\begin{compactitem}
\item The proposed duality between orthosymplectic and unitary quivers, with trivial 1-form symmetries, is likely to be extended to a duality of orthosymplectic and unitary quivers involving non-trivial 1-form symmetries.
\item Generating functions, which are sensitive to higher-form symmetries, are the Coulomb branch Hilbert series and also the superconformal index as well as the A-twisted index. See for instance \cite{Beratto:2021xmn} for the analysis with the superconformal index and \cite{Eckhard:2019jgg} for the observation that the Witten index is sensitive to 1-form symmetries.
\end{compactitem}
A systematic study requires an at least partially refined evaluation of the supersymmetric indices or Hilbert series, which is beyond the scope of the present paper. We leave this for future work.

\section{Summary and discussions}
 \label{sec:summary-open}
The central motivation for this paper has been the question whether two magnetic quivers that describe the same moduli space are dual as honest 3d $\Ncal=4$ theories.

As a first probe, the superconformal index has been computed for the unitary magnetic quivers as well as the orthosymplectic magnetic quiver.
The results of \S \ref{sec:UOSp-3d} show that the indices for the two different 3d mirror quivers are compatible with each other in the given order of the perturbative evaluation. Coincidentally, these results are also valid for the finite coupling magnetic quivers for the 5d $\sprm(k)$ theories of \S \ref{sec:5d_Sp}. Moreover, in \S \ref{sec:UOSp-5d} the superconformal indices for the two infinite coupling magnetic quivers of 5d $\sprm(k)$ have been computed perturbatively. Again, in the given order of evaluation, the results are compatible with one another. As a consistency check, the Coulomb limit and the Higgs limit of the index results agree with the known Coulomb branch and Higgs branch Hilbert series, respectively.  Also, we perform the identification of some 0-form symmetries, including that from gauging the 1-form symmetry, in the Coulomb branch Hilbert series between the two magnetic quivers  in \S \ref{sec:refinement}.

As a second probe, half BPS line defects have been introduced into the two classes of magnetic quivers. The 3d setup of \S \ref{sec:3d_Sp} served as a playground to familiarise oneself with Wilson and vortex lines because the entire setting admits a brane realisation in Type IIB. In \S \ref{sec:Wilson_mirrors} a pattern has been presented as a dictionary to match Wilson line defects between the two 3d $\Ncal=4$ $\sprm(k)$ mirror theories. Computationally, evidence is provided by matching the B-twisted indices with Wilson line insertion. Inclusion of vortex lines in the two mirrors has been considered in \S \ref{sec:vortex_in_mirrors}, based on the mirror symmetry of Wilson and vortex line discussed in \S \ref{sec:mirror_Wilson_vortex}.
Moving on to the magnetic quivers for the 5d $\sprm(k)$ theories at infinite coupling  limit, the inclusion of line defects is solely based on quantum field theory rather than underlying brane configurations. As a first conceptual step, the allowed representations for the Wilson lines have been detailed in \S \ref{sec:Wilson_unframed}. Thereafter, an intriguing matching pattern between the Wilson lines in the two different types of magnetic quivers has been uncovered in \S \ref{sec:Wilson_unframed_MagQuiv}. Again, the evidence stems from explicitly computing B-twisted indices.

The results clearly indicate that the unitary and orthosymplectic magnetic quivers agree not only on their moduli spaces, but also have equal superconformal indices, and they admit matching Wilson line defects. In \eqref{single-Wilson}, we demonstrated the agreement for a Wilson line with higher representations in the $E_4$ quivers. It would be intriguing to find a rule of the agreement for Wilson lines with higher representations in the two magnetic quivers.  In \S\ref{sec:vortex_in_mirrors}, we propose the matching of vortex defects in the unitary and orthosymplectic quivers mirror dual to the fundamental Wilson line in $\sprm(k)$ SQCD. Nevertheless, there is a large class of vortex-type line defects that preserve the A-type supercharge \cite{Dimofte:2019zzj}. Finding the correspondence of vortex line defects in the two magnetic quivers in \S\ref{sec:5d_Sp} is also a consequential task left for future work.  In order to further study A-type line defects in the proposed duality, it would be desirable to, firstly, improve the computability of A-twisted indices and, secondly, clarify the vortex line defects in orthosymplectic quivers.

Other known cases of a duality between unitary and orthosymplectic quivers include, for instance, different 3d mirrors of certain Argyres-Douglas theories $(A_{4m−1},D_3) \cong  (A_{4m−1},A_3)$ as well as $(D_3,D_{2n+2}) \cong (A_3,D_{2n+2}) $ \cite{Carta:2021whq,Carta:2021dyx}, due to the $A_3\cong D_3 $ Lie algebra isomorphism. So far, the proposed duality has been checked by the match of Higgs/Coulomb Hilbert series in \cite{Carta:2021whq,Carta:2021dyx}, but it would be desirable to extend the analysis along the lines of this paper.

Let us discuss another interesting open problem. In \S\ref{sec:UOSp-3d},  the unitary and orthosymplectic mirror theories in \eqref{eq:U_mirror_Nf=2k+1} and \eqref{eq:OSp_mirror_Nf=2k+1}, respectively, for $\sprm(1)$ theory with 3 flavours are obtained via mirror symmetry.  At the same time, the unitary $A_3$ quiver theory in \eqref{eq:U_mirror_Nf=2k+1} (for $k=1$) is mirror dual to $\urm(1)$ theory with 4 flavours. The coincidence of the $A_3$ and $D_3$ unitary quiver has been observed long ago in \cite{Kapustin:1998fa}, and recently revisited from the viewpoint of $S^3$ partition functions \cite[\S4.1]{Dey:2021rxw}. As a result, we have a web of the following 3d $\Ncal =4$ dualities.
\begin{equation}
\raisebox{-.5\height}{
\begin{tikzpicture}
	\begin{pgfonlayer}{nodelayer}
		\node [style=bluegauge] (0) at (3, -0.25-4) {};
		\node [style=redgauge] (1) at (4.25, -0.25-4) {};
		\node [style=redgauge] (4) at (1.75, -0.25-4) {};
		\node [style=none] (5) at (3, -0.75-4) {2};
		\node [style=none] (6) at (4.25, -0.75-4) {2};
		\node [style=none] (9) at (1.75, -0.75-4) {2};
		\node [style=flavorRed] (10) at (3, 0.75-4) {};
		\node [style=none] (11) at (3, 1.25-4) {2};
		\node [style=gauge3] (12) at (-5.25, -0.25-4) {};
		\node [style=gauge3] (13) at (-4.25, -0.25-4) {};
		\node [style=gauge3] (14) at (-3.25, -0.25-4) {};
		\node [style=none] (17) at (-5.25, 1.25-4) {1};
		\node [style=none] (18) at (-3.25, 1.25-4) {1};
		\node [style=none] (19) at (-5.25, -0.75-4) {1};
		\node [style=none] (20) at (-4.25, -0.75-4) {1};
		\node [style=none] (21) at (-3.25, -0.75-4) {1};
		\node [style=none] (22) at (-2.25, 0-4) {};
		\node [style=none] (23) at (1, 0) {};
		\node [style=blankflavor] (24) at (-5.25, 0.75-4) {};
		\node [style=blankflavor] (25) at (-3.25, 0.75-4) {};
	\end{pgfonlayer}
	\begin{pgfonlayer}{edgelayer}
		\draw (4) to (0);
		\draw (0) to (1);
		\draw (10) to (0);
		\draw (12) to (14);
		\draw (24) to (12);
		\draw (25) to (14);
	\end{pgfonlayer}
	\begin{pgfonlayer}{nodelayer}
		\node [style=gauge3] (0) at (-5, -4+4) {};
		\node [style=flavour2] (1) at (-3.5, -4+4) {};
		\node [style=none] (2) at (-5, -4.5+4) {1};
		\node [style=none] (3) at (-3.5, -4.5+4) {4};
	\end{pgfonlayer}
	\begin{pgfonlayer}{edgelayer}
		\draw (0) to (1);
	\end{pgfonlayer}
    \node [style=bluegauge] (5) at (2.2, -4+4) {};
		\node [style=flavorRed] (6) at (3.7, -4+4) {};
		\node [style=none] (7) at (2.2, -4.5+4) {2};
		\node [style=none] (8) at (3.7, -4.5+4) {6};
	\begin{pgfonlayer}{edgelayer}
		\draw (5) to (6);
	\end{pgfonlayer}
		\node [style=none] (24) at (-2.25, -4) {};
		\node [style=none] (25) at (1,-4) {};
		\node [style=none] (26) at (-4.25,-1) {};
		\node [style=none] (27) at (-4.25,-2.25) {};
		\node [style=none] (28) at (3,-1){};
		\node [style=none] (29) at (3,-2.25){};
    \node [style=none] (30) at (-2.25,-2.5){};
    \node [style=none] (31) at (1,-1){};
		\draw [style=<->] (24.center) to node [above] {U/OSp duality}  (25.center);
		\draw [style=<->] (26.center) to node [left] {MS${}_1$} (27.center);
		\draw [style=<->] (29.center) to node [left] {MS${}_3$} (28.center);
   	\draw [style=<->] (30.center) to (31.center);
   	\node at (-1.1,-1.6) {MS${}_2$};
\end{tikzpicture}
}\label{MS3}
\end{equation}

The superconformal indices of these theories agree, see Table \ref{tab:SCI_Sp_mirrors}. Now let us consider the $S^3$ partition functions of these theories.
The $S^3$ partition function of the SQED is given in\cite[\S2.2]{Benvenuti:2011ga}
\begin{align}\label{SQED-S3}
Z_{(1)-[4]}^{S^3}=&\int \! ds \frac{e^{2\pi i(\xi_1-\xi_2) s}}{\ch(s-m_1)\ch(s-m_2)\ch(s-m_3)\ch(s-m_4)}\cr
 =&\frac{i}{i^3(e^{\pi(\xi_1-\xi_2)}-e^{-\pi(\xi_1-\xi_2)})}\sum_{i=1}^4\frac{e^{2\pi im_i(\xi_1-\xi_2)}}{\prod_{j\neq i}\sh(m_i-m_j)}~,
\end{align}
where $\xi_1-\xi_2$ is the FI parameter and $m_i$ ($i=1,\ldots,4$) are the mass parameters.
The $S^3$ partition function of the unitary $A_3$ quiver theory is given by
\begin{align}\label{A3-S3}
    Z_{A_3}^{S^3}=&\int \! \prod_{j=1}^3 ds_j \frac{e^{2\pi i\sum_{j=1}^3(\xi_j-\xi_{j+1}) s_i}}{\ch(m_1-s_1)\ch(s_1-s_2)\ch(s_2-s_3)\ch(s_3-m_2)}\cr
    =&\int  \prod_{j=1}^3 ds_j  \prod_{k=1}^4dt_k\frac{e^{2\pi i\sum_{j=1}^3(\xi_j-\xi_{j+1}) s_i} e^{-2\pi i(t_1(m_1-s_1)+t_2(s_1-s_2)+t_3(s_2-s_3)+t_4(s_3-m_2))}}{\prod_{k=1}^4\ch(t_k)}\cr
  =& e^{2\pi i (-m_1\xi_1+m_2\xi_4)} \int \! dt_1 \frac{e^{2\pi i(m_1-m_2) t_1}}{\ch(t_1-\xi_1)\ch(t_1-\xi_2)\ch(t_1-\xi_3)\ch(t_1-\xi_4)} \,,
\end{align}
where $\xi_i,-\xi_{i+1}$ ($i=1,2,3$)  is the FI parameter of the $i$-th $\urm(1)$ gauge node from the left, and $m_{1,2}$ are the mass parameters of the flavours.
From the second to the third line, performing $ds_j$ ($j=1,2,3$) integrals, we obtain the delta functions that impose the relations
$$t_{2} = t_{1}+\xi_{1}-\xi_{2}~,\quad t_{3} = t_{1}+\xi_{1}-\xi_{3}~,\quad  t_{4} = t_{1}+\xi_{1}-\xi_{4}~ ,$$
and we then integrate over $dt_{k}$ ($k=2,3,4$). It is straightforward to verify that the last line is the same integral as in \eqref{SQED-S3} up to a factor under the exchange of the FI and mass parameters, which shows the mirror symmetry MS${}_1$ in \eqref{MS3}. To compute the sphere partition function of the $\sprm(1)$ with 3 flavours, we introduce an UV regulator $\xi$ and take the limit $\xi\to 0$ by using the L'Hospital rule:
\begin{equation}\label{SQCD-S3}
  Z_{\textrm{SQCD}}^{S^3}=\lim_{\xi\to0}\int ds \frac{e^{2\pi i \xi s}\sh^2(2s)}{\prod_{i=1}^3\ch(\pm s-m_i)}= \sum_{i=1}^3\frac{m_i\sh(2m_i)}{\prod_{j\neq i}\sh(m_j\pm m_i)}~.
\end{equation}
This is equal to the limit $m_i\to0$ of \eqref{A3-S3} with the substitution $\xi_4=\xi_1+\xi_2+\xi_3$, up to the same mirror parameter changes ($\xi_i\leftrightarrow m_i$). This confirms the mirror symmetry MS${}_2$ in \eqref{MS3}. Note that it is necessary to take the limit $\xi\to 0$ in \eqref{SQCD-S3} for this equality. Therefore, the regulator $\xi$ does not capture the emergent $\sorm(2)$ symmetry at IR.
To show the mirror symmetry  MS${}_3$ and the unitary/orthosymplectic duality, we take the limit $\xi_j\to0$
\begin{align}\label{A3-S3-limit}
  \lim_{\xi_j\to0}  Z_{A_3}^{S^3}=\frac{(m_1-m_2)\left(1+(m_1-m_2)^{2}\right) }{6\sh(m_1-m_2)}~.
\end{align}
It is also straightforward to compute the $S^3$ partition function of the orthosymplectic quiver theory by introducing regulators
\begin{align}
    Z_{\textrm{OSp}}^{S^3}=\lim_{\xi_i,\eta\to0}\int dsdt_1dt_2 \frac{e^{2\pi i(\xi_1 t_1+\xi_2 t_2 +\eta s)}\sh^2(2s)}{\ch(\pm s-m)\prod_{i=1,2}\ch(t_i\pm s)}
    =\frac{m\left(1+4 m^{2}\right) }{3\sh(2m)}~.
\end{align}
This agrees with \eqref{A3-S3-limit} with the identification of the mass parameters
\begin{equation}\label{mass-id}
  m=\frac{m_1-m_2}{2}~.
\end{equation}

Next, let us consider line operators dual to a Wilson loop in the SQED with 4 flavours.  The Wilson loop with charge $q$ results in the shift $\xi_1\to \xi_1-iq$ in \eqref{SQED-S3}.
The dual vortex in the unitary $A_3$ quiver theory is described as supersymmetric quantum mechanics that couples to two adjacent 3d $\urm(1)$ nodes (including both gauge and flavour nodes). One configuration is illustrated in \eqref{MS3-line}, which is also equivalent to their hopping-dual configurations. Note that the directions of the arrows for 1d chiral are for $q<0$ in the unitary $A_3$ quiver of \eqref{MS3-line}, and the directions become opposite for $q>0$. In fact, it is straightforward to see that this vortex produces the shift $m_1\to m_1-iq$ in the $S^3$ partition function. Likewise, the vortex in the orthosymplectic quiver of \eqref{MS3-line} becomes the dual configuration when $q$ is an even negative integer due to \eqref{mass-id}. Namely, it introduces the shift $m\to m+i|q|/2$ in the $S^3$ partition function.  We do not know a description of the dual vortex defect in the orthosymplectic quiver for odd $q$.
Moreover, for any value of charge $q$, it remains an open problem to describe a dual B-type defect in the $\sprm(1)$ SQCD with 3 flavours because the theory lacks an explicit FI parameter.
\begin{equation}
\raisebox{-.5\height}{
\begin{tikzpicture}
	\begin{pgfonlayer}{nodelayer}
		\node [style=bluegauge,label=north west:{2}] (0) at (3, -0.25-4) {};
		\node [style=flavorRed,label=above:{2}] (1) at (4.25, -0.25-4) {};
		\node [style=redgauge,label=above:{2}] (4) at (1.75, -0.25-4) {};
		\node [style=redgauge] (10) at (3, 0.75-4) {};
		\node [style=none] (11) at (3, 1.25-4) {2};
		\node [style=gauge3,label=left:{1}] (12) at (-5.25, -0.25-4) {};
		\node [style=gauge3,label=above:{1}] (13) at (-4.25, -0.25-4) {};
		\node [style=gauge3,label=right:{1}] (14) at (-3.25, -0.25-4) {};
		\node [style=none] (17) at (-5.25, 1.25-4) {1};
		\node [style=none] (18) at (-3.25, 1.25-4) {1};
		\node [style=none] (22) at (-2.25, 0-4) {};
		\node [style=none] (23) at (1, 0) {};
		\node [style=blankflavor] (24) at (-5.25, 0.75-4) {};
		\node [style=blankflavor] (25) at (-3.25, 0.75-4) {};
	\end{pgfonlayer}
  \node (SQM1) [defect,label=below:{SQM},label=right:{$\frac{|q|}{2}$}]  at (-4.75+8.5,-5.75) {};
  \begin{scope}[decoration={markings,mark =at position 0.5 with {\arrow{stealth}}}]
\draw[postaction={decorate},color=red] (1)--(SQM1) ;
  \draw[postaction={decorate},color=red] (SQM1)--(0);
\end{scope}
    \draw[dashed] (-6+8.5,-5)--(-3.5+8.5,-5);
    \node at (-6.5+11.5,-5.5) {1d};
    \node at (-6.5+11.5,-4.5) {3d};
	\begin{pgfonlayer}{edgelayer}
		\draw (4) to (0);
		\draw (0) to (1);
		\draw (10) to (0);
		\draw (12) to (14);
		\draw (24) to (12);
		\draw (25) to (14);
	\end{pgfonlayer}
	\begin{pgfonlayer}{nodelayer}
		\node [style=gauge3,label=above:{$\mathcal{W}^q$}] (0) at (-5, -4+4) {};
		\node [style=flavour2] (1) at (-3.5, -4+4) {};
		\node [style=none] (2) at (-5, -4.5+4) {1};
		\node [style=none] (3) at (-3.5, -4.5+4) {4};
	\end{pgfonlayer}
	\begin{pgfonlayer}{edgelayer}
		\draw (0) to (1);
	\end{pgfonlayer}
    \node [style=bluegauge] (5) at (2.2, -4+4) {};
		\node [style=flavorRed] (6) at (3.7, -4+4) {};
		\node [style=none] (7) at (2.2, -4.5+4) {2};
		\node [style=none] (8) at (3.7, -4.5+4) {6};
	\begin{pgfonlayer}{edgelayer}
		\draw (5) to (6);
	\end{pgfonlayer}
		\node [style=none] (24) at (-2.25, -4) {};
		\node [style=none] (25) at (1,-4) {};
		\node [style=none] (26) at (-4.25,-1) {};
		\node [style=none] (27) at (-4.25,-2.25) {};
		\node [style=none] (28) at (3,-1){};
		\node [style=none] (29) at (3,-2.25){};
    \node [style=none] (30) at (-2.25,-2.5){};
    \node [style=none] (31) at (1,-1){};
		\draw [style=<->] (24.center) to node [above] {U/OSp duality} node [below] {$q<0,q:$even} (25.center);
		\draw [style=<->] (26.center) to node [left] {MS${}_1$} (27.center);
		\draw [style=<->] (29.center) to node [left] {MS${}_3$} (28.center);
   	\draw [style=<->] (30.center) to (31.center);
   	\node at (-1.1,-1.6) {MS${}_2$};
    \node (SQM) [defect,label=below:{SQM},label=right:{$|q|$}]  at (-4.75,-5.75) {};
    \begin{scope}[decoration={markings,mark =at position 0.5 with {\arrow{stealth}}}]
  \draw[postaction={decorate},color=red] (12)--(SQM) ;
    \draw[postaction={decorate},color=red] (SQM)--(13);
  \end{scope}
      \draw[dashed] (-6,-5)--(-3.5,-5);
      \node at (-6.5,-5.5) {1d};
      \node at (-6.5,-4.5) {3d};
\end{tikzpicture}
}\label{MS3-line}
\end{equation}
As seen in \S\ref{sec:mirror_Wilson_vortex} and Appendix \ref{app:sphere}, we propose the vortex configurations in the two mirror theories dual to the fundamental Wilson line of the $\sprm(1)$ SQCD for $N_f>3$.
However, we do not know how to describe A-type line defects mirror dual to the fundamental Wilson line of the $\sprm(1)$ SQCD with 3 flavours. Only with 3 flavours, we cannot manipulate the integrand \eqref{SQCD-S3} of the $\sprm(1)$ SQCD into the integrand \eqref{A3-S3} of the unitary mirror quiver since $\sh^2$ in the numerator of the integrand \eqref{SQCD-S3} cannot be removed by the Cauchy determinant formula. This is different from the computations in \eqref{k1-vortex}. Hence, it is interesting to find a description of A-type line defects in the two mirror theories dual to the fundamental Wilson line of the $\sprm(1)$ SQCD with 3 flavours. Supersymmetric enhancement from 1d $\Ncal=2$ to $\Ncal=4$ may be a potential approach for this problem. We leave these open problems to future research.

\paragraph{Acknowledgements.}
We would like to thank Jin Chen, Stefano Cremonesi, Anindya Dey, Tudor Dimofte, Dongmin Gang, Amihay Hanany, Ken Kikuchi, Sung-Soo Kim, and Futoshi Yagi for discussion and correspondence.
 S.N. would like to thank Yau Center, Tsinghua University for the warm hospitality where part of the work was carried out and preliminary results were presented. The research of S.N. is supported by the National Science Foundation of China (No.12050410234) and Fudan University Original Project (No. IDH1512092/002).
 M.S. and Z.Z. are grateful for the warm hospitality of Fudan University, Department of Physics during various stages of this work.
M.S. was further supported by the National Natural Science Foundation of China (grant no.\ 11950410497), and the China Postdoctoral Science Foundation (grant no.\ 2019M650616).
%
\appendix
\section{Background material}
\label{app:brane}
In this appendix, some basics about Type IIB brane configurations for 3d $\Ncal=4$ theories are summarised. In addition, some notations are laid out.

\subsection{\texorpdfstring{Brane realisation of 3d $\Ncal=4$ theories}{Brane realisation of 3d N=4 theories}}
The starting point is the D3-D5-NS5 setup of \cite{Hanany:1996ie} in which the branes occupy space-time dimension as summarised in Table \ref{tab:branes}. Each brane individually breaks half the supercharges; however, the arrangement is such that any two of the three branes imply that the third brane can be added without breaking supersymmetry further. Hence, the system has 8 supercharges. The crucial feature of the D3-D5-NS5 setup \cite{Hanany:1996ie} is that the system breaks the 10d space-time symmetry
\begin{align}
    \sorm(1,9) \to \sorm(1,2) \times \sorm(3)_{3,4,5} \times \sorm(3)_{7,8,9}
    \label{eq:breaking_Poincare}
\end{align}
and the rotational symmetries  $\sorm(3)_{3,4,5}\subset \surm(2)_C$ and $\sorm(3)_{7,8,9}\subset \surm(2)_H$ realise the 3d $\Ncal=4$ R-symmetry $\sorm(4)_R \cong \surm(2)_C \times \surm(2)_H$ geometrically.

Starting from the Type IIB brane setup D3-D5-NS5, the resulting classes of 3d $\Ncal=4$ quiver gauge theories can be enriched by including orientifolds and orbifold planes. There are three types to consider: O3, O5, and ON planes, and the reader is referred to \cite{Uranga:1998uj,Gimon:1996rq,Hanany:1997gh,Kapustin:1998fa,Hanany:1999sj,Feng:2000eq,Gaiotto:2008ak} for details.

\begin{table}[ht]
    \centering
    \begin{tabular}{ccccccccccc}
    \toprule
 IIB & $0$ & $1$ & $2$ & $3$ & $4$ & $5$ & $6$ & $7$ & $8$ & $9$ \\ \midrule
 NS5/ON & $\times$ & $\times$& $\times$& $\times$& $\times$& $\times$ \\
 D3/O3 &  $\times$ & $\times$& $\times$ & & & & $\times$ \\
 D5/O5 &  $\times$ & $\times$& $\times$ & & & & & $\times$  & $\times$ & $\times$  \\
 & \multicolumn{3}{c}{$\leftarrow \R^{1,2} \rightarrow$}  & \multicolumn{3}{|c|}{$\underbrace{\leftarrow \R^3_{3,4,5} \rightarrow}_{\circlearrowleft \surm(2)_C}$} & & \multicolumn{3}{|c}{$\underbrace{\leftarrow \R^3_{7,8,9} \rightarrow}_{\circlearrowleft \surm(2)_H}$}\\
\bottomrule
 \end{tabular}
    \caption{Space-time occupation of the D3-D5-NS5 setup of \cite{Hanany:1996ie}. Likewise, O3 or O5 planes occupy the same directions as D3 or D5 branes, respectively; while ON planes are parallel to NS5 branes.}
    \label{tab:branes}
\end{table}

The graphical notation for the different branes and orientifolds is as follows:
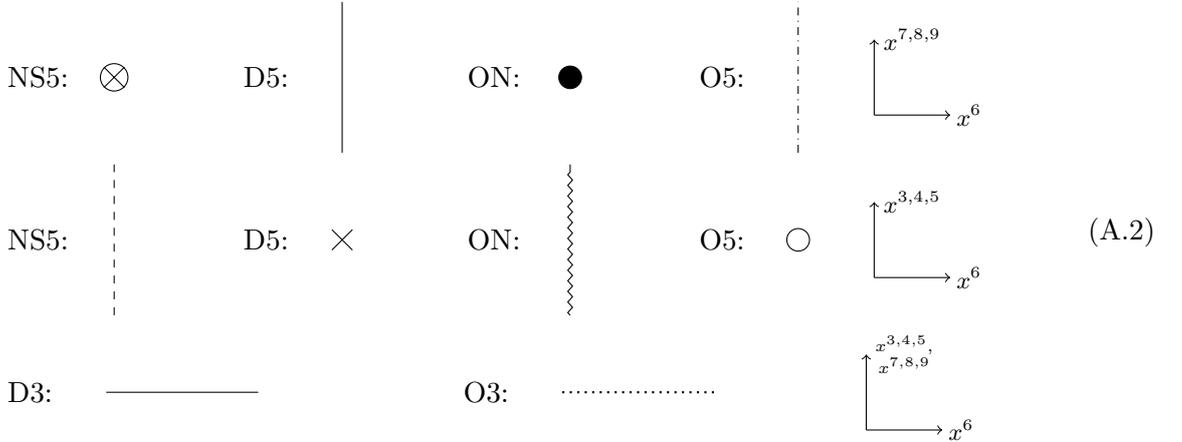
\begin{align}
    &\raisebox{-.5\height}{
        \begin{tikzpicture}
             \node at (0,0) {NS5:};
             \ns{1,0}
             \node at (3,0) {D5:};
             \draw (4,-1)--(4,1);
             \node at (6,0) {ON:};
             \node at (7,0) [defect] {};
             \node at (9,0) {O5:};
             \draw[dashdotted] (10,-1)--(10,1);
            \draw[->] (11,-0.5)--(12,-0.5);
            \draw[->] (11,-0.5)--(11,0.5);
            \node at (12.25,-0.5) {\footnotesize{$x^6$}};
            \node at (11.5,0.5) {\footnotesize{$x^{7,8,9}$}};
        \end{tikzpicture}
    } \notag \\
    &\raisebox{-.5\height}{
        \begin{tikzpicture}
             \node at (0,0) {NS5:};
             \draw[dashed] (1,-1)--(1,1);
             \node at (3,0) {D5:};
             \dfive{4,0}
            \node at (6,0) {ON:};
            \draw [line join=round,decorate, decoration={zigzag, segment length=4,amplitude=.9,post=lineto,post length=2pt}] (7,-1) to (7,1);
             \node at (9,0) {O5:};
            \node at (10,0) [gauge] {};
            \draw[->] (11,-0.5)--(12,-0.5);
            \draw[->] (11,-0.5)--(11,0.5);
            \node at (12.25,-0.5) {\footnotesize{$x^6$}};
            \node at (11.5,0.5) {\footnotesize{$x^{3,4,5}$}};
        \end{tikzpicture}
    }\label{notation} \\
    &\raisebox{-.5\height}{
        \begin{tikzpicture}
             \node at (0,0) {D3:};
             \draw (1,0)--(3,0);
             \node at (6,0) {O3:};
             \draw[dotted,thick] (7,0)--(9,0);
            \draw[->] (11,-0.5)--(12,-0.5);
            \draw[->] (11,-0.5)--(11,0.5);
            \node at (12.25,-0.5) {\footnotesize{$x^6$}};
            \node at (11.5,0.5) {\footnotesize{$\substack{x^{3,4,5}, \\ x^{7,8,9}}$}};
        \end{tikzpicture}
    } \notag
\end{align}

\subsection{Quiver gauge theories}
A Quiver diagram, composed of nodes and edges, encodes a 3d $\Ncal=4$ field theory as follows:
\begin{compactitem}
\item Gauge nodes $\bigcirc$ denotes dynamical vector multiplets, whiled flavour nodes $\Box$ denote background vector multiplets. The notation used in this paper is
\begin{align}
    \raisebox{-.5\height}{
\begin{tikzpicture}
    \node [gauge,label=below:{$k$}]{};
\end{tikzpicture}}
\; \longleftrightarrow \; \urm(k)
\qquad
    \raisebox{-.5\height}{
\begin{tikzpicture}
    \node [gaugeSO,label=below:{$k$}]{};
\end{tikzpicture}}
\; \longleftrightarrow \; \sorm(k)
 \qquad
    \raisebox{-.5\height}{
\begin{tikzpicture}
    \node [gaugeSp,label=below:{$2k$}]{};
\end{tikzpicture}}
\; \longleftrightarrow \; \sprm(k)
\end{align}
and the same colour-coding is used for flavour nodes.
\item An edge between two nodes corresponds to a hypermultiplet $H=(X,Y^\dagger)$, with $X,Y$ two $\Ncal=2$ chiral multiplets. In a unitary quiver, an edge represents
\begin{align}
    \raisebox{-.5\height}{
\begin{tikzpicture}
    \node (g1) [gauge,label=below:{$k_1$}]{};
    \node (g2) [gauge,right of=g1,label=below:{$k_2$}]{};
    \draw (g1)--(g2);
\end{tikzpicture}}
\quad \longleftrightarrow  \quad
\text{bifundamental hyper: } \quad
H=(X,Y^\dagger)\in \mathbf{k_1}\otimes \overline{\mathbf{k_2}}
\end{align}
\textit{i.e.}\ each chiral transforms as bifundamental
$ X\in \mathbf{k_1}\otimes \overline{\mathbf{k_2}}$ and
$Y\in \overline{\mathbf{k_1}}\otimes \mathbf{k_2}$. On the other hand, in an orthosymplectic quiver, an edge stands for
\begin{align}
    \raisebox{-.5\height}{
\begin{tikzpicture}
    \node (g1) [gaugeSO,label=below:{$k_1$}]{};
    \node (g2) [gaugeSp,right of=g1,label=below:{$2k_2$}]{};
    \draw (g1)--(g2);
\end{tikzpicture}}
\quad \longleftrightarrow  \quad
\text{half-hypermultiplet:} \quad
h=(X,JY)\in \mathbf{k_1}\otimes \mathbf{2k_2}
\end{align}
but the individual chirals \emph{do not} transform as $ \mathbf{k_1}\otimes \mathbf{2k_2}$. Only the use of the $\sprm(k_2)$ invariant tensor $J$ allows to define $h=(X,JY)$ that transforms suitably.
\end{compactitem}

An important concept is the classification of 3d $\Ncal=4$ theories as \emph{good}, \emph{ugly}, or \emph{bad} based on the \emph{balance} of the gauge groups \cite{Gaiotto:2008ak}. For a quiver with a $G=\urm(n),\sorm(n), \sprm(n)$ gauge node which is connected to $N_f$ fundamental flavours, the node is \emph{good} if
\begin{align}
\label{eq:def_balance}
    \text{for} \;\; \urm(n): \; N_f \geq 2n
    \,, \qquad
    \text{for} \;\; \sorm(n): \; N_f \geq n-1
    \,, \qquad
    \text{for} \;\; \sprm(n):  \; N_f \geq  n+1 \,.
\end{align}
If the equality holds, the node is called \emph{balanced} and Coulomb branch global symmetry is expected to be enhanced to some non-abelian group due to monopole operators. If the inequality is strict, the node is referred to as \emph{over-balanced}.

\subsection{Brane realisation of line defects}
3d $\Ncal=4$ SCFTs admit two physically distinct classes of half-BPS line defects that are supported on a straight line in a flat space. The preserved symmetries distinguish these. As detailed in \cite{Assel:2015oxa}, superconformal line defects with support on a time-like line are invariant under either a $\urm(1,1|2)_W$ or $\urm(1,1|2)_V$ subalgebra of the full 3d $\Ncal=4 $ superconformal algebra $\mathrm{OSp}(4|4)$.

Given a UV Lagrangian description of a 3d $\Ncal=4$ SCFT, a line defect can be made invariant under half of the supercharges of the 3d $\Ncal=4$ Poincar\'e supersymmetry algebra of the UV theory.

There are two inequivalent 1d $\Ncal=4$ supersymmetric quantum mechanics (SQM) subalgebras of the 3d $\Ncal=4$ Poincar\'e subalgebra that can be preserved by a line defect.
$\mathrm{SQM}_W$ preserves $\urm(1)_C\times \surm(2)_H$, while $\mathrm{SQM}_V$ preserves $\surm(2)_C\times \urm(1)_H$.

In a 3d $\Ncal=4$ gauge theory, two classes of line defects can be defined, one of which preserves $\mathrm{SQM}_W$, while the other preserves $\mathrm{SQM}_V$. These line defects are then expected to flow to superconformal defects in the IR, which preserve  $\urm(1,1|2)_W$ or $\urm(1,1|2)_V$, respectively.

Both classes of supersymmetric line defect can be realised in a UV 3d $\Ncal=4$ theory as follows: a certain 1d $\Ncal=4$ SQM theory is coupled to the bulk 3d $\Ncal=4$ theory. Such a coupling is canonically realised by gauging the flavour symmetries of the SQM theory with 3d $\Ncal=4$ vector multiplets. In addition, superpotential couplings between the defect and bulk fields need to be specified in the construction.
For this work, the relevant line defects are realised in brane configurations.

\paragraph{Brane realisation of a Wilson line defect.}
A Wilson line defect can be introduced by an F1 string stretched along one of the $x^{3,4,5}$ directions; to be specific, say F1 is extended along the $x^5$ direction. On one side, the F1 ends on a D3 brane stack. The inclusion of the F1 breaks the rotational symmetries in \eqref{eq:breaking_Poincare} further to
\begin{align}
    \sorm(3)_{3,4,5} \times \sorm(3)_{7,8,9} \to \sorm(2)_{3,4} \times \sorm(3)_{7,8,9} \subset \urm(1)_C \times \surm(2)_{H}
\end{align}
which is compatible with the $\mathrm{SQM}_W$ subalgebra.

On the other end, the F1 can be terminated on a D5 or D5${}^\prime$ brane without breaking any further supersymmetries, see Figure \ref{fig:Wilson_branes}. Depending on D5 or D5${}^\prime$, the representation of the Wilson line changes \cite{Assel:2015oxa}, as reviewed now. (See also \cite{Yamaguchi:2006tq,Gomis:2006sb,Gomis:2006im} for an earlier brane realisation of Wilson loops.)

\begin{figure}[ht]
    \centering
\begin{subfigure}{0.6\textwidth}
\centering
    \begin{tabular}{ccccccccccc}
    \toprule
 IIB & $0$ & $1$ & $2$ & $3$ & $4$ & $5$ & $6$ & $7$ & $8$ & $9$ \\ \midrule
 NS5 & $\times$ & $\times$& $\times$& $\times$& $\times$& $\times$ \\
 D3 &  $\times$ & $\times$& $\times$ & & & & $\times$ \\
 D5 &  $\times$ & $\times$& $\times$ & & & & & $\times$  & $\times$ & $\times$  \\
 F1 &  $\times$ & & & & & $\times$ & & & &  \\
 D5${}^\prime$ &  $\times$  & & & $\times$& $\times$ & & & $\times$  & $\times$ & $\times$  \\
 & \multicolumn{3}{c}{$\leftarrow \R^{1,2} \rightarrow$}  & \multicolumn{2}{|c|}{$\underbrace{ \R^2_{3,4}}_{\circlearrowleft \urm(1)_C}$} & & & \multicolumn{3}{|c}{$\underbrace{\leftarrow \R^3_{7,8,9} \rightarrow}_{\circlearrowleft \surm(2)_H}$}\\
\bottomrule
 \end{tabular}
 \caption{}
\end{subfigure}
\begin{subfigure}{0.38\textwidth}
\centering
\begin{tikzpicture}
    \draw[dashed] (0,-1)--(0,1) (4,-1)--(4,1);
    \draw (0,0.05)--(4,0.05) (0,0.15)--(4,0.15) (0,-0.15)--(4,-0.15);
    \node at (2,-0.05) {$\ldots$};
    \draw [red,line join=round,decorate, decoration={zigzag, segment length=4,amplitude=.9,post=lineto,post length=2pt}] (2,0.05) -- (2,1);
    \dfive{2,1}
    \node at (2.3,0.5) {\footnotesize{F1}};
    \node at (1,-0.4) {\footnotesize{$N$ D3}};
    \node at (2.75,1.1) {\footnotesize{D5/D5${}^\prime$}};
    \draw[->] (3-0.25,2.15)--(4-0.25,2.15);
    \draw[->] (3-0.25,2.15)--(3-0.25,3.15);
    \node at (4.2-0.25,2.15) {\footnotesize{$x^6$}};
    \node at (3.3-0.25,3.6-0.35) {\footnotesize{$x^5$}};
\end{tikzpicture}
\caption{}
\end{subfigure}
    \caption{Brane realisation of Wilson line defects.}
    \label{fig:Wilson_branes}
\end{figure}

Suppose an F1 is stretched between a stack of $N$ D3s and a D5${}^\prime$ brane. The quantisation of the open string states between the D3 and the D5${}^\prime$ gives rise to a 1d complex Fermion in the bifundamental representation of $\urm(1)\times\urm(N)$, with mass $m \sim x^5$ set by the location of the D5${}^\prime$.
Considering $k$ F1s and integrating out the Fermions gives rise to the insertion of a supersymmetric Wilson loop operator in the $k$-th antisymmetric representation $\Lambda^k$ of $\urm(N)$. In terms of branes, the weights of $\Lambda^k$ are realised by the different ways the $k$ F1s can be suspended between the D3 and D5${}^\prime$. The s-rule dictates that there can only be $0$ or $1$ F1 stretched between the same pair of D3 and D5${}^\prime$ branes.

Analogously, the quantisation of an F1 between a stack of $N$ D3s and a D5 brane gives rise to a heavy hypermultiplet in the bifundamental representation of $\urm(1) \times \urm(N)$. Considering $k$ F1s and integrating out the heavy hypermultiplets yields the insertion of a supersymmetric Wilson loop operator in the symmetric representation $\mathcal{S}^k$ of $\urm(N)$. The weights of $\mathcal{S}^k$ are manifest in the brane configuration due to the s-rule. The number of F1s between the same pair of a D3 and a D5 is not restricted.

\paragraph{Brane realisation of a vortex line defect.}
A vortex line defect, as a natural candidate for a 3d mirror of a Wilson line, is constructed from a D1 string stretched along one of the $x^{7,8,9}$ directions; to be specific, D1 is extended along the $x^9$ direction. On one end, the D-string ends on a stack of D3 branes. The addition of the D1 breaks the rotational symmetries in \eqref{eq:breaking_Poincare} further to
\begin{align}
    \sorm(3)_{3,4,5} \times \sorm(3)_{7,8,9} \to \sorm(3)_{3,4,5} \times \sorm(2)_{7,8} \subset \surm(2)_C \times \urm(1)_H
\end{align}
which is compatible with the $\mathrm{SQM}_V$ subalgebra.

\begin{figure}[ht]
    \centering
\begin{subfigure}{0.6\textwidth}
\centering
    \begin{tabular}{ccccccccccc}
    \toprule
 IIB & $0$ & $1$ & $2$ & $3$ & $4$ & $5$ & $6$ & $7$ & $8$ & $9$ \\ \midrule
 NS5 & $\times$ & $\times$& $\times$& $\times$& $\times$& $\times$ \\
 D3 &  $\times$ & $\times$& $\times$ & & & & $\times$ \\
 D5 &  $\times$ & $\times$& $\times$ & & & & & $\times$  & $\times$ & $\times$  \\
 D1 &  $\times$ & & & & & & & & & $\times$ \\
 NS5${}^\prime$ &  $\times$  & & & $\times$& $\times$ & $\times$ & & $\times$  & $\times$ &   \\
 & \multicolumn{3}{c}{$\leftarrow \R^{1,2} \rightarrow$}  & \multicolumn{3}{|c|}{$\underbrace{\leftarrow \R^3_{3,4,5} \rightarrow}_{\circlearrowleft \surm(2)_C}$}
  & & \multicolumn{2}{|c|}{$\underbrace{ \R^2_{7,8}}_{\circlearrowleft \urm(1)_H}$} & \\
\bottomrule
 \end{tabular}
 \caption{}
\end{subfigure}
\begin{subfigure}{0.38\textwidth}
\centering
\begin{tikzpicture}
    \draw (0,0.05)--(4,0.05) (0,0.15)--(4,0.15) (0,-0.15)--(4,-0.15);
    \node at (2,-0.05) {$\cdots$};
    \ns{0,0}
    \ns{4,0}
    \draw[red] (2,0.05) -- (2,1);
    \ns{2,1}
    \node at (2.3,0.5) {\footnotesize{D1}};
    \node at (1,-0.4) {\footnotesize{$N$ D3}};
    \node at (3,1.1) {\footnotesize{NS5/NS5${}^\prime$}};
    \draw[->] (3-0.25,2.15)--(4-0.25,2.15);
    \draw[->] (3-0.25,2.15)--(3-0.25,3.15);
    \node at (4.2-0.25,2.15) {\footnotesize{$x^6$}};
    \node at (3.3-0.25,3.6-0.35) {\footnotesize{$x^9$}};
\end{tikzpicture}
\caption{}
\label{subfig:vortex_brane_pic}
\end{subfigure}
    \caption{Brane realisation of vortex line defects.}
    \label{fig:Vortex_branes}
\end{figure}
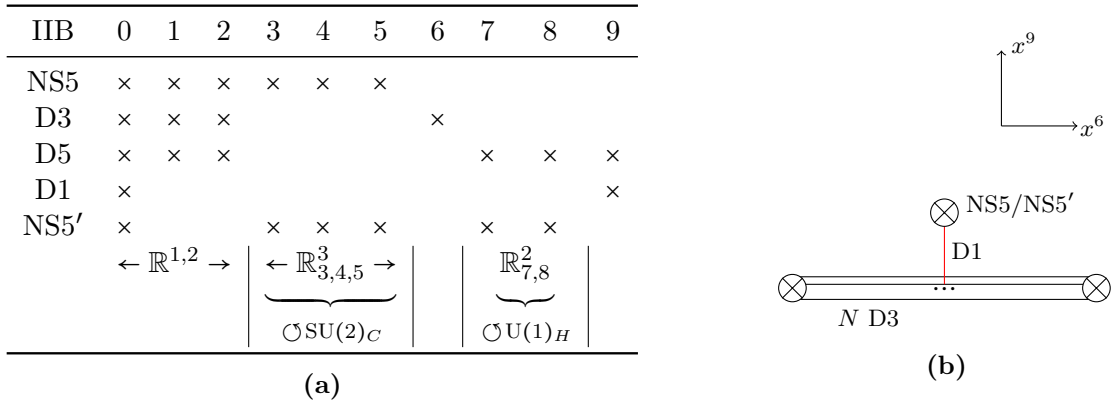

On the other end, the D1 can end on an NS5 or NS5${}^\prime$ brane without breaking supersymmetry further, see Figure \ref{fig:Vortex_branes}. Depending on the type of 5-brane, the properties of the 1d defect theory changes. (See also \cite{Alday:2009fs} for a brane realisation of codimension 2 defects in 4d, based on \cite{Hanany:1997vm}.)

To derive the 1d $\Ncal=4$ gauge theory description of the vortex defect, it is convenient to align the defect branes, \textit{i.e.}\ the D1s and the additional 5-branes, with either one of the nearest NS5 branes in the main brane configuration. As evident from Figure \ref{subfig:vortex_brane_pic}, there are, in general, two NS branes that are closest. This gives rise to a 1d theory defined by moving the defect to the nearest NS5 brane on the left-hand side and another 1d theory by moving to the right-hand side.

Gauge theory parameters have an interpretation in the brane setup. The distance of the NS5/NS5${}^\prime$ in $x^9$ direction gives the inverse gauge coupling. The relative distance in $x^6$ direction yields the FI parameter for the 1d theory. Therefore, Figure \ref{subfig:vortex_brane_pic} can be understood as the FI $>0$ deformation of the 1d theory associated to the left NS5, or as the FI $<0$ deformation of the 1d theory associated to the right NS5.

According to \cite{Assel:2015oxa}, the 1d $\Ncal=4$ quiver gauge theory is read off by the following rules:
\begin{compactitem}
\item A stack of $k$ D1s between an NS5 and an NS5${}^\prime$ yields a $\urm(k)$ vector multiplet.
\item A stack of $k$ D1s between two NS5s of two NS5${}^\prime$ branes yields a $\urm(k)$ vector multiplet and an adjoint chiral multiplet.
\item an NS5 or NS5${}^\prime$ with $k_1$ D1s ending from one side and $k_2$ D1s ending from the other side gives rise to a chiral multiplet in the bifundamental representation $\boldsymbol{k_1} \otimes \overline{\boldsymbol{k_2}}$ of $\urm(k_1) \times \urm(k_2)$ and a chiral multiplet in the anti-bifundamental representation $\overline{\boldsymbol{k_1}} \otimes \boldsymbol{k_2}$.
\item For $k$ D1s ending on an NS5 in the main brane configuration, which has $N_L$ D3s ending from the left and $N_R$ D3s ending from the right, we associate a bifundamental $\boldsymbol{N_R} \otimes \overline{\boldsymbol{k}}$ chiral multiplet of $\urm(N_R)\times \urm(k)$ and a bifundamental $\boldsymbol{k} \otimes \overline{\boldsymbol{N_L}}$ chiral multiplet of $\urm(k)\times \urm(N_L)$.
\end{compactitem}
To exemplify the notation relevant for this paper, consider the following brane setup:
\begin{align}
    \raisebox{-.5\height}{
    \begin{tikzpicture}
    \draw (-2,0.15)--(2,0.15) (-2,0.05)--(2,0.05)
    (-2,-0.15)--(2,-0.15);
    \node at (-1,-0.05) {$\ldots$};
    \node at (1,-0.05) {$\ldots$};
    \draw[red] (0,0)--(0,1);
    \ns{0,0}
    \ns{0,1}
    \ns{-2,0}
    \ns{2,0}
    \node at (-1,-0.35) {\footnotesize{$N_L$ D3}};
    \node at (1,-0.35) {\footnotesize{$N_R$ D3}};
    \node[red] at (0.3,0.5) {\footnotesize{D1}};
    \node at (0,1.5) {\footnotesize{ NS5/NS5${}^\prime$}};
    %
    \node (g1) [gauge,label=below:{\footnotesize{$\urm(N_L)$}}] at (6,0) {};
    \node (g2) [gauge,label=below:{\footnotesize{$\urm(N_R)$}}] at (8,0) {};
    \node (SQM) [defect,label=above:{\footnotesize{$\urm(1)$}}] at (7,1) {};
    \draw (g1)--(g2) (5,0)--(g1) (g2)--(9,0);
    \begin{scope}[decoration={markings,mark =at position 0.5 with {\arrow{stealth}}}]
    \draw[postaction={decorate},color=red] (g1) -- (SQM);
    \draw[postaction={decorate},color=red] (SQM) -- (g2);
    \draw[dashed] (5.5,0.45)--(8.5,0.45);
    \node at (9,0.75) {\footnotesize{1d}};
    \node at (9,0.25) {\footnotesize{3d}};
     \end{scope}
    \end{tikzpicture}
        }
\end{align}
such that the arrow of the chiral multiplet points towards the gauge node in which it transforms in the fundamental representation.

Adding an O3 plane along the stack of D3 branes affects 3d low-energy theory by projecting the gauge groups to orthogonal and symplectic groups. In contrast, the gauge group of the 1d SQM is not affected by the orientifold projection as the D1 branes are suspended between NS5 branes without crossing the orientifold.


\section{Superconformal index}
\label{app:superconf_index}
In \S\ref{sec:3d_Sp}, we study the two theories mirror dual to the 3d $\Ncal=4$ $\sprm(k)$ SQCD with $N_f$ flavours. Following the computational recipe in \S\ref{sec:SCI}, we calculate the superconformal indices of the 3d unitary and orthosymplectic mirror quivers in \S\ref{sec:UOSp-3d}, respectively, and check the agreement perturbatively. Here, we summarise the matching results of the superconformal indices in Table \ref{tab:SCI_Sp_mirrors}.
\begin{table}[ht]
    \centering
    \ra{1.75}
    \begin{tabular}{ccl}\toprule
    $\boldsymbol{k}$  & $\boldsymbol{N_f}$    & \textbf{superconformal index}  \\ \midrule
      $1$   & $3$ & $\begin{aligned}
1&+\sqrt{\mathfrak{q}}\left(\frac{15}{\mathfrak{t}^{2}}+\mathfrak{t}^{2}\right)+\mathfrak{q}\left(-17+\frac{84}{\mathfrak{t}^{4}}+3 \mathfrak{t}^{4}\right)
+\mathfrak{q}^{3 / 2}\left(\frac{300}{\mathfrak{t}^{6}}-\frac{173}{\mathfrak{t}^{2}}-2 \mathfrak{t}^{2}+3 \mathfrak{t}^{6}\right)\\
&+\mathfrak{q}^{2}\left(138+\frac{825}{\mathfrak{t}^{8}}-\frac{707}{\mathfrak{t}^{4}}-2 \mathfrak{t}^{4}+5 \mathfrak{t}^{8}\right)+\cdots
\end{aligned}$
\\ \midrule
      $1$   & $4$ & $\begin{aligned}
     1&+\frac{28 \sqrt{\mathfrak{q}}}{\mathfrak{t}^{2}}+\mathfrak{q}\left(-29+\frac{300}{\mathfrak{t}^{4}}+2 \mathfrak{t}^{4}\right)+\mathfrak{q}^{3 / 2}\left(\frac{1925}{\mathfrak{t}^{6}}-\frac{649}{\mathfrak{t}^{2}}-\mathfrak{t}^{2}+\mathfrak{t}^{6}\right) \\ &+\mathfrak{q}^{2}\left(376+\frac{8918}{\mathfrak{t}^{8}}-\frac{5643}{\mathfrak{t}^{4}}-\mathfrak{t}^{4}+3 \mathfrak{t}^{8}\right)+\cdots
\end{aligned}$
\\ \midrule
      $1$   & $5$ & $\begin{aligned}
    1&+\frac{45 \sqrt{\mathfrak{q}}}{\mathfrak{t}^{2}}+\mathfrak{q}\left(-46+\frac{770}{\mathfrak{t}^{4}}+\mathfrak{t}^{4}\right)+\mathfrak{q}^{3 / 2}\left(\frac{7644}{\mathfrak{t}^{6}}-\frac{1714}{\mathfrak{t}^{2}}+\mathfrak{t}^{6}\right)\\
    &+\mathfrak{q}^{2}\left(988+\frac{52920}{\mathfrak{t}^{8}}-\frac{24574}{\mathfrak{t}^{4}} -\mathfrak{t}^{4}+2 \mathfrak{t}^{8}\right)+\cdots
\end{aligned}$
\\ \midrule
      $1$   & $6$ & $\begin{aligned}1&+\frac{66 \sqrt{\mathfrak{q}}}{\mathfrak{t}^{2}}+\mathfrak{q}\left(-67+\frac{1638}{\mathfrak{t}^{4}}+\mathfrak{t}^{4}\right)+4 \mathfrak{q}^{3/2}\left(\frac{5775}{\mathfrak{t}^{6}}-\frac{929}{\mathfrak{t}^{2}}\right)\\ &+\mathfrak{q}^{2}\left(\frac{222156}{t^{8}}-\frac{78299}{t^{4}}+2143 +2 t^{8}\right)+\cdots\end{aligned}$
      \\ \midrule
      $2$   & $4$ & $\begin{aligned}
1&+\sqrt{\mathfrak{q}}\left( \frac{28}{\mathfrak{t}^{2}}+3 \mathfrak{t}^{2}\right)
+\mathfrak{q}\left(\frac{335}{\mathfrak{t}^{4}}+52 +6 \mathfrak{t}^{4}\right)
+\mathfrak{q}^{3 / 2}\left(\frac{2492}{\mathfrak{t}^{6}}+\frac{104}{\mathfrak{t}^{2}} +48 \mathfrak{t}^{2}+13 \mathfrak{t}^{6}\right) \\  &+\mathfrak{q}^{2}\left(\frac{13524}{\mathfrak{t}^{8}}- \frac{2524}{ \mathfrak{t}^{4}}-96 +4 0 \mathfrak{t}^{4}+23 \mathfrak{t}^{8}\right)+\cdots
\end{aligned}$
\\ \midrule
      $2$   & $5$ & $\begin{aligned}
1+\sqrt{\mathfrak{q}}\left(\frac{45}{\mathfrak{t}^{2}}+\mathfrak{t}^{2}\right)+\mathfrak{q}\left(-2+\frac{980}{\mathfrak{t}^{4}}+3\mathfrak{t}^{4}\right)+\cdots~.
\end{aligned}$
\\ \midrule
      $2$   & $6$ & $\begin{aligned}1+\frac{66\sqrt{\mathfrak{q}}}{\mathfrak{t}^{2}}+\cdots\end{aligned}$
      \\\bottomrule
    \end{tabular}
    \caption{Superconformal index for the unitary and orthosymplectic mirrors of $\sprm(k)$ with $N_f$ flavours.}
    \label{tab:SCI_Sp_mirrors}
\end{table}

\clearpage
\section{Twisted indices}
\label{app:twisted_indices}
Twisted indices for 3d $\Ncal=4$ theories have been introduced in \S\ref{sec:twisted_ind_def}. In this appendix, computational results supporting the statements in the main text are presented.
\subsection{\texorpdfstring{A-twisted index $\sprm(1)$,  $N_f=3,4,5,6$ flavours and vortex defect}{A-twisted index Sp1, 3,4,5,6 flavours and vortex defect}}
\label{app:A-twist_Sp1}\label{app:A-twisted}
Analogously to \cite{Assel:2015oxa}, the vortex line is introduced via a 1d SQM coupled to the 3d theory as in \eqref{fig:vortex}. The A-twisted indices of the $\sprm(1)$ theory with $N_f=3,\ldots,6$ in presence of vortex lines are presented in Table \ref{tab:vortx_SU2_results}. The notation for the splitting $L +(N_f-L)$ indicates that $L$ flavours couple to the 1d SQM.

As a remark, comparing B-twisted indices of the mirror theories with Wilson line defects and the A-twisted indices in Table \ref{tab:vortx_SU2_results}, we need to take care of some prefactors. In \S\ref{sec:Wilson_mirrors}, the B-twisted index has been computed via the Higgs branch Hilbert series with insertion of a Wilson line. The twisted index and the Hilbert series match up to overall prefactors; see \cite{Closset:2016arn}. Likewise, the computations in Table \ref{tab:vortx_SU2_results} display the A-twisted index, which again is related to a Coulomb branch Hilbert series by overall factors.

\begin{table}[ht]
    \begin{subtable}[t]{0.48\textwidth}
    \centering
      \ra{1.2}
    \begin{tabular}{cl} \toprule
      Vortex line   &   A-twisted index  \\ \midrule
      3+0   &$-\frac{ t (1+t^2)}{(-1+t^2)^2 (1+t)^2}$ \\
      2+1   &$-\frac{2 t^2}{(-1+t)^2 (1+t)^2}$  \\
      1+2  &$-\frac{t(1+t^2)}{(-1+t^2)^2}$\\
      0+3  &$\frac{(1+t^2)^2}{2(-1+t^2)^2}$  \\ \bottomrule
    \end{tabular}
    \caption{$N_f=3$}
    \label{tab:vortex_SU2_Nf=3}
    \end{subtable}
    \hfill
    \begin{subtable}[t]{0.48\textwidth}
    \centering
      \ra{1.2}
    \begin{tabular}{cl} \toprule
      Vortex line   &   A-twisted index  \\ \midrule
      4+0  &$-\frac{t^2 (1+t^2)}{(-1+t^2)^2 (1+t^2)}$ \\
      3+1  &$\frac{2t^3}{(-1+t^2)^2(1+t^2)}$ \\
      2+2   &$\frac{t^2}{(-1+t)^2(1+t)^2}$  \\
      1+3   &$\frac{t(1+t^4)}{(-1+t)^2(1+t)^2(1+t^2)}$ \\
      0+4   &$\frac{-(1+t^4)}{2(-1+t^2)^2}$ \\\bottomrule
    \end{tabular}
    \caption{$N_f=4$}
    \label{tab:vortex_SU2_Nf=4}
    \end{subtable}
    \\
    \begin{subtable}[t]{0.48\textwidth}
    \centering
      \ra{1.2}
    \begin{tabular}{cl} \toprule
      Vortex line   &   A-twisted index  \\ \midrule
      5+0  &$\frac{t^3(1+t^2)}{(-1+t)^2(1+t)^2(1-t+t^2)(1+t+t^2)}$ \\
      4+1  &$-\frac{2t^4}{(-1+t)^2(1+t)^2(1-t+t^2)(1+t+t^2)}$ \\
      3+2   &$\frac{t^3(1+t^2)}{(-1+t)^2(1+t)^2(1-t+t^2)(1+t+t^2)}$  \\
      2+3   &$-\frac{t^2(1+t^4)}{(-1+t)^2(1+t)^2(1-t+t^2)(1+t+t^2)}$ \\
      1+4  &$\frac{t(1+t^2)(1-t^2+t^4)}{(-1+t)^2(1+t)^2(1-t+t^2)(1+t+t^2)}$\\
      0+5  &$-\frac{(1+t^2)^2(1-t^2+t^4)}{2(-1+t)^2(1+t)^2(1-t+t^2)(1+t+t^2)}$\\ \bottomrule
    \end{tabular}
    \caption{$N_f=5$}
    \label{tab:vortex_SU2_Nf=5}
    \end{subtable}
    \hfill
    \begin{subtable}[t]{0.48\textwidth}
      \centering
        \ra{1.2}
    \begin{tabular}{cl} \toprule
      Vortex line   &   A-twisted index  \\ \midrule
      6+0  &$-\frac{t^4 (1+t^2)}{(-1+t)^2(1+t)^2 (1+t^2)(1+t^4)}$ \\
      5+1  &$\frac{2t^5}{(-1+t)^2(1+t)^2(1+t^2)(1+t^4)}$ \\
      4+2   &$-\frac{t^4}{(-1+t)^2(1+t)^2(1+t^4)}$  \\
      3+3   &$\frac{t^3}{(-1+t)^2(1+t)^2(1+t^2)}$ \\
      2+4  &$-\frac{t^2(1-t^2+t^4)}{(-1+t)^2(1+t)^2(1+t^4)}$\\
      1+5  &$\frac{t(1+t^8)}{(-1+t)^2(1+t)^2(1+t^2)(1+t^4)}$\\
      0+6  &$-\frac{1+t^8}{2(-1+t)^2(1+t)^2(1+t^4)}$\\ \bottomrule
    \end{tabular}
    \caption{$N_f=6$}
    \label{tab:vortex_SU2_Nf=6}
    \end{subtable}
    \caption{A-twisted index for $\sprm(1)$ with $N_f$ flavours in the presence of a vortex line defect characterised by a splitting $L+(N_f -L)$ and the fundamental representation of $\sprm(1)$.}
    \label{tab:vortx_SU2_results}
\end{table}

\subsection{\texorpdfstring{B-twisted index for mirrors of $\sprm(k)$ and Wilson defect}{B-twisted index for mirrors of Sp(k) and Wilson defect}}
This appendix provides the computational results for the B-twisted index for the 3d mirrors of $\sprm(k)$ SQCD in the presence of Wilson line defects, see Tables \ref{Sp1Nf3WilsonHS}--\ref{Sp2Nf6WilsonHS}.

\begin{table}[ht]
    \centering
    \ra{1.6}
 \scalebox{1}{   \begin{tabular}{ccl} \toprule
 \multirow{2}{*}{
\raisebox{+2\height}{\begin{tikzpicture}
	\begin{pgfonlayer}{nodelayer}
		\node [style=none] (12) at (1, -3) {\color{red}{A}};
		\node [style=none] (13) at (2.25, -3) {\color{red}{B}};
		\node [style=none] (16) at (3.5, -3) {\color{red}{C}};
		\node [style=gauge3] (20) at (1, -2) {};
		\node [style=none] (21) at (1, -2.5) {1};
		\node [style=gauge3] (22) at (2.25, -2) {};
		\node [style=none] (23) at (2.25, -2.5) {1};
		\node [style=gauge3] (24) at (3.5, -2) {};
		\node [style=none] (25) at (3.5, -2.5) {1};
		\node [style=blankflavor] (26) at (1, -1) {};
		\node [style=none] (27) at (1, -0.5) {1};
		\node [style=blankflavor] (28) at (3.5, -1) {};
		\node [style=none] (29) at (3.5, -0.5) {1};
	\end{pgfonlayer}
	\begin{pgfonlayer}{edgelayer}
		\draw (26) to (20);
		\draw (28) to (24);
		\draw (24) to (20);
	\end{pgfonlayer}
\end{tikzpicture}
}
}
&&\\
&A,C   & $\frac{t}{(1-t)^2(1+t)^2}$\\
 &B   &$\frac{2 t^2}{(-1+t)^2 (1+t)^2(1+t^2)}$ \\
&&\\
\bottomrule
    \end{tabular}}
    \caption{$\sprm(1)$ with $N_f=3$. B-twisted index with Wilson lines added at the labelled position in the quiver. }
    \label{Sp1Nf3WilsonHS}
\end{table}

\begin{table}[ht]
    \centering
    \ra{1.9}
 \scalebox{1}{   \begin{tabular}{ccl} \toprule
 \multirow{2}{*}{
\raisebox{+2\height}{\begin{tikzpicture}
	\begin{pgfonlayer}{nodelayer}
		\node [style=gauge3] (0) at (4.25, 1.25) {};
		\node [style=gauge3] (1) at (5.25, 0.5) {};
		\node [style=gauge3] (2) at (5.25, 2.25) {};
		\node [style=none] (5) at (4.25, 0.75) {2};
		\node [style=none] (6) at (5.25, 0) {1};
		\node [style=none] (7) at (5.25, 2.75) {1};
		\node [style=gauge3] (10) at (3, 1.25) {};
		\node [style=none] (11) at (3, 0.75) {1};
		\node [style=none] (12) at (3, 0.25) {\color{red}{A}};
		\node [style=none] (13) at (4.25, 0.25) {\color{red}{B}};
		\node [style=none] (16) at (5.75, 2.25) {\color{red}{C}};
		\node [style=none] (17) at (5.75, 0.5) {\color{red}{D}};
		\node [style=blankflavor] (18) at (4.25, 2.5) {};
		\node [style=none] (19) at (4.25, 3) {1};
	\end{pgfonlayer}
	\begin{pgfonlayer}{edgelayer}
		\draw (0) to (2);
		\draw (1) to (0);
		\draw (18) to (0);
		\draw (0) to (10);
	\end{pgfonlayer}
\end{tikzpicture}
}
}
&&\\
&A,C,D   & $\frac{ t^2}{(-1+t)^2 (1+t)^2 \left(1+t^2\right)}$\\
 &B   &$\frac{t}{(-1+t)^2 (1+t)^2}$ \\
&& \\
\bottomrule
    \end{tabular}}
    \caption{$\sprm(1)$ with $N_f=4$. B-twisted index with Wilson lines added at the labelled position in the quiver. }
    \label{Sp1Nf4WilsonHS}
\end{table}

\begin{table}[]
    \centering
      \ra{2}
 \scalebox{1}{   \begin{tabular}{ccl} \toprule
\multirow{4}{*}{\raisebox{-.5\height}{ \begin{tikzpicture}
	\begin{pgfonlayer}{nodelayer}
		\node [style=gauge3] (0) at (3, 0) {};
		\node [style=gauge3] (1) at (4, -0.75) {};
		\node [style=gauge3] (2) at (4, 1) {};
		\node [style=gauge3] (3) at (1.75, 0) {};
		\node [style=none] (4) at (1.75, -0.5) {2};
		\node [style=none] (5) at (3, -0.5) {2};
		\node [style=none] (6) at (4, -1.25) {1};
		\node [style=none] (7) at (4, 1.5) {1};
		\node [style=none] (8) at (0.5, -0.5) {1};
		\node [style=gauge3] (9) at (0.5, 0) {};
		\node [style=blankflavor] (11) at (1.75, 1) {};
		\node [style=none] (12) at (1.75, 1.5) {1};
		\node [style=none] (13) at (0.5, -1) {\color{red}{A}};
		\node [style=none] (14) at (1.75, -1) {\color{red}{B}};
		\node [style=none] (15) at (3, -1) {\color{red}{C}};
		\node [style=none] (16) at (4.5, 1) {\color{red}{D}};
		\node [style=none] (17) at (4.5, -0.75) {\color{red}{E}};
	\end{pgfonlayer}
	\begin{pgfonlayer}{edgelayer}
		\draw (0) to (2);
		\draw (1) to (0);
		\draw (0) to (3);
		\draw (3) to (9);
		\draw (11) to (3);
	\end{pgfonlayer}
\end{tikzpicture}}}
     & A & $\frac{t^2 \left(1-t^8\right)}{(1-t^4)^2(1-t^6)}$ \\
      &     B   & $\frac{t (1-t^8)}{(1-t^2)(1-t^4)(1-t^6)}$  \\
        &       C   &  $\frac{t^2 (1-t^4)}{(1-t^2)^2(1-t^6)}$\\
      & D, E   & $\frac{t^3}{(1-t^2)(1-t^6)}$ \\ \bottomrule
\end{tabular}}
    \caption{$\sprm(1)$ with $N_f=5$. B-twisted index with Wilson lines added at the labelled position in the quiver. This is also the $E_5$ family for $k=1$.}
    \label{Sp1Nf5WilsonHS}
\end{table}

\begin{table}[ht]
    \centering
    \ra{2}
 \scalebox{1}{   \begin{tabular}{ccl} \toprule
 \multirow{5}{*}{
\raisebox{-.5\height}{\begin{tikzpicture}
	\begin{pgfonlayer}{nodelayer}
		\node [style=gauge3] (58) at (-0.75, 1.25) {};
		\node [style=gauge3] (59) at (0.25, 0.5) {};
		\node [style=gauge3] (60) at (0.25, 2.25) {};
		\node [style=gauge3] (61) at (-2, 1.25) {};
		\node [style=none] (62) at (-2, 0.75) {2};
		\node [style=none] (63) at (-0.75, 0.75) {2};
		\node [style=none] (64) at (0.25, 0) {1};
		\node [style=none] (65) at (0.25, 2.75) {1};
		\node [style=gauge3] (66) at (-3.25, 1.25) {};
		\node [style=none] (67) at (-3.25, 0.75) {2};
		\node [style=gauge3] (68) at (-4.5, 1.25) {};
		\node [style=none] (69) at (-4.5, 0.75) {1};
		\node [style=none] (72) at (-4.5, 0.25) {\color{red}{A}};
		\node [style=none] (73) at (-3.25, 0.25) {\color{red}{B}};
		\node [style=none] (74) at (-2, 0.25) {\color{red}{C}};
		\node [style=none] (75) at (-0.75, 0.25) {\color{red}{D}};
		\node [style=none] (76) at (0.75, 2.25) {\color{red}{E}};
		\node [style=none] (77) at (0.75, 0.5) {\color{red}{F}};
		\node [style=blankflavor] (78) at (-3.25, 2.5) {};
		\node [style=none] (79) at (-3.25, 3) {1};
	\end{pgfonlayer}
	\begin{pgfonlayer}{edgelayer}
		\draw (58) to (60);
		\draw (59) to (58);
		\draw (58) to (61);
		\draw (66) to (68);
		\draw (66) to (61);
		\draw (78) to (66);
	\end{pgfonlayer}
\end{tikzpicture}}
}
&A   & $\frac{t^2 \left(t^4-t^2+1\right)}{(t-1)^2 (t+1)^2 \left(t^2+1\right) \left(t^4+1\right)}$ \\
&B   & $\frac{t \left(t^4-t^2+1\right)}{(t-1)^2 (t+1)^2 \left(t^4+1\right)}$ \\
&C  & $\frac{t^2}{(t-1)^2 (t+1)^2 \left(t^2+1\right)}$ \\
&D   &$\frac{t^3}{(t-1)^2 (t+1)^2 \left(t^4+1\right)}$ \\
&E,F   & $\frac{t^4}{(t-1)^2 (t+1)^2 \left(t^2+1\right) \left(t^4+1\right)}$ \\ \bottomrule
    \end{tabular}}
    \caption{$\sprm(1)$ with $N_f=6$. B-twisted index with Wilson lines added at the labelled position in the quiver. }
    \label{Sp1Nf6WilsonHS}
\end{table}

\begin{table}[ht]
    \centering
      \ra{1.5}
 \scalebox{1}{   \begin{tabular}{ccl} \toprule
\multirow{5}{*}{
\raisebox{0\height}{\begin{tikzpicture}
	\begin{pgfonlayer}{nodelayer}
		\node [style=gauge3] (0) at (4.25, 0) {};
		\node [style=gauge3,label=above:{\color{red}{D}}] (1) at (5.25, -0.75) {};
		\node [style=gauge3,label=below:{\color{red}{C}}] (2) at (5.25, 1) {};
		\node [style=gauge3] (3) at (3, 0) {};
		\node [style=none] (4) at (3, -0.5) {1};
		\node [style=none] (5) at (4.25, -0.5) {2};
		\node [style=none] (6) at (5.25, -1.25) {1};
		\node [style=none] (7) at (5.25, 1.5) {2};
		\node [style=none] (52) at (3, -1) {\color{red}{A}};
		\node [style=none] (53) at (4.25, -1) {\color{red}{B}};
		\node [style=flavour2] (58) at (6.25, 1) {};
		\node [style=none] (60) at (6.75, 1) {2};
	\end{pgfonlayer}
	\begin{pgfonlayer}{edgelayer}
		\draw (0) to (2);
		\draw (1) to (0);
		\draw (0) to (3);
		\draw (2) to (58);
	\end{pgfonlayer}
\end{tikzpicture}
}
}
&&\\
&A,D  & $\frac{2 t^3 \left(1+t^2\right) \left(1+t^4\right)}{(-1+t)^4 (1+t)^4 \left(1-t+t^2\right)^2 \left(1+t+t^2\right)^2}$ \\
&B   & $\frac{2 t^2 \left(1+t^2\right)^2 \left(1+t^4\right)}{(-1+t)^4 (1+t)^4 \left(1-t+t^2\right)^2 \left(1+t+t^2\right)^2}$ \\
&C   & $\frac{2 t \left(1+t^2\right) \left(1+t^4\right)^2}{(-1+t)^4 (1+t)^4 \left(1-t+t^2\right)^2 \left(1+t+t^2\right)^2}$ \\
&&\\
 \bottomrule
    \end{tabular}}
    \caption{The Higgs branch of $\sprm(2)$ with $N_f=4$ consists of two identical cones. The Coulomb branch of this quiver is one of the two cones. B-twisted index with Wilson lines at the labelled position in the quiver. }
    \label{Sp2Nf4WilsonHS}
\end{table}

\begin{table}[ht]
    \centering
      \ra{1.75}
 \scalebox{1}{   \begin{tabular}{ccl} \toprule
\multirow{5}{*}{
\raisebox{.5\height}{ \begin{tikzpicture}
	\begin{pgfonlayer}{nodelayer}
		\node [style=gauge3] (0) at (4.25, 0) {};
		\node [style=gauge3,label=above:{\color{red}{E}}] (1) at (5.25, -0.75) {};
		\node [style=gauge3,label=below:{\color{red}{D}}] (2) at (5.25, 1) {};
		\node [style=gauge3] (3) at (3, 0) {};
		\node [style=none] (4) at (3, -0.5) {2};
		\node [style=none] (5) at (4.25, -0.5) {3};
		\node [style=none] (6) at (5.25, -1.25) {2};
		\node [style=none] (7) at (5.25, 1.5) {2};
		\node [style=gauge3] (8) at (1.75, 0) {};
		\node [style=none] (9) at (1.75, -0.5) {1};
		\node [style=none] (52) at (1.75, -1) {\color{red}{A}};
		\node [style=none] (53) at (3, -1) {\color{red}{B}};
		\node [style=none] (54) at (4.25, -1) {\color{red}{C}};
		\node [style=flavour2] (58) at (6.25, 1) {};
		\node [style=flavour2] (59) at (6.25, -0.75) {};
		\node [style=none] (60) at (6.75, 1) {1};
		\node [style=none] (61) at (6.75, -0.75) {1};
	\end{pgfonlayer}
	\begin{pgfonlayer}{edgelayer}
		\draw (0) to (2);
		\draw (1) to (0);
		\draw (0) to (3);
		\draw (8) to (3);
		\draw (2) to (58);
		\draw (1) to (59);
	\end{pgfonlayer}
\end{tikzpicture}}
}
&A   & $\frac{2 t^4 \left(1-t^2+t^4\right)}{(-1+t)^4 (1+t)^4 \left(1+t^2\right)^2 \left(1+t^4\right)}$ \\
&B   & $\frac{2 t^3 \left(1-t^2+t^4\right)}{(-1+t)^4 (1+t)^4 \left(1+t^2\right) \left(1+t^4\right)}$ \\
&C   & $\frac{2 t^2 \left(1-t+t^2\right) \left(1+t+t^2\right) \left(1-t^2+t^4\right)}{(-1+t)^4 (1+t)^4 \left(1+t^2\right)^2 \left(1+t^4\right)}$\\
&D,E   & $\frac{t \left(1-t^2+t^4\right)}{(-1+t)^4 (1+t)^4 \left(1+t^2\right)}$ \\
     \bottomrule\end{tabular}}
    \caption{$\sprm(2)$ with $N_f=5$. B-twisted index with Wilson lines at the labelled position in the quiver. }
    \label{Sp2Nf5WilsonHS}
\end{table}

\begin{table}[ht]
    \centering
      \ra{1.75}
 \scalebox{1}{   \begin{tabular}{ccl} \toprule
\multirow{5}{*}{
\raisebox{-.5\height}{ \begin{tikzpicture}
	\begin{pgfonlayer}{nodelayer}
		\node [style=gauge3] (0) at (4.25, 0) {};
		\node [style=gauge3] (1) at (5.25, -0.75) {};
		\node [style=gauge3] (2) at (5.25, 1) {};
		\node [style=gauge3] (3) at (3, 0) {};
		\node [style=none] (4) at (3, -0.5) {3};
		\node [style=none] (5) at (4.25, -0.5) {4};
		\node [style=none] (6) at (5.25, -1.25) {2};
		\node [style=none] (7) at (5.25, 1.5) {2};
		\node [style=gauge3] (8) at (1.75, 0) {};
		\node [style=none] (9) at (1.75, -0.5) {2};
		\node [style=gauge3] (10) at (0.5, 0) {};
		\node [style=none] (13) at (0.5, -0.5) {1};
		\node [style=blankflavor] (49) at (4.25, 1) {};
		\node [style=none] (51) at (4.25, 1.5) {1};
		\node [style=none] (52) at (0.5, -1) {\color{red}{A}};
		\node [style=none] (53) at (1.75, -1) {\color{red}{B}};
		\node [style=none] (54) at (3, -1) {\color{red}{C}};
		\node [style=none] (55) at (4.25, -1) {\color{red}{D}};
		\node [style=none] (56) at (5.75, 1) {\color{red}{E}};
		\node [style=none] (57) at (5.75, -0.75) {\color{red}{F}};
	\end{pgfonlayer}
	\begin{pgfonlayer}{edgelayer}
		\draw (0) to (2);
		\draw (1) to (0);
		\draw (0) to (3);
		\draw (8) to (10);
		\draw (8) to (3);
		\draw (49) to (0);
	\end{pgfonlayer}
\end{tikzpicture}}
}
&A   & $\frac{t^4 \left(t^8+1\right)}{(t-1)^4 (t+1)^4 \left(t^2+1\right)^2 \left(t^2-t+1\right) \left(t^2+t+1\right) \left(t^4+1\right)}$ \\
&B   & $\frac{t^3 \left(t^8+1\right)}{(t-1)^4 (t+1)^4 \left(t^2+1\right) \left(t^2-t+1\right) \left(t^2+t+1\right) \left(t^4+1\right)}$ \\
&C   & $\frac{t^2 \left(t^8+1\right)}{(t-1)^4 (t+1)^4 \left(t^2+1\right)^2 \left(t^4+1\right)}$ \\
&D   & $\frac{t \left(t^8+1\right)}{(t-1)^4 (t+1)^4 \left(t^2+1\right) \left(t^2-t+1\right) \left(t^2+t+1\right)}$ \\
&E, F   & $\frac{t^2 \left(t^8+1\right)}{(t-1)^4 (t+1)^4 \left(t^2+1\right)^2 \left(t^2-t+1\right) \left(t^2+t+1\right)}$ \\ \bottomrule
    \end{tabular}}
    \caption{$\sprm(2)$ with $N_f=6$. B-twisted index with Wilson lines at the labelled position in the quiver. }
    \label{Sp2Nf6WilsonHS}
\end{table}

\FloatBarrier

\subsection{Wilson lines for exceptional families}
\label{choiceofU1}
For $\sprm(k)$ SQCD theories, the unitary 3d mirror can be read off from a D3-D5-NS5 brane set up in the presence of O5 orientifold planes. The NS5 branes carry non-dynamical degrees of freedom and are therefore seen as flavour nodes in the quiver. On the other hand, $E_{6,7,8}$ unitary quivers are obtained from  $(p,q)$ 5-brane, $[p,q]$ 7-brane settings where all 5-branes carry dynamical degrees of freedom. As a result, magnetic quivers obtained from brane webs are all unframed. For unitary unframed quivers, a choice needs to be made as to where to gauge an overall $\urm(1)$ that corresponds to fixing the centre of mass of the brane system. For two different choices of ungauging, a Wilson line placed at the same position in the quiver can result in two different B-twisted indices.

 We wish to ungauge on an $\urm(1)$ so that the resulting quiver has an explicit flavour group. Otherwise, as described in \S\ref{sec:Wilson_unframed_MagQuiv}, a Wilson line in the fundamental representation of a single gauge group in an unframed quiver is not invariant under the discrete 1-form symmetry and the B-twisted index is zero. In this paper, we focused on the $E_n$ exceptional families with $n \leq 7$. For $k=1$, all unitary quivers are affine Dynkin diagrams of $\mathfrak{e}_n$. Due to the symmetry of these quivers, ungauging any of the $\urm(1)$s will give the same B-twisted index. For $k>1$, the quivers are no longer affine Dynkin diagrams but with one of the tails extended. This extended tail is often called a $T[\surm(N)]$ tail as it is a sequence of nodes with increasing ranks (in steps of one)  that begins with $\urm(1)$ and ends with  $\urm(N)$ for some $N$. As we see in \S\ref{sec:3d_Sp}, a Wilson line on the  $\urm(1)$ in the long $T[\surm(N)]$ tail is always required in order to match with a Wilson line on a $\sorm(\mathrm{even})$ node in the orthosymplectic counterpart. Hence, we will not ungauge the $\urm(1)$ node in the $T[\surm(N)]$ tail. As a result, for the $E_7$ family, the $\urm(1)$ node to be ungauged is the one away from the extended tail. For $E_n$ and $n\leq 6$, there are two $\urm(1)$ nodes away from the extended tail and, since the quiver has reflection symmetry, we can choose either one of them to be ungauged. With these choices of ungauging, we can find a nice matching with the Wilson lines on the unframed orthosymplectic quivers as shown in \S\ref{sec:Wilson_unframed_MagQuiv}. Examples of B-twisted indices with Wilson lines in this ungauging scheme at various gauge groups are given in Table \ref{E3WilsonHSk1}-\ref{E7WilsonHSk1}.

\begin{table}[ht]
    \centering
      \ra{1.2}
 \scalebox{1}{   \begin{tabular}{ccc} \toprule
 \raisebox{-.5\height}{
 \begin{tikzpicture}
	\begin{pgfonlayer}{nodelayer}
		\node [style=gauge3] (5) at (0, 0) {};
		\node [style=none] (10) at (0, -0.5) {1};
		\node [style=none] (11) at (1, -0.5) {1};
		\node [style=gauge3] (31) at (1, 0) {};
		\node [style=blankflavor] (33) at (1, 1) {};
		\node [style=none] (34) at (1, 1.5) {1};
		\node [style=blankflavor] (35) at (0, 1) {};
		\node [style=none] (36) at (0, 1.5) {1};
		\node [style=none] (37) at (0, -1) {\color{red}{A}};
		\node [style=none] (38) at (1, -1) {\color{red}{B}};
	\end{pgfonlayer}
	\begin{pgfonlayer}{edgelayer}
		\draw (5) to (31);
		\draw (33) to (31);
		\draw (35) to (5);
	\end{pgfonlayer}
\end{tikzpicture}}
&
      A, B   & $\frac{t}{(-1+t)^2 \left(1+t+t^2\right)}$ \\ \bottomrule
    \end{tabular}}
    \caption{$E_3$ family for $k=1$. B-twisted index with Wilson lines at the labelled position in the quiver.}
    \label{E3WilsonHSk1}
\end{table}

\begin{table}[ht]
    \centering
    \ra{2}
 \scalebox{1}{   \begin{tabular}{ccc} \toprule
\multirow{4}{*}{\raisebox{-.5\height}{\begin{tikzpicture}
	\begin{pgfonlayer}{nodelayer}
		\node [style=gauge3] (0) at (3, 0) {};
		\node [style=gauge3] (1) at (4, -0.75) {};
		\node [style=gauge3] (2) at (4, 1) {};
		\node [style=gauge3] (3) at (1.75, 0) {};
		\node [style=none] (4) at (1.75, -0.5) {2};
		\node [style=none] (5) at (3, -0.5) {2};
		\node [style=none] (6) at (4, -1.25) {1};
		\node [style=none] (7) at (4, 1.5) {1};
		\node [style=none] (8) at (0.5, -0.5) {1};
		\node [style=gauge3] (9) at (0.5, 0) {};
		\node [style=blankflavor] (14) at (1.75, 1) {};
		\node [style=none] (15) at (2.25, 1) {1};
		\node [style=none] (16) at (0.5, -1) {\color{red}{A}};
		\node [style=none] (17) at (1.75, -1) {\color{red}{B}};
		\node [style=none] (18) at (3, -1) {\color{red}{C}};
		\node [style=none] (19) at (4.5, 1) {\color{red}{D}};
		\node [style=none] (20) at (4.5, -0.75) {\color{red}{E}};
		\node [style=blankflavor] (21) at (0.5, 1) {};
		\node [style=none] (22) at (1, 1) {1};
	\end{pgfonlayer}
	\begin{pgfonlayer}{edgelayer}
		\draw (0) to (2);
		\draw (1) to (0);
		\draw (0) to (3);
		\draw (3) to (9);
		\draw (9) to (21);
		\draw (14) to (3);
	\end{pgfonlayer}
\end{tikzpicture}}}
&   A   &  $\frac{t \left(1+t^4\right) \left(1+t^3-t^4+t^5+t^8\right)}{(-1+t)^4 (1+t)^2 \left(1+t^2\right) \left(1-t+t^2\right) \left(1+t+t^2\right) \left(1+t+t^2+t^3+t^4+t^5+t^6\right)}$ \\
     &  B   & $\frac{t \left(1+t^4\right) \left(1+t^3+t^6\right)}{(-1+t)^4 (1+t)^2 \left(1-t+t^2\right) \left(1+t+t^2\right) \left(1+t+t^2+t^3+t^4+t^5+t^6\right)}$ \\
     &   C   &  $\frac{t^2 \left(1+t^2\right) \left(1+t^3+t^6\right)}{(-1+t)^4 (1+t)^2 \left(1-t+t^2\right) \left(1+t+t^2\right) \left(1+t+t^2+t^3+t^4+t^5+t^6\right)}$ \\
      &   D, E   & $\frac{t^3 \left(1+t^3+t^6\right)}{(-1+t)^4 (1+t)^2 \left(1-t+t^2\right) \left(1+t+t^2\right) \left(1+t+t^2+t^3+t^4+t^5+t^6\right)}$ \\ \bottomrule
    \end{tabular}}
    \caption{$E_3$ family for $k=2$. B-twisted index with Wilson lines at the labelled position in the quiver. }
    \label{E3WilsonHSk2}
\end{table}

\begin{table}[]
    \centering
    \ra{1.2}
 \scalebox{0.9}{   \begin{tabular}{c|c} \toprule

         & \begin{tikzpicture}
	\begin{pgfonlayer}{nodelayer}
		\node [style=gauge3] (0) at (-1, 0) {};
		\node [style=gauge3] (1) at (-2, 0) {};
		\node [style=gauge3] (2) at (-3, 0) {};
		\node [style=gauge3] (3) at (-1, 1) {};
		\node [style=gauge3] (5) at (0, 0) {};
		\node [style=none] (7) at (-1, -0.5) {4};
		\node [style=none] (8) at (-2, -0.5) {3};
		\node [style=none] (9) at (-3, -0.5) {2};
		\node [style=none] (10) at (0, -0.5) {3};
		\node [style=none] (11) at (1, -0.5) {1};
		\node [style=none] (12) at (-1, 1.5) {2};
		\node [style=gauge3] (31) at (1, 0) {};
		\node [style=gauge3] (33) at (-4, 0) {};
		\node [style=none] (34) at (1, 1.5) {1};
		\node [style=none] (35) at (-4, -0.5) {1};
		\node [style=blankflavor] (36) at (1, 1) {};
		\node [style=blankflavor] (37) at (0, 1) {};
		\node [style=none] (38) at (0, 1.5) {1};
		\node [style=none] (39) at (-4, -1) {\color{red}{A}};
		\node [style=none] (40) at (-3, -1) {\color{red}{B}};
		\node [style=none] (41) at (-2, -1) {\color{red}{C}};
		\node [style=none] (42) at (-1, -1) {\color{red}{D}};
		\node [style=none] (43) at (0, -1) {\color{red}{E}};
		\node [style=none] (44) at (1, -1) {\color{red}{F}};
		\node [style=none] (45) at (-1.5, 1) {\color{red}{G}};
	\end{pgfonlayer}
	\begin{pgfonlayer}{edgelayer}
		\draw (3) to (0);
		\draw (0) to (5);
		\draw (2) to (1);
		\draw (1) to (0);
		\draw (5) to (31);
		\draw (33) to (2);
		\draw (31) to (36);
		\draw (37) to (5);
	\end{pgfonlayer}
\end{tikzpicture}
 \\ \midrule
      A   & $\dfrac{(1-t)\left(\begin{array}{c} t^5 (1+2 t+3 t^2+4 t^3+5 t^4+6 t^5+8 t^6+10 t^7+12 t^8+13 t^9+14 t^{10}+15 t^{11}+17 t^{12}\\+18 t^{13}+19 t^{14}+19 t^{15}+19 t^{16}+18 t^{17}+17 t^{18}+15
  t^{19}+14 t^{20}+13 t^{21}\\+12 t^{22}+10 t^{23}+8 t^{24}+6 t^{25}+5 t^{26}+4 t^{27}+3 t^{28}+2 t^{29}+t^{30}) \end{array}\right)}{(1-t^2)(1-t^3)(1-t^6)(1-t^7)(1-t^8)(1-t^{10})(1-t^{11})
    } $\\
      B   &  $\dfrac{(1-t)(1-t^4)\left(\begin{array}{c} t^4 (1+2 t+3 t^2+4 t^3+5 t^4+6 t^5+8 t^6+10 t^7+12 t^8+13 t^9+14 t^{10}+15 t^{11}+17 t^{12}\\+18 t^{13}+19 t^{14}+19 t^{15}+19 t^{16}+18 t^{17}+17 t^{18}+15
  t^{19}+14 t^{20}+13 t^{21}\\+12 t^{22}+10 t^{23}+8 t^{24}+6 t^{25}+5 t^{26}+4 t^{27}+3 t^{28}+2 t^{29}+t^{30})            \end{array} \right)}{(1-t^2)^2(1-t^3)(1-t^6)(1-t^7)(1-t^8)(1-t^{10})(1-t^{11})
    } $ \\
        C   &  $\dfrac{(1-t)\left(\begin{array}{c}  t^3 (1+2 t+3 t^2+4 t^3+5 t^4+6 t^5+8 t^6+10 t^7+12 t^8+13 t^9+14 t^{10}+15 t^{11}+17 t^{12}\\+18 t^{13}+19 t^{14}+19 t^{15}+19 t^{16}+18 t^{17}+17 t^{18}+15
  t^{19}+14 t^{20}+13 t^{21}\\+12 t^{22}+10 t^{23}+8 t^{24}+6 t^{25}+5 t^{26}+4 t^{27}+3 t^{28}+2 t^{29}+t^{30})          \end{array} \right)}{(1-t^2)^2(1-t^3)(1-t^7)(1-t^8)(1-t^{10})(1-t^{11})
    } $ \\
         D   & $\dfrac{(1-t)\left(\begin{array}{c}      t^2 (1+2 t+3 t^2+4 t^3+5 t^4+6 t^5+8 t^6+10 t^7+12 t^8+13 t^9+14 t^{10}+15 t^{11}+17 t^{12}\\+18 t^{13}+19 t^{14}+19 t^{15}+19 t^{16}+18 t^{17}+17 t^{18}+15
  t^{19}+14 t^{20}+13 t^{21}\\+12 t^{22}+10 t^{23}+8 t^{24}+6 t^{25}+5 t^{26}+4 t^{27}+3 t^{28}+2 t^{29}+t^{30})      \end{array} \right)}{(1-t^2)^2(1-t^3)(1-t^6)(1-t^7)(1-t^{10})(1-t^{11})
    } $  \\
          E   & $\dfrac{(1-t)\left(\begin{array}{c}   t (1-t^2+t^4) (1+2 t+3 t^2+4 t^3+5 t^4+6 t^5+8 t^6+10 t^7+12 t^8+13 t^9+14 t^{10}+15 t^{11}+17 t^{12}\\+18 t^{13}+19 t^{14}+19 t^{15}+19 t^{16}+18
  t^{17}+17 t^{18}+15 t^{19}+14 t^{20}+13 t^{21}\\+12 t^{22}+10 t^{23}+8 t^{24}+6 t^{25}+5 t^{26}+4 t^{27}+3 t^{28}+2 t^{29}+t^{30})       \end{array} \right)}{(1-t^2)^2(1-t^3)(1-t^7)(1-t^8)(1-t^{10})(1-t^{11})
    } $ \\
          F   & $\dfrac{  t (1-t) \left(1-t^3+t^6\right) \left(\begin{array}{c} 1+2 t+3 t^2+5 t^3+6 t^4+8 t^5+9 t^6+10 t^7+12 t^8+13 t^9+14 t^{10}+15 t^{11}\\+16 t^{12}+17 t^{13}+17 t^{14}+17 t^{15}+17 t^{16}+17
  t^{17}+16 t^{18}+15 t^{19}+14 t^{20}+13\\ t^{21}+12 t^{22}+10 t^{23}+9 t^{24}+8 t^{25}+6 t^{26}+5 t^{27}+3 t^{28}+2 t^{29}+t^{30}    \end{array} \right)}{\left(1-t^3\right) \left(1-t^4\right)^2 \left(1-t^6\right) \left(1-t^7\right) \left(1-t^{10}\right) \left(1-t^{11}\right)
    } $  \\
             G   &  $\dfrac{(1-t)t^3 \left(\begin{array}{c} (1+2 t+3 t^2+4 t^3+5 t^4+6 t^5+8 t^6+10 t^7+12 t^8+13 t^9+14 t^{10}\\+15 t^{11}+17 t^{12}+18 t^{13}+19 t^{14}+19 t^{15}+19 t^{16}+18 t^{17}+17 t^{18}+15 t^{19}\\+14 t^{20}+13 t^{21}+12
   t^{22}+10 t^{23}+8 t^{24}+6 t^{25}+5 t^{26}+4 t^{27}+3 t^{28}+2 t^{29}+t^{30})       \end{array} \right)}{(1-t^2)(1-t^3)(1-t^4)(1-t^6)(1-t^7)(1-t^{10})(1-t^{11})
  } $ \\ \bottomrule
    \end{tabular}}
    \caption{$E_3$ family for $k=3$. B-twisted index with Wilson lines at the labelled position in the quiver. }
    \label{E3WilsonHSk3}
\end{table}

\begin{table}[]
    \centering
      \ra{3}
 \scalebox{1}{   \begin{tabular}{ccl} \toprule
\multirow{2}{*}{ \raisebox{-.5\height}{
\begin{tikzpicture}
	\begin{pgfonlayer}{nodelayer}
		\node [style=gauge3] (0) at (-3, 0) {};
		\node [style=gauge3] (1) at (-4, 0) {};
		\node [style=gauge3] (2) at (-5, 0) {};
		\node [style=none] (7) at (-3, -0.5) {1};
		\node [style=none] (8) at (-4, -0.5) {1};
		\node [style=none] (9) at (-5, -0.5) {1};
		\node [style=gauge3] (33) at (-6, 0) {};
		\node [style=none] (35) at (-6, -0.5) {1};
		\node [style=blankflavor] (36) at (-6, 1) {};
		\node [style=blankflavor] (37) at (-3, 1) {};
		\node [style=none] (38) at (-3, 1.5) {1};
		\node [style=none] (39) at (-6, 1.5) {1};
		\node [style=none] (40) at (-6, -1) {\color{red}{A}};
		\node [style=none] (41) at (-5, -1) {\color{red}{B}};
		\node [style=none] (42) at (-4, -1) {\color{red}{C}};
		\node [style=none] (43) at (-3, -1) {\color{red}{D}};
	\end{pgfonlayer}
	\begin{pgfonlayer}{edgelayer}
		\draw (2) to (1);
		\draw (1) to (0);
		\draw (33) to (2);
		\draw (36) to (33);
		\draw (37) to (0);
	\end{pgfonlayer}
\end{tikzpicture} } }
        &       A, D   & $\frac{t \left(1-t+t^2\right)}{(-1+t)^2 \left(1+t+t^2+t^3+t^4\right)}$ \\
      & B, C   &$\frac{t^2}{(-1+t)^2 \left(1+t+t^2+t^3+t^4\right)}$ \\ \bottomrule
    \end{tabular}}
    \caption{$E_4$ family for $k=1$. B-twisted index with Wilson lines at the labelled position in the quiver.}
    \label{E4WilsonHSk1}
\end{table}

\begin{table}[]
    \centering
      \ra{2}
 \scalebox{1}{   \begin{tabular}{ccl} \toprule
\multirow{6}{*}{ \raisebox{-.5\height}{    \begin{tikzpicture}
	\begin{pgfonlayer}{nodelayer}
		\node [style=gauge3] (0) at (-1, 0) {};
		\node [style=gauge3] (1) at (-2, 0) {};
		\node [style=gauge3] (2) at (-3, 0) {};
		\node [style=gauge3] (3) at (-1, 1) {};
		\node [style=gauge3] (5) at (0, 0) {};
		\node [style=none] (7) at (-1, -0.5) {3};
		\node [style=none] (8) at (-2, -0.5) {2};
		\node [style=none] (9) at (-3, -0.5) {1};
		\node [style=none] (10) at (0, -0.5) {2};
		\node [style=none] (11) at (1, -0.5) {1};
		\node [style=none] (12) at (-0.5, 1) {2};
		\node [style=none] (13) at (-0.5, 2) {1};
		\node [style=gauge3] (31) at (1, 0) {};
		\node [style=blankflavor] (32) at (-1, 2) {};
		\node [style=blankflavor] (33) at (1, 1) {};
		\node [style=none] (34) at (1, 1.5) {1};
		\node [style=none] (35) at (-3, -1) {\color{red}{A}};
		\node [style=none] (36) at (-2, -1) {\color{red}{B}};
		\node [style=none] (37) at (-1, -1) {\color{red}{C}};
		\node [style=none] (38) at (0, -1) {\color{red}{D}};
		\node [style=none] (39) at (1, -1) {\color{red}{E}};
		\node [style=none] (40) at (-1.5, 1) {\color{red}{F}};
	\end{pgfonlayer}
	\begin{pgfonlayer}{edgelayer}
		\draw (3) to (0);
		\draw (0) to (5);
		\draw (2) to (1);
		\draw (1) to (0);
		\draw (32) to (3);
		\draw (5) to (31);
		\draw (33) to (31);
	\end{pgfonlayer}
\end{tikzpicture}}}
&   A   & $\frac{t^4 \left(1+t+t^2+t^3+t^4+t^5+t^6+t^7+t^8+t^9+t^{10}+t^{11}+t^{12}\right)}{(-1+t)^4 (1+t)^2 \left(1+t^2\right) \left(1-t+t^2\right) \left(1+t+t^2\right)^2
  \left(1+t+t^2+t^3+t^4\right) \left(1+t^3+t^6\right)}$ \\
  &    B   & $\frac{t^3 \left(1+t+t^2+t^3+t^4+t^5+t^6+t^7+t^8+t^9+t^{10}+t^{11}+t^{12}\right)}{(-1+t)^4 (1+t)^2 \left(1-t+t^2\right) \left(1+t+t^2\right)^2 \left(1+t+t^2+t^3+t^4\right)
  \left(1+t^3+t^6\right)}$ \\
  &     C   & $\frac{t^2 \left(1+t+t^2+t^3+t^4+t^5+t^6+t^7+t^8+t^9+t^{10}+t^{11}+t^{12}\right)}{(-1+t)^4 (1+t)^2 \left(1+t^2\right) \left(1+t+t^2\right) \left(1+t+t^2+t^3+t^4\right)
  \left(1+t^3+t^6\right)}$ \\
     &   D   & $\frac{t^2 \left(1+t+t^3+t^4+t^5+t^6+t^7+t^8+t^9+t^{10}+t^{11}+t^{13}+t^{14}\right)}{(-1+t)^4 (1+t)^2 \left(1-t+t^2\right) \left(1+t+t^2\right)^2 \left(1+t+t^2+t^3+t^4\right)
  \left(1+t^3+t^6\right)}$ \\
    &      E   &  $\frac{t \left(1+t^2+t^3+t^5+t^7+t^8+t^9+t^{10}+t^{11}+t^{13}+t^{15}+t^{16}+t^{18}\right)}{(-1+t)^4 (1+t)^2 \left(1+t^2\right) \left(1-t+t^2\right) \left(1+t+t^2\right)^2
  \left(1+t+t^2+t^3+t^4\right) \left(1+t^3+t^6\right)}$\\
     &      F   & $\frac{t \left(1-t^2+t^4\right) \left(1+t^2+t^3+t^4+2 t^5+t^6+2 t^7+t^8+t^9+t^{10}+t^{12}\right)}{(-1+t)^4 (1+t)^2 \left(1-t+t^2\right) \left(1+t+t^2\right)^2
  \left(1+t+t^2+t^3+t^4\right) \left(1+t^3+t^6\right)}$ \\ \bottomrule
    \end{tabular}}
    \caption{$E_4$ family for $k=2$. B-twisted index with Wilson lines at the labelled position in the quiver. }
    \label{E4WilsonHSk2}
\end{table}

\begin{table}[]
    \centering
      \ra{2}
 \scalebox{1}{   \begin{tabular}{ccl} \toprule
\multirow{4}{*}{
\raisebox{-.5\height}{\begin{tikzpicture}
	\begin{pgfonlayer}{nodelayer}
		\node [style=gauge3] (0) at (0, 0) {};
		\node [style=gauge3] (1) at (-1, 0) {};
		\node [style=gauge3] (2) at (-2, 0) {};
		\node [style=gauge3] (3) at (0, 1) {};
		\node [style=gauge3] (5) at (1, 0) {};
		\node [style=none] (7) at (0, -0.5) {3};
		\node [style=none] (8) at (-1, -0.5) {2};
		\node [style=none] (9) at (-2, -0.5) {1};
		\node [style=none] (10) at (1, -0.5) {2};
		\node [style=none] (11) at (2, -0.5) {1};
		\node [style=none] (12) at (0.5, 1) {2};
		\node [style=none] (13) at (0.5, 2) {1};
		\node [style=gauge3] (31) at (2, 0) {};
		\node [style=blankflavor] (32) at (0, 2) {};
		\node [style=none] (33) at (-2, -1) {\color{red}{A}};
		\node [style=none] (34) at (-1, -1) {\color{red}{B}};
		\node [style=none] (35) at (0, -1) {\color{red}{C}};
		\node [style=none] (36) at (1, -1) {\color{red}{D}};
		\node [style=none] (37) at (2, -1) {\color{red}{E}};
		\node [style=none] (38) at (-0.5, 1) {\color{red}{F}};
	\end{pgfonlayer}
	\begin{pgfonlayer}{edgelayer}
		\draw (3) to (0);
		\draw (0) to (5);
		\draw (2) to (1);
		\draw (1) to (0);
		\draw (32) to (3);
		\draw (5) to (31);
	\end{pgfonlayer}
\end{tikzpicture}}
}
& A, E & $\frac{t^4}{(1-t^4)(1-t^6)}$ \\
& B, D   &  $\frac{t^3}{(1-t^2)(1-t^6)}$ \\
& C   & $\frac{t^2}{(1-t^2)(1-t^4)}$ \\
&  F   & $\frac{t (1-t^{12})}{(1-t^4)(1-t^6)^2}$ \\ \bottomrule
    \end{tabular}}
    \caption{$E_6$ family for $k=1$. B-twisted index with Wilson lines at the labelled position in the quiver. }
    \label{E6WilsonHSk1}
\end{table}

\begin{table}[]
    \centering
      \ra{1.75}
\begin{tabular}{ccl} \toprule
    \multirow{7}{*}{
    \raisebox{-.5\height}{\begin{tikzpicture}
	\begin{pgfonlayer}{nodelayer}
		\node [style=gauge3] (0) at (-1, 0) {};
		\node [style=gauge3] (1) at (-2, 0) {};
		\node [style=gauge3] (2) at (-3, 0) {};
		\node [style=gauge3] (3) at (-1, 1) {};
		\node [style=gauge3] (5) at (0, 0) {};
		\node [style=none] (7) at (-1, -0.5) {4};
		\node [style=none] (8) at (-2, -0.5) {3};
		\node [style=none] (9) at (-3, -0.5) {2};
		\node [style=none] (10) at (0, -0.5) {3};
		\node [style=none] (11) at (1, -0.5) {2};
		\node [style=none] (12) at (-0.5, 1) {2};
		\node [style=gauge3] (31) at (1, 0) {};
		\node [style=gauge3] (33) at (-4, 0) {};
		\node [style=none] (34) at (2, -0.5) {1};
		\node [style=none] (35) at (-4, -0.5) {1};
		\node [style=blankflavor] (36) at (2, 0) {};
		\node [style=none] (37) at (-4, -1) {\color{red}{A}};
		\node [style=none] (38) at (-3, -1) {\color{red}{B}};
		\node [style=none] (39) at (-2, -1) {\color{red}{C}};
		\node [style=none] (40) at (-1, -1) {\color{red}{D}};
		\node [style=none] (41) at (0, -1) {\color{red}{E}};
		\node [style=none] (42) at (1, -1) {\color{red}{F}};
		\node [style=none] (43) at (-1, 1.5) {\color{red}{G}};
	\end{pgfonlayer}
	\begin{pgfonlayer}{edgelayer}
		\draw (3) to (0);
		\draw (0) to (5);
		\draw (2) to (1);
		\draw (1) to (0);
		\draw (5) to (31);
		\draw (33) to (2);
		\draw (31) to (36);
	\end{pgfonlayer}
\end{tikzpicture}}
}
 & A   & $\frac{t^6}{(1-t^6)(1-t^8)}$ \\
 & B  &  $\frac{t^5(1-t^4)}{(1-t^2)(1-t^6)(1-t^8)}$\\
 &  C &  $\frac{t^4}{(1-t^2)(1-t^8)}$\\
 &  D & $\frac{t^3}{(1-t^2)(1-t^6)}$ \\
 &  E &  $\frac{t^2 (1-t^{12})}{(1-t^4)(1-t^6)(1-t^8)}$ \\
 &  F & $\frac{t(1-t^{20})}{(1-t^6)(1-t^8)(1-t^{10})}$ \\
 &  G &  $\frac{t^4}{(1-t^4)(1-t^6)}$\\ \bottomrule
    \end{tabular}
    \caption{$E_7$ family for $k=1$. B-twisted index with Wilson lines at the labelled position in the quiver. }
    \label{E7WilsonHSk1}
\end{table}


\FloatBarrier
\section{Sphere partition function}
\label{app:sphere}
In this appendix, we provide the computations and results of $S^3$ partition functions both with and without line defects for 3d $\Ncal=4$ theories relevant in this paper.

\subsection{\texorpdfstring{$\sprm(k)$ SQCDs and their mirrors}{Sp(k) SQCDs and their mirrors}}

\paragraph{$\sprm(k)$ SQCD.} The $S^3$ partition function for $\sprm(k)$ SQCD with $N_f$ fundamental hypermultiplets is given by
\begin{align}\label{eq:Sp-SQCD-S3}
Z^{S^3}_{\sprm(k),N_f}(m)=&\lim_{\xi\to0} \int [d\mathbf{s}] \frac{e^{2\pi i\xi\sum_{i=1}^k s_i}\prod_{\alpha\in \Delta}\sh(\alpha\cdot s)}{\prod_{j=1}^{N_f}\prod_{w\in \Box}\ch(w\cdot s-m_j)}\cr
=&\sum_{I\in C_{k}^{N_f}}\prod_{j=1}^k\frac{ m_{I_j}\sh(2m_{I_j})}{\prod_{\ell\not \in I} \sh(m_{\ell}\pm m_{I_j})}~.
\end{align}
where $I$ runs over all combinations $C_{k}^{N_{f}}$ of $k$ different integers in $\left\{1, \ldots, N_{f}\right\}$. From the first to the second line, we take poles at \begin{equation}\label{eq:poles}s_j=\pm m_{I_j}+i\frac{2k_j+1}{2}~,\quad k_j\in \mathbb{Z}_{\ge0}\end{equation}
and their permutations $S_k$ on $I$. Then, we take the limit $\xi\to 0$ by using L'Hospital's rule to obtain the second line.

\paragraph{Vortex loops in $\sprm(k)$ SQCD.} The partition function of the $\urm(1)$ supersymmetric quantum mechanics for a vortex defect of type $N_{f}=L+\left(N_{f}-L\right)$ in \eqref{fig:vortex} is given by
\begin{equation}\label{eq:vortex-SQM}
  Z_{(L,N_f-L),\mathbf{2k}}(z)=\sum_{w_i\in \Box} \prod_{j\neq i}\frac{\sh(w_i\cdot s-w_j\cdot s+iz)}{\sh(w_i\cdot s-w_j\cdot s) }\prod_{\ell=1}^L  \frac{\ch(w_i\cdot s-m_\ell)}{\ch(w_i\cdot s-m_\ell+iz) }~.
\end{equation}
Note that the fundamental representation of $\sprm(k)$ is $2k$-dimensional so that \eqref{eq:vortex-SQM} is the sum of $2k$ terms.
Due to the factors $\ch(w_i\cdot s-m_k+iz)$ in the denominator, some of the poles in \eqref{eq:poles} are shifted by $-i$. The explicit evaluations of the residues yield the $S^3$ partition function with the vortex loop
\begin{align}\label{eq:S3-vortex}
Z^{S^3}_{(L,N_f-L),\mathbf{2k}}=&\lim_{\xi\to0,z\to1} \int [d\mathbf{s}] \frac{e^{2\pi i\xi\sum_{i=1}^k s_i}\prod_{\alpha\in \Delta}\sh(\alpha\cdot s)}{\prod_{j=1}^{N_f}\prod_{w\in \Box}\ch(w\cdot s-m_j)}   Z_{(L,N_f-L),\mathbf{2k}}(z) ~,\cr
=&(2k-L)\cdot Z^{S^3}_{\sprm(k),N_f}(m)+\sum_{j=1}^L Z^{S^3}_{\sprm(k),N_f}(m_j\to m_j-i)~.
\end{align}

\begin{equation}\label{fig:k=1Dtypemirror}
    \begin{tikzpicture}
  \node (g4) [gauge,label=below:{\footnotesize{$1$}},label=above:{\footnotesize{$\xi_1-\xi_2$}}] {};
  \node (g5) [gauge,right of =g4,label=below:{\footnotesize{${2}$}}] {};
  \node (g6) [gauge,right of =g5,label=below:{\footnotesize{${2}$}},label=above:{\footnotesize{$\xi_3-\xi_4$}}] {};
  \node (g7) [right of =g6] {$\ldots$};
  \node (g8) [gauge,right of =g7,label=below:{\footnotesize{$2$}}] {};
  \node (g9) [gauge,below right of =g8,label=below:{\footnotesize{$1$}},label=right:{\footnotesize{$\xi_{N_f-1}-\xi_{N_f}$}}] {};
  \node (g10) [gauge,above right of =g8,label=below:{\footnotesize{$1$}},label=right:{\footnotesize{$\xi_{N_f-1}+\xi_{N_f}$}}] {};
  \node (f1) [flavour,above of=g5, label=left:{\footnotesize{$1$}}] {};
	\draw  (g4)--(g5) (g5)--(g6) (g6)--(g7) (g7)--(g8) (g8)--(g9) (g8)--(g10) (g5)--(f1);
    \node (h2) [below of =g5] {\footnotesize{$\xi_{2}-\xi_{3}$}};
	\end{tikzpicture}
    \end{equation}
The mirror unitary quiver is the $D_{N_f}$-type quiver \eqref{fig:k=1Dtypemirror} and the equivalence of the $S^3$ partition functions can be derived for $k=1$. Note that \eqref{eq:S3-vortex} at $k=1$ is
\begin{align}\label{eq:S3-vortex-k1}
Z^{S^3}_{(L,N_f-L),\mathbf{2}}=\sum_{j=1}^{L}\frac{(2m_j-i) \sh(2m_j)}{\prod_{\ell\neq j}\sh(m_\ell\pm m_j)}+\sum_{k=L+1}^{N_f}\frac{(2m_k) \sh(2m_k)}{\prod_{\ell\neq k}\sh(m_\ell\pm m_k)}~.
\end{align}
The derivation is almost the same as in \cite[\S3.2]{Benvenuti:2011ga} so that we refer the reader to it for more detail.

\paragraph{$D$-type quiver.} The $S^3$ partition function of the $D$-type quiver is
\begin{align}
      Z^{S^3}_{D}=&\frac{1}{2^{N_f-3} } \int  dz^{(1)} dz_{\pm}  \prod_{j=2}^{N_f-2}  d z^{(j)}_1 d z^{(j)}_2  \frac{   e^{2\pi i (\xi_1-\xi_{2})z^{(1)}} e^{2\pi i \sum_{j=2}^{N_f-2}( \xi_j-\xi_{j+1})(z^{(j)}_1+z^{(j)}_2 )}   e^{2\pi i (\xi_{N_f-1}\pm \xi_{N_f})z_{\pm}} }{\prod_{i=1,2} \ch(z^{(2)}_i-z^{(1)}) \ch(z^{(2)}_i-m)  \ch(z^{(N_f-2)}_i-z_\pm) }
      \\ & \times  \frac{\prod_{j=2}^{N_f-2} \sh^2(z^{(j)}_1-z^{(j)}_2)}{\prod_{j=2}^{N_f-3}\prod_{k,\ell=1,2} \ch(z^{(j)}_k-z^{(j+1)}_\ell )}\cr
      =& \frac{1}{2}
       \frac{1}
       { \sh(\xi_1-\xi_{2}) \sh(\xi_{N_f-1}-\xi_{N_f})}
       \int  dz_{+}  \prod_{j=2}^{N_f-2}  d z^{(j)}_1 d z^{(j)}_2  \frac{e^{2\pi i \sum_{j=2}^{N_f-2}( \xi_j-\xi_{j+1})(z^{(j)}_1+z^{(j)}_2 )}   e^{2\pi i (\xi_{N_f-1}+\xi_{N_f})z_{+}} }{\prod_{i=1,2} \ch(z^{(2)}_i-m)  \ch(z^{(N_f-2)}_i-z_+)}\cr
  & \times
       \frac{(e^{2 \pi i (\xi_1-\xi_{2}) z^{(2)}_1} -  e^{2 \pi i (\xi_1-\xi_{2}) z^{(2)}_2} )
       (e^{2 \pi i (\xi_{N_f-1}-\xi_{N_f}) z^{(N_f-2)}_1} -  e^{2 \pi i (\xi_{N_f-1}-\xi_{N_f}) z^{(N_f-2)}_2} )}{\prod_{j=2}^{N_f-3}\prod_{i=1,2} \ch(z^{(j)}_i-z^{(j+1)}_i )}~.
\end{align}
From the first and the second line, we apply the Cauchy determinant formula and use the $T[\surm(2)]$ formula. Now we  take the Fourier transform of each ch
\begin{align}\label{Fourier}
Z^{S^3}_{D}=&\frac{1}{2}
 \frac{1}
 { \sh(\xi_1-\xi_{2}) \sh(\xi_{N_f-1}-\xi_{N_f})}
 \int  dz_{+}  \prod_{j=2}^{N_f-2}  d z^{(j)}_1 d z^{(j)}_2 e^{2\pi i \sum_{j=2}^{N_f-2}( \xi_j-\xi_{j+1})(z^{(j)}_1+z^{(j)}_2 )}   e^{2\pi i (\xi_{N_f-1}+\xi_{N_f})z_{+}} \cr
 & (e^{2 \pi i (\xi_1-\xi_{2}) z^{(2)}_1} -  e^{2 \pi i (\xi_1-\xi_{2}) z^{(2)}_2} )
 (e^{2 \pi i (\xi_{N_f-1}-\xi_{N_f}) z^{(N_f-2)}_1} -  e^{2 \pi i (\xi_{N_f-1}-\xi_{N_f}) z^{(N_f-2)}_2} )\cr
 & \int \frac{d s_{1} d s_{2}}{\ch s_{1} \ch s_{2}} e^{2 \pi i(s_{1}(z_{1}^{(2)}-m)+s_{2}(z_{2}^{(2)}-m))} \int \frac{d p_{1} d p_{2}}{\ch p_{1} \ch p_{2}} e^{2 \pi i(p_{1}(z_{1}^{(N_f-2)}-z_+)+p_{2}(z_{2}^{(N_f-2)}-z_+))} \cr
 & \int  \prod_{j=2}^{N_f-3} \frac{ d t_{1}^{(j)} d t_{2}^{(j)}}{\ch t_{1}^{(j)} \ch t_{2}^{(j)}} e^{2 \pi i \sum_{j=2}^{N_f-3}(t_{1}^{(j)}(z_{1}^{(j)}-z_{1}^{(j+1)})+t_{2}^{(j)}(z_{2}^{(j)}-z_{2}^{(j+1)}))}~.
\end{align}
Expanding the second line, we have four terms and let us focus on the term proportional to $e^{2 \pi i (\xi_1-\xi_{2}) z^{(2)}_1+ 2 \pi i(\xi_{N_f-1}-\xi_{N_f}) z^{(N_f-2)}_1}$.
Performing the integrals over the variables $z_{+}$ $z^{(j)}_i$ ($i=1,2$ $j=2,\ldots,N_f-2$), we obtain delta functions which impose the relations
\begin{align}\label{eq:deltafunction}
s_{2} =-\xi_{2}-s_{1}~,\quad p_{1} =\xi_{N_f}-s_{1}~,\quad  p_{2}= \xi_{N_f-1}+s_{1}~,\quad   t_1^{(j)}=s_1-\xi_{j+1}~, \quad   t_2^{(j)}=-s_1-\xi_{j+1}~,
\end{align}
where we shift $s_1\to s_1-\xi_1$. The other three terms impose similar relations where $\xi_1\leftrightarrow\xi_2$ and $\xi_{N_f-1}\leftrightarrow\xi_{N_f}$. Finally, we obtain
\begin{align}
Z^{S^3}_{D}=&\frac{e^{2\pi i(\xi_1+\xi_2)m}}{\sh(\xi_2-\xi_1)\sh(\xi_{N_f-1}-\xi_{N_f})} \int d s_1 \frac{1}{\prod^{N_f-2}_{j=2} \ch(s_1\pm\xi_{j})  }\cr &\times\left(\frac{1}{\ch(s_1-\xi_1) \ch(s_1+\xi_2)}-\frac{1}{\ch(s_1+\xi_1) \ch(s_1-\xi_2)}\right)\frac{1}{\ch(s_1-\xi_{N_f-1}) \ch(s_1+\xi_{N_f})}~\cr
=&e^{2\pi i(\xi_1+\xi_2)m}\sum_{i=1}^{N_f}\frac{\xi_i \sh(2\xi_i)}{\prod_{k\neq i}\sh(\xi_k\pm \xi_i)}~.
\end{align}
Taking the limit $m\to 0$, it agrees with the $S^3$ partition function of the $\sprm(1)$ SQCD with $N_f$ flavours under the exchange of the FI and mass parameters $\xi_i \leftrightarrow m_i$.

\paragraph{Wilson loops in $D$-type quiver.} Now let us include a fundamental Wilson loop in the $L$-th gauge node from the left in \eqref{fig:k=1Dtypemirror}. For $L=1$, the fundamental Wilson loop amounts to the shift $\xi_1\to \xi_1-i$. Hence, the sum of the Wilson loop at the left $\urm(1)$ gauge group \eqref{eq:Sp1_split_1_n-1_a} and the flavour Wilson loop \eqref{eq:Sp1_split_1_n-1_b} is
\begin{align}\label{eq:Wilson1}
Z^{S^3}_{\mathcal{W} \text{ at } [1,0,\ldots,0,0]}+Z^{S^3}_{D}
=\frac{(2\xi_{1}-i) \sh(2\xi_{1})}{\prod_{\ell\neq 1}\sh(\xi_\ell\pm \xi_{1})}+\sum_{j=2}^{N_f}\frac{(2\xi_j) \sh(2\xi_j)}{\prod_{\ell\neq j}\sh(\xi_\ell\pm \xi_j)}~,
\end{align}
which agrees with the vortex \eqref{eq:S3-vortex-k1} of the $1+(N_f-1)$ splitting under $\xi\leftrightarrow m$.
For $2\le L\le N_f-2$, the relations in \eqref{eq:deltafunction} are modified so that
\begin{align}\label{eq:Wilson2}
&Z^{S^3}_{\mathcal{W} \text{ at } [0,\ldots,0,1,0,\ldots,0]}\cr
=&\frac{1}{\sh(\xi_2-\xi_1)\sh(\xi_{N_f-1}-\xi_{N_f})} \int d s_1 \frac{1}{\prod^{N_f-2}_{k=L+1} \ch(s_1\pm\xi_{k}) \prod^{L}_{j=2} \ch(s_1-\xi_{j}-i)\ch(s_1+\xi_{j})  }\cr &\times\left(\frac{1}{\ch(s_1-\xi_1-i) \ch(s_1+\xi_2)}-\frac{1}{\ch(s_1+\xi_1) \ch(s_1-\xi_2-i)}\right)\frac{1}{\ch(s_1-\xi_{N_f-1}) \ch(s_1+\xi_{N_f})}~\cr
&+\frac{1}{\sh(\xi_2-\xi_1)\sh(\xi_{N_f-1}-\xi_{N_f})} \int d s_1 \frac{1}{\prod^{N_f-2}_{k=L+1} \ch(s_1\pm\xi_{k}) \prod^{L}_{j=2} \ch(s_1-\xi_{j})\ch(s_1+\xi_{j}-i)  }\cr &\times\left(\frac{1}{\ch(s_1-\xi_1) \ch(s_1+\xi_2-i)}-\frac{1}{\ch(s_1+\xi_1-i) \ch(s_1-\xi_2)}\right)\frac{1}{\ch(s_1-\xi_{N_f-1}) \ch(s_1+\xi_{N_f})}~\cr
=&\sum_{j=1}^{L}\frac{(2\xi_j-i) \sh(2\xi_j)}{\prod_{\ell\neq j}\sh(\xi_\ell\pm \xi_j)}+\sum_{k=L+1}^{N_f}\frac{(2\xi_k) \sh(2\xi_k)}{\prod_{\ell\neq k}\sh(\xi_\ell\pm \xi_k)}~.
\end{align}
This is equal to \eqref{eq:S3-vortex-k1} under  $\xi\leftrightarrow m$.
The fundamental Wilson loop at the spinor node with the FI parameter $\xi_{N_f-1}-\xi_{N_f}$ gives rise to the shift
\begin{equation}
  \xi_j\to\xi_j-\frac{i}{2}~(j=1,\ldots,N_f-1)~,\quad  \xi_{N_f}\to\xi_{N_f}+\frac{i}{2}~.
\end{equation}
The fundamental Wilson loop at the other spinor node with the FI parameter $\xi_{N_f-1}+\xi_{N_f}$ gives rise to the shift
\begin{equation}
  \xi_j\to\xi_j-\frac{i}{2}~,\quad   (j=1,\ldots,N_f)~.
\end{equation}
Therefore, the sum of the fundamental Wilson loop at spinor nodes \eqref{eq:Sp1_split_n-1_1_a} and \eqref{eq:Sp1_split_n-1_1_b} is
\begin{align}\label{eq:Wilson3}
Z^{S^3}_{\mathcal{W} \text{ at } [0,\ldots,1,0]}+Z^{S^3}_{\mathcal{W} \text{ at } [0,\ldots,0,1]}
=\sum_{j=1}^{N_f-1}\frac{(2\xi_j-i) \sh(2\xi_j)}{\prod_{\ell\neq j}\sh(\xi_\ell\pm \xi_j)}+\frac{(2\xi_{N_f}) \sh(2\xi_{N_f})}{\prod_{\ell\neq N_f}\sh(\xi_\ell\pm \xi_{N_f})}~,
\end{align}
which corresponds to the vortex of the $(N_f-1)+1$ splitting in the mirror dual theory. The Wilson loop with multiplicity two \eqref{eq:Sp1_split_n_0_a} at the node with the FI parameter $\xi_{N_f-1}+\xi_{N_f}$ is
\begin{align}\label{eq:Wilson4}
2\cdot Z^{S^3}_{\mathcal{W} \text{ at } [0,\ldots,0,1]}
=\sum_{j=1}^{N_f}\frac{(2\xi_j-i) \sh(2\xi_j)}{\prod_{\ell\neq j}\sh(\xi_\ell\pm \xi_j)}~,
\end{align}
corresponding to the vortex of $N_f+0$ splitting in the mirror dual theory.

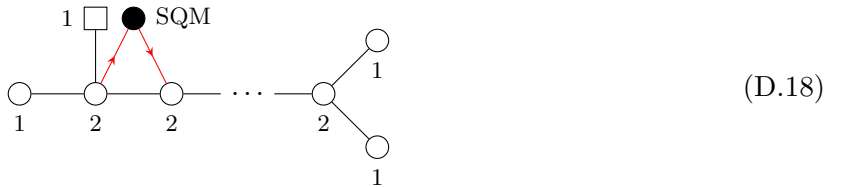
\begin{align}\label{D-vortex}
    \raisebox{-.5\height}{
    \begin{tikzpicture}
	\node (g4) [gauge,label=below:{\footnotesize{$1$}}] {};
	\node (g5) [gauge,right of =g4,label=below:{\footnotesize{2}}] {};
	\node (g6) [gauge,right of =g5,label=below:{\footnotesize{2}}] {};
	\node (g7) [right of =g6] {$\ldots$};
	\node (g8) [gauge,right of =g7,label=below:{\footnotesize{2}}] {};
	\node (g9) [gauge,below right of =g8,label=below:{\footnotesize{1}}] {};
	\node (g10) [gauge,above right of =g8,label=below:{\footnotesize{1}}] {};
	\node (f1) [flavour,above of=g5, label=left:{\footnotesize{$1$}}] {};
	\draw  (g4)--(g5) (g5)--(g6)
	(g6)--(g7) (g7)--(g8) (g8)--(g9) (g8)--(g10) (g5)--(f1);
%
    \node (SQM) at (1.5,1) [defect,label=right:{\footnotesize{SQM}}] {};
    \begin{scope}[decoration={markings,mark =at position 0.5 with {\arrow{stealth}}}]
    \draw[postaction={decorate},color=red] (g5) -- (SQM);
    \draw[postaction={decorate},color=red] (SQM) -- (g6);
     \end{scope}
	\end{tikzpicture}
    }
    \end{align}

\paragraph{Vortex loops in $D$-type quiver.}
Let us evaluate the vortex contribution \eqref{D-vortex}.
\begin{equation}
  Z^{\mathrm{SQM}}=\sum_{\ell =1,2}\frac{\sh(z_1^{(2)}-z_2^{(2)}-(-1)^\ell iz)}{\sh(z_{1}^{(2)}-z_2^{(2)})} \prod_{j=1,2}\frac{\ch(z_\ell^{(2)}-z_j^{(3)})}{\ch(z_\ell^{(2)}-z_j^{(3)}+iz)}~.
\end{equation}
This contribution amounts to a multiplication of a factor
$
(e^{-2\pi t_1^{(2)}z}+ e^{-2\pi t_2^{(2)}z})
$
to the integrand of \eqref{Fourier}.
Since the relations \eqref{eq:deltafunction} stay as they are, the vortex expectation value is
\begin{align}\label{k1-vortex}
Z^{S^3}_{D,\Vcal}=&\frac{1}{\sh(\xi_2-\xi_1)\sh(\xi_{N_f-1}-\xi_{N_f})} \int d s_1 \frac{(e^{2\pi s_1}+ e^{-2\pi s_1})}{\prod^{N_f-2}_{j=2} \ch(s_1\pm\xi_{j})  }\cr &\times\left(\frac{1}{\ch(s_1-\xi_1) \ch(s_1+\xi_2)}-\frac{1}{\ch(s_1+\xi_1) \ch(s_1-\xi_2)}\right)\frac{1}{\ch(s_1-\xi_{N_f-1}) \ch(s_1+\xi_{N_f})}~\cr
=&\sum_{i=1}^{N_f}\frac{\xi_i \sh(2\xi_i)(e^{2\pi \xi_i}+ e^{-2\pi \xi_i})}{\prod_{k\neq i}\sh(\xi_k\pm \xi_i)}~.
\end{align}
It is easy to see that it is equivalent to the expectation value of the fundamental Wilson loop in the $\sprm(1)$ SQCD. The derivation of the hopping duality is given in \cite[\S5.6]{Assel:2015oxa} so that the vortex defects \eqref{vortex1}, \eqref{vortex2}, \eqref{vortex3} are equivalent.

\paragraph{$D$-type quiver of higher rank.}

Let us briefly sketch the idea of how to obtain a closed-form expression of the $S^3$ partition function of the $D$-type quiver in \eqref{eq:unitary-mirror} mirror to the $\sprm(k)$ SQCD with $N_f$ flavours. The left tail in \eqref{eq:unitary-mirror} is the $T[\urm(2k)]$ theory and its $S^3$ partition function is obtained in \cite{Benvenuti:2011ga}. Also, the contribution from the spinor nodes can be read off from the  $S^3$ partition function of the $\urm(k)$ SQCD with $2k$ flavours \eqref{eq:U-SQCD-FI}. Consequently, we have
\begin{align}
  Z=&\frac{1}{(k!)^{N_{f}-2k-1}}\int \prod_{j=1}^{N_{f}-2k-1}d\mathbf{z}^{(j)}\frac{e^{2 \pi i \sum_{j=1}^{N_{f}-2k-1}\left(\xi_{2k-1+j}-\xi_{2k+j}\right)(\sum_{k=1}z_{k}^{(j)})}\prod_{j=1}^{N_{f}-2k-1}\prod_{n\neq \ell} \sh^{2}(z_{n}^{(j)}-z_{\ell}^{(j)})}{\prod_{n=1}^{2k}\ch(z_{n}^{(1)}-m)\prod_{j=1}^{N_{f}-2k-2} \prod_{n, \ell=1}^{2k} \ch(z_{n}^{(j)}-z_{\ell}^{(j+1)})}\cr
  & Z^{S^3}_{T[\urm(2k)]}(\mathbf{z}^{(1)},\mathbf{\xi})Z^{S^3}_{\urm(k),N_f}(z^{(N_{f}-2k-1)},\xi_{N_f-1}-\xi_{N_f})Z^{S^3}_{\urm(k),N_f}(z^{(N_{f}-2k-1)},\xi_{N_f-1}+\xi_{N_f})~.
\end{align}
Then, applying the Cauchy determinant formulas repeatedly, we have
\begin{align}
  Z=&\frac{1}{\sh^k(\xi_{N_f-1}\pm\xi_{N_f})\prod_{n\neq \ell}^{2k} \sh(\xi_n-\xi_\ell)}\int \prod_{j=1}^{N_{f}-2k-1}d\mathbf{z}^{(j)}\frac{e^{2 \pi i \sum_{j=1}^{N_{f}-2k-1}\left(\xi_{2k-1+j}-\xi_{2k+j}\right)(\sum_{k=1}z_{k}^{(j)})}}{\prod_{n=1}^{2k}\ch(z_{n}^{(1)}-m)\prod_{j=1}^{N_{f}-2k-2} \prod_{n=1}^{2k} \ch(z_{n}^{(j)}-z_{n}^{(j+1)})}\cr &\sum_{J\in C_{k}^{N_f}}\frac{ e^{2\pi i( \xi_{N_f-1}-\xi_{N_f})(\sum_{j=1}^km_{J_j})}}{\prod_{j=1}^k\prod_{\ell\not \in I} \sh(m_{\ell}- m_{I_j})} \sum_{I\in C_{k}^{N_f}}\frac{ e^{2\pi i( \xi_{N_f-1}+\xi_{N_f})(\sum_{j=1}^km_{I_j})}}{\prod_{j=1}^k\prod_{\ell\not \in I} \sh(m_{\ell}- m_{I_j})}~.
\end{align}
For any choice of $I,J\in C_{k}^{N_f}$, we can further apply the Cauchy determinant formula so that there is no sh in the numerator of the integrand. As before, using Fourier transformations of $1/\ch$ and $1/\sh$, we can obtain delta-functions by integrating the gauge fugacities. By performing Fourier integrals further, the formula simplifies to a combination of products of the $\urm(1)$ integral formula \cite[\S2.2]{Benvenuti:2011ga}. During the manipulation, we use the identity $\sh(x)=-i\ch(x+\frac{i}{2})$. In this way, we can evaluate the $S^3$ partition function of the $D$-type quiver of higher ranks, and we verify that it is equal to \eqref{eq:Sp-SQCD-S3} up to a factor under $\xi_i\leftrightarrow m_i$ when $k=2$.

\subsection{Other theories}
Let us summarise closed-form expressions of sphere partition functions for familiar 3d $\Ncal=4$ theories here. The $S^3$ partition function of SQCDs for other gauge groups can be evaluated as in \eqref{eq:Sp-SQCD-S3}.
\paragraph{$\urm(k)$ SQCD.}

The $S^3$ partition function of $\urm(k)$ SQCD with $N_f$ flavours with an FI parameter $\xi$ is
\begin{align}\label{eq:U-SQCD-FI}
Z^{S^3}_{\urm(k),N_f}(m,\xi)
=&\sum_{I\in C_{k}^{N_f}}\frac{ e^{2\pi i \xi(\sum_{j=1}^km_{I_j})}}{\sh^k(\xi)\prod_{j=1}^k\prod_{\ell\not \in I} \sh(m_{\ell}- m_{I_j})}~.
\end{align}
If we turn off the FI parameter, it becomes
\begin{align}\label{eq:U-SQCD}
Z^{S^3}_{\urm(k),N_f}(m)=\lim_{\xi\to0} Z^{S^3}_{\urm(k),N_f}(m,\xi)
=\sum_{I\in C_{k}^{N_f}}\frac{ (\sum_{j=1}^km_{I_j})^k}{\prod_{j=1}^k\prod_{\ell\not \in I} \sh(m_{\ell}- m_{I_j})}~.
\end{align}

\paragraph{$\sorm(2k)$ SQCD.}
Since $\sorm(k)$ SQCD is not endowed with an FI parameter, we introduce a regulator and take its zero limit as in \eqref{eq:Sp-SQCD-S3}. The resulting partition function is
\begin{align}
Z^{S^3}_{\sorm(2k),N_f}(m)
=&\sum_{I\in C_{k}^{N_f}}\frac{ (\sum_{j=1}^km_{I_j})^k}{\prod_{j=1}^k \sh(m_{I_j})\prod_{\ell\not \in I} \sh(m_{\ell}\pm m_{I_j})}~.
\end{align}

\paragraph{$T[G]$ theory.}
We also calculate the $S^3$ partition function for the $T[G]$ theory \cite{Gaiotto:2008ak}
\begin{equation}\label{eq:TG-S3}
Z^{S^3}_{T[G]}=\textrm{const}\prod_{\alpha\in \Delta^+}\frac{\alpha\cdot m}{\sh(\alpha\cdot m)}~.
\end{equation}
For $A$ type, a closed-form expression with FI parameters is obtained in \cite{Benvenuti:2011ga}. We take the limit that the FI parameters are zero, and the constant is equal to
$$\frac1{G(1+N)}=\frac{1}{\prod_{k=1}^{N-1}k!}~,$$
where $G(x)$ is the Barnes gamma function.
For $C$ and $D$ types, we can also evaluate the $S^3$ partition function by introducing regulators repeatedly. We have checked \eqref{eq:TG-S3} up to $T[\sorm(6)]$.

%
\FloatBarrier
 \bibliographystyle{JHEP}
 {\small{
 \bibliography{references}

\providecommand{\href}[2]{#2}\begingroup\raggedright\begin{thebibliography}{10}

\bibitem{Cabrera:2018jxt}
S.~Cabrera, A.~Hanany and F.~Yagi, \emph{{Tropical Geometry and Five
  Dimensional Higgs Branches at Infinite Coupling}},
  \href{https://doi.org/10.1007/JHEP01(2019)068}{\emph{JHEP} {\bfseries 01}
  (2019) 068} [\href{https://arxiv.org/abs/1810.01379}{{\ttfamily
  1810.01379}}].

\bibitem{Cabrera:2019izd}
S.~Cabrera, A.~Hanany and M.~Sperling, \emph{{Magnetic quivers, Higgs branches,
  and 6d $N$=(1,0) theories}}, \href{https://doi.org/10.1007/JHEP07(2019)137,
  10.1007/JHEP06(2019)071}{\emph{JHEP} {\bfseries 06} (2019) 071}
  [\href{https://arxiv.org/abs/1904.12293}{{\ttfamily 1904.12293}}].

\bibitem{Bourget:2019aer}
A.~Bourget, S.~Cabrera, J.~F. Grimminger, A.~Hanany, M.~Sperling, A.~Zajac
  et~al., \emph{{The Higgs mechanism --- Hasse diagrams for symplectic
  singularities}}, \href{https://doi.org/10.1007/JHEP01(2020)157}{\emph{JHEP}
  {\bfseries 01} (2020) 157}
  [\href{https://arxiv.org/abs/1908.04245}{{\ttfamily 1908.04245}}].

\bibitem{Bourget:2019rtl}
A.~Bourget, S.~Cabrera, J.~F. Grimminger, A.~Hanany and Z.~Zhong, \emph{{Brane
  Webs and Magnetic Quivers for SQCD}},
  \href{https://doi.org/10.1007/JHEP03(2020)176}{\emph{JHEP} {\bfseries 03}
  (2020) 176} [\href{https://arxiv.org/abs/1909.00667}{{\ttfamily
  1909.00667}}].

\bibitem{Cabrera:2019dob}
S.~Cabrera, A.~Hanany and M.~Sperling, \emph{{Magnetic Quivers, Higgs Branches,
  and 6d N=(1,0) Theories -- Orthogonal and Symplectic Gauge Groups}},
  \href{https://doi.org/10.1007/JHEP02(2020)184}{\emph{JHEP} {\bfseries 02}
  (2020) 184} [\href{https://arxiv.org/abs/1912.02773}{{\ttfamily
  1912.02773}}].

\bibitem{Bourget:2020gzi}
A.~Bourget, J.~F. Grimminger, A.~Hanany, M.~Sperling and Z.~Zhong,
  \emph{{Magnetic Quivers from Brane Webs with O5 Planes}},
  \href{https://doi.org/10.1007/JHEP07(2020)204}{\emph{JHEP} {\bfseries 07}
  (2020) 204} [\href{https://arxiv.org/abs/2004.04082}{{\ttfamily
  2004.04082}}].

\bibitem{Bourget:2020asf}
A.~Bourget, J.~F. Grimminger, A.~Hanany, M.~Sperling, G.~Zafrir and Z.~Zhong,
  \emph{{Magnetic quivers for rank 1 theories}},
  \href{https://doi.org/10.1007/JHEP09(2020)189}{\emph{JHEP} {\bfseries 09}
  (2020) 189} [\href{https://arxiv.org/abs/2006.16994}{{\ttfamily
  2006.16994}}].

\bibitem{Closset:2020scj}
C.~Closset, S.~Schafer-Nameki and Y.-N. Wang, \emph{{Coulomb and Higgs Branches
  from Canonical Singularities: Part 0}},
  \href{https://doi.org/10.1007/JHEP02(2021)003}{\emph{JHEP} {\bfseries 02}
  (2021) 003} [\href{https://arxiv.org/abs/2007.15600}{{\ttfamily
  2007.15600}}].

\bibitem{Akhond:2020vhc}
M.~Akhond, F.~Carta, S.~Dwivedi, H.~Hayashi, S.-S. Kim and F.~Yagi,
  \emph{{Five-brane webs, Higgs branches and unitary/orthosymplectic magnetic
  quivers}}, \href{https://doi.org/10.1007/JHEP12(2020)164}{\emph{JHEP}
  {\bfseries 12} (2020) 164}
  [\href{https://arxiv.org/abs/2008.01027}{{\ttfamily 2008.01027}}].

\bibitem{vanBeest:2020kou}
M.~van Beest, A.~Bourget, J.~Eckhard and S.~Schafer-Nameki, \emph{{(Symplectic)
  Leaves and (5d Higgs) Branches in the Poly(go)nesian Tropical Rain Forest}},
  \href{https://doi.org/10.1007/JHEP11(2020)124}{\emph{JHEP} {\bfseries 11}
  (2020) 124} [\href{https://arxiv.org/abs/2008.05577}{{\ttfamily
  2008.05577}}].

\bibitem{Bourget:2020mez}
A.~Bourget, S.~Giacomelli, J.~F. Grimminger, A.~Hanany, M.~Sperling and
  Z.~Zhong, \emph{{S-fold magnetic quivers}},
  \href{https://doi.org/10.1007/JHEP02(2021)054}{\emph{JHEP} {\bfseries 02}
  (2021) 054} [\href{https://arxiv.org/abs/2010.05889}{{\ttfamily
  2010.05889}}].

\bibitem{VanBeest:2020kxw}
M.~Van~Beest, A.~Bourget, J.~Eckhard and S.~Sch\"afer-Nameki, \emph{{(5d
  RG-flow) Trees in the Tropical Rain Forest}},
  \href{https://doi.org/10.1007/JHEP03(2021)241}{\emph{JHEP} {\bfseries 03}
  (2021) 241} [\href{https://arxiv.org/abs/2011.07033}{{\ttfamily
  2011.07033}}].

\bibitem{Eckhard:2020jyr}
J.~Eckhard, S.~Sch\"afer-Nameki and Y.-N. Wang, \emph{{Trifectas for T$_{N}$ in
  5d}}, \href{https://doi.org/10.1007/JHEP07(2020)199}{\emph{JHEP} {\bfseries
  07} (2020) 199} [\href{https://arxiv.org/abs/2004.15007}{{\ttfamily
  2004.15007}}].

\bibitem{Closset:2020afy}
C.~Closset, S.~Giacomelli, S.~Schafer-Nameki and Y.-N. Wang, \emph{{5d and 4d
  SCFTs: Canonical Singularities, Trinions and S-Dualities}},
  \href{https://doi.org/10.1007/JHEP05(2021)274}{\emph{JHEP} {\bfseries 05}
  (2021) 274} [\href{https://arxiv.org/abs/2012.12827}{{\ttfamily
  2012.12827}}].

\bibitem{Akhond:2021knl}
M.~Akhond, F.~Carta, S.~Dwivedi, H.~Hayashi, S.-S. Kim and F.~Yagi,
  \emph{{Factorised 3d $\mathcal{N}=4$ orthosymplectic quivers}},
  \href{https://doi.org/10.1007/JHEP05(2021)269}{\emph{JHEP} {\bfseries 05}
  (2021) 269} [\href{https://arxiv.org/abs/2101.12235}{{\ttfamily
  2101.12235}}].

\bibitem{Bourget:2021xex}
A.~Bourget, J.~F. Grimminger, A.~Hanany, R.~Kalveks, M.~Sperling and Z.~Zhong,
  \emph{{Folding Orthosymplectic Quivers}},
  \href{https://arxiv.org/abs/2107.00754}{{\ttfamily 2107.00754}}.

\bibitem{Akhond:2021ffo}
M.~Akhond and F.~Carta, \emph{{Magnetic quivers from brane webs with
  O7$^+$-planes}},  \href{https://arxiv.org/abs/2107.09077}{{\ttfamily
  2107.09077}}.

\bibitem{vanBeest:2021xyt}
M.~van Beest and S.~Giacomelli, \emph{{Connecting 5d Higgs Branches via
  Fayet-Iliopoulos Deformations}},
  \href{https://arxiv.org/abs/2110.02872}{{\ttfamily 2110.02872}}.

\bibitem{Bourget:2021csg}
A.~Bourget, J.~F. Grimminger, M.~Martone and G.~Zafrir, \emph{{Magnetic quivers
  for rank 2 theories}},  \href{https://arxiv.org/abs/2110.11365}{{\ttfamily
  2110.11365}}.

\bibitem{Sperling:2021fcf}
M.~Sperling and Z.~Zhong, \emph{{Balanced B and D-type orthosymplectic quivers
  -- Magnetic quivers for product theories}},
  \href{https://arxiv.org/abs/2111.00026}{{\ttfamily 2111.00026}}.

\bibitem{DelZotto:2014kka}
M.~Del~Zotto and A.~Hanany, \emph{{Complete Graphs, Hilbert Series, and the
  Higgs branch of the 4d $\mathcal{N} =$ 2 $(A_n,A_m)$ SCFTs}},
  \href{https://doi.org/10.1016/j.nuclphysb.2015.03.017}{\emph{Nucl. Phys. B}
  {\bfseries 894} (2015) 439}
  [\href{https://arxiv.org/abs/1403.6523}{{\ttfamily 1403.6523}}].

\bibitem{Cremonesi:2015lsa}
S.~Cremonesi, G.~Ferlito, A.~Hanany and N.~Mekareeya, \emph{{Instanton
  Operators and the Higgs Branch at Infinite Coupling}},
  \href{https://doi.org/10.1007/JHEP04(2017)042}{\emph{JHEP} {\bfseries 04}
  (2017) 042} [\href{https://arxiv.org/abs/1505.06302}{{\ttfamily
  1505.06302}}].

\bibitem{Ferlito:2017xdq}
G.~Ferlito, A.~Hanany, N.~Mekareeya and G.~Zafrir, \emph{{3d Coulomb branch and
  5d Higgs branch at infinite coupling}},
  \href{https://doi.org/10.1007/JHEP07(2018)061}{\emph{JHEP} {\bfseries 07}
  (2018) 061} [\href{https://arxiv.org/abs/1712.06604}{{\ttfamily
  1712.06604}}].

\bibitem{Mekareeya:2017jgc}
N.~Mekareeya, K.~Ohmori, Y.~Tachikawa and G.~Zafrir, \emph{{E$_{8}$ instantons
  on type-A ALE spaces and supersymmetric field theories}},
  \href{https://doi.org/10.1007/JHEP09(2017)144}{\emph{JHEP} {\bfseries 09}
  (2017) 144} [\href{https://arxiv.org/abs/1707.04370}{{\ttfamily
  1707.04370}}].

\bibitem{Hanany:2018uhm}
A.~Hanany and N.~Mekareeya, \emph{{The small E$_{8}$ instanton and the Kraft
  Procesi transition}},
  \href{https://doi.org/10.1007/JHEP07(2018)098}{\emph{JHEP} {\bfseries 07}
  (2018) 098} [\href{https://arxiv.org/abs/1801.01129}{{\ttfamily
  1801.01129}}].

\bibitem{Hanany:2018vph}
A.~Hanany and G.~Zafrir, \emph{{Discrete Gauging in Six Dimensions}},
  \href{https://doi.org/10.1007/JHEP07(2018)168}{\emph{JHEP} {\bfseries 07}
  (2018) 168} [\href{https://arxiv.org/abs/1804.08857}{{\ttfamily
  1804.08857}}].

\bibitem{Hanany:2018cgo}
A.~Hanany and M.~Sperling, \emph{{Discrete quotients of 3-dimensional $
  \mathcal{N}=4 $ Coulomb branches via the cycle index}},
  \href{https://doi.org/10.1007/JHEP08(2018)157}{\emph{JHEP} {\bfseries 08}
  (2018) 157} [\href{https://arxiv.org/abs/1807.02784}{{\ttfamily
  1807.02784}}].

\bibitem{Intriligator:1996ex}
K.~A. Intriligator and N.~Seiberg, \emph{{Mirror symmetry in three-dimensional
  gauge theories}},
  \href{https://doi.org/10.1016/0370-2693(96)01088-X}{\emph{Phys. Lett. B}
  {\bfseries 387} (1996) 513}
  [\href{https://arxiv.org/abs/hep-th/9607207}{{\ttfamily hep-th/9607207}}].

\bibitem{Hanany:1996ie}
A.~Hanany and E.~Witten, \emph{{Type IIB superstrings, BPS monopoles, and
  three-dimensional gauge dynamics}},
  \href{https://doi.org/10.1016/S0550-3213(97)00157-0,
  10.1016/S0550-3213(97)80030-2}{\emph{Nucl. Phys.} {\bfseries B492} (1997)
  152} [\href{https://arxiv.org/abs/hep-th/9611230}{{\ttfamily
  hep-th/9611230}}].

\bibitem{Seiberg:1996nz}
N.~Seiberg and E.~Witten, \emph{{Gauge dynamics and compactification to
  three-dimensions}},  in \emph{{Conference on the Mathematical Beauty of
  Physics (In Memory of C. Itzykson)}}, pp.~333--366, 6, 1996,
  \href{https://arxiv.org/abs/hep-th/9607163}{{\ttfamily hep-th/9607163}}.

\bibitem{Hanany:1999sj}
A.~Hanany and A.~Zaffaroni, \emph{{Issues on orientifolds: On the brane
  construction of gauge theories with SO(2n) global symmetry}},
  \href{https://doi.org/10.1088/1126-6708/1999/07/009}{\emph{JHEP} {\bfseries
  07} (1999) 009} [\href{https://arxiv.org/abs/hep-th/9903242}{{\ttfamily
  hep-th/9903242}}].

\bibitem{Feng:2000eq}
B.~Feng and A.~Hanany, \emph{{Mirror symmetry by O3 planes}},
  \href{https://doi.org/10.1088/1126-6708/2000/11/033}{\emph{JHEP} {\bfseries
  11} (2000) 033} [\href{https://arxiv.org/abs/hep-th/0004092}{{\ttfamily
  hep-th/0004092}}].

\bibitem{Assel:2015oxa}
B.~Assel and J.~Gomis, \emph{{Mirror Symmetry And Loop Operators}},
  \href{https://doi.org/10.1007/JHEP11(2015)055}{\emph{JHEP} {\bfseries 11}
  (2015) 055} [\href{https://arxiv.org/abs/1506.01718}{{\ttfamily
  1506.01718}}].

\bibitem{Dey:2021jbf}
A.~Dey, \emph{{Line Defects in Three Dimensional Mirror Symmetry beyond Linear
  Quivers}},  \href{https://arxiv.org/abs/2103.01243}{{\ttfamily 2103.01243}}.

\bibitem{Gaiotto:2014kfa}
D.~Gaiotto, A.~Kapustin, N.~Seiberg and B.~Willett, \emph{{Generalized Global
  Symmetries}}, \href{https://doi.org/10.1007/JHEP02(2015)172}{\emph{JHEP}
  {\bfseries 02} (2015) 172} [\href{https://arxiv.org/abs/1412.5148}{{\ttfamily
  1412.5148}}].

\bibitem{Bourget:2020xdz}
A.~Bourget, J.~F. Grimminger, A.~Hanany, R.~Kalveks, M.~Sperling and Z.~Zhong,
  \emph{{Magnetic Lattices for Orthosymplectic Quivers}},
  \href{https://doi.org/10.1007/JHEP12(2020)092}{\emph{JHEP} {\bfseries 12}
  (2020) 092} [\href{https://arxiv.org/abs/2007.04667}{{\ttfamily
  2007.04667}}].

\bibitem{Benvenuti:2006qr}
S.~Benvenuti, B.~Feng, A.~Hanany and Y.-H. He, \emph{{Counting BPS Operators in
  Gauge Theories: Quivers, Syzygies and Plethystics}},
  \href{https://doi.org/10.1088/1126-6708/2007/11/050}{\emph{JHEP} {\bfseries
  11} (2007) 050} [\href{https://arxiv.org/abs/hep-th/0608050}{{\ttfamily
  hep-th/0608050}}].

\bibitem{Feng:2007ur}
B.~Feng, A.~Hanany and Y.-H. He, \emph{{Counting gauge invariants: The
  Plethystic program}},
  \href{https://doi.org/10.1088/1126-6708/2007/03/090}{\emph{JHEP} {\bfseries
  03} (2007) 090} [\href{https://arxiv.org/abs/hep-th/0701063}{{\ttfamily
  hep-th/0701063}}].

\bibitem{Gray:2008yu}
J.~Gray, A.~Hanany, Y.-H. He, V.~Jejjala and N.~Mekareeya, \emph{{SQCD: A
  Geometric Apercu}},
  \href{https://doi.org/10.1088/1126-6708/2008/05/099}{\emph{JHEP} {\bfseries
  05} (2008) 099} [\href{https://arxiv.org/abs/0803.4257}{{\ttfamily
  0803.4257}}].

\bibitem{Cremonesi:2013lqa}
S.~Cremonesi, A.~Hanany and A.~Zaffaroni, \emph{{Monopole operators and Hilbert
  series of Coulomb branches of $3d$ $\mathcal{N} = 4$ gauge theories}},
  \href{https://doi.org/10.1007/JHEP01(2014)005}{\emph{JHEP} {\bfseries 01}
  (2014) 005} [\href{https://arxiv.org/abs/1309.2657}{{\ttfamily 1309.2657}}].

\bibitem{Bullimore:2015lsa}
M.~Bullimore, T.~Dimofte and D.~Gaiotto, \emph{{The Coulomb Branch of 3d
  ${\mathcal{N}= 4}$ Theories}},
  \href{https://doi.org/10.1007/s00220-017-2903-0}{\emph{Commun. Math. Phys.}
  {\bfseries 354} (2017) 671}
  [\href{https://arxiv.org/abs/1503.04817}{{\ttfamily 1503.04817}}].

\bibitem{Grimminger:2020dmg}
J.~F. Grimminger and A.~Hanany, \emph{{Hasse diagrams for 3d $ \mathcal{N} $ =
  4 quiver gauge theories \textemdash{} Inversion and the full moduli space}},
  \href{https://doi.org/10.1007/JHEP09(2020)159}{\emph{JHEP} {\bfseries 09}
  (2020) 159} [\href{https://arxiv.org/abs/2004.01675}{{\ttfamily
  2004.01675}}].

\bibitem{Bourget:2021siw}
A.~Bourget, J.~F. Grimminger, A.~Hanany, M.~Sperling and Z.~Zhong,
  \emph{{Branes, Quivers, and the Affine Grassmannian}},
  \href{https://arxiv.org/abs/2102.06190}{{\ttfamily 2102.06190}}.

\bibitem{Razamat:2014pta}
S.~S. Razamat and B.~Willett, \emph{{Down the rabbit hole with theories of
  class $ \mathcal{S} $}},
  \href{https://doi.org/10.1007/JHEP10(2014)099}{\emph{JHEP} {\bfseries 10}
  (2014) 099} [\href{https://arxiv.org/abs/1403.6107}{{\ttfamily 1403.6107}}].

\bibitem{Closset:2016arn}
C.~Closset and H.~Kim, \emph{{Comments on twisted indices in 3d supersymmetric
  gauge theories}}, \href{https://doi.org/10.1007/JHEP08(2016)059}{\emph{JHEP}
  {\bfseries 08} (2016) 059}
  [\href{https://arxiv.org/abs/1605.06531}{{\ttfamily 1605.06531}}].

\bibitem{Kapustin:2009kz}
A.~Kapustin, B.~Willett and I.~Yaakov, \emph{{Exact Results for Wilson Loops in
  Superconformal Chern-Simons Theories with Matter}},
  \href{https://doi.org/10.1007/JHEP03(2010)089}{\emph{JHEP} {\bfseries 03}
  (2010) 089} [\href{https://arxiv.org/abs/0909.4559}{{\ttfamily 0909.4559}}].

\bibitem{Dimofte:2019zzj}
T.~Dimofte, N.~Garner, M.~Geracie and J.~Hilburn, \emph{{Mirror symmetry and
  line operators}}, \href{https://doi.org/10.1007/JHEP02(2020)075}{\emph{JHEP}
  {\bfseries 02} (2020) 075}
  [\href{https://arxiv.org/abs/1908.00013}{{\ttfamily 1908.00013}}].

\bibitem{Aharony:2013hda}
O.~Aharony, N.~Seiberg and Y.~Tachikawa, \emph{{Reading between the lines of
  four-dimensional gauge theories}},
  \href{https://doi.org/10.1007/JHEP08(2013)115}{\emph{JHEP} {\bfseries 08}
  (2013) 115} [\href{https://arxiv.org/abs/1305.0318}{{\ttfamily 1305.0318}}].

\bibitem{Okazaki:2019ony}
T.~Okazaki, \emph{{Mirror symmetry of 3D $\mathcal{N}=4$ gauge theories and
  supersymmetric indices}},
  \href{https://doi.org/10.1103/PhysRevD.100.066031}{\emph{Phys. Rev. D}
  {\bfseries 100} (2019) 066031}
  [\href{https://arxiv.org/abs/1905.04608}{{\ttfamily 1905.04608}}].

\bibitem{Rozansky:1996bq}
L.~Rozansky and E.~Witten, \emph{{HyperKahler geometry and invariants of three
  manifolds}}, \href{https://doi.org/10.1007/s000290050016}{\emph{Selecta
  Math.} {\bfseries 3} (1997) 401}
  [\href{https://arxiv.org/abs/hep-th/9612216}{{\ttfamily hep-th/9612216}}].

\bibitem{Nekrasov:2014xaa}
N.~A. Nekrasov and S.~L. Shatashvili, \emph{{Bethe/Gauge correspondence on
  curved spaces}}, \href{https://doi.org/10.1007/JHEP01(2015)100}{\emph{JHEP}
  {\bfseries 01} (2015) 100} [\href{https://arxiv.org/abs/1405.6046}{{\ttfamily
  1405.6046}}].

\bibitem{Gukov:2015sna}
S.~Gukov and D.~Pei, \emph{{Equivariant Verlinde formula from fivebranes and
  vortices}}, \href{https://doi.org/10.1007/s00220-017-2931-9}{\emph{Commun.
  Math. Phys.} {\bfseries 355} (2017) 1}
  [\href{https://arxiv.org/abs/1501.01310}{{\ttfamily 1501.01310}}].

\bibitem{Benini:2015noa}
F.~Benini and A.~Zaffaroni, \emph{{A topologically twisted index for
  three-dimensional supersymmetric theories}},
  \href{https://doi.org/10.1007/JHEP07(2015)127}{\emph{JHEP} {\bfseries 07}
  (2015) 127} [\href{https://arxiv.org/abs/1504.03698}{{\ttfamily
  1504.03698}}].

\bibitem{Benini:2016hjo}
F.~Benini and A.~Zaffaroni, \emph{{Supersymmetric partition functions on
  Riemann surfaces}}, {\emph{Proc. Symp. Pure Math.} {\bfseries 96} (2017) 13}
  [\href{https://arxiv.org/abs/1605.06120}{{\ttfamily 1605.06120}}].

\bibitem{Gukov:2020lqm}
S.~Gukov, P.-S. Hsin, H.~Nakajima, S.~Park, D.~Pei and N.~Sopenko,
  \emph{{Rozansky-Witten geometry of Coulomb branches and logarithmic knot
  invariants}}, \href{https://doi.org/10.1016/j.geomphys.2021.104311}{\emph{J.
  Geom. Phys.} {\bfseries 168} (2021) 104311}
  [\href{https://arxiv.org/abs/2005.05347}{{\ttfamily 2005.05347}}].

\bibitem{Benvenuti:2011ga}
S.~Benvenuti and S.~Pasquetti, \emph{{3D-partition functions on the sphere:
  exact evaluation and mirror symmetry}},
  \href{https://doi.org/10.1007/JHEP05(2012)099}{\emph{JHEP} {\bfseries 05}
  (2012) 099} [\href{https://arxiv.org/abs/1105.2551}{{\ttfamily 1105.2551}}].

\bibitem{Goddard:1976qe}
P.~Goddard, J.~Nuyts and D.~I. Olive, \emph{{Gauge Theories and Magnetic
  Charge}}, \href{https://doi.org/10.1016/0550-3213(77)90221-8}{\emph{Nucl.
  Phys. B} {\bfseries 125} (1977) 1}.

\bibitem{Kapustin:2005py}
A.~Kapustin, \emph{{Wilson-'t Hooft operators in four-dimensional gauge
  theories and S-duality}},
  \href{https://doi.org/10.1103/PhysRevD.74.025005}{\emph{Phys. Rev. D}
  {\bfseries 74} (2006) 025005}
  [\href{https://arxiv.org/abs/hep-th/0501015}{{\ttfamily hep-th/0501015}}].

\bibitem{Hanany:2016ezz}
A.~Hanany and M.~Sperling, \emph{{Coulomb branches for rank 2 gauge groups in
  3d $ \mathcal{N}=4 $ gauge theories}},
  \href{https://doi.org/10.1007/JHEP08(2016)016}{\emph{JHEP} {\bfseries 08}
  (2016) 016} [\href{https://arxiv.org/abs/1605.00010}{{\ttfamily
  1605.00010}}].

\bibitem{Tong:2017oea}
D.~Tong, \emph{{Line Operators in the Standard Model}},
  \href{https://doi.org/10.1007/JHEP07(2017)104}{\emph{JHEP} {\bfseries 07}
  (2017) 104} [\href{https://arxiv.org/abs/1705.01853}{{\ttfamily
  1705.01853}}].

\bibitem{Dey:2020hfe}
A.~Dey, \emph{{Three dimensional mirror symmetry beyond ADE quivers and
  Argyres-Douglas theories}},
  \href{https://doi.org/10.1007/JHEP07(2021)199}{\emph{JHEP} {\bfseries 07}
  (2021) 199} [\href{https://arxiv.org/abs/2004.09738}{{\ttfamily
  2004.09738}}].

\bibitem{Gaiotto:2008ak}
D.~Gaiotto and E.~Witten, \emph{{S-Duality of Boundary Conditions In N=4 Super
  Yang-Mills Theory}},
  \href{https://doi.org/10.4310/ATMP.2009.v13.n3.a5}{\emph{Adv. Theor. Math.
  Phys.} {\bfseries 13} (2009) 721}
  [\href{https://arxiv.org/abs/0807.3720}{{\ttfamily 0807.3720}}].

\bibitem{Assel:2018exy}
B.~Assel and S.~Cremonesi, \emph{{The Infrared Fixed Points of 3d
  $\mathcal{N}=4$ $USp(2N)$ SQCD Theories}},
  \href{https://doi.org/10.21468/SciPostPhys.5.2.015}{\emph{SciPost Phys.}
  {\bfseries 5} (2018) 015} [\href{https://arxiv.org/abs/1802.04285}{{\ttfamily
  1802.04285}}].

\bibitem{Kapustin:1998fa}
A.~Kapustin, \emph{{D(n) quivers from branes}},
  \href{https://doi.org/10.1088/1126-6708/1998/12/015}{\emph{JHEP} {\bfseries
  12} (1998) 015} [\href{https://arxiv.org/abs/hep-th/9806238}{{\ttfamily
  hep-th/9806238}}].

\bibitem{Hanany:2016gbz}
A.~Hanany and R.~Kalveks, \emph{{Quiver Theories for Moduli Spaces of Classical
  Group Nilpotent Orbits}},
  \href{https://doi.org/10.1007/JHEP06(2016)130}{\emph{JHEP} {\bfseries 06}
  (2016) 130} [\href{https://arxiv.org/abs/1601.04020}{{\ttfamily
  1601.04020}}].

\bibitem{Cabrera:2017ucb}
S.~Cabrera, A.~Hanany and Z.~Zhong, \emph{{Nilpotent orbits and the Coulomb
  branch of $T^\sigma (G)$ theories: special orthogonal vs orthogonal gauge
  group factors}}, \href{https://doi.org/10.1007/JHEP11(2017)079}{\emph{JHEP}
  {\bfseries 11} (2017) 079}
  [\href{https://arxiv.org/abs/1707.06941}{{\ttfamily 1707.06941}}].

\bibitem{Cabrera:2018ldc}
S.~Cabrera, A.~Hanany and R.~Kalveks, \emph{{Quiver Theories and Formulae for
  Slodowy Slices of Classical Algebras}},
  \href{https://doi.org/10.1016/j.nuclphysb.2018.12.022}{\emph{Nucl. Phys. B}
  {\bfseries 939} (2019) 308}
  [\href{https://arxiv.org/abs/1807.02521}{{\ttfamily 1807.02521}}].

\bibitem{Callan:1997kz}
C.~G. Callan and J.~M. Maldacena, \emph{{Brane death and dynamics from the
  Born-Infeld action}},
  \href{https://doi.org/10.1016/S0550-3213(97)00700-1}{\emph{Nucl. Phys. B}
  {\bfseries 513} (1998) 198}
  [\href{https://arxiv.org/abs/hep-th/9708147}{{\ttfamily hep-th/9708147}}].

\bibitem{Hama:2010av}
N.~Hama, K.~Hosomichi and S.~Lee, \emph{{Notes on SUSY Gauge Theories on
  Three-Sphere}}, \href{https://doi.org/10.1007/JHEP03(2011)127}{\emph{JHEP}
  {\bfseries 03} (2011) 127} [\href{https://arxiv.org/abs/1012.3512}{{\ttfamily
  1012.3512}}].

\bibitem{Hama:2011ea}
N.~Hama, K.~Hosomichi and S.~Lee, \emph{{SUSY Gauge Theories on Squashed
  Three-Spheres}}, \href{https://doi.org/10.1007/JHEP05(2011)014}{\emph{JHEP}
  {\bfseries 05} (2011) 014} [\href{https://arxiv.org/abs/1102.4716}{{\ttfamily
  1102.4716}}].

\bibitem{Intriligator:1997pq}
K.~A. Intriligator, D.~R. Morrison and N.~Seiberg, \emph{{Five-dimensional
  supersymmetric gauge theories and degenerations of Calabi-Yau spaces}},
  \href{https://doi.org/10.1016/S0550-3213(97)00279-4}{\emph{Nucl. Phys. B}
  {\bfseries 497} (1997) 56}
  [\href{https://arxiv.org/abs/hep-th/9702198}{{\ttfamily hep-th/9702198}}].

\bibitem{Brunner:1997gk}
I.~Brunner and A.~Karch, \emph{{Branes and six-dimensional fixed points}},
  \href{https://doi.org/10.1016/S0370-2693(97)00935-0}{\emph{Phys. Lett. B}
  {\bfseries 409} (1997) 109}
  [\href{https://arxiv.org/abs/hep-th/9705022}{{\ttfamily hep-th/9705022}}].

\bibitem{Bergman:2015dpa}
O.~Bergman and G.~Zafrir, \emph{{5d fixed points from brane webs and
  O7-planes}}, \href{https://doi.org/10.1007/JHEP12(2015)163}{\emph{JHEP}
  {\bfseries 12} (2015) 163}
  [\href{https://arxiv.org/abs/1507.03860}{{\ttfamily 1507.03860}}].

\bibitem{Chacaltana:2012ch}
O.~Chacaltana, J.~Distler and Y.~Tachikawa, \emph{{Gaiotto duality for the
  twisted $A_{2N-1}$ series}},
  \href{https://doi.org/10.1007/JHEP05(2015)075}{\emph{JHEP} {\bfseries 05}
  (2015) 075} [\href{https://arxiv.org/abs/1212.3952}{{\ttfamily 1212.3952}}].

\bibitem{Chacaltana:2011ze}
O.~Chacaltana and J.~Distler, \emph{{Tinkertoys for the $D_N$ series}},
  \href{https://doi.org/10.1007/JHEP02(2013)110}{\emph{JHEP} {\bfseries 02}
  (2013) 110} [\href{https://arxiv.org/abs/1106.5410}{{\ttfamily 1106.5410}}].

\bibitem{Beratto:2021xmn}
E.~Beratto, N.~Mekareeya and M.~Sacchi, \emph{{Zero-form and one-form
  symmetries of the ABJ and related theories}},
  \href{https://arxiv.org/abs/2112.09531}{{\ttfamily 2112.09531}}.

\bibitem{Eckhard:2019jgg}
J.~Eckhard, H.~Kim, S.~Schafer-Nameki and B.~Willett, \emph{{Higher-Form
  Symmetries, Bethe Vacua, and the 3d-3d Correspondence}},
  \href{https://doi.org/10.1007/JHEP01(2020)101}{\emph{JHEP} {\bfseries 01}
  (2020) 101} [\href{https://arxiv.org/abs/1910.14086}{{\ttfamily
  1910.14086}}].

\bibitem{Carta:2021whq}
F.~Carta, S.~Giacomelli, N.~Mekareeya and A.~Mininno, \emph{{Conformal
  manifolds and 3d mirrors of Argyres-Douglas theories}},
  \href{https://doi.org/10.1007/JHEP08(2021)015}{\emph{JHEP} {\bfseries 08}
  (2021) 015} [\href{https://arxiv.org/abs/2105.08064}{{\ttfamily
  2105.08064}}].

\bibitem{Carta:2021dyx}
F.~Carta, S.~Giacomelli, N.~Mekareeya and A.~Mininno, \emph{{Conformal
  Manifolds and 3d Mirrors of $(D_n,D_m)$ Theories}},
  \href{https://arxiv.org/abs/2110.06940}{{\ttfamily 2110.06940}}.

\bibitem{Dey:2021rxw}
A.~Dey, \emph{{Higgs Branches of Argyres-Douglas theories as Quiver
  Varieties}},  \href{https://arxiv.org/abs/2109.07493}{{\ttfamily
  2109.07493}}.

\bibitem{Uranga:1998uj}
A.~M. Uranga, \emph{{Towards mass deformed N=4 SO(n) and Sp(k) gauge theories
  from brane configurations}},
  \href{https://doi.org/10.1016/S0550-3213(98)00370-8}{\emph{Nucl. Phys. B}
  {\bfseries 526} (1998) 241}
  [\href{https://arxiv.org/abs/hep-th/9803054}{{\ttfamily hep-th/9803054}}].

\bibitem{Gimon:1996rq}
E.~G. Gimon and J.~Polchinski, \emph{{Consistency conditions for orientifolds
  and d manifolds}},
  \href{https://doi.org/10.1103/PhysRevD.54.1667}{\emph{Phys. Rev. D}
  {\bfseries 54} (1996) 1667}
  [\href{https://arxiv.org/abs/hep-th/9601038}{{\ttfamily hep-th/9601038}}].

\bibitem{Hanany:1997gh}
A.~Hanany and A.~Zaffaroni, \emph{{Branes and six-dimensional supersymmetric
  theories}}, \href{https://doi.org/10.1016/S0550-3213(98)00355-1}{\emph{Nucl.
  Phys. B} {\bfseries 529} (1998) 180}
  [\href{https://arxiv.org/abs/hep-th/9712145}{{\ttfamily hep-th/9712145}}].

\bibitem{Yamaguchi:2006tq}
S.~Yamaguchi, \emph{{Wilson loops of anti-symmetric representation and
  D5-branes}}, \href{https://doi.org/10.1088/1126-6708/2006/05/037}{\emph{JHEP}
  {\bfseries 05} (2006) 037}
  [\href{https://arxiv.org/abs/hep-th/0603208}{{\ttfamily hep-th/0603208}}].

\bibitem{Gomis:2006sb}
J.~Gomis and F.~Passerini, \emph{{Holographic Wilson Loops}},
  \href{https://doi.org/10.1088/1126-6708/2006/08/074}{\emph{JHEP} {\bfseries
  08} (2006) 074} [\href{https://arxiv.org/abs/hep-th/0604007}{{\ttfamily
  hep-th/0604007}}].

\bibitem{Gomis:2006im}
J.~Gomis and F.~Passerini, \emph{{Wilson Loops as D3-Branes}},
  \href{https://doi.org/10.1088/1126-6708/2007/01/097}{\emph{JHEP} {\bfseries
  01} (2007) 097} [\href{https://arxiv.org/abs/hep-th/0612022}{{\ttfamily
  hep-th/0612022}}].

\bibitem{Alday:2009fs}
L.~F. Alday, D.~Gaiotto, S.~Gukov, Y.~Tachikawa and H.~Verlinde, \emph{{Loop
  and surface operators in N=2 gauge theory and Liouville modular geometry}},
  \href{https://doi.org/10.1007/JHEP01(2010)113}{\emph{JHEP} {\bfseries 01}
  (2010) 113} [\href{https://arxiv.org/abs/0909.0945}{{\ttfamily 0909.0945}}].

\bibitem{Hanany:1997vm}
A.~Hanany and K.~Hori, \emph{{Branes and N=2 theories in two-dimensions}},
  \href{https://doi.org/10.1016/S0550-3213(97)00754-2}{\emph{Nucl. Phys. B}
  {\bfseries 513} (1998) 119}
  [\href{https://arxiv.org/abs/hep-th/9707192}{{\ttfamily hep-th/9707192}}].

\end{thebibliography}\endgroup
 }}

\end{document}